\documentclass{aastex631}
\usepackage{dirtytalk}

\usepackage{booktabs} 
\usepackage{graphicx} 
\usepackage{multirow}
\usepackage{algorithmic}
\usepackage{algorithm}
\usepackage{float}
\definecolor{darkgreen}{RGB}{0,100,0} 
\usepackage{array}
\usepackage{ulem}
\usepackage{amsmath}
\usepackage{hyperref}
\usepackage{pifont}
\newcommand{\xmark}{\ding{55}}%

\usepackage{subcaption} 

\begin{document}


\title{Review of Machine Learning Models for Solar Energetic Particle Prediction} 

\author[0000-0002-0786-7307]{Spiridon Kasapis}
\affiliation{Department of Astrophysical Sciences, Princeton University, Princeton, NJ, USA}
\affiliation{Computational Physics Branch, NASA Ames Research Center, Moffett Field, CA, USA}

\author[0000-0001-8045-2709]{Pouya Hosseinzadeh}
\affiliation{Department of Computer Science, Utah State University, Logan, UT, USA}

\author[0000-0002-3787-1622]{Kathryn Whitman}
\affiliation{Space Radiation Analysis Group, NASA Johnson Space Center, Houston, TX, USA}
\affiliation{KBR Wyle Services, LLC, TX, USA}

\author[0000-0002-4996-0753]{Ricky Egeland}
\affiliation{Space Radiation Analysis Group, NASA Johnson Space Center, Houston, TX, USA}

\author[0000-0001-6913-1330]{Manolis Georgoulis}
\affiliation{Johns Hopkins Applied Physics Lab, 11100 Johns Hopkins Rd, Laurel, MD 20723, United States}
\affiliation{Research Center for Astronomy and Applied Mathematics of the Academy of Athens, 4 Soranou Efesiou Street, Athens 11527, Greece}

\author[0000-0002-8164-5948]{Angelos Vourlidas} 
\affiliation{Johns Hopkins Applied Physics Lab, 11100 Johns Hopkins Rd, Laurel, MD 20723, United States}

\author[0000-0002-9479-8644]{Athanasios Papaioannou}
\affiliation{Institute for Astronomy, Astrophysics, Space Applications and Remote Sensing, National Observatory of Athens, Athens, Greece}

\author[0000-0003-1192-0868]{Eleni Lavasa}
\affiliation{Institute for Astronomy, Astrophysics, Space Applications and Remote Sensing, National Observatory of Athens, Athens, Greece}

\author[0000-0002-5162-8821]{Anastasios Anastasiadis}
\affiliation{Institute for Astronomy, Astrophysics, Space Applications and Remote Sensing, National Observatory of Athens, Athens, Greece}

\author[0000-0002-8252-9869]{Giorgos Giannopoulos}
\affiliation{Institute for Astronomy, Astrophysics, Space Applications and Remote Sensing, National Observatory of Athens, Athens, Greece}

\author[0000-0002-4716-0840]{Andrés Muñoz-Jaramillo}
\affiliation{Southwest Research Institute, Boulder, CO, USA}

\author[0000-0003-1258-0308]{Bala Poduval}
\affiliation{Space Science Center, University of New Hampshire, Durham, NH, USA}

\author[0000-0003-4144-2270]{Irina N. Kitiashvili}
\affiliation{Computational Physics Branch, NASA Ames Research Center, Moffett Field, CA, USA}

\author[0000-0003-0364-4883]{Alexander G. Kosovichev}
\affiliation{Computational Physics Branch, NASA Ames Research Center, Moffett Field, CA, USA}
\affiliation{Department of Physics, New Jersey Institute of Technology, Newark, NJ, USA}

\author[0000-0002-4001-1295]{Viacheslav Sadykov}
\affiliation{Physics and Astronomy Department, Georgia State University, Atlanta, GA, USA}

\author[0000-0001-5693-6383]{Soukaina Filali Boubrahimi}
\affiliation{Department of Computer Science, Utah State University, Logan, UT, USA}

\author[0009-0005-5373-5515]{Tate T. Hutchins}
\affiliation{Department of Computer Science, Princeton University, Princeton, NJ, USA}
\affiliation{Department of Astrophysical Sciences, Princeton University, Princeton, NJ, USA}

\author[0000-0001-7952-8032]{Hameedullah A. Farooki}
\affiliation{Department of Astrophysical Sciences, Princeton University, Princeton, NJ, USA}

\author[0000-0002-7341-2992]{Manuel E. Cuesta}
\affiliation{Department of Astrophysical Sciences, Princeton University, Princeton, NJ, USA}

\author[0000-0003-0412-1064]{Leng Y. Khoo}
\affiliation{Department of Astrophysical Sciences, Princeton University, Princeton, NJ, USA}

\author[0000-0003-2847-7110]{Sungmin Pak}
\affiliation{Department of Astrophysical Sciences, Princeton University, Princeton, NJ, USA}

\author[0009-0006-8032-4380]{Robert Czarnota}
\affiliation{Department of Mathematics, Rowan University, Glassboro, NJ, USA}
\affiliation{Department of Astrophysical Sciences, Princeton University, Princeton, NJ, USA}

\author[0000-0002-8111-1444]{Jamie S. Rankin}
\affiliation{Department of Astrophysical Sciences, Princeton University, Princeton, NJ, USA}

\author[0000-0003-2685-9801]{Jamey Szalay}
\affiliation{Department of Astrophysical Sciences, Princeton University, Princeton, NJ, USA}

\author[0000-0002-3093-458X]{Mitchell M. Shen}
\affiliation{Department of Astrophysical Sciences, Princeton University, Princeton, NJ, USA}

\author[0000-0002-7655-6019]{Georgios Livadiotis}
\affiliation{Department of Astrophysical Sciences, Princeton University, Princeton, NJ, USA}

\author[0000-0002-9246-996X]{Zigong Xu}
\affiliation{Division of Physics Mathematics and Astronomy, California Institute of Technology, Pasadena, CA, USA}

\author[0000-0001-6160-1158]{David J. McComas}
\affiliation{Department of Astrophysical Sciences, Princeton University, Princeton, NJ, USA}

\author[0000-0002-8483-519X]{Nikolaos Sarlis}
\affiliation{Department of Physics, National and Kapodistrian University of Athens, Athens, Greece}

\author[0000-0002-5189-5612]{Dionissios Hristopulos}
\affiliation{School of Electrical and Computer Engineering, Technical University of Crete, Chania, Greece}

\author[0000-0003-1572-8734]{Arik Posner}
\affiliation{Space Radiation Analysis Group, NASA Johnson Space Center, Houston, TX, USA}

\author[0000-0001-9254-9057]{Alec J. Engell}
\affiliation{NextGen Federal Systems, Morgantown, WV, USA}

\author[0000-0001-5560-9102]{Mohammed AbuBakr Ali}
\affiliation{National Authority for Remote Sensing and Space Science, Cairo, Egypt}

\author[0000-0002-8939-0759]{Ali G. A. Abdelkawy}
\affiliation{Department of Astronomy and Meteorology, Faculty of Science, Al-Azhar University, Cairo, Egypt}

\author{Abdelrazek M. K. Shaltout}
\affiliation{Department of Astronomy and Meteorology, Faculty of Science, Al-Azhar University, Cairo, Egypt}

\author{M. M. Beheary}
\affiliation{Department of Astronomy and Meteorology, Faculty of Science, Al-Azhar University, Cairo, Egypt}

\author[0000-0002-1604-3326]{Christina O. Lee}
\affiliation{Space Sciences Lab, University of California, Berkeley, CA, USA}

\author[0000-0003-1162-5842]{Sigiava Aminalragia-Giamini}
\affiliation{Space Applications and Research Consultancy, Athens, Greece}

\author[0000-0002-5191-0149]{Constantinos Papadimitriou}
\affiliation{Space Applications and Research Consultancy, Athens, Greece}
\affiliation{Department of Physics, National and Kapodistrian University of Athens, Athens, Greece}

\author[0000-0001-9716-608X]{Ingmar Sandberg}
\affiliation{Space Applications and Research Consultancy, Athens, Greece}

\author[0000-0002-4381-3197]{Savvas Raptis}
\affiliation{Johns Hopkins Applied Physics Lab, 11100 Johns Hopkins Rd, Laurel, MD 20723, United States}

\author[0000-0002-9303-7835]{Shah Muhammad Hamdi}
\affiliation{Department of Computer Science, Utah State University, Logan, UT, USA}

\author[0000-0001-5481-4534]{Monica Laurenza}
\affiliation{Institute for Space Astrophysics and Planetology, Via del Fosso del Cavaliere 100, 00133, Rome, Italy}

\author[0000-0002-6303-5329]{Mirko Stumpo}
\affiliation{Institute for Space Astrophysics and Planetology, Via del Fosso del Cavaliere 100, 00133, Rome, Italy}

\author[0000-0003-1080-3424]{Sumanth A. Rotti}
\affiliation{Department of Physics and Astronomy, Georgia State University, Atlanta, GA 30303, USA}
\affiliation{Aryabhatta Research Institute of Observational Sciences (ARIES), Manora Peak, Nainital-263001, Uttarakhand, India}

\author[0009-0001-5404-8689]{India Jackson}
\affiliation{Physics and Astronomy Department, Georgia State University, Atlanta, GA, USA}

\author[0000-0003-3196-3822]{Aatiya Ali}
\affiliation{Physics and Astronomy Department, Georgia State University, Atlanta, GA, USA}

\author[0000-0001-9854-8100]{Atilim Gunes Baydin}
\affiliation{Department of Computer Science, Oxford University, Oxford, England}

\author[0000-0002-3737-9283]{Nathan Schwadron}
\affiliation{Space Science Center, University of New Hampshire, Durham, NH, USA}
\affiliation{Department of Astrophysical Sciences, Princeton University, Princeton, NJ, USA}

\author[0000-0002-5014-7022]{Subhamoy Chatterjee}
\affiliation{Southwest Research Institute, San Antonio, TX, USA}

\author[0000-0001-9323-1200]{Maher A. Dayeh}
\affiliation{Southwest Research Institute, San Antonio, TX, USA}

\author[0000-0003-2846-2453]{Gelu M. Nita}
\affiliation{Department of Physics, New Jersey Institute of Technology, Newark, NJ, USA}

\author{Patrick M. O’Keefe}
\affiliation{Computer Science Department, New Jersey Institute of Technology, Newark, NJ, USA}

\author[0009-0002-2545-8037]{Chun Jie Chong}
\affiliation{Computer Science Department, New Jersey Institute of Technology, Newark, NJ, USA}

\author[0009-0002-1860-6265]{Paul Kosovich}
\affiliation{Department of Physics, New Jersey Institute of Technology, Newark, NJ, USA}

\author[0000-0002-3364-7463]{Russell D. Marroquin}
\affiliation{Department of Physics, University of California San Diego, La Jolla, CA 92093, USA}

\author[0000-0002-9799-9265]{Berkay Aydin}
\affiliation{Department of Computer Science, Georgia State University, Atlanta, GA 30303, USA}

\author[0000-0001-8078-6856]{Petrus C. Martens}
\affiliation{Department of Physics and Astronomy, Georgia State University, Atlanta, GA 30303, USA}

\author[0000-0002-4453-9097]{Lulu Zhao}
\affiliation{Department of Climate and Space Sciences and Engineering, University of Michigan, Ann Arbor, MI, USA}

\author[0000-0002-9516-8134]{Yang Chen}
\affiliation{Department of Statistics, University of Michigan, Ann Arbor, MI, USA}

\author[0000-0001-9414-9890]{Yian Yu}
\affiliation{Department of Statistics, University of Michigan, Ann Arbor, MI, USA}

\author[0000-0002-5662-9604]{Monica G. Bobra}
\affiliation{Office of Data and Innovation, State of California, Sacramento, CA}

\author[0000-0003-0472-9408]{Ward Manchester}
\affiliation{Department of Climate and Space Sciences and Engineering, University of Michigan, Ann Arbor, MI, USA}

\author[0000-0001-9360-4951]{Tamas Gombosi}
\affiliation{Department of Climate and Space Sciences and Engineering, University of Michigan, Ann Arbor, MI, USA}

\author[0000-0003-3529-8743]{Ming Zhang}
\affiliation{Department of Electrical Engineering and Computer Science, Florida Institute of Technology, Melbourne, FL, USA}

\author[0000-0002-1293-1246]{Jesse Torres}
\affiliation{Department of Electrical Engineering and Computer Science, Florida Institute of Technology, Melbourne, FL, USA}

\author[0000-0002-3878-4205]{Philip K. Chan}
\affiliation{Department of Electrical Engineering and Computer Science, Florida Institute of Technology, Melbourne, FL, USA}

\author[0000-0001-9333-6539]{Mohamed Nedal}
\affiliation{Astronomy and Astrophysics Section, School of Cosmic Physics, Dublin Institute for Advanced Studies, DIAS Dunsink Observatory, Dublin D15 XR2R, Ireland}

\author[0000-0002-6591-4482]{Kamen Kozarev}
\affiliation{Institute of Astronomy of the Bulgarian Academy of Sciences, Sofia, Bulgaria}

\author[0000-0001-6855-5799]{Peijin Zhang}
\affiliation{Center for Solar-Terrestrial Research, New Jersey Institute of Technology, Newark, NJ 07102, USA}
\affiliation{Cooperative Programs for the Advancement of Earth System Science, University Corporation for Atmospheric Research, Boulder, CO, USA}

\author[0000-0002-6202-8565]{Kimberly Moreland}
\affiliation{Southwest Research Institute, San Antonio, TX, USA}
\affiliation{CIRES, University of Colorado Boulder, Boulder, CO, USA}
\affiliation{Space Weather Prediction Center, NOAA, Boulder, CO, USA}
\affiliation{Department of Physics and Astronomy, College of Science, The University of Texas at San Antonio, San Antonio, TX, USA}

\author[0000-0003-2595-3185]{Hazel M. Bain}
\affiliation{Space Weather Prediction Center, National Oceanic and Atmospheric Administration, Boulder, CO, USA}

\author[0000-0003-0508-4912]{Samuel Hart}
\affiliation{Southwest Research Institute, San Antonio, TX, USA}
\affiliation{The University of Texas at San Antonio, San Antonio, TX, USA}

\author[0000-0001-7514-6571]{Michael J. Starkey}
\affiliation{Southwest Research Institute, San Antonio, TX, USA}

\author[0000-0002-8768-1819]{Alan G. Ling}
\affiliation{Atmospheric and Environmental Research, Inc., MA, USA}

\author[0000-0002-7102-5032]{Simone Benella}
\affiliation{Institute for Space Astrophysics and Planetology, Via del Fosso del Cavaliere 100, 00133, Rome, Italy}

\begin{abstract}

Solar energetic particle (SEP) events have attracted increasing attention due to their significant radiation hazards for aviation, spacecraft electronics, and human missions beyond Earth’s magnetosphere. From a scientific perspective, SEP events are intriguing because they arise from a set of physical processes extending from the solar surface and corona through the heliosphere, offering insight into particle acceleration and transport mechanisms that are widely applicable across astrophysics. Therefore, advancing our ability to understand and predict SEP events is essential both for deepening our knowledge of such mechanisms and for safeguarding space technologies and exploration. Traditionally, researchers have modeled SEPs using physics-based simulations and empirical methods. More recently, machine learning (ML) has emerged as a new tool for understanding and predicting SEP events. The purpose of this manuscript is to review the currently available ML models for SEP prediction, identify the datasets used for training, compare their architectures, inputs, and outputs, and, based on these insights, outline good practices and recommendations for future research.

\end{abstract}



\section{Introduction} \label{sec:Introduction}

The heliosphere, the region dominated by the Sun’s influence, can be seen as a complex system of interconnected subsystems \citep{engelbrecht2022theory, cohen2026imap}. Within this domain, the constantly changing magnetic activity of the Sun shapes what is known as space weather \citep[e.g.,][]{temmer2021space,gopalswamy2022sun} in combination with the galaxy's hazards via galactic cosmic rays \citep{rankin2022galactic}. Space weather disturbances, particularly their more hazardous aspects, are driven by highly energetic solar events such as flares, coronal mass ejections (CMEs), and solar energetic particle (SEP) events \citep[e.g.,][]{papaioannou2016solar,buzulukova2022space}. These high-energy phenomena are critical contributors to technological disruptions in space and on Earth, underscoring the importance of understanding and predicting them to mitigate their potentially severe impacts \citep{georgoulis2024prediction}.

SEP events are characterized by the rapid acceleration and release of high-energy electrons \citep[e.g.,][]{MitchellEA2025ApJ_EnergeticElectronDelay_Type3Bursts}, protons and heavier ions \citep{McComasEA2019Nature_PSPISOIS_FirstResults,CohenEA2021AandA_He_to_H_abundance_inside_1au_PSPISOIS,CohenEA2021AandA_ISOISobservations_29Nov2020_SEPs,pak2025species} into the heliosphere \citep{desai2016large,reames2021solar}. These particles, accelerated by transient solar phenomena such as solar flares and CMEs, exhibit energies spanning from keV to GeV \citep{reames2013two}. Once energized, SEPs propagate along interplanetary magnetic field lines, creating complex spatial and temporal distributions influenced by the physical characteristics of their sources, e.g., diffusive shock acceleration \citep{AxfordEA1977ICRC_CosRayAccel_Shock,Bell1978MNRAS_ShockAccel_CosRay,BlandfordOstriker1978ApJL_ParticleAccel_Shock,drury1983introduction}, and by their transport through the interplanetary medium \citep{zank2015diffusive,ChhiberEA2021AandA_FLRW_SEPs_PSP,SubashchandarEA2025ApJL_SpatialDiffCoeff_PerpPara_PSP,CuestaEA2025ApJL_DiffusionCoefficients_SEPfitting_QLT}. 

SEP events pose significant challenges to space weather forecasting due to their rapid onset and variability and potential impact on both technological systems and human health \citep[e.g.,][]{tobiska2015advances,mishev2015computation,miroshnichenko2018retrospective}. SEPs are central to space weather concerns as they represent a major radiation hazard for astronauts, particularly during extravehicular activities, and for passengers and crew on high-latitude flights \citep{cucinotta2013safe,tobiska2015advances}. These high-energy particles can also damage satellite electronics, disrupt communication systems, and impair navigation and power infrastructure on Earth \citep{schrijver2015understanding}. For interplanetary missions, SEP events present critical risks, especially for astronauts on the Moon or Mars, where the lack of a planetary magnetic field and thin or non-existent atmospheres offer limited shielding from solar radiation \citep{zeitlin2013measurements}. As the National Aeronautics and Space Administration (NASA) and other space agencies plan long-term missions to the Moon and Mars, accurate forecasting of SEP events has become imperative for ensuring mission success and the safety of human explorers \citep{neukart2024towards,creech2022artemis}.

Despite decades of research, forecasting SEP events with high confidence remains an open challenge. Traditional forecasting approaches span from physics-based models ---which simulate particle acceleration and transport processes, but often require substantial computational resources--- to empirical and statistical models that provide faster predictions, but rely heavily on historical correlations \citep{whitman2023review}. Recently, machine learning (ML) has emerged as a powerful alternative tool, capable of uncovering nonlinear patterns across diverse solar and heliospheric datasets \citep{aminalragia2021solar,neukart2024towards,kasapis2025reconstructing}. ML has created a new area of heliophysics research \citep{nita2020machine,berger2021machine} and a new community of researchers \citep{camporeale2020ml,narock2022supporting}. This growing body of work includes a recent review of empirical and physics-based models that also highlights the expanding role of ML techniques in space‑weather and specifically in SEP forecasting \citep{2025SSRv..221...82P}. ML methods hold promise for improving both the accuracy and the speed of SEP forecasts, particularly as the volume of space-based observations continues to grow.

The application of ML to the prediction of SEPs faces several challenges, including the rarity of SEP events (typically an SEP event is defined as $\geq10$ MeV particles surpassing a 10 pfu limit\footnote{\url{https://www.swpc.noaa.gov/products/goes-proton-flux}}) and in particular the high-energy ones \citep{Waterfall_2023}, the severe class imbalance between SEP and non-SEP cases (which becomes more pronounced as the particle's energy increase), and the need to integrate heterogeneous and often sparse data sources such as proton fluxes, solar images, flare catalogs, and active region (AR) properties \citep{chatterjee2024mempsep,hosseinzadeh2025end, 2025SSRv..221...82P}. In addition, because most ML‑based SEP prediction efforts remain proof‑of‑concept, researchers have adopted widely varying validation metrics, input data choices, target definitions, and modeling setups, which makes meaningful comparison between studies difficult. Overcoming these limitations requires careful and more coordinated model design, innovative data augmentation strategies, and the incorporation of domain knowledge to ensure both predictive skill and physical interpretability \citep{lavasa2021assessing,sadykov2021prediction,ali2024predicting,hosseinzadeh2024improving}.

In this paper, we provide a comprehensive review of ML models developed for SEP prediction. Twenty four approaches are categorized based on their model architectures, input data, and output predictions (Section~\ref{sec:Overview_and_Categorization} and Appendix~\ref{app:Description_of_Models}). The datasets commonly used to train and validate these ML models (Section~\ref{sec:Datasets} and Appendix~\ref{app:Datasets}) are identified and their performance and limitations are discussed (Section~\ref{sec:Discussion}). An effort is made to systematically compare existing models, assess the current state of the field, identify open challenges, and outline opportunities for advancing SEP forecasting to operational use. Moving beyond cataloging existing SEP forecasting approaches \citep[as already done by][]{whitman2023review}, the current review is focused on ML applications and synthesizes common trends, limitations, and emerging directions identifying both current capabilities and remaining challenges that need to be addressed before SEP ML forecasting becomes completely operational. By clarifying or capturing the state of the field and highlighting opportunities for future progress, this work aims to lead the development of next-generation SEP prediction systems that combine predictive skill, physical interpretability and operational reliability. Ultimately, this work provides a cartography of the current state of SEP‑prediction research using ML and to guide future efforts toward more reliable, interpretable, and operationally useful approaches for space‑weather forecasting, as discussed in Section~\ref{sec:Conclusions}.


\vspace{19pt}

\section{Overview and Categorization of ML Models} \label{sec:Overview_and_Categorization}

\begin{table}[h] 
\centering
\begin{tabular}{|c|c|p{4cm}|p{3cm}|c|c|}
\hline
 \textbf{Section}  & \textbf{Model }               & \textbf{References  }               & \textbf{Access Links} & \textbf{Type }                   & \textbf{Complexity} \\ \hline
\ref{sec:XGBoost}  & XGBoost              & \cite{ali2024predicting}                & \href{https://sun.njit.edu/SEP3/datasets.html}{SEP List} and \href{http://www.ncei.noaa.gov/data/goes-space-environment-monitor/access/avg/}{Data}   & Gradient Boosting       & 2          \\ \hline
\ref{sec:STSF} & STSF                 & \cite{rotti2024short}, \cite{rotti2024precise}              &   \href{https://github.com/sumanth-ra23/SEP-Predictions}{Model}  and  \href{https://dataverse.harvard.edu/dataset.xhtml?persistentId=doi:10.7910/DVN/DZYLHK}{Data}              & Time Series Classifier  & 4          \\ \hline
\ref{sec:SHARP-SMARP}  & SMARP-SHARP        & \cite{kasapis2022interpretable}, \cite{kasapis2024forecasting}         &     \href{github.com/skasapis/SEP_Pred_SMARP-SHARP}{Model}, \href{https:/ngdc.noaa.gov/stp/satellite/goes/doc/SPE.txt}{SEP List} and Data             & Linear SVM              & 7          \\ \hline
\ref{sec:AA} & AA                   & \cite{lavasa2021assessing}             &  \href{https://github.com/SolarML/SEP-ML}{Model and Data}                 & Random Forest           & 8          \\ \hline
\ref{sec:ESPERTA} & ESPERTA              & \cite{laurenza2009technique,laurenza2018short,laurenza2024upgrades}, \cite{alberti2017solar, alberti2019forecasting}, \cite{stumpo2021open}, \cite{benella2023statistical}   &   N/A                & Logistic Regression     &      12      \\ \hline
\ref{sec:UMASEP} & UMASEP             & \cite{nunez2011predicting}            &  \href{https://ccmc.gsfc.nasa.gov/scoreboards/sep/}{CCMC SEP Scoreboard}                 &   Regression Tree Ensemble                      &     20       \\ \hline
\ref{sec:UMASOD} & UMASOD               & \cite{nunez2020predicting}               &        N/A           &   Decision Tree        &    30        \\ \hline
\ref{sec:MS_SEP} & MS-SEP             & \cite{ali2025forecasting}             &     \href{https://github.com/SolarML/SEP-ML}{Model}  and \href{ftp://ftp.swpc.noaa.gov/pub/indices/events/}{Data 1}, \href{https://cdaw.gsfc.nasa.gov/CME_list/}{Data 2},  \href{ftp://ftp.swpc.noaa.gov/pub/indices/SPE.txt}{Data 3}     &    Random Forest                     &     52       \\ \hline
\ref{sec:CART}  & CART & \cite{boubrahimi2017prediction}         &     N/A              & Decision Tree (CART)    & 61         \\ \hline
\ref{sec:RH} & RH     & \cite{o2024random}             &    \href{https://sun.njit.edu/SEP3/datasets.html}{SPE Catalog}               &      Random Ensemble of NNs          &     202       \\ \hline
\ref{sec:SSEP-Survival}  & SSEP    & \cite{martens2024advancing}, \cite{DVN/GXY9MZ_2024}             &  \href{https://github.com/indiajacksonphd}{Model} and \href{https://doi.org/10.7910/DVN/GXY9MZ}{Data}                 & Random Survival Forests & 300        \\ \hline
\ref{sec:SEP-C} & SEP-C            & \cite{torres2022machine}               &    \href{https://zenodo.org/records/12832882}{Model} and  \href{https://cdaw.gsfc.nasa.gov/CME_list/radio/waves_type2.html}{Data 1}, \href{https://cdaw.gsfc.nasa.gov/CME_list/}{Data 2}            &      Neural Network                   &     780       \\ \hline
\ref{sec:CANN} & CANN           & \cite{sadykov2021prediction}            &    \href{https://sun.njit.edu/SEP3/datasets.html}{SPE Catalog and Model}               &   Custom Architecture NN   &     1,243       \\ \hline
\ref{sec:SEP-E} & SEP-E            &  \cite{torres2025machine}               &    \href{https://zenodo.org/records/12832882}{Model} and \href{https://cdaw.gsfc.nasa.gov/CME_list/radio/waves_type2.html}{Data 1}, \href{https://cdaw.gsfc.nasa.gov/CME_list/}{Data 2}            &      Neural Network                   &     1,530       \\ \hline
\ref{sec:SPRINTS} & SPRINTS              & \cite{engell2017sprints}           &  \href{https://ccmc.gsfc.nasa.gov/scoreboards/sep/}{Outputs}                & MLP                     & 5,401      \\ \hline
\ref{sec:TSF}  & TSF                  & \cite{hosseinzadeh2024improving}        &  \href{https://github.com/pouyahosseinzadeh/Solar-Energetic-Particle-Event-Prediction-Data-Augmentation}{Model}                 & Time Series Forest      & 15,000     \\ \hline
\ref{sec:UDM}  & UDM                  & \cite{hosseinzadeh2024toward}        &    \href{https://github.com/pouyahosseinzadeh/High-Impact-SEP-Prediction---Space-Weather}{Model}               & Time Series Forest      & 15,548     \\ \hline
\ref{sec:UNSPELL} & UNSPELL              & \cite{aminalragia2021solar} &  \href{https://www.ncei.noaa.gov/data/goes-space-environment-monitor/access/science/xrs/}{Data 1}, \href{https://www.ngdc.noaa.gov/stp/space-weather/solar-data/solar-features/solar-flares/x-rays/goes/xrs/}{Data 2}              & NN Ensemble             & 81,120 \\ \hline
\ref{sec:TS-HOG-TB}  & TS-HOG-TB            & \cite{hosseinzadeh2025end}        &    \href{https://github.com/pouyahosseinzadeh/Solar-Energetic-Particle-Event-Prediction-Ensemble-TS-HOG-TB}{Model}               & Ensemble Method         & 100,000    \\ \hline
\ref{sec:SEPNET}  & SEPNET            & \cite{yu2025solar}        &    \href{https://github.com/yuyian/SEP-Prediction}{Model}               & Transformer         & 130,000    \\ \hline
\ref{sec:BiLSTM-SEP} & BiLSTM-SEP           & \cite{nedal2023forecasting}               &   \href{https://gitlab.com/iahelio/mosaiics/sep-lstm/}{Model} and \href{https://omniweb.gsfc.nasa.gov)}{Data 1}, \href{(https://satdat.ngdc.noaa.gov/sem/goes/data/avg}{Data 2}, \href{https://www.sidc.be/silso/home}{Data 3}              & BiLSTM NN               & 333,699    \\ \hline
\ref{sec:MEMPSEP} & MEMPSEP              & \cite{chatterjee2024mempsep}, \cite{dayeh2024mempsep}, \cite{moreland2024mempsep}          &   \href{https://zenodo.org/records/11201195}{Model}  and \href{https://zenodo.org/records/10044865}{Data}              & Convolutional NN                     & 6,092,617  \\ \hline
\ref{sec:PSPSP} & PSPSP              &  \cite{hutchins2026}          &   \href{https://github.com/thutch17/PSP-SEP-Event-Prediction}{Code and Data}              & Neural Network                     & 13,814,081  \\ \hline
\ref{sec:EPREM-S} &    EPREM-S                  & \cite{baydin2023surrogate}            & \href{https://zenodo.org/records/10038847}{Model} and \href{https://zenodo.org/records/10109868}{Data}                 &        Feed-Forward NN                 &      285,881,344      \\ \hline

\end{tabular}
\caption{List of ML models that predict SEP events. The first two columns reference the Appendix~\ref{app:Description_of_Models} subsection of the models' description and their name. The following columns include references to the models, links to the associated code and data (if available), and the models' \textit{Type} and \textit{Complexity} descriptors. Models are summarized in this table ---and in the manuscript as a whole--- in order of complexity (number of trainable parameters), beginning from the least complex models. A list of acronyms (including the acronym names of the models) along with their definitions is available in Appendix~\ref{app:Acronyms}.}
\label{tab:all_models}
\end{table}


To facilitate a systematic comparison of ML models for SEP prediction, we organize the descriptions of 24 models (Table~\ref{tab:all_models}) identified in the English literature into three major categories: \textit{Architecture}, \textit{Input}, and \textit{Output} (Table~\ref{tab:Qualitative_Table}). Each category contains several subfields that capture aspects of the model’s design. The {\it Architecture} section includes descriptors such as algorithm type and complexity; {\it Input} covers data shape, physical type, historical depth, diversity, class imbalance, and sample characteristics; and {\it Output} describes the prediction format, triggering mechanism, and forecast window. This type of categorization is followed throughout Appendix~\ref{app:Description_of_Models}, where single-page model descriptions are provided, based on the submissions offered by the authors of each model. The quantitative and qualitative values for each model based on such a categorization are prescribed in Tables~\ref{tab:XGBoost}-\ref{tab:EPREM-S}. These values were again submitted by the modelers through responding to the form presented in Appendix~\ref{app:Questionnaire}.

The \textit{Architecture} subgroup captures the structural and computational characteristics of each SEP prediction system. It includes the \textit{Model Type}, a categorical descriptor indicating the algorithm class ---such as Support Vector Machines (SVMs), random forests, deep Neural Networks (NNs), or Long Short-Term Memory (LSTM) methods--- which defines the model’s learning strategy and architecture. Complementing this is the \textit{Model Complexity}, a numerical value representing the total number of trainable parameters, i.e., the number of internal weights adjusted during training (often referred to as ``model weights''). This scalar metric reflects the model’s depth and capacity, offering insight into its expressiveness, computational cost, and potential for overfitting. In practice, this index spans several orders of magnitude: shallow models such as SVMs or random forests typically involve only tens to a few hundred trainable parameters, whereas modern deep NNs can contain tens of millions of parameters.

The \textit{Input} subgroup captures the structure and physical nature of the data used to train the SEP prediction models. The \textit{Input Shape} is a categorical descriptor that defines the dimensionality of each individual sample—whether it is point-like (0D), time series (1D), spectra (1D), or imagery (2D). This classification reflects the format of a single event, not the overall dataset structure. The \textit{Input Type} is also categorical and refers to the physical quantity represented in the input, such as magnetic fields, extreme ultraviolet (EUV) or X-ray imagery, electric fields, white-light observations, radio measurements, and others. Beyond structure and physical type, several quantitative subcategories describe the statistical and temporal properties of the input dataset. \textit{Input History} refers to the total time span (in years) covered by the training data, \textit{Input Diversity} captures the total number of positive (SEP-producing) and negative events (non SEP-producing) used for training, while \textit{Input Imbalance} quantifies the rarity of SEP-producing events as a fraction (0 to 1) of the total sample set. Additionally, \textit{Input Sample Size} measures the data volume (in bytes) of a single event, and \textit{Input Sample Coverage} indicates the temporal duration (in hours) represented by each input sample. These metrics are useful for understanding the dataset richness, bias, and the temporal resolution of each model’s learning process.

Lastly, the \textit{Output} subgroup describes the nature and format of the predictions produced by each SEP model. The \textit{Output Prediction} is a categorical descriptor that defines the kind of prediction made, such as binary classification (SEP vs. non-SEP, all-clear warnings), regression of time-series quantities (onset or peak time forecasts) and probability estimates. Models may fall into multiple categories depending on how their outputs are processed or interpreted. The \textit{Output Type} qualitative category (Triggered vs. Continuous) further distinguishes models based on their operational logic: triggered models (often referred to in literature as post-eruptive) issue predictions in response to specific solar events (e.g., flares or CMEs), while continuous models provide ongoing forecasts based on ambient solar or heliospheric conditions (often referred to as pre-eruptive). Finally, \textit{Forecast Window}, a quantitative metric, defines the forecast window, the time span over which the prediction is valid or expected to occur, such as SEP onset within the next 7 hours or an all-clear prediction for the next 24 hours. 

The qualitative descriptors of each corresponding model that exists in the current literature are prescribed in Table~\ref{tab:Qualitative_Table} and Figure~\ref{fig:Qualitative_Plot} of Section~\ref{sec:Discussion}. The quantitative descriptors are used to create Figures~\ref{fig:Quantitative_Plot} and \ref{fig:Time_Coverage} in an attempt to compare the different models with each other and map the state of current research in the field. In Appendix~\ref{app:Description_of_Models}, each model and its descriptors are summarized in more detail.


\section{Datasets for SEP Prediction Using ML} \label{sec:Datasets}

ML applications for SEP prediction rely fundamentally on the quality, relevance, and structure of the datasets used for training and evaluation. Unlike traditional modeling approaches, ML methods require large volumes of labeled data that capture the complexity of SEP-related phenomena, including solar activity indicators, particle flux measurements, and contextual heliospheric conditions. The choice of dataset directly influences model performance, generalizability, and interpretability. Subsections~\ref{sec:MEMPSEP-III_Dataset}-\ref{sec:CLEAR_Dataset} of Appendix~\ref{app:Datasets} summarize the key datasets that have been developed and used for SEP prediction using ML, highlighting their characteristics, input features, target variables, and typical use cases. A list of these five datasets, along with the links to access them, can be found in Table~\ref{tab:sep_datasets}. Note that all models presented in this manuscript are supervised, therefore they require datasets that provide event labeling. Future studies could utilize these datasets, without their labels, to explore the capabilities of unsupervised ML approaches for SEP prediction.

\begin{table}[htbp]
\centering
\caption{List of datasets that were created with the purpose of being used for SEP prediction. The title of the relevant publication, the associated Digital Object Identifier (DOI), the respective Appendix~\ref{app:Datasets} dataset description and useful links to access the datasets (if available) are included.}
\label{tab:sep_datasets}
\begin{tabular}{|p{3.5cm}|p{5.5cm}|p{4.5cm}|p{2.5cm}|}
\hline
\textbf{Developer \& Title} & \textbf{Publication Title} & \textbf{Description and DOI} & \textbf{Dataset Access} \\
\hline
Kimberly Moreland \, \, \, \, \, \, \, \, \, (MEMPSEP-III Dataset) & MEMPSEP-III. A Machine Learning-Oriented Multivariate Data Set for Forecasting the Occurrence and Properties of Solar Energetic Particle Events Using a Multivariate Ensemble Approach & Appendix~\ref{sec:MEMPSEP-III_Dataset} \, \, \, \, \, \, \, \, \, \, \, \,  \href{https://agupubs.onlinelibrary.wiley.com/doi/epdf/10.1029/2023SW003765}{10.1029/2023SW003765} & \href{https://zenodo.org/records/10044865}{Zenodo} \\
\hline
Pouya Hosseinzadeh \, \, \, (MTS-SEP Dataset) & Improving Solar Energetic Particle Event Prediction through Multivariate Time Series Data Augmentation & Appendix~\ref{sec:MTS-SEP_Dataset} \, \, \, \, \, \, \, \, \, \, \, \, \href{https://iopscience.iop.org/article/10.3847/1538-4365/ad1de0/pdf}{10.3847/1538-4365/ad1de0} & \href{https://github.com/pouyahosseinzadeh/Solar-Energetic-Particle-Event-Prediction-Data-Augmentation}{GitHub} \\
\hline
Sumanth A. Rotti \, \, \, \, \, \, (GSEP Dataset) & Integrated Geostationary Solar Energetic Particle Events Catalog: GSEP & Appendix~\ref{sec:GSEP_Dataset} \, \, \, \, \, \, \, \, \, \, \, \, \href{https://iopscience.iop.org/article/10.3847/1538-4365/ac87ac/pdf}{10.3847/1538-4365/ac87ac} \href{https://iopscience.iop.org/article/10.3847/1538-4365/acdace/pdf}{10.3847/1538-4365/acdace} & \href{https://dataverse.harvard.edu/dataset.xhtml?persistentId=doi:10.7910/DVN/DZYLHK}{Harvard Dataverse} \\
\hline
Paul Kosovich \, \, \, (SHARP-SMARP Dataset) & Time series of magnetic field parameters of merged MDI and HMI space-weather active region patches as potential tool for solar flare forecasting & Appendix~\ref{sec:SMARP-SHARP_Dataset} \, \, \, \, \, \, \, \, \, \, \, \, \href{https://iopscience.iop.org/article/10.3847/1538-4357/ad60c3/pdf}{/10.3847/1538-4357/ad60c3/} & \href{https://drive.google.com/file/d/1pIMvvw3GXZkuQ6WIFIhisauWMP2uUIaC/view?usp=sharing}{Google Drive} \\ \hline
Kathryn Whitman \, \, \, (CLEAR Dataset) & CLEAR SEP Benchmark Dataset & Appendix~\ref{sec:CLEAR_Dataset} \, \, \, \, \, \, \, \, \, \, \, \, \href{https://ccmc.gsfc.nasa.gov/swxcoe/clear/}{CLEAR Benchmark Website} & \href{https://ccmc.gsfc.nasa.gov/swxcoe/clear/clear_data/ClearBenchmarkDataset_Description_20251214.pdf}{Description} and  \href{https://ccmc.gsfc.nasa.gov/swxcoe/clear/benchmark.php}{Dataset}\\
\hline
\end{tabular}
\end{table}

\begin{figure}[h]  
    \centering
    \includegraphics[width=\textwidth]{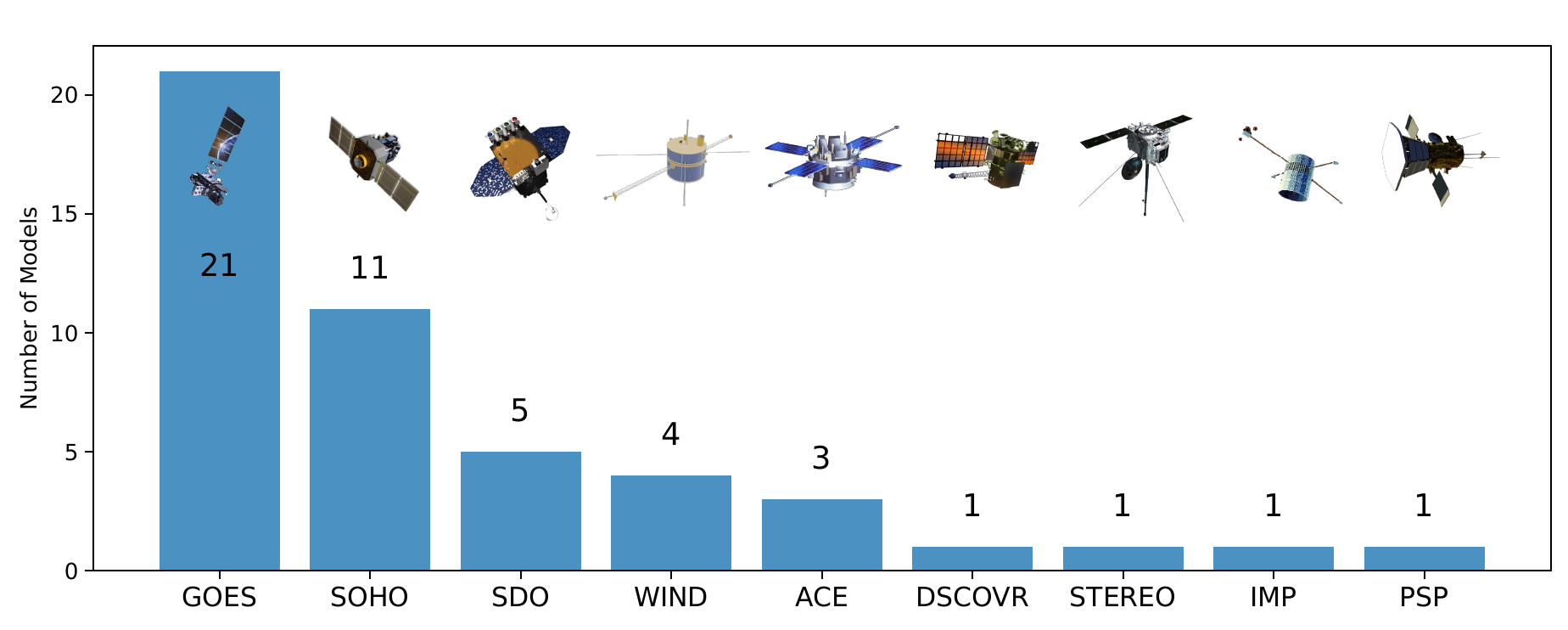} 
    \caption{Histogram illustrating the number of SEP forecasting models (from 24 total) that utilized space-based observations from each respective spacecraft.} 
    \label{fig:Satellites}
\end{figure}

It should be noted that this list is not exhaustive, but rather a list of datasets curated for the specific task of SEP prediction. A majority of the works in Table~\ref{tab:all_models} and Appendix~\ref{app:Description_of_Models} have used data from various sources for which although extensive processing might have taken place, they have not been published as peer reviewed publications. To avoid limiting the datasets list to peer reviewed publications, Table~\ref{tab:all_models} contains (if available from the modelers) links to their datasets, repositories and code. In summary, the 24 studies have used data from nine different spacecraft as seen in Figure~\ref{fig:Satellites}, with an overwhelming majority (21/24) using the National Oceanic and Atmospheric Administration (NOAA) Geostationary Operational Environmental Satellite (GOES) series \citep{rodriguez2010east, hu2022calibration, sellers1996design}, many times complemented with data from other satellites such as the Solar and Heliospheric Observatory \citep[SOHO;][]{domingo1995soho, domingo1995soho2}, the Solar Dynamics Observatory \citep[SDO;][]{pesnell2012solar}, the Wind \citep{von1995energetic} spacecraft, the Advanced Composition Explorer \citep[ACE;][]{stone1998advanced}, the Deep Space Climate Observatory \citep[DSCOVR;][]{burt2012deep} and the Solar TErrestrial RElations Observatory \citep[STEREO;][]{kaiser2008stereo}. The BiLSTM-SEP and PSPSP models have also used data from the Interplanetary Monitoring Platform \citep[IMP;][]{simunac2004solar} and the Parker Solar Probe \citep[PSP;][]{fox2016solar, raouafi2023parker}. Some studies such as MEMPSEP, SEPNET and ESPERTA have also used ground-based datasets and relevant event catalogs such as the Space Weather Database Of Notifications, Knowledge, Information \href{https://kauai.ccmc.gsfc.nasa.gov/DONKI/}{(DONKI)} developed at the Community Coordinated Modeling Center (CCMC\footnote{\url{https://ccmc.gsfc.nasa.gov/}}) and the Low-Frequency Array \citep[LOFAR;][]{van2013lofar}.

The majority of missions from which data are used for ML-aided SEP predictions reside either in Earth orbits ---such as GOES in geostationary orbit and SDO in inclined geosynchronous orbit--- or at the Sun–Earth Lagrange 1 (L1) point, where SOHO, WIND, ACE, and DSCOVR operate. These spacecraft therefore primarily sample the space environment relevant to geoeffective events. A geoeffective SEP event generates particles that reach Earth with sufficient intensity and energy to produce measurable impacts on the near‑Earth space environment. Only a handful of studies (PSPSEP and MEMPSEP) use data from spacecraft located away from the Earth–L1 system, such as the PSP, which operates in a highly eccentric heliocentric orbit deep in the inner heliosphere and STEREO, which is in a heliocentric orbit near Earth's orbit. These are also the only studies that have used imagery as inputs to their models (Table~\ref{tab:Qualitative_Table}, 2D column). Since most satellites in Earth orbit or at L1 downlink their measurements in near‑real‑time (e.g., SDO, SOHO, WIND) and some even in real‑time (GOES, ACE, DSCOVR), the majority of existing SEP‑prediction studies could in principle evolve toward operational forecasting using near‑real‑time inputs. What constitutes an operational model, and the criteria for determining operational feasibility, are discussed in Section~\ref{sec:Paths_for_Future_Research}. Despite the availability of multiple SEP and solar datasets, inconsistencies in event definitions, preprocessing pipelines, and temporal coverage ---stemming from the differences in the modelers' dataset selection--- remain significant barriers to fair model comparison, as discussed later in this manuscript. Continued efforts toward dataset harmonization and shared validation resources will be essential for advancing ML-based SEP forecasting.


\section{Discussion} \label{sec:Discussion}

The past decade, the heliophysics community has explored a broad set of ML models, which have different architectures, use different data, and are at varying levels of maturity. Predicting SEP events using ML is an emerging field, and most studies are proofs of concept, compared to other physics-based and empirical models that are well-established and already provide real-time operational forecasts \citep{whitman2023review}. Table~\ref{tab:Qualitative_Table} provides a community‑driven overview of existing SEP ML models, summarizing their architectures, inputs, and outputs based on the grouping described in Section~\ref{sec:Overview_and_Categorization} and the detailed questionnaire completed by model developers and presented in Appendix~\ref{app:Questionnaire}. While this study makes every effort to compile a comprehensive list of models developed to date, it reflects only the English‑language literature and the state of the field prior to 2026. As with any rapidly evolving research area, new ML‑based SEP prediction approaches continued to appear as this work was being written. Consequently, the conclusions drawn here should be interpreted as a snapshot of an active and continually developing domain rather than a definitive or static inventory.

\begin{table}[h] 
\centering
\caption{Qualitative taxonomy of the SEP ML models. This table presents a summary of the qualitative values listed in Tables~\ref{tab:XGBoost}-\ref{tab:EPREM-S}, for the three model aspects discussed in Section~\ref{sec:Discussion}: a) \textit{Architecture} (Section~\ref{sec:Model_Architectures}), b) \textit{Inputs} (Section~\ref{sec:Inputs_Comparison}) and \textit{Outputs} (Section~\ref{sec:Outputs_and_Testing}).}
\label{tab:Qualitative_Table}
\begin{tabular}{|llllll|lllllllllllll|lllll|}
\hline
\multicolumn{6}{|c|}{Architecture}  & \multicolumn{13}{c|}{Inputs}  & \multicolumn{5}{c|}{Output}  \\ \hline

\multicolumn{1}{|l|}{Model Name}  & \multicolumn{5}{c|}{Type}  & \multicolumn{3}{c|}{Shape}  & \multicolumn{10}{c|}{Type}  & \multicolumn{3}{c|}{Prediction} & \multicolumn{2}{c|}{Type}  \\ \hline

\multicolumn{1}{|l|}{} & \rotatebox{90}{Neural Network} & \rotatebox{90}{Forest/Decision Tree} & \rotatebox{90}{Linear/Logistic Regression} & \rotatebox{90}{Support Vector Machine} & \rotatebox{90}{Ensemble} & \rotatebox{90}{0D} & \rotatebox{90}{1D} & \rotatebox{90}{2D} & \rotatebox{90}{Soft X-ray} & \rotatebox{90}{Proton Flux} & \rotatebox{90}{Solar Wind} & \rotatebox{90}{Space-Based Radio} & \rotatebox{90}{Ground-Based Radio} & \rotatebox{90}{Electron Flux} & \rotatebox{90}{Flare Location} & \rotatebox{90}{Magnetic Fields} & \rotatebox{90}{Coronagraphs} & \rotatebox{90}{EUV Imagery} & \rotatebox{90}{Classification} & \rotatebox{90}{Probability} & \rotatebox{90}{Regression (Time-Series)} & \rotatebox{90}{Triggered} & \rotatebox{90}{Continuous} \\ \hline

\multicolumn{1}{|l|}{\ref{sec:XGBoost} XGBoost}     & \multicolumn{1}{l|}{}  & \multicolumn{1}{l|}{\xmark} & \multicolumn{1}{l|}{}  & \multicolumn{1}{l|}{}  &  \xmark  & \multicolumn{1}{l|}{}   & \multicolumn{1}{l|}{\xmark}   & \multicolumn{1}{l|}{}   & \multicolumn{1}{l|}{\xmark}   & \multicolumn{1}{l|}{\xmark}  & \multicolumn{1}{l|}{}  & \multicolumn{1}{l|}{}   & \multicolumn{1}{l|}{} & \multicolumn{1}{l|}{}  & \multicolumn{1}{l|}{} & \multicolumn{1}{l|}{}  & \multicolumn{1}{l|}{}  &   & \multicolumn{1}{l|}{\xmark}   & \multicolumn{1}{l|}{} & \multicolumn{1}{l|}{}  & \multicolumn{1}{l|}{}  &   \xmark   \\ \hline

\multicolumn{1}{|l|}{\ref{sec:STSF} STSF}     & \multicolumn{1}{l|}{}   & \multicolumn{1}{l|}{\xmark}  & \multicolumn{1}{l|}{}  & \multicolumn{1}{l|}{}    &    \xmark   & \multicolumn{1}{l|}{}   & \multicolumn{1}{l|}{\xmark}   & \multicolumn{1}{l|}{}   & \multicolumn{1}{l|}{\xmark}  & \multicolumn{1}{l|}{\xmark}  & \multicolumn{1}{l|}{}  & \multicolumn{1}{l|}{}  & \multicolumn{1}{l|}{}  & \multicolumn{1}{l|}{}  & \multicolumn{1}{l|}{} & \multicolumn{1}{l|}{} & \multicolumn{1}{l|}{}  &   & \multicolumn{1}{l|}{\xmark}  & \multicolumn{1}{l|}{}  & \multicolumn{1}{l|}{}  & \multicolumn{1}{l|}{}   &   \xmark   \\ \hline

\multicolumn{1}{|l|}{\ref{sec:SHARP-SMARP} SMARP-SHARP}  & \multicolumn{1}{l|}{}  & \multicolumn{1}{l|}{}  & \multicolumn{1}{l|}{\xmark}  & \multicolumn{1}{l|}{\xmark}    &          & \multicolumn{1}{l|}{\xmark}   & \multicolumn{1}{l|}{}   & \multicolumn{1}{l|}{}   & \multicolumn{1}{l|}{}  & \multicolumn{1}{l|}{}  & \multicolumn{1}{l|}{}  & \multicolumn{1}{l|}{}  & \multicolumn{1}{l|}{}  & \multicolumn{1}{l|}{}  & \multicolumn{1}{l|}{} & \multicolumn{1}{l|}{\xmark} & \multicolumn{1}{l|}{}  &   & \multicolumn{1}{l|}{\xmark}   & \multicolumn{1}{l|}{\xmark}  & \multicolumn{1}{l|}{}   & \multicolumn{1}{l|}{\xmark}  &  \\ \hline

\multicolumn{1}{|l|}{\ref{sec:AA} AA} & \multicolumn{1}{l|}{}  & \multicolumn{1}{l|}{\xmark}  & \multicolumn{1}{l|}{}  & \multicolumn{1}{l|}{}   &   & \multicolumn{1}{l|}{\xmark}   & \multicolumn{1}{l|}{}   & \multicolumn{1}{l|}{}   & \multicolumn{1}{l|}{\xmark}  & \multicolumn{1}{l|}{}  & \multicolumn{1}{l|}{}  & \multicolumn{1}{l|}{}   & \multicolumn{1}{l|}{} & \multicolumn{1}{l|}{}  & \multicolumn{1}{l|}{}  & \multicolumn{1}{l|}{}   & \multicolumn{1}{l|}{\xmark}  &   & \multicolumn{1}{l|}{\xmark}  & \multicolumn{1}{l|}{}  & \multicolumn{1}{l|}{}   & \multicolumn{1}{l|}{\xmark}   &   \\ \hline

\multicolumn{1}{|l|}{\ref{sec:ESPERTA} ESPERTA}  & \multicolumn{1}{l|}{} & \multicolumn{1}{l|}{} & \multicolumn{1}{l|}{\xmark} & \multicolumn{1}{l|}{} & \multicolumn{1}{l|}{} & \multicolumn{1}{l|}{\xmark} & \multicolumn{1}{l|}{}  & \multicolumn{1}{l|}{}  & \multicolumn{1}{l|}{\xmark}  & \multicolumn{1}{l|}{} & \multicolumn{1}{l|}{} & \multicolumn{1}{l|}{\xmark}  & \multicolumn{1}{l|}{\xmark} & \multicolumn{1}{l|}{}  & \multicolumn{1}{l|}{\xmark}  & \multicolumn{1}{l|}{} & \multicolumn{1}{l|}{} &  & \multicolumn{1}{l|}{\xmark} & \multicolumn{1}{l|}{\xmark} & \multicolumn{1}{l|}{}  & \multicolumn{1}{l|}{\xmark}  & \\ \hline

\multicolumn{1}{|l|}{\ref{sec:UMASEP} UMASEP} & \multicolumn{1}{l|}{}  & \multicolumn{1}{l|}{\xmark}  & \multicolumn{1}{l|}{\xmark} & \multicolumn{1}{l|}{} & \multicolumn{1}{l|}{\xmark} & \multicolumn{1}{l|}{} & \multicolumn{1}{l|}{\xmark}  & \multicolumn{1}{l|}{}  & \multicolumn{1}{l|}{\xmark}  & \multicolumn{1}{l|}{\xmark}  & \multicolumn{1}{l|}{}  & \multicolumn{1}{l|}{}  & \multicolumn{1}{l|}{} & \multicolumn{1}{l|}{}  & \multicolumn{1}{l|}{}  & \multicolumn{1}{l|}{}  & \multicolumn{1}{l|}{}  &  & \multicolumn{1}{l|}{\xmark} & \multicolumn{1}{l|}{} & \multicolumn{1}{l|}{\xmark}  & \multicolumn{1}{l|}{}  & \xmark \\ \hline

\multicolumn{1}{|l|}{\ref{sec:UMASOD} UMASOD}  & \multicolumn{1}{l|}{} & \multicolumn{1}{l|}{\xmark}  & \multicolumn{1}{l|}{}  & \multicolumn{1}{l|}{}   &   & \multicolumn{1}{l|}{} & \multicolumn{1}{l|}{\xmark}  & \multicolumn{1}{l|}{}  & \multicolumn{1}{l|}{\xmark}   & \multicolumn{1}{l|}{} & \multicolumn{1}{l|}{}  & \multicolumn{1}{l|}{\xmark}  & \multicolumn{1}{l|}{}  & \multicolumn{1}{l|}{} & \multicolumn{1}{l|}{}  & \multicolumn{1}{l|}{} & \multicolumn{1}{l|}{}  &   & \multicolumn{1}{l|}{\xmark}  & \multicolumn{1}{l|}{} & \multicolumn{1}{l|}{}  & \multicolumn{1}{l|}{\xmark} &  \\ \hline

\multicolumn{1}{|l|}{\ref{sec:MS_SEP} MS-SEP} & \multicolumn{1}{l|}{} & \multicolumn{1}{l|}{\xmark} & \multicolumn{1}{l|}{} & \multicolumn{1}{l|}{}  &  & \multicolumn{1}{l|}{} &\multicolumn{1}{l|}{\xmark} & \multicolumn{1}{l|}{} & \multicolumn{1}{l|}{\xmark} & \multicolumn{1}{l|}{} & \multicolumn{1}{l|}{} & \multicolumn{1}{l|}{\xmark}  & \multicolumn{1}{l|}{} & \multicolumn{1}{l|}{} & \multicolumn{1}{l|}{} & \multicolumn{1}{l|}{} & \multicolumn{1}{l|}{\xmark} &   & \multicolumn{1}{l|}{\xmark}  & \multicolumn{1}{l|}{} & \multicolumn{1}{l|}{} & \multicolumn{1}{l|}{\xmark} &  \\ \hline

\multicolumn{1}{|l|}{\ref{sec:CART} CART}  & \multicolumn{1}{l|}{}  & \multicolumn{1}{l|}{\xmark} & \multicolumn{1}{l|}{} & \multicolumn{1}{l|}{} &  & \multicolumn{1}{l|}{} & \multicolumn{1}{l|}{\xmark}   & \multicolumn{1}{l|}{}   & \multicolumn{1}{l|}{\xmark}   & \multicolumn{1}{l|}{\xmark} & \multicolumn{1}{l|}{}  & \multicolumn{1}{l|}{}  & \multicolumn{1}{l|}{}  & \multicolumn{1}{l|}{}  & \multicolumn{1}{l|}{}  & \multicolumn{1}{l|}{}  & \multicolumn{1}{l|}{}  &   & \multicolumn{1}{l|}{\xmark} & \multicolumn{1}{l|}{} & \multicolumn{1}{l|}{}  & \multicolumn{1}{l|}{\xmark}  &   \\ \hline

\multicolumn{1}{|l|}{\ref{sec:RH} RH} & \multicolumn{1}{l|}{\xmark} & \multicolumn{1}{l|}{} & \multicolumn{1}{l|}{} & \multicolumn{1}{l|}{}  &  \xmark & \multicolumn{1}{l|}{\xmark} & \multicolumn{1}{l|}{} & \multicolumn{1}{l|}{}  & \multicolumn{1}{l|}{\xmark} & \multicolumn{1}{l|}{} & \multicolumn{1}{l|}{} & \multicolumn{1}{l|}{}  & \multicolumn{1}{l|}{}  & \multicolumn{1}{l|}{}  & \multicolumn{1}{l|}{\xmark} & \multicolumn{1}{l|}{} & \multicolumn{1}{l|}{} &  & \multicolumn{1}{l|}{\xmark}  & \multicolumn{1}{l|}{} & \multicolumn{1}{l|}{}  & \multicolumn{1}{l|}{\xmark} &  \\ \hline

\multicolumn{1}{|l|}{\ref{sec:SSEP-Survival} SSEP}   & \multicolumn{1}{l|}{}  & \multicolumn{1}{l|}{\xmark}  & \multicolumn{1}{l|}{}  & \multicolumn{1}{l|}{} & \multicolumn{1}{l|}{} & \multicolumn{1}{l|}{\xmark} & \multicolumn{1}{l|}{}  & \multicolumn{1}{l|}{}  & \multicolumn{1}{l|}{}  & \multicolumn{1}{l|}{}  & \multicolumn{1}{l|}{} & \multicolumn{1}{l|}{}  & \multicolumn{1}{l|}{}  & \multicolumn{1}{l|}{} & \multicolumn{1}{l|}{\xmark}  & \multicolumn{1}{l|}{}  & \multicolumn{1}{l|}{}  &  & \multicolumn{1}{l|}{}  & \multicolumn{1}{l|}{\xmark}  & \multicolumn{1}{l|}{}  & \multicolumn{1}{l|}{\xmark} &   \\ \hline

\multicolumn{1}{|l|}{\ref{sec:SEP-C} SEP-C} & \multicolumn{1}{l|}{\xmark}  & \multicolumn{1}{l|}{}  & \multicolumn{1}{l|}{} & \multicolumn{1}{l|}{}   &   & \multicolumn{1}{l|}{\xmark}   & \multicolumn{1}{l|}{}   & \multicolumn{1}{l|}{}   & \multicolumn{1}{l|}{}  & \multicolumn{1}{l|}{\xmark}  & \multicolumn{1}{l|}{\xmark}   & \multicolumn{1}{l|}{}  & \multicolumn{1}{l|}{} & \multicolumn{1}{l|}{}  & \multicolumn{1}{l|}{} & \multicolumn{1}{l|}{} & \multicolumn{1}{l|}{\xmark} &  & \multicolumn{1}{l|}{\xmark}  & \multicolumn{1}{l|}{\xmark} & \multicolumn{1}{l|}{\xmark}  & \multicolumn{1}{l|}{}  &  \xmark \\ \hline

\multicolumn{1}{|l|}{\ref{sec:CANN} CANN}  & \multicolumn{1}{l|}{\xmark} & \multicolumn{1}{l|}{} & \multicolumn{1}{l|}{}  & \multicolumn{1}{l|}{}  & \multicolumn{1}{l|}{}  &  \multicolumn{1}{l|}{\xmark} & \multicolumn{1}{l|}{} & \multicolumn{1}{l|}{}  & \multicolumn{1}{l|}{\xmark}  & \multicolumn{1}{l|}{\xmark}  & \multicolumn{1}{l|}{} & \multicolumn{1}{l|}{}  & \multicolumn{1}{l|}{\xmark} & \multicolumn{1}{l|}{} & \multicolumn{1}{l|}{}  & \multicolumn{1}{l|}{\xmark} & \multicolumn{1}{l|}{}  &  & \multicolumn{1}{l|}{\xmark}  & \multicolumn{1}{l|}{\xmark} & \multicolumn{1}{l|}{} & \multicolumn{1}{l|}{}  &  \xmark \\ \hline

\multicolumn{1}{|l|}{\ref{sec:SEP-E} SEP-E} & \multicolumn{1}{l|}{\xmark}  & \multicolumn{1}{l|}{}  & \multicolumn{1}{l|}{} & \multicolumn{1}{l|}{}   &   & \multicolumn{1}{l|}{}   & \multicolumn{1}{l|}{\xmark}   & \multicolumn{1}{l|}{}   & \multicolumn{1}{l|}{}  & \multicolumn{1}{l|}{\xmark}  & \multicolumn{1}{l|}{}   & \multicolumn{1}{l|}{}  & \multicolumn{1}{l|}{} & \multicolumn{1}{l|}{\xmark}  & \multicolumn{1}{l|}{} & \multicolumn{1}{l|}{} & \multicolumn{1}{l|}{} &  & \multicolumn{1}{l|}{\xmark}  & \multicolumn{1}{l|}{\xmark} & \multicolumn{1}{l|}{\xmark}  & \multicolumn{1}{l|}{}  &  \xmark \\ \hline

\multicolumn{1}{|l|}{\ref{sec:SPRINTS} SPRINTS}  & \multicolumn{1}{l|}{\xmark} & \multicolumn{1}{l|}{} & \multicolumn{1}{l|}{} & \multicolumn{1}{l|}{} &  & \multicolumn{1}{l|}{\xmark} & \multicolumn{1}{l|}{} & \multicolumn{1}{l|}{}  & \multicolumn{1}{l|}{\xmark}   & \multicolumn{1}{l|}{}  & \multicolumn{1}{l|}{} & \multicolumn{1}{l|}{} & \multicolumn{1}{l|}{} & \multicolumn{1}{l|}{}  & \multicolumn{1}{l|}{\xmark}  & \multicolumn{1}{l|}{}  & \multicolumn{1}{l|}{} &  & \multicolumn{1}{l|}{\xmark} & \multicolumn{1}{l|}{\xmark} & \multicolumn{1}{l|}{} & \multicolumn{1}{l|}{\xmark} &  \\ \hline

\multicolumn{1}{|l|}{\ref{sec:TSF} TSF}  & \multicolumn{1}{l|}{} & \multicolumn{1}{l|}{\xmark}  & \multicolumn{1}{l|}{}  & \multicolumn{1}{l|}{}    &  & \multicolumn{1}{l|}{\xmark} & \multicolumn{1}{l|}{\xmark}  & \multicolumn{1}{l|}{}  & \multicolumn{1}{l|}{}  & \multicolumn{1}{l|}{\xmark} & \multicolumn{1}{l|}{} & \multicolumn{1}{l|}{} & \multicolumn{1}{l|}{} & \multicolumn{1}{l|}{}  & \multicolumn{1}{l|}{}  & \multicolumn{1}{l|}{}  & \multicolumn{1}{l|}{}  &   & \multicolumn{1}{l|}{\xmark}  & \multicolumn{1}{l|}{} & \multicolumn{1}{l|}{}  & \multicolumn{1}{l|}{\xmark}  &   \\ \hline

\multicolumn{1}{|l|}{\ref{sec:UDM} UDM}  & \multicolumn{1}{l|}{} & \multicolumn{1}{l|}{\xmark}  & \multicolumn{1}{l|}{} & \multicolumn{1}{l|}{}  &  & \multicolumn{1}{l|}{} & \multicolumn{1}{l|}{\xmark}  & \multicolumn{1}{l|}{}   & \multicolumn{1}{l|}{}  & \multicolumn{1}{l|}{\xmark} & \multicolumn{1}{l|}{} & \multicolumn{1}{l|}{}  & \multicolumn{1}{l|}{} & \multicolumn{1}{l|}{}  & \multicolumn{1}{l|}{}  & \multicolumn{1}{l|}{}  & \multicolumn{1}{l|}{}  & \xmark  & \multicolumn{1}{l|}{\xmark} & \multicolumn{1}{l|}{} & \multicolumn{1}{l|}{}  & \multicolumn{1}{l|}{\xmark}  &   \\ \hline

\multicolumn{1}{|l|}{\ref{sec:UNSPELL} UNSPELL}  & \multicolumn{1}{l|}{\xmark} & \multicolumn{1}{l|}{} & \multicolumn{1}{l|}{} & \multicolumn{1}{l|}{} &  \xmark & \multicolumn{1}{l|}{\xmark} & \multicolumn{1}{l|}{} & \multicolumn{1}{l|}{} & \multicolumn{1}{l|}{\xmark} & \multicolumn{1}{l|}{}  & \multicolumn{1}{l|}{} & \multicolumn{1}{l|}{}  & \multicolumn{1}{l|}{} & \multicolumn{1}{l|}{} & \multicolumn{1}{l|}{\xmark}  & \multicolumn{1}{l|}{}  & \multicolumn{1}{l|}{} &  & \multicolumn{1}{l|}{\xmark}  & \multicolumn{1}{l|}{\xmark} & \multicolumn{1}{l|}{} & \multicolumn{1}{l|}{\xmark}  &  \\ \hline

\multicolumn{1}{|l|}{\ref{sec:TS-HOG-TB} TS-HOG-TB}  & \multicolumn{1}{l|}{}  & \multicolumn{1}{l|}{\xmark}  & \multicolumn{1}{l|}{} & \multicolumn{1}{l|}{\xmark}  & \xmark & \multicolumn{1}{l|}{}  & \multicolumn{1}{l|}{\xmark}  & \multicolumn{1}{l|}{}  & \multicolumn{1}{l|}{}  & \multicolumn{1}{l|}{\xmark} & \multicolumn{1}{l|}{} & \multicolumn{1}{l|}{}  & \multicolumn{1}{l|}{}  & \multicolumn{1}{l|}{}  & \multicolumn{1}{l|}{}  & \multicolumn{1}{l|}{}  & \multicolumn{1}{l|}{}  & \xmark & \multicolumn{1}{l|}{\xmark}  & \multicolumn{1}{l|}{}  & \multicolumn{1}{l|}{}  & \multicolumn{1}{l|}{\xmark}  &  \\ \hline

\multicolumn{1}{|l|}{\ref{sec:SEPNET} SEPNET}  & \multicolumn{1}{l|}{\xmark}  & \multicolumn{1}{l|}{}  & \multicolumn{1}{l|}{} & \multicolumn{1}{l|}{}  &  & \multicolumn{1}{l|}{}  & \multicolumn{1}{l|}{\xmark}  & \multicolumn{1}{l|}{}  & \multicolumn{1}{l|}{\xmark}  & \multicolumn{1}{l|}{} & \multicolumn{1}{l|}{} & \multicolumn{1}{l|}{}  & \multicolumn{1}{l|}{}  & \multicolumn{1}{l|}{}  & \multicolumn{1}{l|}{}  & \multicolumn{1}{l|}{\xmark}  & \multicolumn{1}{l|}{}  &  & \multicolumn{1}{l|}{\xmark}  & \multicolumn{1}{l|}{\xmark}  & \multicolumn{1}{l|}{}  & \multicolumn{1}{l|}{}  & \xmark \\ \hline

\multicolumn{1}{|l|}{\ref{sec:BiLSTM-SEP} BiLSTM-SEP}  & \multicolumn{1}{l|}{\xmark}  & \multicolumn{1}{l|}{}  & \multicolumn{1}{l|}{}  & \multicolumn{1}{l|}{}  &  & \multicolumn{1}{l|}{} & \multicolumn{1}{l|}{\xmark} & \multicolumn{1}{l|}{} & \multicolumn{1}{l|}{\xmark}  & \multicolumn{1}{l|}{\xmark}  & \multicolumn{1}{l|}{\xmark}  & \multicolumn{1}{l|}{} & \multicolumn{1}{l|}{\xmark} & \multicolumn{1}{l|}{} & \multicolumn{1}{l|}{} & \multicolumn{1}{l|}{\xmark}  & \multicolumn{1}{l|}{}  &   & \multicolumn{1}{l|}{\xmark} & \multicolumn{1}{l|}{\xmark} & \multicolumn{1}{l|}{\xmark} & \multicolumn{1}{l|}{} & \xmark \\ \hline

\multicolumn{1}{|l|}{\ref{sec:MEMPSEP} MEMPSEP} & \multicolumn{1}{l|}{\xmark}  & \multicolumn{1}{l|}{} & \multicolumn{1}{l|}{}  & \multicolumn{1}{l|}{}  & \multicolumn{1}{l|}{\xmark} & \multicolumn{1}{l|}{\xmark} & \multicolumn{1}{l|}{\xmark} & \multicolumn{1}{l|}{\xmark} & \multicolumn{1}{l|}{\xmark}  & \multicolumn{1}{l|}{}  & \multicolumn{1}{l|}{\xmark}  & \multicolumn{1}{l|}{\xmark} & \multicolumn{1}{l|}{} & \multicolumn{1}{l|}{\xmark}  & \multicolumn{1}{l|}{}  & \multicolumn{1}{l|}{\xmark}  & \multicolumn{1}{l|}{\xmark}  & \xmark & \multicolumn{1}{l|}{\xmark} & \multicolumn{1}{l|}{\xmark} & \multicolumn{1}{l|}{\xmark}  & \multicolumn{1}{l|}{\xmark}  &  \\ \hline

\multicolumn{1}{|l|}{\ref{sec:PSPSP} PSPSP} & \multicolumn{1}{l|}{\xmark}  & \multicolumn{1}{l|}{}  & \multicolumn{1}{l|}{} & \multicolumn{1}{l|}{} & \multicolumn{1}{l|}{} & \multicolumn{1}{l|}{} & \multicolumn{1}{l|}{\xmark}  & \multicolumn{1}{l|}{\xmark}  & \multicolumn{1}{l|}{}  & \multicolumn{1}{l|}{\xmark}  & \multicolumn{1}{l|}{\xmark}  & \multicolumn{1}{l|}{}  & \multicolumn{1}{l|}{} & \multicolumn{1}{l|}{}  & \multicolumn{1}{l|}{}  & \multicolumn{1}{l|}{}  & \multicolumn{1}{l|}{}  & \xmark & \multicolumn{1}{l|}{\xmark} & \multicolumn{1}{l|}{} & \multicolumn{1}{l|}{\xmark}  & \multicolumn{1}{l|}{}  & \xmark  \\ \hline

\multicolumn{1}{|l|}{\ref{sec:EPREM-S} EPREM-S} & \multicolumn{1}{l|}{\xmark} & \multicolumn{1}{l|}{} & \multicolumn{1}{l|}{} & \multicolumn{1}{l|}{} &  & \multicolumn{1}{l|}{}  & \multicolumn{1}{l|}{\xmark}  & \multicolumn{1}{l|}{}  & \multicolumn{1}{l|}{} & \multicolumn{1}{l|}{\xmark} & \multicolumn{1}{l|}{} & \multicolumn{1}{l|}{} & \multicolumn{1}{l|}{} & \multicolumn{1}{l|}{}  & \multicolumn{1}{l|}{}  & \multicolumn{1}{l|}{} & \multicolumn{1}{l|}{}  &   & \multicolumn{1}{l|}{}  & \multicolumn{1}{l|}{} & \multicolumn{1}{l|}{}  & \multicolumn{1}{l|}{}  & \xmark \\ \hline

\end{tabular}
\end{table}

Our review compares three key aspects of these works: the inputs, the ML models' architectures trained on those inputs, and their outputs. SEP prediction approaches employ a wide range of model architectures, often driven by available input data and the desired output. Our first goal is to map the complexity of the different ML model architectures (Architecture column of Table~\ref{tab:Qualitative_Table} and Section~\ref{sec:Model_Architectures}). Furthermore, by comparing the inputs (Input column of Figure~\ref{tab:Qualitative_Table} and Section~\ref{sec:Inputs_Comparison}), we can understand which space- or ground-based observations the community has used, which physical quantities have proven most useful, which datasets are more accessible or easier to work with, and whether there is valuable information that has been overlooked for SEP forecasting tasks. However, comparing the models' outputs (Output column of Table~\ref{tab:Qualitative_Table} and Section~\ref{sec:Outputs_and_Testing}) highlights the current status of the field: most studies explore different forecasting setups that produce different outputs. As a result, there is no standardized prediction output across models, making it difficult to compare results directly, since different studies aim to predict fundamentally different quantities. Some works predict the expected SEP onset time at Earth (e.g., SSEP, MEMPSEP), others provide all-clear predictions (e.g., BiLSTM-SEP, CANN), and several estimate the probability that a flare will produce an SEP (e.g., SHARP-SMARP, TSF, UDM, AA). Certain approaches produce entirely different types of outputs, such as inner-heliosphere particle intensity profiles (e.g., PSPSP) or surrogate physics-based simulations (e.g., EPREM-S). Additionally, some studies perform pre-eruptive forecasting while others rely on post-eruptive, triggered prediction setups. These diverse output targets make direct comparison between models inherently challenging or even impossible. Sections~\ref{sec:Model_Architectures}-\ref{sec:Paths_for_Future_Research} examine the current landscape by comparing model architectures, inputs and outputs and then outline paths towards improved comparability and operational readiness.

\subsection{Model Architectures} \label{sec:Model_Architectures}

Table~\ref{tab:Qualitative_Table} presents a summary of the qualitative responses submitted by SEP modelers with regards to the model's architecture, input and outputs. The questionnaire researchers responded to is available in Appendix~\ref{app:Questionnaire}. The values checked within Table~\ref{tab:Qualitative_Table} for each model are also those presented in Tables~\ref{tab:XGBoost}-\ref{tab:EPREM-S}. The left-most group of columns in Table~\ref{tab:Qualitative_Table} (Architecture) presents a summary of the model types the community has used. 

Many models (11/24) use some type of forest or decision tree architecture. The type ranges from a gradient-boosted decision tree architecture (i.e., XGBoost) to the Random Forest, a bagging ensemble classifier built on decision trees (AA) and to the Classification and Regression Tree (CART) model, which are simple decision trees that use splitting criteria to identify feature thresholds that best separate SEP and non-SEP events. Such methods are appealing because they perform well with relatively small datasets, are robust to noise and imbalance, and offer easier interpretability through feature importance analysis. Therefore, these remain competitive baseline approaches for SEP prediction. Similarly, NNs are also a popular type of model (11/24). A wide variety of NN architectures have been used. For example, BiLSTM-SEP uses a Bidirectional LSTM network, a type of Recurrent Neural Network (RNN), the MEMPSEP model uses Convolutional Neural Networks (CNNs), SEPNET uses Transformers and approaches such as UNSPELL and RH use ensembles of NNs. Seven of the models (7/24) use some type of ensemble architecture, a method where multiple models are combined to produce a single, usually better, prediction than any individual model could achieve on its own. Fewer models (5/24) have used lower-complexity architectures, such as regression models or SVMs. Nonetheless, they remain valuable and can act as useful baselines or exploratory tools for identifying useful information about physical parameters.

\begin{figure}[h] 
    \centering 
    \includegraphics[width=\textwidth]{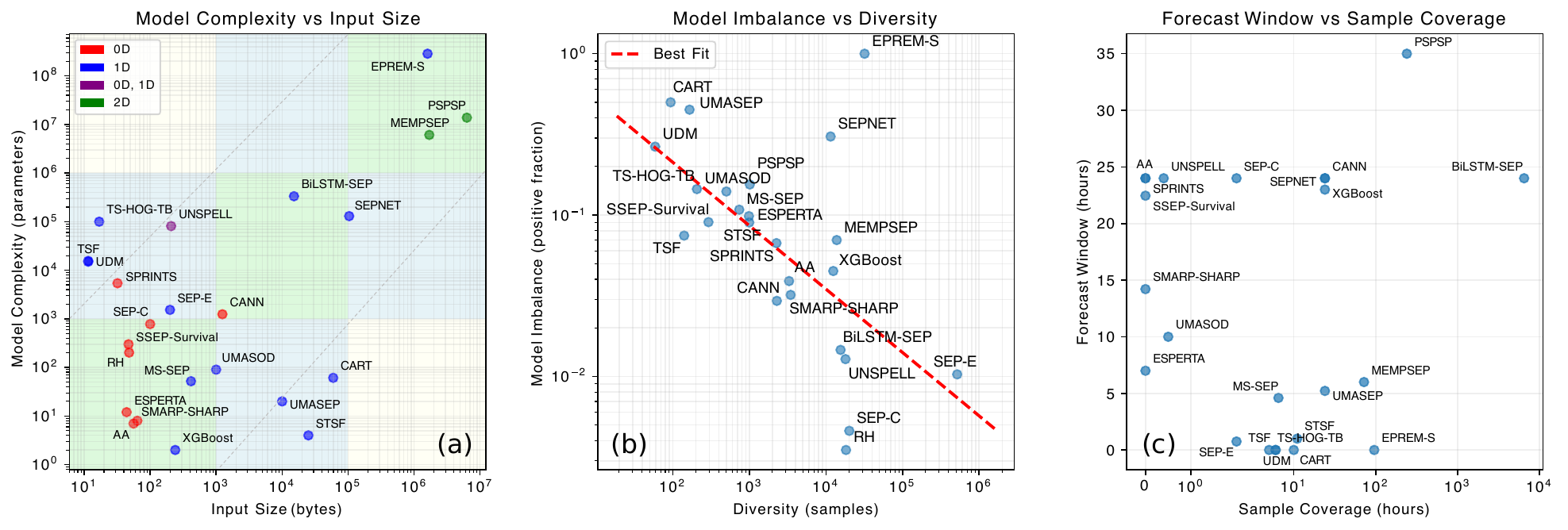} 
        \caption{Plots of the quantitative categorization submissions of the SEP modelers tabulated in Tables~\ref{tab:XGBoost}-\ref{tab:EPREM-S}. Panel (a) presents the model complexity against the input size in bytes. Panel (b) presents the input's imbalance against the input diversity. Panel (c) presents the forecast window against the sample coverage.} 
        \label{fig:Quantitative_Plot} 
\end{figure}

The models examined in this review differ not only in architecture but also in complexity. Here, \textit{Model Complexity} refers to the number of free parameters (trainable weights) contained within each model. Figure~\ref{fig:Quantitative_Plot}a illustrates this complexity as a function of \textit{Input Size}. For visualization, models are grouped into three categories: low complexity (1–1,000 parameters; bottom row), medium complexity (1,000–100,000 parameters; middle row), and high complexity ($\geq 100,000$ parameters; top row). A clear observation from this distribution is that, because ML-based SEP prediction is still a nascent field, most models (12/24) fall into the low-complexity range. Eight studies use models in the medium-complexity range (majority of them using a low input size, with the exception of CANN, BiLSTM-SEP and SEPNET), and only three employ deeper architectures with more than $10^5$ trainable parameters. For context, modern large-scale ML models such as Large Language Models (LLMs; ChatGPT, Copilot, Claude, Gemini, etc.) contain billions of trainable parameters ($\geq 10^9$), while the heliophysics foundation model SuryaFM \citep{roy2025surya} has $3.7 \times 10^8$ parameters ---comparable to the EPREM-S surrogate model. This contrast highlights a clear opportunity for future research: the development and exploration of more complex, higher-capacity ML models.

Based on this complexity categorization, Figure~\ref{fig:Quantitative_Plot}a is divided into nine regions. The green regions highlight groups of models whose complexity and input size align ---low complexity with small input size, medium complexity with moderate input size, and high complexity with large input size. Input size reflects the amount of information available to each model during training. Most studies (14/24) fall within these green regions (and 18/24 are within the gray dotted trend lines), supporting the intuitive assumption that larger input sizes often require deeper model architectures. One might also expect that deeper architectures would yield better performance. However, among the high‑complexity models, only MEMPSEP provides predictions of geoeffective SEP events, as EPREM-S and PSPSP do not (their uniqueness is further discussed in Sections~\ref{sec:PSPSP} and \ref{sec:EPREM-S}), therefore this assumption cannot be tested yet. The overwhelming majority (22/24) of models presented here predict geoeffective events, with the exception of EPREM-S and PSPSP which either do not provide a prediction (EPREM-S) or predict particles within the inner heliosphere and not at Earth (PSPSP).

In the medium‑complexity and medium‑input‑size category, only two models appear: BiLSTM-SEP and CANN. When comparing these groups to the lower‑complexity, lower‑input‑size models, the results in Table~\ref{tab:results} (which will be discussed in further detail later in the text) do not show improved performance. In fact, BiLSTM-SEP, SEPNET and MEMPSEP report lower True Skill Score (TSS) and Heidke Skill Score (HSS) scores than their simpler counterparts. Overall, no clear trend emerges linking model complexity to predictive performance. This is further illustrated in Table~\ref{tab:results}, where models are ordered by complexity, yet no increasing performance trend is observed across any of the five evaluation metrics. The white regions in the top-left and bottom-right of Figure~\ref{fig:Quantitative_Plot}a represent areas where one would not expect models to appear (and indeed none do): small input sizes do not require models with a large number of trainable parameters, and conversely, low‑complexity models are generally unsuitable for processing large, information‑rich datasets such as 2D inputs.

Although there is no clear evidence that increasing model complexity alone leads to improved predictive performance, this should not discourage the community from pursuing more complex architectures. As discussed in detail in Section~\ref{sec:Outputs_and_Testing}, comparing model results at this early stage of ML-based SEP prediction is not reliable, since most reported scores correspond to different forecast windows, different prediction targets, or entirely different validation setups. What can be confirmed from our analysis is that larger input sizes (i.e., more relevant information) tend to correspond to, and often require, higher model complexity. Therefore, this study cannot conclusively determine whether more complex architectures yield better predictive performance. Nonetheless, higher-complexity models remain underexplored as seen in Table~\ref{tab:Qualitative_Table} and since performance improvement comes from richer input data, future progress will likely depend on combining more informative datasets with higher complexity models capable of levereging them effectively. At the current stage of the field, modelers should recognize the trade-off between training time and development difficulty on one hand, and the potential benefits of incorporating richer information on the other. As with the other qualitative characteristics in Table~\ref{tab:Qualitative_Table}, the distribution of model architectures used across the literature is summarized in Figure~\ref{fig:Qualitative_Plot}b.

In summary, classical ML methods remain competitive for SEP prediction when data volumes are limited. Their strengths lie in robustness and interpretability, though performance improvements often require richer physical inputs rather than algorithmic complexity. Ensemble and tree-based methods provide stable performance under class imbalance and remain attractive for operational deployment due to their robustness and moderate computational requirements. Time-series architectures show promise for improving onset-time prediction by exploiting temporal evolution of solar and heliospheric measurements, though they remain constrained by limited event statistics. Deep neural architectures enable integration of complex data sources such as imagery and multivariate time-series, but their potential is currently limited by data scarcity rather than model capability. Overall, differences in predictive performance across models appear to depend more strongly on input richness and data quality than on architectural sophistication alone. 

\subsection{Inputs Comparison} \label{sec:Inputs_Comparison}

\begin{figure}[h] 
    \centering 
    \includegraphics[width=\textwidth]{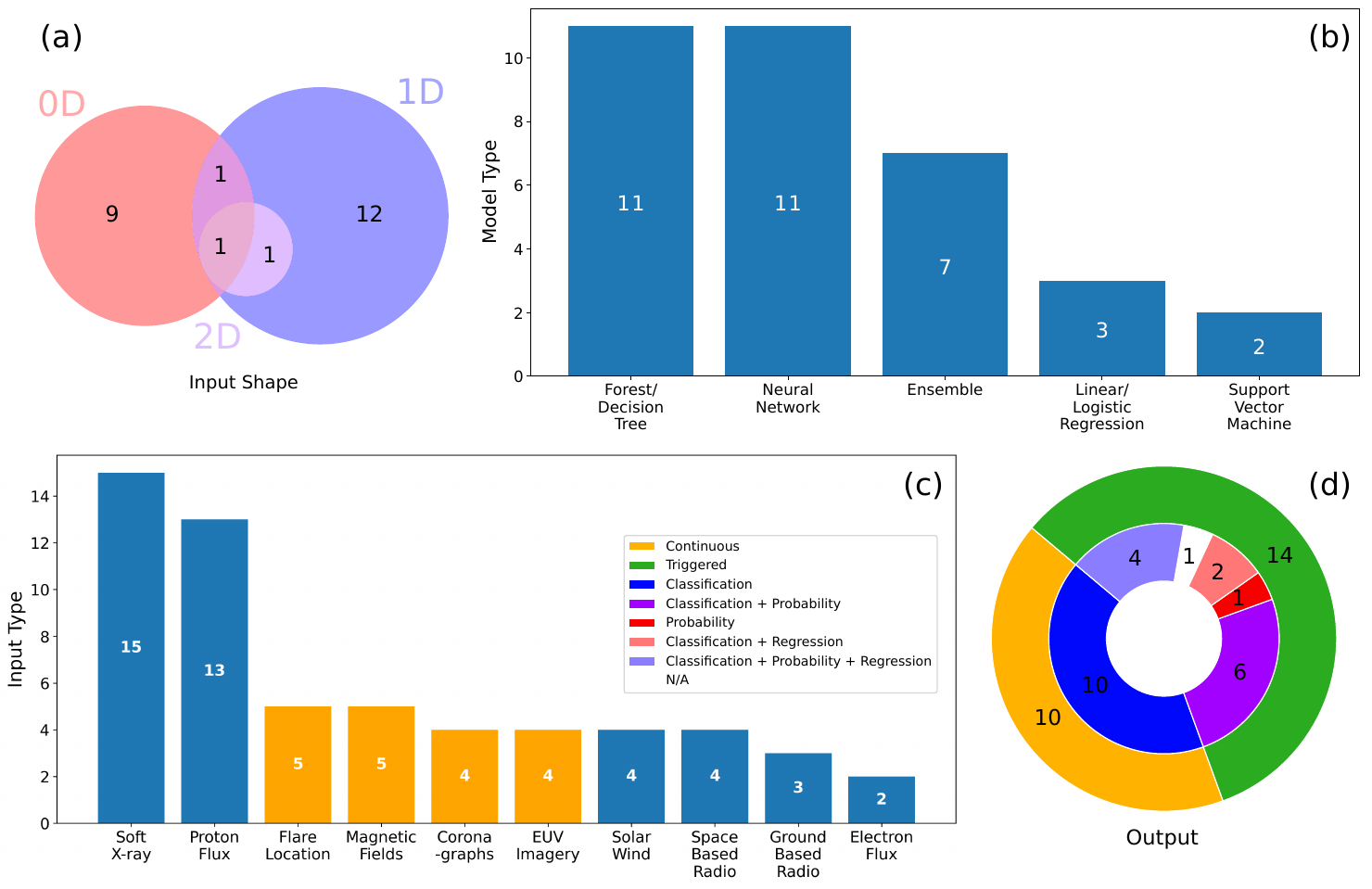} 
    \caption{Summary plots of the quantitative classifications presented in Table~\ref{tab:Qualitative_Table}. The Venn diagram at (a) and histogram at (c) summarize the \textit{Input} columns (Shape and Type), the histogram at (b) the \textit{Architecture} columns and the pie chart at (d) the \textit{Output} columns (Prediction for the inner circle and Type for the outer circle). In the Input Shape and Output plots, each model is shown separately, whereas in the Input Type and Model Type histograms, methods that employ multiple types contribute to all relevant histogram bars.} 
    \label{fig:Qualitative_Plot}
\end{figure}

The different model architectures above have been trained on inputs that vary in type, shape, size, and coverage, obtained primarily from the space missions illustrated in Figure~\ref{fig:Satellites}. The middle columns of Table~\ref{tab:Qualitative_Table} (Input columns) prescribe the shape and type categories to which the inputs of each model belong. As seen in the Venn diagram of Figure~\ref{fig:Qualitative_Plot}a, most models (12/24) use some type of time series input, followed by single-point inputs (9/24) often related to a progenitor event, such as flares and CMEs (triggered prediction). Only two models, MEMPSEP and PSPSP use 2D data, such as sequences of full‐disc line‐of‐sight magnetograms from the Michelson Doppler Imager \citep[MDI/SOHO;][]{scherrer1995solar} and the Helioseismic and Magnetic Imager \citep[HMI/SDO;][]{scherrer2012helioseismic} or full-disk EUV imagery from the Atmospheric Imaging Assembly \citep[AIA/SDO;][]{lemen2012atmospheric}, in addition to 1D quantities. Note that there are a number of studies that use more than one input, and these inputs in some cases are of different shape (TSF, MEMPSEP and PSPSP).

In terms of the type of observations used as input, ten different physical quantities have been used to train ML models. These include solar EUV imagery or magnetograms and their derivatives such as the Space-Weather HMI Active Region Patches (SHARP) and the Space-Weather MDI Active Region Patches (SMARP) data \citep{bobra2014helioseismic,bobra2021smarps}. Derivatives of observations captured by the Large Angle and Spectrometric Coronagraph \citep[LASCO;][]{brueckner1995large}
instrument, such as CME lists, have also been used to train the AA, SEP-C, MEMPSEP, ESPERTA and MS-SEP models. In terms of timeline or point data, studies have used soft X-ray, proton and electron flux, ground and space-based radio data and solar wind parameters or flare location information. The histogram of Figure~\ref{fig:Qualitative_Plot}c shows the number of models using the various input types. The overwhelming majority uses soft X-ray (15/24) and Proton Flux (13/24) data from the GOES satellites (Figure~\ref{fig:Satellites}) because they are available in near real-time (1 minute cadence\footnote{\url{https://www.swpc.noaa.gov/products/goes-x-ray-flux}}), extend over a long time period (since 1984) spanning multiple Solar Cycles (SCs), are obtained from well calibrated instruments ---such as the X-Ray Sensor \citep[XRS;][]{chamberlin2009next, woods2024goes}--- and are used in operational decision-making by federal agencies and the commercial sector. More specifically, GOES data are used for making operational decisions at NASA and by other operational communities since they provide operationally-supported measurements which include a 24/7 support, a primary and secondary backup in case of failure, and the detectors do not saturate during high intensity periods when operations are most likely to be affected. The blue histogram bars in Figure~\ref{fig:Qualitative_Plot}c indicate in-situ measurements, while the orange bars indicate remote-sensing observations. Note that in Table~\ref{tab:Qualitative_Table} and Figure~\ref{fig:Qualitative_Plot}c the Magnetic Fields \textit{Inputs} Type refers to both in-situ and remote observations although only Bi-LSTM and MEMPSEP use in-situ magnetic field observations (MEMPSEP uses both). As expected due to the nature of the SEP prediction problem, most studies use some type of in-situ measurement while a smaller number of studies complement their analysis with data captured from remote observations.

A common problem in SEP prediction, especially when using data-hungry ML models, is the absence of positive events (SEP occurrences)-- what is commonly known as data imbalance. More specifically, flare-based studies encounter the flare imbalance problem where the overwhelming majority of flares do not produce SEPs (negative), while for models that use time-series the majority of their data is labeled as ``quiet times" due to the rarity of significant space weather events. Here, we express the imbalance in each model as a qualitative parameter in Tables~\ref{tab:XGBoost}-\ref{tab:EPREM-S}. The imbalance parameter is defined as the ratio of positive events over the total number of events, therefore, it is always an integer between 0 and 1. Figure~\ref{fig:Quantitative_Plot}b shows the diversity of each model (number of total samples used; the number of SEP and non-SEP occurrences together) against the imbalance parameter. As expected, models trained on larger number of samples (more diverse) are highly imbalanced (their imbalance ratio is lower), while models trained on a smaller number of samples are less imbalanced (imbalance ratios closer to 1). EPREM-S is again a special case where the model sees only SEP occurrences.

Interestingly, two models that deviate strongly from the trend in Figure~\ref{fig:Quantitative_Plot}b (the red dotted best‑fit line) ---SEP‑C and RH--- are also ones that report the highest Probability of Detection (POD $\geq0.9$) and TSS ($\geq0.9$), but at the same time the lowest HSS ($\leq0.25$), as shown in Table~\ref{tab:results} and will be further discussed in the text that follows. Both studies train their models on a very small number of positive events compared to other works with a similar total number of samples (e.g., BiLSTM-SEP and UNSPELL). This type of performance is common in highly imbalanced scenarios in which many correct negatives result in high POD but do not necessarily represent skill in predicting rare events \citep{pierce1884,whitman2026}. In rare‑event settings, TSS can remain high when most positive events are captured (high POD), because the False Alarm Rate (FAR) is also very high (for SEP-C, a FAR of 0.88 is reported), making the models non-useful for operational predictions, where lots of false alarms become an inconvenience. This might also occur because correct rejections dominate the confusion matrix, causing TSS to track POD closely. Therefore, having the highest TSS and POD does not imply that SEP‑C and RH are the best-performing models, nor does it suggest that studies should avoid training on larger sets of positive events. Rather, it demonstrates that high scores can be misleading when viewed in isolation ---especially in rare‑event regimes--- and must be interpreted alongside other metrics such as HSS, which in this case clearly indicates limited overall skill. Recall that multiplying the model imbalance (Figure~\ref{fig:Quantitative_Plot}b; y‑axis) with the diversity (x‑axis) yields the number of positive events available for training each model.

In future studies, data augmentation and models that can produce realistic SEP distributions should be explored as a remedy to imbalance. Imbalance mitigation through data augmentation \citep{bahri2023shapelet}, has been applied to SEP prediction only in the study by \citet{hosseinzadeh2024improving} (see Table~\ref{tab:sep_datasets}). The results of the relevant model (TSF) appear promising: the model reports high TSS and HSS values of 0.80 and 0.90, respectively. However, this is only a single data point, so conclusions should be made with caution until more studies reproduce similar results. It should also be noted that even with augmentation, TSF still exhibits relatively low imbalance compared to other studies with a similar number of samples. Nevertheless, the community should explore this direction further, as data augmentation has already shown performance improvements in flare prediction tasks \citep{li2025evaluating, wen2024class, grim2024solar}. Beyond this, no major trends or correlations between dataset diversity, imbalance, and the reported model performance can be identified from the existing literature, mainly due to the difficulty in comparing fairly the models' results. 

\begin{figure}[h]  
    \centering 
    \includegraphics[width=\textwidth]{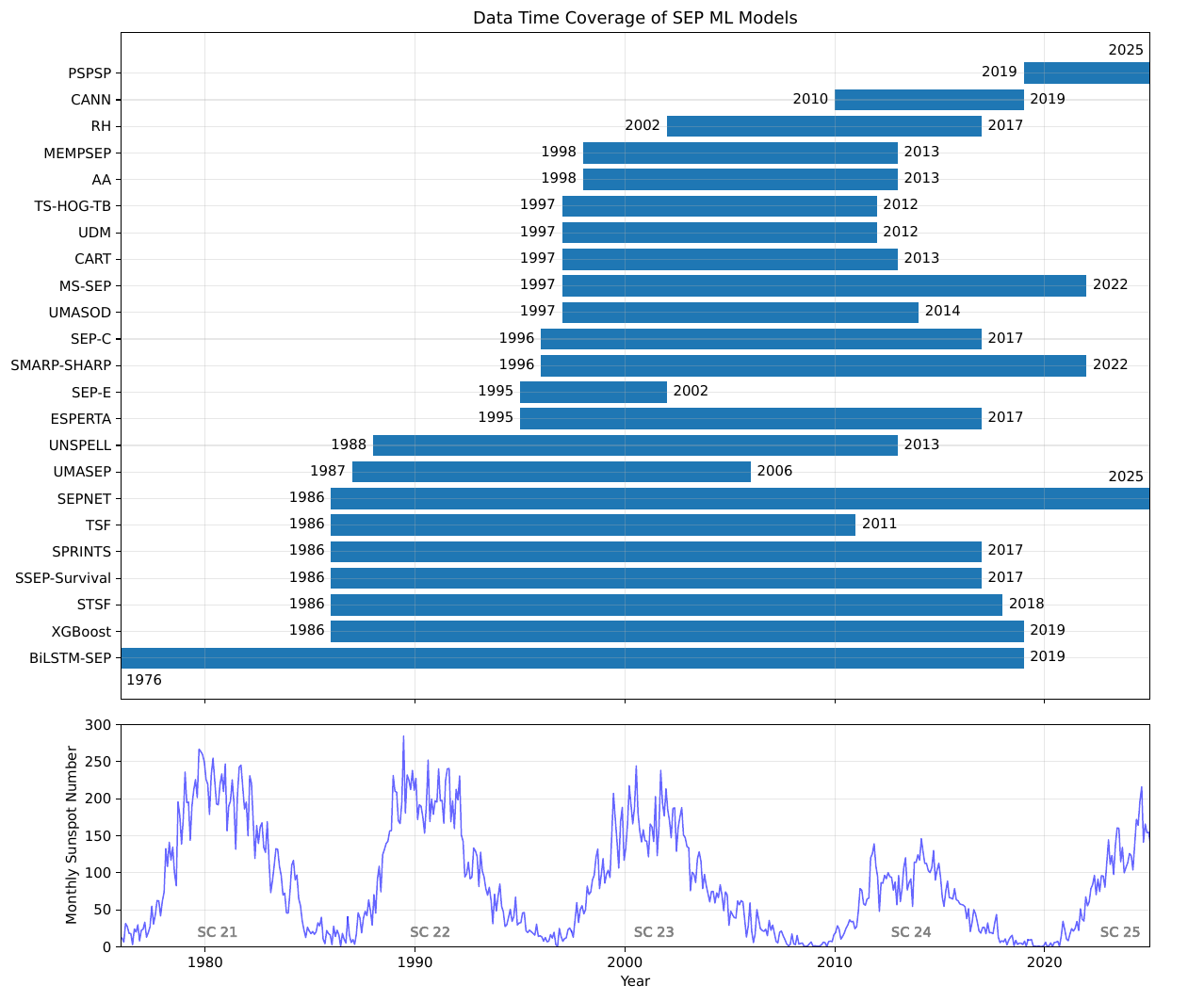} 
    \caption{Time coverage of the reviewed SEP prediction models (start and end years, per study) shown alongside the monthly sunspot number, illustrating how each model's training and evaluation data align with the SC variability.} 
    \label{fig:Time_Coverage}
\end{figure}

Lastly, an important input parameter is the input size, which in this discussion is particularly relevant because larger inputs generally require more complex models to process them effectively. For this reason, input size is plotted against model complexity in Figure~\ref{fig:Quantitative_Plot}a. The colors of the points indicate the data dimensionality used by each model (0D, 1D, 0D+1D, or 2D), following the classification in Table~\ref{tab:Qualitative_Table}. As expected, models using 2D data exhibit large input sizes and correspondingly rely on more complex architectures. Since input size reflects the amount of information processed per sample, it is reasonable to hypothesize that models with larger input sizes should achieve better performance. However, the current results presented in Table~\ref{tab:results} do not support this expectation. Models in the mid‑range of input sizes (BiLSTM-SEP, CANN, CART, and STSF) report some of the lowest POD values (0.73 for CART and 0.61 for BiLSTM-SEP, where available), yet at the same time achieve high F1 scores (e.g., 0.82 for CART) and notably low FAR (0.09 for BiLSTM-SEP). Attention should also be given to models with small input sizes but medium complexity—such as TS-HOG-TB, TSF, UDM, SPRINTS, and UNSPELL. Despite using relatively limited inputs, these models train more parameters and achieve performance comparable to models with substantially larger input sizes. These architectures (particularly NNs and ensemble forests) appear capable of extracting meaningful information even from smaller datasets and warrant further exploration. Nevertheless, due to the inconsistent use of validation metrics across studies, it remains difficult to draw strong conclusions about the relationship between input size, complexity, and performance at this stage. This difficulty is discussed extensively in Section~\ref{sec:Outputs_and_Testing}.

This review covers studies that were published beginning in 2011 \citep{nunez2011predicting}, during the ascending phase of SC 24. Since then, twenty three more models have been published using the more recent data, up to the maximum of SC 25. Therefore, different models have used data from different periods of solar activity (Figure~\ref{fig:Time_Coverage}). Most of the models discussed here use data from SCs 23 and 24, while nine studies have used historical data from SC 22. Only the BiLSTM-SEP model includes data from SC 21. Note that although most publications cited in this work describe models that appear static in time, there are models such as UMASEP and SEPNET that are being continuously updated and their data now spans up to the most recent data of SC 25, similar to the PSPSP model. Since most observatories (Figure~\ref{fig:Satellites}) providing data for these models remain operational ---although some, such as ACE, now produce data of questionable quality \citep{regnault202020}--- the majority of models could in principle be updated with the most recent observations, but doing so remains at the discretion of the developers.

\subsection{Outputs and Testing} \label{sec:Outputs_and_Testing}

Although all models described in this manuscript have been developed for achieving the same goal ---predicting SEP events, their predictions are different, given that they have different inputs and are developed to forecast different output quantities. The differences between the outputs are captured on the rightmost columns of Table~\ref{tab:Qualitative_Table} (Output columns) where the predictions are classified in two groups: the \textit{Prediction} group (Classification, Probability, and Regression) and the \textit{Type} group (Triggered or Continuous). More information on the definitions of these categorizations is included in Section~\ref{sec:Overview_and_Categorization} and the questionnaire in Appendix~\ref{app:Questionnaire}. The prediction \textit{Types} are mutually exclusive: a triggered model (in which a prediction is issued based on an event such as a flare or CME; post-eruptive) cannot be continuous (in which predictions are issued continuously without dependence on an event; pre-eruptive), and vice versa. On the other hand, the \textit{Prediction} categorization can be often mutually inclusive, meaning that a model that outputs probabilities can also provide a classification given a defined threshold. 

The nested pie chart of Figure~\ref{fig:Qualitative_Plot}d summarizes the nature of the predictions provided by the models in this review. There is an almost even split between continuous (10/24) and triggered (14/24) predictions (outer pie chart). The inner pie chart shows that the overwhelming majority (22/24) of models output an event classification, with ten of them also providing a probability. All the regression models use a threshold for an SEP event in order to provide a classification. Only a minority of models do not provide an event classification (SSEP and EPREM-S). For instance, EPREM-S is a NN trained on synthetic data, not built to predict the occurrence of SEP events (though it has the capability), the SSEP model outputs a function that indicates the probability that an SEP event has not occurred at a given, post-flare, time. Another difference among the model outputs is the forecast window, which is plotted against the sample coverage Figure~\ref{fig:Quantitative_Plot}c. Although we expect that smaller forecast windows should produce better results (Table~\ref{tab:results}), even this expectation is not obvious here, underlying once again the difficulty of comparing between the different model results in this stage of the community's efforts. 

Since most models provide some type of binary classification, whether this is an ``All Clear" or a binary yes/no for SEP events ($>10$ MeV), the community has used a wide variety of evaluation metrics that capture different aspects of the predictive capabilities of each model. Table~\ref{tab:Metrics} presents a summary of all the metrics that have been used by the community to validate their ML models. More specifically, the different columns represent the eight most used metrics: the True Skill Score \citep[TSS;][]{doswell1990summary}, the Heidke Skill Score (HSS), the Accuracy (ACC), the Probability of Detection \citep[POD or Recall;][]{wehling2011probability}, the False Alarm Rate \citep[FAR;][]{macmillan1985detection}, the F1 score \citep[F1;][]{lipton2014thresholding}, the Precision (Prec) and the Area Under the Curve \citep[AUC;][]{muschelli2020roc}. Here we need to note that although most of the metrics are derivatives of a contingency table, the AUC is different in that it captures the best performing threshold introduced in order to turn an output probability into a binary classification. The last column of Table~\ref{tab:Metrics} presents other, less popular metrics used for model evaluation. The FAR measures the fraction of all non-events that were incorrectly predicted as events (\(\text{False\ Alarms}/\text{Total\ Non-Events}\)). On the other hand, the False Alarm Ratio (FAR\textsuperscript{*}) measures the fraction of all "yes" forecasts that were wrong (\(\text{False\ Alarms}/\text{Total\ Forecasted\ Events}\)). Note that herein, FAR indicates the False Alarm Rate whereas the FAR\textsuperscript{*} is the False Alarm Ratio. In practice, the FAR\textsuperscript{*} is often the more challenging metric to optimize, as it penalizes every incorrect ‘yes’ forecast directly and therefore reflects the forecaster’s precision more strictly than the FAR. The last column of Table~\ref{tab:Metrics} presents a list of other, less utilized by the community metrics, such as the Pearson Correlation Coefficient \citep[PCC;][]{benesty2009pearson, kasapis2023turning}, the Root Mean Squared Error \citep[RMSE;][]{hodson2022root}, the $R^2$ Score \citep{ash1999r2}, the Kaplan-Meier estimate \citep[KM;][]{goel2010understanding}, Balanced Accuracy (BA) and others. A list of acronyms, which includes the aforementioned metrics, is available in Appendix~\ref{app:Acronyms}.

\begin{table}[h] 
\centering
\caption{Summary of metrics used to validate the SEP ML models' outputs.}
\label{tab:Metrics}
\begin{tabular}{|l|c|c|c|c|c|c|c|c|c|}
\hline
Model & TSS & HSS & ACC & POD & FAR & F1 & Prec & AUC & Other Metrics \\ \hline
\ref{sec:XGBoost} XGBoost     &  \xmark   &  \xmark   &     &  \xmark   &     &    &     &     &       \\ \hline
\ref{sec:STSF} STSF     &  \xmark   & \xmark    &     &     &     & \xmark   &     &  \xmark   &  MCC, GSS     \\ \hline
\ref{sec:SHARP-SMARP} SMARP-SHARP     &  \xmark   &  \xmark   & \xmark    &  \xmark   &  \xmark   &  \xmark  &    &    &     \\ \hline
\ref{sec:AA} AA      &  \xmark   & \xmark    &     &  \xmark   &  \xmark   & \xmark   &     &     &       \\ \hline
\ref{sec:ESPERTA} ESPERTA    &     &     &     &   \xmark  &  \xmark   & \xmark   &     &     &    CSI   \\ \hline
\ref{sec:UMASEP} UMASEP    &     &     &     &  \xmark   &  \xmark   &    &     &     &       \\ \hline
\ref{sec:UMASOD} UMASOD    &     &     &     &  \xmark   &  \xmark   &    &     &  \xmark   &       \\ \hline
\ref{sec:MS_SEP} MS-SEP     & \xmark    &  \xmark   &     & \xmark    & \xmark    &  \xmark  &     &     &       \\ \hline
\ref{sec:CART} CART     &     &     &  \xmark   &  \xmark   &     & \xmark   &  \xmark   &  \xmark   &       \\ \hline
\ref{sec:RH} RH    &  \xmark   &  \xmark   &  \xmark   &  \xmark   &   &   &  \xmark   &  \xmark   & BA \\ \hline
\ref{sec:SSEP-Survival} SSEP      &     &     &     &     &     &    &     &     &  KM Est, Cox PH \& Param. Models  \\ \hline
\ref{sec:SEP-C} SEP-C    &  \xmark   &  \xmark   &     &  \xmark   & \xmark    &  \xmark  &  \xmark   &  &    \\ \hline
\ref{sec:CANN} CANN     &  \xmark   &  \xmark   &     &     &     &    &     &  \xmark   &       \\ \hline
\ref{sec:SEP-E} SEP-E    &      &     &     &  \xmark   &     &  \xmark  &  \xmark   &  &  MAE  \\ \hline
\ref{sec:SPRINTS} SPRINTS   &     &  \xmark   &     &  \xmark   &  \xmark   &    &     &     &       \\ \hline
\ref{sec:TSF} TSF    &     &     &  \xmark   &     &     &  \xmark  &     &     &       \\ \hline
\ref{sec:UDM}  UDM   &  \xmark   &  \xmark   &  \xmark   &  \xmark   &     &  \xmark  &  \xmark   &    &    \\ \hline
\ref{sec:UNSPELL} UNSPELL    &  \xmark   &     &     &   \xmark  &   \xmark  &    &     &  \xmark   &       \\ \hline
\ref{sec:TS-HOG-TB} TS-HOG-TB    &  \xmark   &     &     &  \xmark   &     &  \xmark  &   \xmark  &     &       \\ \hline
\ref{sec:SEPNET} SEPNET    &  \xmark    &   \xmark  &  \xmark  &  \xmark &  \xmark &    \xmark &    &  \xmark   &       \\ \hline
\ref{sec:BiLSTM-SEP} BiLSTM-SEP    &  \xmark   &  \xmark   &  \xmark   &  \xmark   & \xmark    &    &  \xmark   &     &  CSI     \\ \hline
\ref{sec:MEMPSEP} MEMPSEP    &  \xmark   &  \xmark   &  \xmark   & \xmark  &   &   &   &  \xmark  &  $R^2$, RMSE, PCC, BS, ECE, FPR  \\ \hline
\ref{sec:MEMPSEP} PSPSP    &  \xmark   &  \xmark   &  \xmark   & \xmark  &  \xmark &  \xmark & \xmark  &    &    \\ \hline
\ref{sec:EPREM-S} EPREM-S    &     &     &     &     &     &    &     &     &  Deep Ensemble (see Appendix~\ref{sec:EPREM-S})    \\ \hline
Total    &  15   &  14   &  9   &  19   &  12   &  13  &  8   &   8  &      \\ \hline
\end{tabular}
\end{table}

Five of the aforementioned metrics ---namely the POD, TSS, HSS, F1 and FAR--- have been used by the majority of the studies. To facilitate comparison across models, Table~\ref{tab:results} summarizes the values of these metrics as reported in the respective publications. However, several difficulties arise when attempting to compare results between studies. Most works generally follow the NOAA SEP event definitions\footnote{\url{https://www.swpc.noaa.gov/products/goes-proton-flux}}, in which proton event alerts are issued for several thresholds and two particle-energy levels, with the $\geq10$ MeV channel aligned with the NOAA Solar Radiation Storm S-scale thresholds ---10, 100, 1000, 10000, and 100000~pfu (S1-S5 thresholds)--- and the $\geq100$ MeV channel being based on a single threshold of 1 pfu. Nevertheless, the specific thresholds adopted by each study vary. For example, models such as CART, TSF, UDM, and TS-HOG-TB use a $\geq 100$ MeV threshold, whereas most studies classify events using the $\geq 10$ MeV criterion. Additionally, some studies do not employ the NOAA definition at all. UNSPELL, for instance, uses the European Space Agency (ESA) Solar Energetic Particle Environment Modeling (SEPEM) reference event list\footnote{\url{http://sepem.eu/help/event_ref.html}}, a standardized catalog of solar proton events (SPEs). Other approaches diverge for methodological reasons: EPREM-S does not perform event prediction, and PSPSP defines a mission-specific threshold tailored to Parker Solar Probe measurements, since it does not aim to predict geoeffective events. When the SEP event definition differs between studies, and when training and testing conditions are not identical, a fair comparison of the reported results becomes impossible.

As is evident in Table~\ref{tab:results}, it is difficult to compare even studies that adopt a common SEP event definition because they often do not evaluate their performance using the same set of metrics. Although having multiple common metrics is important and necessary, as comparison based on a single metric is biased, it is often insufficient for a fair comparison of model predictions. For instance, although the MS-SEP, SMARP, and AA models share three metrics in common (TSS, HSS, and F1), they employ different forecast windows of 5, 14, and 24~hours, respectively. Figure~\ref{fig:Quantitative_Plot} also shows that both the forecast windows and the corresponding input‑sample coverage vary widely across studies, ranging from just a few hours to as long as four days for models such as SPRINTS. A substantial number of studies do not report a forecast window at all, either because they do not issue a prediction in the sense of an advance alert, or because they do not provide a geoeffective event prediction (as is the case for EPREM-S or SSEP). There are also differences in the physical targets and prediction objectives of the models. Forecast horizons and predicted quantities vary considerably. For example, some models attempt to determine whether a space weather event (flare, CME, etc.) will produce an SEP (models marked as Triggered in Table~\ref{tab:Qualitative_Table}), whereas others provide all-clear predictions (CANN) or estimate particle fluxes like BiLSTM-SEP (which may then be reduced to a binary outcome). Regardless of these methodological differences, most studies ultimately reduce their output to a binary yes/no prediction and report a set of common metrics, which are summarized in Table~\ref{tab:results}.

\begin{table}[h]
\centering
\large
\caption{Summary of the validation results for the 24 SEP ML models, as obtained from the relevant publications linked in Table~\ref{tab:all_models}. The FAR results with asterisk\textsuperscript{*} indicate the False Alarm Ratio, not Rate. As mentioned in the text, direct comparison between models, based on the validation measures listed, should be done with caution as the models have been applied with different training and testing setups. EPREM-S adopted a deep learning approach for determining uncertainties of the EPREM-S outputs rather than applying traditional skill scores (MSE of 0.07 was obtained). SSEP did not use for validation any of the five listed metrics, but evaluated performance using the Kaplan–Meier (KM) Estimate and the Cox Proportional Hazards (Cox PH). For studies that have quantified their results' uncertainties, metrics are shown as the mean $\pm$ one standard deviation across k cross‑validation folds.}
\label{tab:results}
\begin{tabular}{|l|c|c|c|c|c|c|c|}
\hline
Model                                 & POD           &   TSS         & HSS           & F1            & FAR        & SEPVAL  \\ \hline
\ref{sec:XGBoost} XGBoost             & $0.81$        & $0.72\pm0.02$ & $0.42$        &               &            &        \\ \hline
\ref{sec:STSF} STSF                   &               & $0.85$        & $0.88$        &               &            &        \\ \hline
\ref{sec:SHARP-SMARP} SMARP-SHARP     &               & $0.39\pm0.19$ & $0.37\pm0.19$ & $0.67\pm0.11$ & $0.30\pm0.14$\textsuperscript{*} &        \\ \hline
\ref{sec:AA} AA                       & $0.76\pm0.06$ & $0.75\pm0.05$ & $0.69\pm0.04$ & $0.70\pm0.04$ & $0.34\pm0.10$        & Yes       \\ \hline
\ref{sec:ESPERTA} ESPERTA             & $0.88$        &               &               & $0.77$        & $0.32$    &        \\ \hline
\ref{sec:UMASEP} UMASEP               & $0.81$        &               &               &               & $0.34$    & Yes       \\ \hline
\ref{sec:UMASOD} UMASOD               & $0.85$        &               &               &               & $0.85$\textsuperscript{*}            &        \\ \hline
\ref{sec:MS_SEP} MS-SEP               & $0.85\pm0.08$ & $0.78\pm0.07$ & $0.71\pm0.03$ & $0.75\pm0.03$ & $0.31\pm0.08$           &        \\ \hline
\ref{sec:CART} CART                   & $0.73$        &               &               & $0.82$        &            &        \\ \hline
\ref{sec:RH} RH                       & $0.96\pm0.00$  & $0.94\pm0.01$ & $0.17\pm0.01$ &               &            &        \\ \hline
\ref{sec:SSEP-Survival} SSEP          &               &               &               &               &            &        \\ \hline
\ref{sec:SEP-C} SEP-C                 & $0.92$        & $0.91\pm0.04$ & $0.25\pm0.06$ & $0.25\pm0.06$ & $0.88$\textsuperscript{*}           &        \\ \hline
\ref{sec:CANN} CANN                   &               & $0.82\pm0.01$ & $0.38\pm0.03$ &               &            &        \\ \hline
\ref{sec:SEP-E} SEP-E                 & $0.70$         &               &               & $0.76$ &            &        \\ \hline
\ref{sec:SPRINTS} SPRINTS             & $0.86$        &               &               &               & $0.37$     & Yes        \\ \hline
\ref{sec:TSF} TSF                     &               & $0.80$        & $0.90$        &               &            &        \\ \hline
\ref{sec:UDM}  UDM                    &               &               &               & $0.79$        &            &        \\ \hline
\ref{sec:UNSPELL} UNSPELL             & $0.86$        & $0.78$        &               &               & $0.08$     & Yes       \\ \hline
\ref{sec:TS-HOG-TB} TS-HOG-TB         & $0.81$        &               &               & $0.80$        &            &        \\ \hline
\ref{sec:SEPNET} SEPNET               & $0.64$        & $0.43$        & $0.42$        & $0.71$        & $0.23$\textsuperscript{*}     &  Yes      \\ \hline
\ref{sec:BiLSTM-SEP} BiLSTM-SEP       & $0.62$        & $0.53$        & $0.73$        &               & $0.09$     &        \\ \hline
\ref{sec:MEMPSEP} MEMPSEP             & $0.83$        & $0.63$        & $0.60$        &               &         & Yes       \\ \hline
\ref{sec:MEMPSEP} PSPSP               & $0.70\pm0.17$ & $0.43\pm0.14$ & $0.31\pm0.12$ & $0.46\pm0.10$ & $0.64\pm0.10$ &        \\ \hline
\ref{sec:EPREM-S} EPREM-S & \multicolumn{6}{|c|}{Uncertainty estimation as explained in the caption and Appendix~\ref{sec:EPREM-S}.} \\ \hline
\end{tabular}
\end{table}

Despite these inconsistencies, a general comparison across studies is still possible using the metrics in Table~\ref{tab:results}. For example, the SMARP–SHARP model shows relatively low performance (TSS: $0.39 \pm 0.19$, HSS: $0.37 \pm 0.19$). This suggests that relying on a low‑complexity model (SVM) together with only point‑data (0D), such as features derived from magnetogram data, likely provides insufficient information for reliable SEP prediction. Although its TSS and HSS values are modest, the model achieves a reasonably low FAR\textsuperscript{*} ($0.30\pm0.14$), indicating that while it may struggle to correctly predict positive events, it does not tend to over-predict them ---making it a conservative model. Nevertheless, such low-complexity models (e.g., XGBoost, STSF, AA, etc.) remain valuable: their interpretability helps identify which parameters in datasets are informative for the SEP prediction task. The AA model, which is similar to the SHARP–SMARP in terms of overall methodological simplicity but employs soft X-ray measurements and coronagraph data, performs substantially better (TSS: $0.75 \pm 0.05$, HSS: $0.69 \pm 0.04$). This improvement can be attributed to the richer physical information content available in these additional data sources. At the other end of the spectrum, MEMPSEP uses the most diverse physical dataset (Magnetic Fields, soft X-ray, Electron Flux, EUV Imagery, Coronagraphs, Space-Based Radio and Solar Wind) and employs a high-complexity ML model (6,092,617 trainable parameters); however, despite its relatively short forecast window (6 hours), it performs well in POD ($0.83$) but not as well in TSS or HSS ($0.63$ and $0.60$). RH and SEP-C have unusually high POD values ($0.96\pm0.0$ and $0.92$). However, this is paired with very high FAR\textsuperscript{*} values ($0.882$; RH does not explicitly report FAR, but its low HSS suggests a similarly high FAR\textsuperscript{*}). In contrast, the BiLSTM-SEP model achieves a lower POD ($0.62$), but also a very low FAR ($0.09$), a combination widely regarded as a strong indicator of a reliable SEP prediction model. Ideally, POD should be high while FAR remains low, but in practice these metrics compete, and modelers navigate the trade‑off between them.

Despite limitations noted above, certain patterns are already observable. For example, models achieving high detection rates often suffer from elevated FAR. This is directly related to the imbalanced problem at hand, as analytically presented in \cite{stumpo2021open} and in \cite{lavasa2021assessing}, with a typical FAR ranging between 0.25-0.45 \citep[see the discussion in][]{2025SSRv..221...82P}. Operational forecasting requires balancing competing objectives depending on user needs. For example, aviation and astronaut safety operations often prioritize minimizing false alarms, whereas scientific monitoring efforts may tolerate higher false alarms in order to avoid missed events. Despite their elevated FAR or reduced POD, ML models can still be operationally valuable because they often provide substantially greater lead times.

Because inconsistent SEP event definitions and non‑uniform training and testing conditions make fair comparison across studies impossible, a community‑standardized validation framework such as the SEPVAL challenge is essential. The Solar Energetic Particle Model Validation Challenge (SEPVAL), organized by the Space Radiation Analysis Group (SRAG) at the NASA Johnson Space Center (JSC) through NASA's Integrated Solar Energetic Proton Event Alert/Warning System (ISEP) collaboration, is an ongoing community-wide effort designed to develop a generalized framework and prescribed methodology to assess the performance of operational and research SEP forecasting models, using standardized event lists, metrics, and evaluation procedures. SEPVAL brings together model developers and space weather end-users to compare SEP predictions against a consistent observational dataset. By providing common input parameters, clearly defined validation periods, and transparent scoring methodologies, SEPVAL aims to identify model strengths and weaknesses, quantify forecast skill, and promote best practices in SEP prediction. SEPVAL provides a platform for more consistent, streamlined, and fair ML model comparisons than what can be achieved through review efforts like this manuscript, effectively addressing the comparison challenges outlined in this work.

To date, 24 SEP models of all types have participated in SEPVAL, including 6 ML models (marked in Table~\ref{tab:results}). A description of the validation methodology developed through SEPVAL and a summary of model performance for participating models is published by \cite{whitman2026}.  The SEPVAL challenge has so far focused on two operational thresholds important to SRAG, the $\geq 10$ MeV that exceeds 10 pfu and the $\geq 100$ MeV that exceeds 1 pfu, for an approximately balanced set of challenge periods comprised of 33 SEP events and 30 non-event periods. All periods are associated with fast CMEs and strong flares, relevant to SRAG operations. Modelers have submitted predictions of all types, including probability of occurrence, binary all clear, peak intensity, fluence, and full-time profiles. Each participating model has a unique set of inputs and outputs which, as has been discussed above, makes it difficult to compare models directly. However, SEPVAL organizers made the decision that the models could be viewed as a group or ensemble and that a mean and median performance derived from this group could be used as a meaningful definition of the state-of-the-art of SEP model performance. In \cite{whitman2026}, the participating SEPVAL models were divided into two groups, pre-eruptive (continuous) and post-eruptive (triggered), and median and top quartile scores were reported for selected metrics for probability, all clear, and peak intensity. Using these metrics as a reference, SEP models may evaluate their predictions for the set of 63 SEPVAL challenge periods and determine whether their performance exceeds the median with the goal of achieving top quartile performance.

\cite{whitman2026} calculated the state-of-the-art median and the top quartile performance for 10 models predicting in real-time on the SEP Scoreboards\footnote{\url{https://ccmc.gsfc.nasa.gov/scoreboards/sep/}}, representing a realistic operational forecasting scenario that includes real-time data latency, data gaps, human-in-the-loop analyses, and true highly-imbalanced climatology. While it may be difficult for models to make a direct comparison to these metrics, they indicate more realistic model performance in an operational setting. If model developers were to run their models in real time for an extended period or perform a simulated real time evaluation, the scores derived from the SEP Scoreboards are most appropriate for comparison.

It should be understood that reliable prediction of SEP events using ML is a complex task. Not only are the generation, transport, and propagation of energetic particles from the photosphere to Earth still active areas of research, but the amount of relevant heliophysics data, although it has increased over the past decades, is still very sparse compared to other problems for which ML has been applied and now outperforms classical methods. It should also be noted that the studies presented in this review represent the very first efforts of the heliophysics community to predict SEP events using the new tool-set provided by ML. Although the community is not yet able to make reliable comparisons between results, nor it is at a stage where operationally robust SEP predictions can be produced, the insights summarized in this paper can help outline several promising paths for future research, which are discussed in the next section.

\section{Paths for Future Research} \label{sec:Paths_for_Future_Research}

Here, some additional observations about the model results can be made, along with recommendations for future research. First, as seen in Table~\ref{tab:results}, many studies do not quantify the uncertainty of their models' performance (e.g., by reporting forecast standard deviations). This implies that either $k$-fold cross‑validation was not performed, or the corresponding variability was not reported. Repeated cross‑validation with different random splits reduces the randomness associated with any single train–test split and yields a more statistically robust and trustworthy estimate of model performance. Therefore, it is strongly recommended that future studies employ repeated $k$-fold cross‑validation and report at least the standard deviation of the model results over the ensemble of $k$-fold splits. Quantification of uncertainty should not be overlooked in ML studies in heliophysics \citep{keegan2025data}.

Furthermore, different studies use different sets of evaluation metrics ---or sometimes only one or two metrics--- making consistent comparison extremely difficult. We recommend that all future studies that ultimately reduce their predictions to a binary yes/no SEP classification, report all five commonly used metrics listed in Table~\ref{tab:results}, together with their standard deviations. POD and FAR provide a complementary picture of event detection performance: POD captures the ability of a model to correctly identify events (sensitivity), while FAR quantifies the frequency of false positives. Together, these two metrics offer a holistic understanding of operational reliability. FAR is a critical metric for operational forecasting; \cite{nunez2011predicting} emphasizes its importance extensively, noting that an effective prediction system should aim to maximize POD while simultaneously minimizing FAR. Meanwhile, TSS is widely used because it adjusts for class imbalance, incorporates both false positives and false negatives, and provides a normalized measure of predictive skill, offering additional insight beyond raw detection rates.

In addition, it would be beneficial for the community to streamline, to the extent possible, the choice of forecast window. As shown in Figure~\ref{fig:Quantitative_Plot}c, the majority of studies aim to predict SEP events within a 24‑hour window. A 24-hour prediction is operationally meaningful for mitigating radiation risks from geoeffective events. Adoption of standarized forecast windows greatly enhances comparability between studies. It is also informative to examine model performance across multiple forecast horizons ---a strategy adopted by studies such as BiLSTM-SEP, SEP-E and MEMPSEP. It is therefore recommended that future work explore at least three representative forecast windows, such as 6~hours, 12~hours, and 24~hours. Studying how performance varies with forecast horizon provides insight into the temporal limitations of the model, the persistence of precursors, and the time window in which predictions are most reliable for practical operational use. 

Lastly, as mentioned previously, the majority of studies define an SEP event using the $\geq 10$ MeV threshold introduced by NOAA. While this definition is widely used, for operational and geoeffective SEP prediction it is important to recognize that higher-energy particles are often the most critical. In particular, $\geq 100$ MeV proton enhancements are especially critical for operational space‑weather decision making, as they directly affect astronaut safety, can increase radiation exposure for high‑altitude aviation, and can disrupt spacecraft operations through communication and navigation disturbances (radio storms) as well as increased atmospheric drag due to upper‑atmosphere inflation. Future work should therefore prioritize forecasting ---rather than only predicting--- particularly with respect to SEP onset time, peak intensity, and total event fluence at locations of interest, other than Earth (e.g., cislunar space or Mars), in order to support NASA exploration activities and other deep-space operational needs. Among the methods reviewed in this study, only the PSPSP model offers some insights about SEP prediction in areas other than near-Earth. Even proof-of-concept ML studies should aim to design their models and outputs as close to operational requirements as possible. Model developers should also explicitly state which operational organizations or user groups their models are intended to support. 

In addition to the above considerations, it is crucial for future SEP prediction studies to address how their models would operate in real-time settings. A key first step is to identify the intended end user and understand what they require from an SEP prediction system. Different organizations ---such as the SRAG, the Space Weather Prediction Center (SWPC), aviation radiation authorities, satellite operators, or mission planners--- have distinct operational needs, and the design of a prediction model should reflect these needs. For example, for the NASA JSC SRAG and agencies concerned with aviation radiation control, reducing false alarms is of the highest priority, as unwarranted alerts carry significant operational cost. For these users, prediction of $\geq 10$ MeV events is useful, but accurate prediction of $\geq 100$ MeV events is especially valuable, given their stronger radiation impact and operational relevance.

Model developers must also consider how the end user intends to use the predictions, in what environment, and under what constraints, following the Research-to-Operations (R2O\footnote{\url{https://bidenwhitehouse.archives.gov/wp-content/uploads/2022/03/03-2022-Space-Weather-R2O2R-Framework.pdf}}) framework. It is important to distinguish between a model that performs well in an operational setting and one that merely runs in real time. Operational readiness requires that the system handle errors, data gaps, and degraded inputs gracefully. If a model fails ---or is unable to provide predictions--- when data are missing, delayed, or corrupted, then it is not operational in practice, regardless of its offline performance. This relates to the broader concept of system robustness and the SWPC readiness levels, where performance testing in real-time conditions is only one (albeit important) component of operational validation.

At present, most ML-based SEP prediction methods are not operational for two main reasons. First, some rely primarily on data that are not available in real time for operational use. For example, data products such as the SMARP-SHARP dataset by \cite{kosovich2024time} (used by the SHARP-SMARP and CANN models), synthetic data (EPREM-S), and even certain high-energy proton measurements (PSPSP) are often unavailable because they are processed manually. Second, although their input data can be accessed in near real-time (all models that use GOES data as input), the readiness of the models themselves is limited. Few studies (e.g., UMASEP) provide the software infrastructure required to deploy their models online, maintain continuous ingestion of real-time measurements, and deliver predictions continuously and without interruptions (real-time space-based observations quite often have data gaps). For a model to be considered operational, it must be supported by automated pipelines, error-handling systems, and ability to run continuously without manual intervention. Complementing the recommendations above, future studies should therefore explicitly address both data availability and model readiness when developing ML-based SEP forecasting systems.

In regard to the inputs, majority of the models (15/24) use soft X-ray measurements from the GOES satellites, due to their low down-link time, ease of use (timelines) and direct relation to space weather activity. On the other hand magnetic field inputs, coronagraphs and electron flux, although they provide meaningful signatures for space weather events, they have been underutilized as they often require more processing and larger models. The community should move towards the use of larger, more complex models (the majority of models in this study have less than 1,000 trainable parameters) that are trained on multiple physical parameters and data streams. As modern ML research has shown, larger models trained on broader and richer datasets consistently outperform smaller ones, especially for complex predictive tasks (see text prediction; e.g., ChatGPT, Claude, Copilot and others). Therefore, while small datasets and lightweight models may be suitable for proof‑of‑concept studies, operational SEP forecasting systems must leverage as many physical parameters and data streams as possible to achieve reliable predictions. 

Future research should also experiment with new inputs and improved datasets or observations. Since February 1st, 2026, the NASA Interstellar Mapping and Acceleration Probe \citep[IMAP;][]{mccomas2018interstellar,mccomas2025interstellar}, located at the L1 point, provides new coordinated and comprehensive observations of the inner heliosphere. In addition to science observations, five in situ instruments on IMAP also make low-latency measurements (e.g. magnetic field, solar wind electrons and protons, energetic protons and electrons) of relevance to space weather operational forecasting. Through the IMAP Active Link for Real-Time \citep[I-ALiRT;][]{lee2025space} space weather data system, these measurements are continuously telemetered in near real-time to Earth. I-ALiRT is based on the data system used for ACE, and therefore it provides similar space weather data products at significantly enhanced cadences, in addition to the new parameters offered. Similarly, new missions such as NOAA's Space weather Observations at L1 to Advance Readiness - 1 (SOLAR-1; formerly called the Space Weather Follow On – Lagrange 1, or SWFO-L1), and the Polarimeter to Unify the Corona and Heliosphere \citep[PUNCH;][]{deforest2022polarimeter} already provide new data relevant to space weather events prediction. 

The majority of geoeffective SEPs originate from solar surface locations that are visible from the Earth-Sun line, where GOES and SDO are stationed. However, \cite{richardson201425} found that on average one quarter of significant SEPs in geospace originate from source locations behind the western limb of the Sun. Most ML models use GOES X-rays or SDO magnetograph observations. While X-ray flares can still be detected, at least partially, if they occur near the limb on the far side, their brightness is reduced as compared to the same event occurring on the near side. Flares occurring farther behind the limb would be entirely missed by GOES. Also critical, currently solar magnetograph measurements cover under 40\% of the visible solar disk. They become inaccurate near the solar limb, with signal degradation starting already at $~60^\circ$ from the observer’s subsolar point. This degradation affects forecasts of a substantial fraction of SEP events in geospace, given that Earth is, on average, magnetically connected to solar longitudes near $W60^\circ$. The combined gaps in accurate observational coverage limit the POD of almost all current forecasting models listed in \cite{whitman2023review} and in this work. In effect, they are an obstacle for many models to become fully operational.

The most effective mitigation of the critical observational gaps near and behind the Sun’s western limb is a mission to Earth-Sun Lagrangian point 4 \citep{bemporad2021possible, posner2021multi, cho2023opening}. The L4 point is gravitationally stable and is located directly over the magnetic footpoint of Earth, therefore allowing for excellent coverage in X-rays and of magnetographic observations in support of SEP forecasting models. X-ray flares can be fully observed from the Eastern limb of the Sun as viewed from Earth to $W150^\circ$, covering essentially all “missed” events identified in \cite{richardson201425}. Magnetograph coverage will extend from $E60^\circ$ to up to $30^\circ$ behind the western limb of the Sun, with the opportunity of stereoscopic views of ARs traversing from the central meridian to $W60^\circ$ if both geospace and L4 locations are equipped with such instrumentation. Thus, a space weather mission to L4 with real-time downlink can elevate SEP forecasting models into operational models for geospace.

In summary, considering the analysis presented in this work and the effort made to compare the community's SEP prediction ML models, it is highly recommended, if the problem setting allows, that future studies follow the list of ``good practices" presented bellow: 

\begin{itemize}
    \setlength{\itemsep}{0pt}
    \setlength{\parskip}{0pt}
    \setlength{\parsep}{0pt}
    \item \textbf{Use of common validation metrics:} A difficulty encountered by the authors of this work while trying to compare the different models, is that modelers use different evaluation metrics for testing. It is recommended that future works will use at least the five metrics outlined in Table~\ref{tab:results}, as they are deemed most useful and are used regularly by the community. This suit of metrics capture well, from different angles, the performance of the model, and adapting to this common validation setup makes one's work much easier to compare to those of the community.
    \item \textbf{Uncertainty quantification:} ML models often tend to overfit on the data they have seen during training and can appear overly confident for some of their predictions. The best way to assess the stability of an ML model, mitigate overconfidence and prove its robust performance, is to cross-evaluate and estimate uncertainty. It is recommended that future research report appropriate measures of uncertainty quantification for the method and the type of task (e.g., regression or classification), such as prediction variance, entropic measures, prediction intervals, etc. Ideally, the reported measures should integrate both aleatoric (due to inherent randomness in the data) and epistemic (due to scarce data or knowledge gaps) uncertainties.
    \item \textbf{Use of common prediction window:} Another difficulty that one will encounter when trying to compare between SEP prediction works, is that oftentimes the scores reported are for different forecasting setups, prediction windows and average waiting times. Figure~\ref{fig:Quantitative_Plot}c shows that a considerable number of studies have used a 24-hour forecasting window. Many studies that report results for a 24-hour forecasting window, also evaluate model performance for windows of different length as well, as all forecast windows are operationally useful. Smaller windows often produce more confident predictions whereas larger windows allow for more reaction time. Therefore, to make one's work easier to compare with the community's, it is recommended that studies assess and report model performance for time windows of different lengths along with the 1-day-ahead (24 hours) prediction.
    \item \textbf{Deeper models on larger datasets:} It has been a decade since the inception of ResNets \citep{he2016deep}, and eight years since Transformers were introduced \citep{vaswani2017attention}. In industry, models with billions of trainable parameters are being trained on terabytes worth of data. With the commissioning of new heliophysics missions, large amounts of new space weather data will be available to the community. In light of these advances, it is recommended that in the future, researchers train larger models with more data than the current state of the art (Figure~\ref{fig:Quantitative_Plot}a). This will allow models to leverage potentially useful physical information in the augmented datasets, hopefully leading to more accurate forecasts.
    \item \textbf{Operational-as-possible validation:} Modelers often choose validation setups that aim to favor their method and make it appear more competitive in the field. However, a fair comparison of results across different studies should take in consideration the operational level of the model. It is therefore recommended that modelers should aim for operational validation settings and model development. A model that performs very well in a simulated environment but fails to perform in real life is less useful than one which can provide real-time predictions but registers worse performance metrics.
    \item \textbf{SEPVAL validation on common events:} SEPVAL is a common validation scheme for operational SEP forecasts, led by the NASA JSC SRAG and the CCMC. It is highly recommended that future works build their models around SEPVAL, in order to validate on the same SEP events as other SEPVAL contributors. SEPVAL is currently the only common validation scheme that exists in the community and ensures comparison between SEP prediction models. 
    \item \textbf{NASA Open Science:} Open sharing of data, information, and knowledge within the scientific community and the wider public accelerates scientific research and understanding. It is recommended that future authors of SEP prediction based on ML approaches, make their data and algorithms publicly available and their experiments easy to reproduce. The best way to ensure reproducibility is to comply with the NASA Open Science\footnote{\url{https://science.nasa.gov/open-science/}} guidelines. 
\end{itemize}

In addition to the above recommendations, this work helps us understand the new research paths that the community can explore. It is important to keep in mind that this review presents the very first 24 ML models the community has developed to tackle the SEP prediction problem; therefore, it is not surprising that there are many unexplored research directions in such a nascent field. Based on the analysis presented in this document, a non‑limiting list of open avenues for future exploration is given below:

\begin{itemize}
    \setlength{\itemsep}{0pt}
    \setlength{\parskip}{0pt}
    \setlength{\parsep}{0pt}
    \item \textbf{Physics informed NNs:} Physics-Informed Neural Networks (PINNs) have proven to be a powerful approach to solving complex, non-linear scientific problems by embedding governing physical laws directly into the ML learning process \citep{raissi2019physics, karniadakis2021physics}. No studies that use PINNs in order to tackle the SEP prediction problem have been identified. Future studies should explore PINNs and other physics-based augmentations in ML models that predict SEP events.
    \item \textbf{Unsupervised models}: All existing work in the literature relies on labeled information (e.g., SEP lists or SEP onset/end times) to supervise model training. However, unsupervised learning offers a powerful alternative by removing the dependence on such externally provided targets, enabling models to discover intrinsic structure, latent representations, or precursor signatures directly from the data itself. Unsupervised ML techniques have already shown promise in heliophysics \citep{woods2021unsupervised, giger2024unsupervised}. Future work should explore SEP prediction using unsupervised approaches, which may reveal new physical insights and reduce biases introduced by manually curated event labels.
    \item \textbf{Inner heliosphere, Mars and cis-lunar environment predictions:} Twenty-two out of the twenty-four studies identified in this research aim to predict geoeffective SEP events, while only PSPSP predicts particle intensities across the heliosphere. With planned manned missions to the Moon and space-based infrastructure extending throughout the heliosphere, predicting SEP events away from Earth becomes increasingly important. Future studies should leverage the data we have from missions further away from the cis-lunar environment, such as the PSP, STEREO-A, and the Solar Orbiter \citep[SolO;][]{muller_solar_2020, marirrodriga2021solar}, in order to predict SEP events in different parts of the heliosphere.
    \item \textbf{New observations:} A number of new heliophysics missions have been commissioned or planned. These missions will, or already do, provide us with new, higher quality data relevant to space weather. It is recommended that future SEP prediction works explore the usage of data from the most recent heliophysics missions such as the IMAP, PUNCH, SOLAR-1 (formerly called SWFO-L1) and Aditya-L1 \citep{tripathi2022aditya} by the Indian Space Research Organisation (ISRO).
    \item \textbf{Real-time data:} Majority of the models discussed in this study are static; they have not been deployed for continuous real-time predictions. An example of real-time operational forecasting is provided by the SEPNET team through the \href{https://mlsw.engin.umich.edu/apps/runSEP}{University of Michigan Machine Learning for Space Weather (MLSW)} website. The SEP forecasting model is run hourly using newly downloaded input features, and users can visualize the predictions on the website, along with the temporal trajectories of the input features and historical predictions. Future work should focus on deploying trained models in a similar fashion, utilizing on-demand, space missions that provide real-time observations, such as SDO, ACE, IMAP (I-ALiRT), STEREO-A, and others.
    \item \textbf{Pre-trained networks:} Most recently, the heliophysics community has trained large models, whose parameters and embeddings can be used as a base for developing new SEP prediction models. Future works could potentially use transfer learning approaches leveraging pre-trained heliophysics NNs or use the embeddings of Foundation Models (FMs) trained on heliophysics data \citep[SuryaFM;][]{roy2025surya,roy2025suryabench} for predicting SEP events.
    \item \textbf{Data augmentation methods:} SEPs are rare events and all models developed suffer from data imbalance. A rather unexplored area of research is the mitigation of the data imbalance problem in SEP prediction through data augmentation methods and production of synthetic but realistic SEP events which can be used for training ML models (see EPREM-S and TSF models in Appendix~\ref{sec:EPREM-S} and \ref{sec:TSF}, respectively).
    \item \textbf{Connecting with other heliophysics predictions}: The applications of ML in heliophysics span from predicting solar flares \citep{jiao2020solar, wang2020predicting, zheng2023comparative}, coronal mass ejections \citep{bobra2016predicting, vourlidas2019predicting, singh2023improving}, the solar sunspot number \citep[SSN]{sierra2024predicting, rodriguez2024hemispheric, qamar2025prediction}, the solar surface flux transport \citep{jeong2025prediction}, and even emerging ARs \citep{kasapis2023predicting, kasapis2025prediction, kosovichev2025structure, tirona2026forecasting}. These predictions are directly related to the production of SEP events. Future work should utilize such ML models to inform their SEP prediction efforts.
\end{itemize}

In summary, SEP prediction using ML remains an emerging field with promising but currently developing capabilities. Progress over the past decade demonstrates the community's growing ability to exploit heliophysical data using modern ML techniques. Continued advances will depend less on algorithmic advance and more on improvements in data integration, evaluation standardization and operational deployment. The insights presented here aim to guide future efforts towards reliable and operationally useful ML-based SEP forecasting systems.

\section{Conclusions} \label{sec:Conclusions}

This document reviews and summarizes (Appendix~\ref{app:Description_of_Models}) more than a decade of research that applies ML to the prediction of SEP events in the English-language literature. This community-wide effort includes descriptions of all identified models, along with tables that capture their quantitative performance and qualitative characteristics. The datasets used ---or created specifically--- for SEP prediction are also compiled and discussed (Section~\ref{sec:Datasets}). Using this consolidated information, we provide a chartography of the current state of the SEP prediction using ML research community, mapping the landscape of model architectures, inputs, and outputs. We also attempt to compare these diverse approaches and their results, highlighting the difficulties and limitations encountered when making such comparisons. Based on this analysis, we outline recommended good practices for future studies and propose new research directions for the community to pursue.

This review compares three core aspects of existing SEP prediction studies ---their inputs, the ML models trained on those inputs, and the outputs they produce. The community employs a wide range of architectures, largely shaped by available data and prediction goals (Section~\ref{sec:Model_Architectures}). By examining the inputs used (Section~\ref{sec:Inputs_Comparison}), we identify which missions and datasets have been most utilized, which observations are more accessible, and which potentially valuable data sources remain underused. Comparison of outputs (Section~\ref{sec:Outputs_and_Testing}) illustrates a key challenge: the lack of standardized forecasting targets across studies, which makes direct comparison difficult. Sections~\ref{sec:Outputs_and_Testing} and \ref{sec:Paths_for_Future_Research} outline a framework to address this issue and suggest concrete directions for future research and progress toward operational ML-aided SEP forecasting.

To ensure that future work is comparable, reliable, and aligned with community and stakeholder's needs, we strongly recommend that new studies adopt the good practices outlined in this document, including the use of common validation metrics, proper uncertainty quantification, consistent prediction windows, larger and more modern ML architectures, operationally realistic validation setups, the use of SEPVAL for validation on common events, and full compliance with NASA Open Science principles. At the same time, our analysis reveals several promising research directions for the community to explore. These include the incorporation of PINNs, prediction efforts beyond geoeffective SEPs and into the broader heliosphere, the use of underutilized datasets and new mission observations, and leveraging pre-trained heliophysics FMs for transfer learning. Together, these recommendations and research avenues provide a clear path for advancing SEP prediction capabilities and strengthening the role of ML in heliophysics.

\section*{Acknowledgments}

This work originated from discussions held during the SEP Monitoring and Forecasting Workshop at Georgia State University, during October 16–19, 2024, for which Dr. Spiridon Kasapis received the AAS SPD Thomas Metcalf Travel Award Report, which we acknowledge along with the NASA AI/ML HECC Expansion Program, and the NASA grants 23-HGIO23\textunderscore2-0077, 20-HSR20\textunderscore2-0037, 80NSSC19K0630, 80NSSC19K0268, 80NSSC20K1870, and 80NSSC22M0162, which also supported Dr. Alexander Kosovichev and Dr. Irina Kitiashvili. This work was also supported, in part, by the IMAP mission as part of NASA's Solar Terrestrial Probes (STP) Program (80GSFC19C0027). Dr. Soukaina Filali Boubrahimi, Shah Muhammad Hamdi, and Pouya Hosseinzadeh were supported in part by funding from the Division of Atmospheric and Geospace Sciences within the Directorate for Geosciences, under NSF awards \#2204363, \#2240022, \#2530946, and \#2301397. Dr. Angelos Vourlidas was supported by the NASA grants 80NSSC22K0970 and 80NSSC23K0412. Dr. Mohamed Nedal acknowledges support by the project "The Origin and Evolution of Solar Energetic Particles”, funded by the European Office of Aerospace Research and Development under award No. FA8655-24-1-7392. Dr. Athanasios Papaioannou, Eleni Lavasa and Dr. Anastasios Anastasiadis  received funding from the European Union’s Horizon Europe programme under grant agreement No 101135044 (SPEARHEAD) [\url{https://spearhead-he.eu/}]. Views and opinions expressed are however those of the author(s) only and do not necessarily reflect those of the European Union or the European Health and Digital Executive Agency (HaDEA). Neither the European Union nor the granting authority can be held responsible for them. Dr. Sumanth A. Rotti was supported by NASA FINESST grant No. 80NSSC21K1388 and SMD grant No. 24-SMDSS24-0045. Dr. Berkay Aydin and Dr. Petrus C. Martens were supported by NASA SWR2O2R grant No. 80NSSC22K0272. Dr. Christina Lee was supported in part by funding from NASA grant No. 80NSSC25K7646.

\section*{Compliance with Ethical Standards}

The authors declare that this work complies with the ethical standards and policies outlined in the \textit{Space Science Reviews} Instructions for Authors regarding compliance with ethical standards. The authors report no competing financial or non-financial interests that could have influenced the work presented in this manuscript. All co-authors have provided written consent to be included as authors and have approved the submitted version of the paper.

\bibliography{sample631}{}

@ARTICLE{MitchellEA2025ApJ_EnergeticElectronDelay_Type3Bursts,
       author = {{Mitchell}, J.~G. and {Christian}, E.~R. and {de Nolfo}, G.~A. and {Cohen}, C.~M.~S. and {Hill}, M.~E. and {Kouloumvakos}, A. and {Labrador}, A.~W. and {Leske}, R.~A. and {McComas}, D.~J. and {McNutt}, R.~L. and {Mitchell}, D.~G. and {Shen}, M. and {Schwadron}, N.~A. and {Wiedenbeck}, M.~E. and {Bale}, S.~D. and {Pulupa}, M.},
        title = "{Delay of Near-relativistic Electrons with Respect to Type III Radio Bursts throughout the Inner Heliosphere}",
      journal = {\apj},
         year = 2025,
        month = feb,
       volume = {980},
       number = {1},
          eid = {96},
        pages = {96},
          doi = {10.3847/1538-4357/adaa7c},
       adsurl = {https://ui.adsabs.harvard.edu/abs/2025ApJ...980...96M}
}

@ARTICLE{CohenEA2021AandA_He_to_H_abundance_inside_1au_PSPISOIS,
       author = {{Cohen}, C.~M.~S. and {Christian}, E.~R. and {Cummings}, A.~C. and {Davis}, A.~J. and {Desai}, M.~I. and {de Nolfo}, G.~A. and {Giacalone}, J. and {Hill}, M.~E. and {Joyce}, C.~J. and {Labrador}, A.~W. and {Leske}, R.~A. and {Matthaeus}, W.~H. and {McComas}, D.~J. and {McNutt}, R.~L. and {Mewaldt}, R.~A. and {Mitchell}, D.~G. and {Mitchell}, J.~G. and {Rankin}, J.~S. and {Roelof}, E.~C. and {Schwadron}, N.~A. and {Stone}, E.~C. and {Szalay}, J.~R. and {Wiedenbeck}, M.~E. and {Vourlidas}, A. and {Bale}, S.~D. and {Pulupa}, M. and {MacDowall}, R.~J.},
        title = "{Parker Solar Probe observations of He/H abundance variations in SEP events inside 0.5 au}",
      journal = {\aap},
         year = 2021,
        month = jun,
       volume = {650},
          eid = {A23},
        pages = {A23},
          doi = {10.1051/0004-6361/202039299},
       adsurl = {https://ui.adsabs.harvard.edu/abs/2021A&A...650A..23C}
}

@ARTICLE{CohenEA2021AandA_ISOISobservations_29Nov2020_SEPs,
       author = {{Cohen}, C.~M.~S. and {Christian}, E.~R. and {Cummings}, A.~C. and {Davis}, A.~J. and {Desai}, M.~I. and {de Nolfo}, G.~A. and {Giacalone}, J. and {Hill}, M.~E. and {Joyce}, C.~J. and {Labrador}, A.~W. and {Leske}, R.~A. and {Matthaeus}, W.~H. and {McComas}, D.~J. and {McNutt}, R.~L. and {Mewaldt}, R.~A. and {Mitchell}, D.~G. and {Mitchell}, J.~G. and {Rankin}, J.~S. and {Roelof}, E.~C. and {Schwadron}, N.~A. and {Stone}, E.~C. and {Szalay}, J.~R. and {Wiedenbeck}, M.~E. and {Vourlidas}, A. and {Bale}, S.~D. and {Pulupa}, M. and {MacDowall}, R.~J.},
        title = "{PSP/IS{\ensuremath{\odot}}IS observations of the 29 November 2020 solar energetic particle event}",
      journal = {\aap},
         year = 2021,
        month = dec,
       volume = {656},
          eid = {A29},
        pages = {A29},
          doi = {10.1051/0004-6361/202140967},
       adsurl = {https://ui.adsabs.harvard.edu/abs/2021A&A...656A..29C}
}

@ARTICLE{McComasEA2019Nature_PSPISOIS_FirstResults,
       author = {{McComas}, D.~J. and {Christian}, E.~R. and {Cohen}, C.~M.~S. and {Cummings}, A.~C. and {Davis}, A.~J. and {Desai}, M.~I. and {Giacalone}, J. and {Hill}, M.~E. and {Joyce}, C.~J. and {Krimigis}, S.~M. and {Labrador}, A.~W. and {Leske}, R.~A. and {Malandraki}, O. and {Matthaeus}, W.~H. and {McNutt}, R.~L. and {Mewaldt}, R.~A. and {Mitchell}, D.~G. and {Posner}, A. and {Rankin}, J.~S. and {Roelof}, E.~C. and {Schwadron}, N.~A. and {Stone}, E.~C. and {Szalay}, J.~R. and {Wiedenbeck}, M.~E. and {Bale}, S.~D. and {Kasper}, J.~C. and {Case}, A.~W. and {Korreck}, K.~E. and {MacDowall}, R.~J. and {Pulupa}, M. and {Stevens}, M.~L. and {Rouillard}, A.~P.},
        title = "{Probing the energetic particle environment near the Sun}",
      journal = {\nat},
         year = 2019,
        month = dec,
       volume = {576},
        pages = {223-227},
          doi = {10.1038/s41586-019-1811-1},
       adsurl = {https://ui.adsabs.harvard.edu/abs/2019Natur.576..223M}
}

@InProceedings{AxfordEA1977ICRC_CosRayAccel_Shock,
  author    = {{Axford}, W.~I. and {Leer}, E. and {Skadron}, G.},
  booktitle = {International Cosmic Ray Conference},
  title     = {{The Acceleration of Cosmic Rays by Shock Waves}},
  year      = {1977},
  month     = jan,
  pages     = {132},
  series    = {International Cosmic Ray Conference},
  volume    = {11},
  adsurl    = {https://ui.adsabs.harvard.edu/abs/1977ICRC...11..132A},
}

@ARTICLE{BlandfordOstriker1978ApJL_ParticleAccel_Shock,
       author = {{Blandford}, R.~D. and {Ostriker}, J.~P.},
        title = "{Particle acceleration by astrophysical shocks.}",
      journal = {\apjl},
         year = 1978,
        month = apr,
       volume = {221},
        pages = {L29-L32},
          doi = {10.1086/182658},
       adsurl = {https://ui.adsabs.harvard.edu/abs/1978ApJ...221L..29B}
}

@ARTICLE{Bell1978MNRAS_ShockAccel_CosRay,
       author = {{Bell}, A.~R.},
        title = "{The acceleration of cosmic rays in shock fronts - I.}",
      journal = {\mnras},
         year = 1978,
        month = jan,
       volume = {182},
        pages = {147-156},
          doi = {10.1093/mnras/182.2.147},
       adsurl = {https://ui.adsabs.harvard.edu/abs/1978MNRAS.182..147B}
}

@ARTICLE{SubashchandarEA2025ApJL_SpatialDiffCoeff_PerpPara_PSP,
       author = {{Subashchandar}, Nibuna S.~M. and {Zhao}, Lingling and {Shalchi}, Andreas and {Zank}, Gary and {Le Roux}, Jakobus and {Li}, Hui and {Zhu}, Xingyu and {Silwal}, Ashok and {Guzman}, Juan G. Alonso},
        title = "{Parallel and Perpendicular Diffusion of Energetic Particles in the Near-Sun Solar Wind Observed by Parker Solar Probe}",
      journal = {\apjl},
         year = 2025,
        month = oct,
       volume = {991},
       number = {2},
          eid = {L30},
        pages = {L30},
          doi = {10.3847/2041-8213/ae063f},
       adsurl = {https://ui.adsabs.harvard.edu/abs/2025ApJ...991L..30S}
}

@ARTICLE{CuestaEA2025ApJL_DiffusionCoefficients_SEPfitting_QLT,
       author = {{Cuesta}, M.~E. and {Fraschetti}, F. and {Livadiotis}, G. and {Farooki}, H.~A. and {Shen}, M.~M. and {Khoo}, L.~Y. and {Szalay}, J.~R. and {Rankin}, J.~S. and {McComas}, D.~J. and {Mitchell}, D.~G. and {Christian}, E.~R. and {Mitchell}, J.~G. and {Berland}, G.~D. and {Cohen}, C.~M.~S. and {Leske}, R.~A. and {Xu}, Z. and {Muro}, G.~D. and {Pecora}, F. and {Ruffolo}, D. and {Matthaeus}, W.~H. and {Giacalone}, J. and {Schwadron}, N.~A. and {Desai}, M.~I. and {Dayeh}, M.~A. and {Bale}, S.~D. and {Stevens}, M.~L. and {Livi}, R.},
        title = "{Distinct Solar Energetic Particle Shock Intensity─Diffusion Coefficient Relationships in the Inner Heliosphere}",
      journal = {\apjl},
         year = 2025,
        month = nov,
       volume = {993},
       number = {1},
          eid = {L15},
        pages = {L15},
          doi = {10.3847/2041-8213/ae109c},
       adsurl = {https://ui.adsabs.harvard.edu/abs/2025ApJ...993L..15C}
}

@ARTICLE{ChhiberEA2021AandA_FLRW_SEPs_PSP,
       author = {{Chhiber}, R. and {Matthaeus}, W.~H. and {Cohen}, C.~M.~S. and {Ruffolo}, D. and {Sonsrettee}, W. and {Tooprakai}, P. and {Seripienlert}, A. and {Chuychai}, P. and {Usmanov}, A.~V. and {Goldstein}, M.~L. and {McComas}, D.~J. and {Leske}, R.~A. and {Szalay}, J.~R. and {Joyce}, C.~J. and {Cummings}, A.~C. and {Roelof}, E.~C. and {Christian}, E.~R. and {Mewaldt}, R.~A. and {Labrador}, A.~W. and {Giacalone}, J. and {Schwadron}, N.~A. and {Mitchell}, D.~G. and {Hill}, M.~E. and {Wiedenbeck}, M.~E. and {McNutt}, R.~L. and {Desai}, M.~I.},
        title = "{Magnetic field line random walk and solar energetic particle path lengths. Stochastic theory and PSP/IS{\ensuremath{\odot}}IS observations}",
      journal = {\aap},
         year = 2021,
        month = jun,
       volume = {650},
          eid = {A26},
        pages = {A26},
          doi = {10.1051/0004-6361/202039816},
archivePrefix = {arXiv},
       eprint = {2011.08329},
 primaryClass = {astro-ph.SR},
       adsurl = {https://ui.adsabs.harvard.edu/abs/2021A&A...650A..26C}
}

@article{kasapis2022interpretable,
  title={Interpretable machine learning to forecast SEP events for solar cycle 23},
  author={Kasapis, Spiridon and Zhao, Lulu and Chen, Yang and Wang, Xiantong and Bobra, Monica and Gombosi, Tamas},
  journal={Space Weather},
  volume={20},
  number={2},
  pages={e2021SW002842},
  year={2022},
  publisher={Wiley Online Library}
}

@article{kasapis2024forecasting,
  title={Forecasting Solar Energetic Particle Events During Solar Cycles 23 and 24 Using Interpretable Machine Learning},
  author={Kasapis, Spiridon and Kitiashvili, Irina N and Kosovich, Paul and Kosovichev, Alexander G and Sadykov, Viacheslav M and O’Keefe, Patrick and Wang, Vincent},
  journal={The Astrophysical Journal},
  volume={974},
  number={1},
  pages={131},
  year={2024},
  publisher={IOP Publishing}
}

@ARTICLE{2025SSRv..221...82P,
       author = {{Papaioannou}, Athanasios and {Strauss}, Roelf Du Toit and {Lario}, David and {Vainio}, Rami and {Wijsen}, Nicolas and {Afanasiev}, Alexander and {Anastasiadis}, Anastasios and {Kouloumvakos}, Athanasios},
        title = "{Predicting Solar Energetic Particles: Solar Storm Watch - Preparing for Space Odyssey}",
      journal = {\ssr},
         year = 2025,
        month = sep,
       volume = {221},
       number = {6},
          eid = {82},
        pages = {82},
          doi = {10.1007/s11214-025-01211-4},
       adsurl = {https://ui.adsabs.harvard.edu/abs/2025SSRv..221...82P}
}

@ARTICLE{Waterfall_2023,
       author = {{Waterfall}, C.~O.~G. and {Dalla}, S. and {Raukunen}, O. and {Heynderickx}, D. and {Jiggens}, P. and {Vainio}, R.},
        title = "{High Energy Solar Particle Events and Their Relationship to Associated Flare, CME and GLE Parameters}",
      journal = {Space Weather},
         year = 2023,
        month = mar,
       volume = {21},
       number = {3},
          eid = {e2022SW003334},
        pages = {e2022SW003334},
          doi = {10.1029/2022SW003334},
archivePrefix = {arXiv},
       eprint = {2303.03935},
 primaryClass = {astro-ph.SR},
       adsurl = {https://ui.adsabs.harvard.edu/abs/2023SpWea..2103334W}
}

@article{kosovich2024time,
  title={Time Series of Magnetic Field Parameters of Merged MDI and HMI Space-weather Active Region Patches as Potential Tool for Solar Flare Forecasting},
  author={Kosovich, Paul A and Kosovichev, Alexander G and Sadykov, Viacheslav M and Kasapis, Spiridon and Kitiashvili, Irina N and O’Keefe, Patrick M and Ali, Aatiya and Oria, Vincent and Granovsky, Samuel and Chong, Chun Jie and others},
  journal={The Astrophysical Journal},
  volume={972},
  number={2},
  pages={169},
  year={2024},
  publisher={IOP Publishing}
}

@article{bobra2021smarps,
  title={SMARPs and SHARPs: Two solar cycles of active region data},
  author={Bobra, Monica G and Wright, Paul J and Sun, Xudong and Turmon, Michael J},
  journal={The Astrophysical Journal Supplement Series},
  volume={256},
  number={2},
  pages={26},
  year={2021},
  publisher={IOP Publishing}
}

@article{ali2024predicting,
  title={Predicting Solar Proton Events of Solar Cycles 22--24 Using GOES Proton and Soft-X-Ray Flux Features},
  author={Ali, Aatiya and Sadykov, Viacheslav and Kosovichev, Alexander and Kitiashvili, Irina N and Oria, Vincent and Nita, Gelu M and Illarionov, Egor and O’Keefe, Patrick M and Francis, Fraila and Chong, Chun-Jie and others},
  journal={The Astrophysical Journal Supplement Series},
  volume={270},
  number={1},
  pages={15},
  year={2024},
  publisher={IOP Publishing}
}

@article{hosseinzadeh2024improving,
  title={Improving solar energetic particle event prediction through multivariate time series data augmentation},
  author={Hosseinzadeh, Pouya and Boubrahimi, Soukaina Filali and Hamdi, Shah Muhammad},
  journal={The Astrophysical Journal Supplement Series},
  volume={270},
  number={2},
  pages={31},
  year={2024},
  publisher={IOP Publishing}
}

@article{hosseinzadeh2025end,
  title={An End-to-end Ensemble Machine Learning Approach for Predicting High-impact Solar Energetic Particle Events Using Multimodal Data},
  author={Hosseinzadeh, Pouya and Boubrahimi, Soukaina Filali and Hamdi, Shah Muhammad},
  journal={The Astrophysical Journal Supplement Series},
  volume={277},
  number={2},
  pages={34},
  year={2025},
  publisher={IOP Publishing}
}

@inproceedings{boubrahimi2017prediction,
  title={On the prediction of> 100 MeV solar energetic particle events using GOES satellite data},
  author={Boubrahimi, Soukaina Filali and Aydin, Berkay and Martens, Petrus and Angryk, Rafal},
  booktitle={2017 IEEE international conference on big data (big data)},
  pages={2533--2542},
  year={2017},
  organization={IEEE}
}

@article{hosseinzadeh2024toward,
  title={Toward enhanced prediction of high-impact solar energetic particle events using multimodal time series data fusion models},
  author={Hosseinzadeh, Pouya and Filali Boubrahimi, Soukaina and Hamdi, Shah Muhammad},
  journal={Space Weather},
  volume={22},
  number={6},
  pages={e2024SW003982},
  year={2024},
  publisher={Wiley Online Library}
}

@article{nedal2023forecasting,
  title={Forecasting solar energetic proton integral fluxes with bi-directional long short-term memory neural networks},
  author={Nedal, Mohamed and Kozarev, Kamen and Arsenov, Nestor and Zhang, Peijin},
  journal={Journal of Space Weather and Space Climate},
  volume={13},
  pages={26},
  year={2023},
  publisher={EDP Sciences}
}

@article{nunez2011predicting,
  title={Predicting solar energetic proton events (E> 10 MeV)},
  author={N{\'u}{\~n}ez, Marlon},
  journal={Space Weather},
  volume={9},
  number={7},
  year={2011},
  publisher={Wiley Online Library}
}

@book{malandraki2018solar,
  title={Solar particle radiation storms forecasting and analysis: The HESPERIA HORIZON 2020 project and beyond},
  author={Malandraki, Olga E and Crosby, Norma B},
  year={2018},
  publisher={Springer Nature}
}

@article{torres2022machine,
  title={A machine learning approach to predicting SEP events using properties of coronal mass ejections},
  author={Torres, Jesse and Zhao, Lulu and Chan, Philip K and Zhang, Ming},
  journal={Space Weather},
  volume={20},
  number={7},
  pages={e2021SW002797},
  year={2022},
  publisher={Wiley Online Library}
}

@article{torres2025machine,
  title={A machine learning approach to predicting SEP proton intensity and events using time series of relativistic electron measurements},
  author={Torres, Jesse and Chan, Philip K and Zhao, Lulu and Zhang, Ming},
  journal={Space Weather},
  volume={23},
  number={2},
  pages={e2024SW003921},
  year={2025},
  publisher={Wiley Online Library}
}

@article{martens2024advancing,
  title={Advancing Solar Energetic Particle Event Prediction through Survival Analysis and Cloud Computing. I. Kaplan--Meier Estimation and Cox Proportional Hazards Modeling},
  author={Jackson, India and Martens, Petrus},
  journal={The Astrophysical Journal Supplement Series},
  volume={272},
  number={2},
  pages={37},
  year={2024},
  publisher={IOP Publishing}
}

@data{DVN/GXY9MZ_2024,
author = {Jackson, India and Petrus Martens},
publisher = {Harvard Dataverse},
title = {{Survival Solar Energetic Particle (SSEP) Dataset}},
UNF = {UNF:6:xn108cK7YYF39LpktDz54g==},
year = {2024},
version = {V1},
doi = {10.7910/DVN/GXY9MZ},
url = {https://doi.org/10.7910/DVN/GXY9MZ}
}

@article{aminalragia2021solar,
  title={Solar energetic particle event occurrence prediction using solar flare soft X-ray measurements and machine learning},
  author={Aminalragia-Giamini, Sigiava and Raptis, Savvas and Anastasiadis, Anastasios and Tsigkanos, Antonis and Sandberg, Ingmar and Papaioannou, Athanasios and Papadimitriou, Constantinos and Jiggens, Piers and Aran, Angels and Daglis, Ioannis A},
  journal={Journal of Space Weather and Space Climate},
  volume={11},
  pages={59},
  year={2021},
  publisher={EDP Sciences}
}

@article{lavasa2021assessing,
  title={Assessing the predictability of solar energetic particles with the use of machine learning techniques},
  author={Lavasa, E and Giannopoulos, Georgios and Papaioannou, Aikaterini and Anastasiadis, Anastasios and Daglis, IA and Aran, Angels and Pacheco, David and Sanahuja, Blai},
  journal={Solar Physics},
  volume={296},
  number={7},
  pages={107},
  year={2021},
  publisher={Springer}
}

@article{drury1983introduction,
  title={An introduction to the theory of diffusive shock acceleration of energetic particles in tenuous plasmas},
  author={Drury, L O'C},
  journal={Reports on Progress in Physics},
  volume={46},
  number={8},
  pages={973},
  year={1983},
  publisher={IOP Publishing}
}

@article{richardson2018prediction,
  title={Prediction of solar energetic particle event peak proton intensity using a simple algorithm based on CME speed and direction and observations of associated solar phenomena},
  author={Richardson, IG and Mays, ML and Thompson, BJ},
  journal={Space Weather},
  volume={16},
  number={11},
  pages={1862--1881},
  year={2018},
  publisher={Wiley Online Library}
}

@article{rotti2024short,
  title={Short-term Classification of Strong Solar Energetic Particle Events Using Multivariate Time-series Classifiers},
  author={Rotti, Sumanth A and Aydin, Berkay and Martens, Petrus C},
  journal={The Astrophysical Journal},
  volume={966},
  number={2},
  pages={165},
  year={2024},
  publisher={IOP Publishing}
}

@article{rotti2024precise,
  title={Precise and Accurate Short-term Forecasting of Solar Energetic Particle Events with Multivariate Time-series Classifiers},
  author={Rotti, Sumanth A and Aydin, Berkay and Martens, Petrus C},
  journal={The Astrophysical Journal},
  volume={974},
  number={2},
  pages={188},
  year={2024},
  publisher={IOP Publishing}
}

@article{rotti2022integrated,
  title={Integrated geostationary solar energetic particle events catalog: GSEP},
  author={Rotti, Sumanth and Aydin, Berkay and Georgoulis, Manolis K and Martens, Petrus C},
  journal={The Astrophysical Journal Supplement Series},
  volume={262},
  number={1},
  pages={29},
  year={2022},
  publisher={IOP Publishing}
}

@article{loning2019sktime,
  title={sktime: A unified interface for machine learning with time series},
  author={L{\"o}ning, Markus and Bagnall, Anthony and Ganesh, Sajaysurya and Kazakov, Viktor and Lines, Jason and Kir{\'a}ly, Franz J},
  journal={arXiv preprint arXiv:1909.07872},
  year={2019}
}

@article{engell2017sprints,
  title={SPRINTS: A framework for solar-driven event forecasting and research},
  author={Engell, AJ and Falconer, DA and Schuh, M and Loomis, J and Bissett, D},
  journal={Space Weather},
  volume={15},
  number={10},
  pages={1321--1346},
  year={2017},
  publisher={Wiley Online Library}
}

@article{chatterjee2024mempsep,
  title={MEMPSEP-I. Forecasting the probability of solar energetic particle event occurrence using a multivariate ensemble of convolutional neural networks},
  author={Chatterjee, Subhamoy and Dayeh, Maher A and Mu{\~n}oz-Jaramillo, Andr{\'e}s and Bain, Hazel M and Moreland, Kimberly and Hart, Samuel},
  journal={Space Weather},
  volume={22},
  number={9},
  pages={e2023SW003568},
  year={2024},
  publisher={Wiley Online Library}
}

@article{dayeh2024mempsep,
  title={MEMPSEP-II. Forecasting the properties of solar energetic particle events using a multivariate ensemble approach},
  author={Dayeh, Maher A and Chatterjee, Subhamoy and Mu{\~n}oz-Jaramillo, Andr{\'e}s and Moreland, Kimberly and Bain, Hazel M and Hart, Samuel T},
  journal={Space Weather},
  volume={22},
  number={9},
  pages={e2023SW003697},
  year={2024},
  publisher={Wiley Online Library}
}

@article{moreland2024mempsep,
  title={MEMPSEP-III. A Machine Learning-Oriented Multivariate Data Set for Forecasting the Occurrence and Properties of Solar Energetic Particle Events Using a Multivariate Ensemble Approach},
  author={Moreland, Kimberly and Dayeh, Maher A and Bain, Hazel M and Chatterjee, Subhamoy and Mu{\~n}oz-Jaramillo, Andr{\'e}s and Hart, Samuel T},
  journal={Space Weather},
  volume={22},
  number={9},
  pages={e2023SW003765},
  year={2024},
  publisher={Wiley Online Library}
}

@article{dayeh2025machine,
  title={A Machine Learning-Ready Data Processing Tool for Near Real-Time Forecasting},
  author={Dayeh, Maher A and Starkey, Michael J and Chatterjee, Subhamoy and Elliott, Heather and Hart, Samuel and Moreland, Kimberly},
  journal={arXiv preprint arXiv:2502.08555},
  year={2025}
}

@article{rankin2022galactic,
  title={Galactic cosmic rays throughout the heliosphere and in the very local interstellar medium},
  author={Rankin, Jamie S and Bindi, Veronica and Bykov, Andrei M and Cummings, Alan C and Della Torre, Stefano and Florinski, Vladimir and Heber, Bernd and Potgieter, Marius S and Stone, Edward C and Zhang, Ming},
  journal={Space Science Reviews},
  volume={218},
  number={5},
  pages={42},
  year={2022},
  publisher={Springer}
}

@article{engelbrecht2022theory,
  title={Theory of cosmic ray transport in the heliosphere},
  author={Engelbrecht, N Eugene and Effenberger, Frederic and Florinski, Vladimir and Potgieter, Marius S and Ruffolo, D and Chhiber, Rohit and Usmanov, Arcadi Vladimirovich and Rankin, Jamie S and Els, Paul L},
  journal={Space Science Reviews},
  volume={218},
  number={4},
  pages={33},
  year={2022},
  publisher={Springer}
}

@article{temmer2021space,
  title={Space weather: The solar perspective: An update to Schwenn (2006)},
  author={Temmer, Manuela},
  journal={Living Reviews in Solar Physics},
  volume={18},
  number={1},
  pages={4},
  year={2021},
  publisher={Springer}
}

@article{rotti2023analysis,
  title={Analysis of SEP events and their possible precursors based on the GSEP Catalog},
  author={Rotti, Sumanth and Martens, Petrus C},
  journal={The Astrophysical Journal Supplement Series},
  volume={267},
  number={2},
  pages={40},
  year={2023},
  publisher={IOP Publishing}
}

@data{gsep_2022,
author = {Rotti, Sumanth and Aydin, Berkay and Georgoulis, Manolis and Martens, Petrus},
publisher = {Harvard Dataverse},
title = {{GSEP Dataset}},
UNF = {UNF:6:DzByXFz8kliDeazIcz8V2Q==},
year = {2022},
version = {V5},
doi = {10.7910/DVN/DZYLHK},
url = {https://doi.org/10.7910/DVN/DZYLHK}
}

@inproceedings{cabello2020fast,
  title={Fast and accurate time series classification through supervised interval search},
  author={Cabello, Nestor and Naghizade, Elham and Qi, Jianzhong and Kulik, Lars},
  booktitle={2020 IEEE International Conference on Data Mining (ICDM)},
  pages={948--953},
  year={2020},
  organization={IEEE}
}

@misc{markussktime, 
    title={sktime/sktime: v0.13.4},
    DOI={10.5281/zenodo.7117735},
    abstractNote={<p>Maintenance release - moves <code>sktime</code> repository to <code>sktime</code> org from <code>alan-turing-institute</code> org.</p>}, 
    publisher={Zenodo}, 
    author={Markus Löning and Franz Király and Tony Bagnall and Matthew Middlehurst and Sajaysurya Ganesh and George Oastler and Jason Lines and Martin Walter and ViktorKaz and Lukasz Mentel and et al.}, 
    year={2022}, 
    month={Sep} }

@article{gopalswamy2022sun,
  title={The Sun and space weather},
  author={Gopalswamy, Nat},
  journal={Atmosphere},
  volume={13},
  number={11},
  pages={1781},
  year={2022},
  publisher={MDPI}
}

@article{papaioannou2016solar,
  title={Solar flares, coronal mass ejections and solar energetic particle event characteristics},
  author={Papaioannou, Athanasios and Sandberg, Ingmar and Anastasiadis, Anastasios and Kouloumvakos, Athanasios and Georgoulis, Manolis K and Tziotziou, Kostas and Tsiropoula, Georgia and Jiggens, Piers and Hilgers, Alain},
  journal={Journal of Space Weather and Space Climate},
  volume={6},
  pages={A42},
  year={2016},
  publisher={EDP Sciences}
}

@article{buzulukova2022space,
  title={Space Weather: From solar origins to risks and hazards evolving in time},
  author={Buzulukova, Natalia and Tsurutani, Bruce},
  journal={Frontiers in Astronomy and Space Sciences},
  volume={9},
  pages={1017103},
  year={2022},
  publisher={Frontiers Media SA}
}

@article{georgoulis2024prediction,
  title={Prediction of solar energetic events impacting space weather conditions},
  author={Georgoulis, Manolis K and Yardley, Stephanie L and Guerra, Jordan A and Murray, Sophie A and Ahmadzadeh, Azim and Anastasiadis, Anastasios and Angryk, Rafal and Aydin, Berkay and Banerjee, Dipankar and Barnes, Graham and others},
  journal={Advances in Space Research},
  year={2024},
  publisher={Elsevier}
}

@article{desai2016large,
  title={Large gradual solar energetic particle events},
  author={Desai, Mihir and Giacalone, Joe},
  journal={Living Reviews in Solar Physics},
  volume={13},
  number={1},
  pages={3},
  year={2016},
  publisher={Springer}
}

@book{reames2021solar,
  title={Solar energetic particles: a modern primer on understanding sources, acceleration and propagation},
  author={Reames, Donald V},
  year={2021},
  publisher={Springer Nature}
}

@article{reames2013two,
  title={The two sources of solar energetic particles},
  author={Reames, Donald V},
  journal={Space Science Reviews},
  volume={175},
  number={1},
  pages={53--92},
  year={2013},
  publisher={Springer}
}

@article{zank2015diffusive,
  title={Diffusive shock acceleration and reconnection acceleration processes},
  author={Zank, GP and Hunana, P and Mostafavi, P and Le Roux, JA and Li, Gang and Webb, GM and Khabarova, O and Cummings, A and Stone, E and Decker, R},
  journal={The Astrophysical Journal},
  volume={814},
  number={2},
  pages={137},
  year={2015},
  publisher={IOP Publishing}
}

@article{tobiska2015advances,
  title={Advances in atmospheric radiation measurements and modeling needed to improve air safety},
  author={Tobiska, W Kent and Atwell, William and Beck, Peter and Benton, Eric and Copeland, Kyle and Dyer, Clive and Gersey, Brad and Getley, Ian and Hands, Alex and Holland, Michael and others},
  journal={Space Weather},
  volume={13},
  number={4},
  pages={202--210},
  year={2015},
  publisher={Wiley Online Library}
}

@article{mishev2015computation,
  title={Computation of dose rate at flight altitudes during ground level enhancements no. 69, 70 and 71},
  author={Mishev, AL and Adibpour, F and Usoskin, IG and Felsberger, E},
  journal={Advances in Space Research},
  volume={55},
  number={1},
  pages={354--362},
  year={2015},
  publisher={Elsevier}
}

@article{miroshnichenko2018retrospective,
  title={Retrospective analysis of GLEs and estimates of radiation risks},
  author={Miroshnichenko, Leonty I},
  journal={Journal of Space Weather and Space Climate},
  volume={8},
  pages={A52},
  year={2018},
  publisher={EDP Sciences}
}

@article{cucinotta2013safe,
  title={How safe is safe enough? Radiation risk for a human mission to Mars},
  author={Cucinotta, Francis A and Kim, Myung-Hee Y and Chappell, Lori J and Huff, Janice L},
  journal={PloS one},
  volume={8},
  number={10},
  pages={e74988},
  year={2013},
  publisher={Public Library of Science San Francisco, USA}
}

@article{schrijver2015understanding,
  title={Understanding space weather to shield society: A global road map for 2015--2025 commissioned by COSPAR and ILWS},
  author={Schrijver, Carolus J and Kauristie, Kirsti and Aylward, Alan D and Denardini, Clezio M and Gibson, Sarah E and Glover, Alexi and Gopalswamy, Nat and Grande, Manuel and Hapgood, Mike and Heynderickx, Daniel and others},
  journal={Advances in Space Research},
  volume={55},
  number={12},
  pages={2745--2807},
  year={2015},
  publisher={Elsevier}
}

@article{zeitlin2013measurements,
  title={Measurements of energetic particle radiation in transit to Mars on the Mars Science Laboratory},
  author={Zeitlin, C and Hassler, DM and Cucinotta, FA and Ehresmann, B and Wimmer-Schweingruber, RF and Brinza, DE and Kang, S and Weigle, G and B{\"o}ttcher, S and B{\"o}hm, E and others},
  journal={science},
  volume={340},
  number={6136},
  pages={1080--1084},
  year={2013},
  publisher={American Association for the Advancement of Science}
}

@article{neukart2024towards,
  title={Towards sustainable horizons: A comprehensive blueprint for Mars colonization},
  author={Neukart, Florian},
  journal={Heliyon},
  volume={10},
  number={4},
  year={2024},
  publisher={Elsevier}
}

@inproceedings{creech2022artemis,
  title={Artemis: an overview of NASA's activities to return humans to the Moon},
  author={Creech, Steve and Guidi, John and Elburn, Darcy},
  booktitle={2022 ieee aerospace conference (aero)},
  pages={1--7},
  year={2022},
  organization={IEEE}
}

@article{whitman2023review,
  title={Review of solar energetic particle prediction models},
  author={Whitman, Kathryn and Egeland, Ricky and Richardson, Ian G and Allison, Clayton and Quinn, Philip and Barzilla, Janet and Kitiashvili, Irina and Sadykov, Viacheslav and Bain, Hazel M and Dierckxsens, Mark and others},
  journal={Advances in Space Research},
  volume={72},
  number={12},
  pages={5161--5242},
  year={2023},
  publisher={Elsevier}
}

@article{whitman2026,
  title={Validation of Solar Energetic Particle Forecasting Models for Space Radiation Operations with {SPHINX} and {VIVID}},
  author={Whitman, Kathryn and Egeland, Ricky and Allison, Clayton and Quinn, Philip and Stegeman, Luke},
  journal={NASA Technical Reports Server},
  number={NASA/TP-20260000463},
  year={2026},
  publisher={NASA}
}

@ARTICLE{pierce1884,
       author = {{Peirce}, C.~S.},
        title = "{The Numerical Measure of the Success of Predictions}",
      journal = {Science},
         year = 1884,
        month = nov,
       volume = {4},
       number = {93},
        pages = {453-454},
          doi = {10.1126/science.ns-4.93.453},
       adsurl = {https://ui.adsabs.harvard.edu/abs/1884Sci.....4..453P}
}

@ARTICLE{sandberg2014,
       author = {{Sandberg}, I. and {Jiggens}, P. and {Heynderickx}, D. and {Daglis}, I.~A.},
        title = "{Cross calibration of NOAA GOES solar proton detectors using corrected NASA IMP-8/GME data}",
      journal = {\grl},
         year = 2014,
        month = jul,
       volume = {41},
       number = {13},
        pages = {4435-4441},
          doi = {10.1002/2014GL060469},
       adsurl = {https://ui.adsabs.harvard.edu/abs/2014GeoRL..41.4435S}
}

@ARTICLE{bruno2017,
       author = {{Bruno}, A.},
        title = "{Calibration of the GOES 13/15 high-energy proton detectors based on the PAMELA solar energetic particle observations}",
      journal = {Space Weather},
         year = 2017,
        month = sep,
       volume = {15},
       number = {9},
        pages = {1191-1202},
          doi = {10.1002/2017SW001672},
archivePrefix = {arXiv},
       eprint = {1708.05641},
 primaryClass = {physics.space-ph},
       adsurl = {https://ui.adsabs.harvard.edu/abs/2017SpWea..15.1191B}
}

@article{baydin2023surrogate,
  title={A surrogate model for studying solar energetic particle transport and the seed population},
  author={Baydin, Atilim Gune{\c{s}} and Poduval, Bala and Schwadron, Nathan A},
  journal={Space Weather},
  volume={21},
  number={12},
  pages={e2023SW003593},
  year={2023},
  publisher={Wiley Online Library}
}

@article{nunez2020predicting,
  title={Predicting> 10 MeV SEP events from solar flare and radio burst data},
  author={N{\'u}{\~n}ez, Marlon and Paul-Pena, Daniel},
  journal={Universe},
  volume={6},
  number={10},
  pages={161},
  year={2020},
  publisher={MDPI}
}

@article{sadykov2021prediction,
  title={Prediction of Solar Proton Events with Machine Learning: Comparison with Operational Forecasts and" All-Clear" Perspectives},
  author={Sadykov, Viacheslav and Kosovichev, Alexander and Kitiashvili, Irina and Oria, Vincent and Nita, Gelu M and Illarionov, Egor and O'Keefe, Patrick and Jiang, Yucheng and Fereira, Sheldon and Ali, Aatiya},
  journal={arXiv preprint arXiv:2107.03911},
  year={2021}
}

@article{bobra2014helioseismic,
  title={The Helioseismic and Magnetic Imager (HMI) vector magnetic field pipeline: SHARPs--space-weather HMI active region patches},
  author={Bobra, Monica G and Sun, Xudong and Hoeksema, J Todd and Turmon, M and Liu, Yang and Hayashi, Keiji and Barnes, Graham and Leka, KD},
  journal={Solar Physics},
  volume={289},
  number={9},
  pages={3549--3578},
  year={2014},
  publisher={Springer}
}

@article{o2024random,
  title={The Random Hivemind: An ensemble deep learning application to the solar energetic particle prediction problem},
  author={O’Keefe, Patrick M and Sadykov, Viacheslav and Kosovichev, Alexander and Kitiashvili, Irina N and Oria, Vincent and Nita, Gelu M and Francis, Fraila and Chong, Chun-Jie and Kosovich, Paul and Ali, Aatiya and others},
  journal={Advances in Space Research},
  volume={74},
  number={12},
  pages={6252--6263},
  year={2024},
  publisher={Elsevier}
}

@article{sadykov2019statistical,
  title={Statistical properties of soft X-ray emission of solar flares},
  author={Sadykov, Viacheslav M and Kosovichev, Alexander G and Kitiashvili, Irina N and Frolov, Alexander},
  journal={The Astrophysical Journal},
  volume={874},
  number={1},
  pages={19},
  year={2019},
  publisher={IOP Publishing}
}

@article{laurenza2009technique,
  title={A technique for short-term warning of solar energetic particle events based on flare location, flare size, and evidence of particle escape},
  author={Laurenza, M and Cliver, EW and Hewitt, J and Storini, M and Ling, AG and Balch, CC and Kaiser, ML},
  journal={Space Weather},
  volume={7},
  number={4},
  year={2009},
  publisher={Wiley Online Library}
}

@article{alberti2017solar,
  title={Solar activity from 2006 to 2014 and short-term forecasts of solar proton events using the ESPERTA model},
  author={Alberti, T and Laurenza, M and Cliver, EW and Storini, M and Consolini, G and Lepreti, Fabio},
  journal={The Astrophysical Journal},
  volume={838},
  number={1},
  pages={59},
  year={2017},
  publisher={IOP Publishing}
}

@article{laurenza2018short,
  title={A short-term ESPERTA-based forecast tool for moderate-to-extreme solar proton events},
  author={Laurenza, MONICA and Alberti, TOMMASO and Cliver, EW},
  journal={The Astrophysical Journal},
  volume={857},
  number={2},
  pages={107},
  year={2018},
  publisher={IOP Publishing}
}

@article{alberti2019forecasting,
  title={Forecasting solar proton events by using the ESPERTA model},
  author={Alberti, T and Laurenza, M and Cliver, EW},
  journal={Nuovo Cimento C},
  volume={42},
  number={1},
  pages={40},
  year={2019}
}

@article{stumpo2021open,
  title={Open issues in statistical forecasting of solar proton events: A machine learning perspective},
  author={Stumpo, Mirko and Benella, Simone and Laurenza, Monica and Alberti, Tommaso and Consolini, Giuseppe and Marcucci, Maria Federica},
  journal={Space Weather},
  volume={19},
  number={10},
  pages={e2021SW002794},
  year={2021},
  publisher={Wiley Online Library}
}

@article{benella2023statistical,
  title={Statistical treatment of solar energetic particle forecasting through supervised learning approaches},
  author={Benella, Simone and Stumpo, Mirko and Laurenza, Monica and Alberti, Tommaso and Consolini, G and Marcucci, MF},
  journal={Proceedings of Science (ECRS)},
  year={2023}
}

@article{laurenza2024upgrades,
  title={Upgrades of the ESPERTA forecast tool for solar proton events},
  author={Laurenza, Monica and Stumpo, Mirko and Zucca, Pietro and Mancini, Mattia and Benella, Simone and Clark, Liam and Alberti, Tommaso and Marcucci, Maria Federica},
  journal={Journal of Space Weather and Space Climate},
  volume={14},
  pages={8},
  year={2024},
  publisher={EDP Sciences}
}

@article{kasapis2025reconstructing,
  title={Reconstructing PSP/ISOIS Pitch Angle Resolved Intensities Using Unsupervised Learning},
  author={Kasapis, Spiridon and Cuesta, Manuel E. and Khoo, Leng Ying and Farooki, Hameedullah A. and Pak, Sungmin and Szalay, Jamey R. and Shen, Mitchell M. and Rankin, Jamie S. and Hristopulos, Dionissios T. and Livadiotis, George and McComas, David J.},
  journal={The Astrophysical Journal},
  year={2025},
  note={Accepted; in production. Manuscript ID: AAS68391R1},
}

@article{narock2022supporting,
  title={Supporting responsible machine learning in heliophysics},
  author={Narock, Ayris and Bard, Christopher and Thompson, Barbara J and Halford, Alexa J and McGranaghan, Ryan M and da Silva, Daniel and Kosar, Burcu and Shumko, Mykhaylo},
  journal={Frontiers in Astronomy and Space Sciences},
  volume={9},
  pages={1064233},
  year={2022},
  publisher={Frontiers Media SA}
}

@article{camporeale2020ml,
  title={Ml-Helio: An emerging community at the intersection between heliophysics and machine learning},
  author={Camporeale, Enrico and Scientific Organizing Committee of ML-Helio},
  journal={Journal of Geophysical Research: Space Physics},
  volume={125},
  number={2},
  pages={e2019JA027502},
  year={2020},
  publisher={Wiley Online Library}
}

@book{berger2021machine,
  title={Machine Learning in Heliophysics},
  author={Berger, Thomas and Camporeale, Enrico and Poduval, Bala and Delouille, Veronique A and Murray, Sophie A},
  year={2021},
  publisher={Frontiers Media SA}
}

@article{nita2020machine,
  title={Machine learning in heliophysics and space weather forecasting: a white paper of findings and recommendations},
  author={Nita, Gelu and Georgoulis, Manolis and Kitiashvili, Irina and Sadykov, Viacheslav and Camporeale, Enrico and Kosovichev, Alexander and Wang, Haimin and Oria, Vincent and Wang, Jason and Angryk, Rafal and others},
  journal={arXiv preprint arXiv:2006.12224},
  year={2020}
}

@article{zheng2023comparative,
  title={Comparative analysis of machine learning models for solar flare prediction},
  author={Zheng, Yanfang and Qin, Weishu and Li, Xuebao and Ling, Yi and Huang, Xusheng and Li, Xuefeng and Yan, Pengchao and Yan, Shuainan and Lou, Hengrui},
  journal={Astrophysics and Space Science},
  volume={368},
  number={7},
  pages={53},
  year={2023},
  publisher={Springer}
}

@article{bobra2016predicting,
  title={Predicting coronal mass ejections using machine learning methods},
  author={Bobra, Monica G and Ilonidis, Stathis},
  journal={The Astrophysical Journal},
  volume={821},
  number={2},
  pages={127},
  year={2016},
  publisher={IOP Publishing}
}

@article{rodriguez2024hemispheric,
  title={Hemispheric sunspot number prediction for solar cycles 25 and 26 using spectral analysis and machine learning techniques},
  author={Rodr{\'\i}guez, Jos{\'e}-V{\'\i}ctor and S{\'a}nchez Carrasco, V{\'\i}ctor Manuel and Rodr{\'\i}guez-Rodr{\'\i}guez, Ignacio and P{\'e}rez Aparicio, Alejandro Jes{\'u}s and Vaquero, Jos{\'e} Manuel},
  journal={Solar Physics},
  volume={299},
  number={8},
  pages={116},
  year={2024},
  publisher={Springer}
}

@article{kasapis2025prediction,
  title={Prediction of Intensity Variations Associated with Emerging Active Regions using Helioseismic Power Maps and Machine Learning},
  author={Kasapis, Spiridon and Kitiashvili, Irina N and Kosovichev, Alexander G and Stefan, John T},
  journal={The Astrophysical Journal Supplement Series},
  volume={280},
  number={2},
  pages={64},
  year={2025},
  publisher={IOP Publishing}
}

@article{roy2025surya,
  title={Surya: Foundation Model for Heliophysics},
  author={Roy, Sujit and Schmude, Johannes and Lal, Rohit and Gaur, Vishal and Freitag, Marcus and Kuehnert, Julian and van Kessel, Theodore and Hegde, Dinesha V and Mu{\~n}oz-Jaramillo, Andr{\'e}s and Jakubik, Johannes and others},
  journal={arXiv preprint arXiv:2508.14112},
  year={2025}
}

@article{roy2025suryabench,
  title={SuryaBench: Benchmark Dataset for Advancing Machine Learning in Heliophysics and Space Weather Prediction},
  author={Roy, Sujit and Hegde, Dinesha V and Schmude, Johannes and Lin, Amy and Gaur, Vishal and Lal, Rohit and Mandal, Kshitiz and Singh, Talwinder and Mu{\~n}oz-Jaramillo, Andr{\'e}s and Yang, Kang and others},
  journal={arXiv preprint arXiv:2508.14107},
  year={2025}
}

@article{ali2025forecasting,
  title={Forecasting solar energetic particles using multi-source data from solar flares, CMEs, and radio bursts with machine learning approaches},
  author={Ali, Mohammed AbuBakr and Abdelkawy, Ali GA and Shaltout, Abdelrazek MK and Beheary, MM},
  journal={Scientific Reports},
  volume={15},
  number={1},
  pages={9546},
  year={2025},
  publisher={Nature Publishing Group UK London}
}

@article{van2013lofar,
  title={LOFAR: The low-frequency array},
  author={van Haarlem, Michael P and Wise, Michael W and Gunst, AW and Heald, George and McKean, John P and Hessels, Jason WT and de Bruyn, A Ger and Nijboer, Ronald and Swinbank, John and Fallows, Richard and others},
  journal={Astronomy \& astrophysics},
  volume={556},
  pages={A2},
  year={2013},
  publisher={EDP Sciences}
}

@article{rodriguez2010east,
  title={The east-west effect in solar proton flux measurements in geostationary orbit: A new GOES capability},
  author={Rodriguez, JV and Onsager, TG and Mazur, JE},
  journal={Geophysical Research Letters},
  volume={37},
  number={7},
  year={2010},
  publisher={Wiley Online Library}
}

@article{hu2022calibration,
  title={Calibration of the GOES 6--16 high-energy proton detectors based on modelling of ground level enhancement energy spectra},
  author={Hu, Shaowen and Semones, Edward},
  journal={Journal of Space Weather and Space Climate},
  volume={12},
  pages={5},
  year={2022},
  publisher={EDP Sciences}
}

@inproceedings{sellers1996design,
  title={Design and calibration of the GOES-8 particle sensors: The EPS and HEPAD},
  author={Sellers, Francis Bach and Hanser, Frederick A},
  booktitle={GOES-8 and Beyond},
  volume={2812},
  pages={353--364},
  year={1996},
  organization={SPIE}
}

@incollection{pesnell2012solar,
  title={The solar dynamics observatory (SDO)},
  author={Pesnell, W Dean and Thompson, B Jꎬ and Chamberlin, PC},
  booktitle={The solar dynamics observatory},
  pages={3--15},
  year={2012},
  publisher={Springer}
}

@article{stone1998advanced,
  title={The advanced composition explorer},
  author={Stone, Edward C and Frandsen, AM and Mewaldt, RA and Christian, ER and Margolies, D and Ormes, JF and Snow, F},
  journal={Space Science Reviews},
  volume={86},
  number={1},
  pages={1--22},
  year={1998},
  publisher={Springer}
}

@article{von1995energetic,
  title={The energetic particles: acceleration, composition, and transport (EPACT) investigation on the wind spacecraft},
  author={Von Rosenvinge, TT and Barbier, LM and Karsch, J and Liberman, R and Madden, MP and Nolan, T and Reames, DV and Ryan, L and Singh, S and Trexel, H and others},
  journal={Space Science Reviews},
  volume={71},
  number={1},
  pages={155--206},
  year={1995},
  publisher={Springer}
}

@inproceedings{burt2012deep,
  title={Deep space climate observatory: The DSCOVR mission},
  author={Burt, Joe and Smith, Bob},
  booktitle={2012 ieee aerospace conference},
  pages={1--13},
  year={2012},
  organization={IEEE}
}

@article{domingo1995soho,
  title={The SOHO mission: an overview},
  author={Domingo, V and Fleck, Bernhard and Poland, Arthur I},
  journal={Solar Physics},
  volume={162},
  number={1},
  pages={1--37},
  year={1995},
  publisher={Springer}
}

@article{domingo1995soho2,
  title={SOHO: the solar and heliospheric observatory},
  author={Domingo, V and Fleck, Bernhard and Poland, AI},
  journal={Space Science Reviews},
  volume={72},
  number={1},
  pages={81--84},
  year={1995},
  publisher={Springer}
}

@article{kaiser2008stereo,
  title={The STEREO mission: An introduction},
  author={Kaiser, Michael L and Kucera, TA and Davila, JM and St. Cyr, OC and Guhathakurta, Madhulika and Christian, Eric},
  journal={Space Science Reviews},
  volume={136},
  number={1},
  pages={5--16},
  year={2008},
  publisher={Springer}
}

@article{scherrer1995solar,
  title={The solar oscillations investigation-Michelson Doppler imager},
  author={Scherrer, P Hꎬ and Bogart, R Sꎬ and Bush, RI and Hoeksema, JTc-a and Kosovichev, AG and Schou, J and Rosenberg, W and Springer, L and Tarbell, TD and Title, A and others},
  journal={Solar Physics},
  volume={162},
  number={1},
  pages={129--188},
  year={1995},
  publisher={Springer}
}

@article{scherrer2012helioseismic,
  title={The helioseismic and magnetic imager (HMI) investigation for the solar dynamics observatory (SDO)},
  author={Scherrer, Philip Hanby and Schou, Jesper and Bush, RI and Kosovichev, AG and Bogart, RS and Hoeksema, JT and Liu, Y and Duvall Jr, TL and Zhao, J and Title, AM and others},
  journal={Solar Physics},
  volume={275},
  number={1},
  pages={207--227},
  year={2012},
  publisher={Springer}
}

@data{hutchins2026,
    author = {Hutchins, Tate and Spiridon Kasapis},
    publisher = {AGU},
    title = {{Solar Energetic Particle Prediction Using Deep Learning and PSP/ISOIS Data}},
    UNF = {UNF:6:xn108cK7YYF39LpktDz54g==},
    year = {2026},
    version = {V1},
    doi = {10.0000/XXX/XXXXXX},
    url = {https://github.com/thutch17/PSP-SEP-Event-Prediction}
    }

@article{doswell1990summary,
  title={On summary measures of skill in rare event forecasting based on contingency tables},
  author={Doswell, CHARL and Davies-Jones, Robert and Keller, David L},
  journal={Weather and forecasting},
  volume={5},
  number={4},
  pages={576--585},
  year={1990}
}

@article{wehling2011probability,
  title={Probability of detection (POD) as a statistical model for the validation of qualitative methods},
  author={Wehling, Paul and LaBudde, Robert A and Brunelle, Sharon L and Nelson, Maria T},
  journal={Journal of AOAC International},
  volume={94},
  number={1},
  pages={335--347},
  year={2011},
  publisher={Oxford University Press}
}

@article{macmillan1985detection,
  title={Detection theory analysis of group data: estimating sensitivity from average hit and false-alarm rates.},
  author={Macmillan, Neil A and Kaplan, Howard L},
  journal={Psychological bulletin},
  volume={98},
  number={1},
  pages={185},
  year={1985},
  publisher={American Psychological Association}
}

@article{lipton2014thresholding,
  title={Thresholding classifiers to maximize F1 score},
  author={Lipton, Zachary C and Elkan, Charles and Narayanaswamy, Balakrishnan},
  journal={stat},
  volume={1050},
  pages={14},
  year={2014}
}

@article{muschelli2020roc,
  title={ROC and AUC with a binary predictor: a potentially misleading metric},
  author={Muschelli III, John},
  journal={Journal of classification},
  volume={37},
  number={3},
  pages={696--708},
  year={2020},
  publisher={Springer}
}

@inproceedings{keegan2025data,
  title={Data-driven solar surface flux transport modeling with uncertainty quantification},
  author={Keegan, Katherine and Bonaventura, Nina and Guzm{\'a}n, Plinio and Karna, Nishu and Hess-Webber, Shea and Kasapis, Spiridon and Jha, Bibhuti Kumar and Mu{\~n}oz-Jaramillo, Andr{\'e}s},
  booktitle={NeurIPS 2025 AI for Science Workshop},
  year={2025}
}

@article{fox2016solar,
  title={The solar probe plus mission: humanity’s first visit to our star},
  author={Fox, NJ and Velli, MC and Bale, SD and Decker, R and Driesman, A and Howard, RA and Kasper, Justin C and Kinnison, J and Kusterer, M and Lario, D and others},
  journal={Space Science Reviews},
  volume={204},
  number={1},
  pages={7--48},
  year={2016},
  publisher={Springer}
}

@article{raouafi2023parker,
  title={Parker solar probe: Four years of discoveries at solar cycle minimum},
  author={Raouafi, Nour E and Matteini, L and Squire, J and Badman, ST and Velli, M and Klein, KG and Chen, CHK and Matthaeus, WH and Szabo, A and Linton, M and others},
  journal={Space Science Reviews},
  volume={219},
  number={1},
  pages={8},
  year={2023},
  publisher={Springer}
}

@article{simunac2004solar,
  title={Solar cycle variations in solar and interplanetary ions observed with Interplanetary Monitoring Platform 8},
  author={Simunac, KDC and Armstrong, TP},
  journal={Journal of Geophysical Research: Space Physics},
  volume={109},
  number={A10},
  year={2004},
  publisher={Wiley Online Library}
}

@article{lakshminarayanan2017simple,
  title={Simple and scalable predictive uncertainty estimation using deep ensembles},
  author={Lakshminarayanan, Balaji and Pritzel, Alexander and Blundell, Charles},
  journal={Advances in neural information processing systems},
  volume={30},
  year={2017}
}

@article{schwadron2010earth,
  title={Earth-Moon-Mars radiation environment module framework},
  author={Schwadron, Nathan A and Townsend, L and Kozarev, Kamen and Dayeh, MA and Cucinotta, Francis and Desai, Mihir and Golightly, Michael and Hassler, D and Hatcher, R and Kim, M-Y and others},
  journal={Space Weather},
  volume={8},
  number={1},
  year={2010},
  publisher={Wiley Online Library}
}

@inproceedings{deforest2022polarimeter,
  title={Polarimeter to unify the corona and heliosphere (punch): Science, status, and path to flight},
  author={DeForest, Craig and Killough, Ronnie and Gibson, Sarah and Henry, Alan and Case, Traci and Beasley, Matthew and Laurent, Glenn and Colaninno, Robin and Waltham, Nick},
  booktitle={2022 IEEE Aerospace Conference (AERO)},
  pages={1--11},
  year={2022},
  organization={IEEE}
}

@article{mccomas2018interstellar,
  title={Interstellar mapping and acceleration probe (IMAP): A new NASA mission},
  author={McComas, DJ and Christian, Eric R and Schwadron, Nathan A and Fox, N and Westlake, J and Allegrini, F and Baker, DN and Biesecker, D and Bzowski, M and Clark, G and others},
  journal={Space science reviews},
  volume={214},
  number={8},
  pages={116},
  year={2018},
  publisher={Springer}
}

@article{mccomas2025interstellar,
  title={Interstellar mapping and acceleration probe: The NASA IMAP mission},
  author={McComas, David J and Christian, Eric R and Schwadron, NA and Gkioulidou, Matina and Allegrini, F and Baker, DN and Bzowski, M and Clark, G and Cohen, CMS and Cohen, I and others},
  journal={Space science reviews},
  volume={221},
  number={8},
  pages={100},
  year={2025},
  publisher={Springer}
}

@article{lee2025space,
  title={Space weather science to enhance forecasting with the NASA IMAP Active Link for Real-Time (I-ALiRT) system},
  author={Lee, Christina O and Christian, Eric R and Sandoval, Laura and Crabtree, Alastair and Desai, Mihir I and Gkioulidou, Matina and Heber, Bernd and Horbury, Timothy and Kistler, Lynn and Knuth, Jenny and others},
  journal={Space Science Reviews},
  volume={221},
  number={8},
  pages={117},
  year={2025},
  publisher={Springer},
  doi={10.1007/s11214-025-01244-9}
}

@inproceedings{he2016deep,
  title={Deep residual learning for image recognition},
  author={He, Kaiming and Zhang, Xiangyu and Ren, Shaoqing and Sun, Jian},
  booktitle={Proceedings of the IEEE conference on computer vision and pattern recognition},
  pages={770--778},
  year={2016}
}

@article{vaswani2017attention,
  title={Attention is all you need},
  author={Vaswani, Ashish and Shazeer, Noam and Parmar, Niki and Uszkoreit, Jakob and Jones, Llion and Gomez, Aidan N and Kaiser, {\L}ukasz and Polosukhin, Illia},
  journal={Advances in neural information processing systems},
  volume={30},
  year={2017}
}

@article{raissi2019physics,
  title={Physics-informed neural networks: A deep learning framework for solving forward and inverse problems involving nonlinear partial differential equations},
  author={Raissi, Maziar and Perdikaris, Paris and Karniadakis, George E},
  journal={Journal of Computational physics},
  volume={378},
  pages={686--707},
  year={2019},
  publisher={Elsevier}
}

@article{karniadakis2021physics,
  title={Physics-informed machine learning},
  author={Karniadakis, George Em and Kevrekidis, Ioannis G and Lu, Lu and Perdikaris, Paris and Wang, Sifan and Yang, Liu},
  journal={Nature Reviews Physics},
  volume={3},
  number={6},
  pages={422--440},
  year={2021},
  publisher={Nature Publishing Group UK London}
}

@article{li2025evaluating,
  title={Evaluating time-series augmentation techniques for deep learning--based solar flare prediction},
  author={Li, Peiyu and Bahri, Omar and Boubrahimi, Souka{\"\i}na Filali and Hamdi, Shah Muhammad},
  journal={The Astrophysical Journal Supplement Series},
  volume={280},
  number={2},
  pages={52},
  year={2025},
  publisher={IOP Publishing}
}

@inproceedings{wen2024class,
  title={Class-Based Time Series Data Augmentation to Mitigate Extreme Class Imbalance for Solar Flare Prediction},
  author={Wen, Junzhi and Angryk, Rafal A},
  booktitle={International Conference on Artificial Intelligence and Soft Computing},
  pages={362--375},
  year={2024},
  organization={Springer}
}

@article{grim2024solar,
  title={Solar flare forecasting based on magnetogram sequences learning with multiscale vision transformers and data augmentation techniques},
  author={Grim, Lu{\'\i}s Fernando L and Gradvohl, Andr{\'e} Leon S},
  journal={Solar Physics},
  volume={299},
  number={3},
  pages={33},
  year={2024},
  publisher={Springer}
}

@inproceedings{bahri2023shapelet,
  title={Shapelet-Preserving Bootstrapping For Time Series Data Augmentation},
  author={Bahri, Omar and Li, Peiyu and Hosseinzadeh, Pouya and Boubrahimi, Souka{\"\i}na Filali and Hamdi, Shah Muhammad},
  booktitle={2023 International Conference on Machine Learning and Applications (ICMLA)},
  pages={453--458},
  year={2023},
  organization={IEEE}
}

@article{pacheco2019analysis,
  title={Analysis and modelling of the solar energetic particle radiation environment in the inner heliosphere in preparation for solar orbiter},
  author={Pacheco Mateo, Daniel},
  year={2019},
  publisher={Universitat de Barcelona}
}

@article{pak2025species,
  title={Species-dependent Variability in the Energy Spectra of Intense Solar Energetic Particle Events Observed by PSP/ISʘIS/EPI-Hi/LET},
  author={Pak, S and Cuesta, ME and Farooki, HA and Khoo, LY and Xu, ZG and Davis, AJ and Leske, RA and Cohen, CMS and McComas, DJ and Shrestha, BL and others},
  journal={The Astrophysical Journal Supplement Series},
  volume={281},
  number={1},
  pages={21},
  year={2025},
  publisher={The American Astronomical Society}
}

@article{kosovichev2025structure,
  title={Structure and dynamics of the Sun’s interior revealed by the Helioseismic and Magnetic Imager},
  author={Kosovichev, Alexander G and Basu, Sarbani and Bekki, Yuto and Buitrago-Casas, Juan Camilo and Chatzistergos, Theodosios and Chen, Ruizhu and Christensen-Dalsgaard, J{\o}rgen and Donea, Alina and Fleck, Bernhard and Fournier, Damien and others},
  journal={Solar Physics},
  volume={300},
  number={5},
  pages={70},
  year={2025},
  publisher={Springer}
}

@article{jiao2020solar,
  title={Solar flare intensity prediction with machine learning models},
  author={Jiao, Zhenbang and Sun, Hu and Wang, Xiantong and Manchester, Ward and Gombosi, Tamas and Hero, Alfred and Chen, Yang},
  journal={Space weather},
  volume={18},
  number={7},
  pages={e2020SW002440},
  year={2020},
  publisher={Wiley Online Library}
}

@article{wang2020predicting,
  title={Predicting solar flares with machine learning: Investigating solar cycle dependence},
  author={Wang, Xiantong and Chen, Yang and Toth, Gabor and Manchester, Ward B and Gombosi, Tamas I and Hero, Alfred O and Jiao, Zhenbang and Sun, Hu and Jin, Meng and Liu, Yang},
  journal={The Astrophysical Journal},
  volume={895},
  number={1},
  pages={3},
  year={2020},
  publisher={The American Astronomical Society}
}

@article{singh2023improving,
  title={Improving the arrival time estimates of coronal mass ejections by using magnetohydrodynamic ensemble modeling, heliospheric imager data, and machine learning},
  author={Singh, Talwinder and Benson, Bernard and Raza, Syed AZ and Kim, Tae K and Pogorelov, Nikolai V and Smith, William P and Arge, Charles N},
  journal={The Astrophysical Journal},
  volume={948},
  number={2},
  pages={78},
  year={2023},
  publisher={The American Astronomical Society}
}

@article{vourlidas2019predicting,
  title={Predicting the geoeffective properties of coronal mass ejections: current status, open issues and path forward},
  author={Vourlidas, A and Patsourakos, S and Savani, NP},
  journal={Philosophical Transactions of the Royal Society A: Mathematical, Physical and Engineering Sciences},
  volume={377},
  number={2148},
  year={2019},
  publisher={The Royal Society}
}

@article{sierra2024predicting,
  title={Predicting sunspot number from topological features in spectral images I: Machine learning approach},
  author={Sierra-Porta, David and Tarazona-Alvarado, Miguel and Acevedo, DD Herrera},
  journal={Astronomy and Computing},
  volume={48},
  pages={100857},
  year={2024},
  publisher={Elsevier}
}

@article{qamar2025prediction,
  title={Prediction of sunspot numbers via Weibull distribution and deep learning},
  author={Qamar, Waqas and Hussain, Majid and Zaheer, M Basit and Akram, Jawaid and Sadiq, Naeem and Uddin, Zaheer},
  journal={Astrophysics and Space Science},
  volume={370},
  number={7},
  pages={68},
  year={2025},
  publisher={Springer}
}

@article{kasapis2023predicting,
  title={Predicting the emergence of solar active regions using machine learning},
  author={Kasapis, Spiridon and Kitiashvili, Irina N and Kosovichev, Alexander G and Stefan, John T and Apte, Bhairavi},
  journal={Proceedings of the International Astronomical Union},
  volume={19},
  number={S365},
  pages={311--319},
  year={2023},
  publisher={Cambridge University Press}
}

@article{jeong2025prediction,
  title={Prediction of the Next Solar Rotation Synoptic Maps Using an Artificial Intelligence--based Surface Flux Transport Model},
  author={Jeong, Hyun-Jin and Jeon, Mingyu and Kim, Daeil and Kim, Youngjae and Baek, Ji-Hye and Moon, Yong-Jae and Choi, Seonghwan},
  journal={The Astrophysical Journal Supplement Series},
  volume={278},
  number={1},
  pages={5},
  year={2025},
  publisher={The American Astronomical Society}
}

@article{tirona2026forecasting,
  title={Forecasting Continuum Intensity for Solar Active Region Emergence Prediction using Transformers},
  author={Tirona, Jonas and Patil, Sarang and Kasapis, Spiridon and Dogan, Eren and Stefan, John and Kitiashvili, Irina N and Kosovichev, Alexander G and Xu, Mengjia},
  journal={arXiv preprint arXiv:2601.13144},
  year={2026}
}

@article{woods2021unsupervised,
  title={Unsupervised machine learning for the identification of preflare spectroscopic signatures},
  author={Woods, Magnus M and Sainz Dalda, Alberto and De Pontieu, Bart},
  journal={The Astrophysical Journal},
  volume={922},
  number={2},
  pages={137},
  year={2021},
  publisher={The American Astronomical Society}
}

@article{giger2024unsupervised,
  title={Unsupervised anomaly detection with Variational AutoEncoders applied to full-disk solar images},
  author={Giger, Marius and Csillaghy, Andr{\'e}},
  journal={Space Weather},
  volume={22},
  number={2},
  pages={e2023SW003516},
  year={2024},
  publisher={Wiley Online Library}
}

@article{lemen2012atmospheric,
  title={The atmospheric imaging assembly (AIA) on the solar dynamics observatory (SDO)},
  author={Lemen, James R and Title, Alan M and Akin, David J and Boerner, Paul F and Chou, Catherine and Drake, Jerry F and Duncan, Dexter W and Edwards, Christopher G and Friedlaender, Frank M and Heyman, Gary F and others},
  journal={Solar Physics},
  volume={275},
  number={1},
  pages={17--40},
  year={2012},
  publisher={Springer}
}

@article{brueckner1995large,
  title={The large angle spectroscopic coronagraph (LASCO) visible light coronal imaging and spectroscopy},
  author={Brueckner, GE and Howard, RA and Koomen, MJ and Korendyke, CM and Michels, DJ and Moses, JD and Socker, DG and Dere, KP and Lamy, PL and Llebaria, A and others},
  journal={Solar Physics},
  volume={162},
  number={1},
  pages={357--402},
  year={1995},
  publisher={Springer}
}

@article{woods2024goes,
  title={GOES-R Series X-Ray Sensor (XRS): 1. Design and pre-flight calibration},
  author={Woods, Thomas N and Eden, Thomas and Eparvier, Francis G and Jones, Andrew R and Woodraska, Donald L and Chamberlin, Phillip C and Machol, Janet L},
  journal={Journal of Geophysical Research: Space Physics},
  volume={129},
  number={11},
  pages={e2024JA032925},
  year={2024},
  publisher={Wiley Online Library}
}

@article{muller_solar_2020,
	title = {The {Solar} {Orbiter} mission: {Science} overview},
	volume = {642},
	copyright = {https://www.edpsciences.org/en/authors/copyright-and-licensing},
	issn = {0004-6361, 1432-0746},
	shorttitle = {The {Solar} {Orbiter} mission},
	url = {https://www.aanda.org/10.1051/0004-6361/202038467},
	doi = {10.1051/0004-6361/202038467},
	language = {en},
	urldate = {2025-10-01},
	journal = {Astronomy \& Astrophysics},
	author = {Müller, D. and St. Cyr, O. C. and Zouganelis, I. and Gilbert, H. R. and Marsden, R. and Nieves-Chinchilla, T. and Antonucci, E. and Auchère, F. and Berghmans, D. and Horbury, T. S. and Howard, R. A. and Krucker, S. and Maksimovic, M. and Owen, C. J. and Rochus, P. and Rodriguez-Pacheco, J. and Romoli, M. and Solanki, S. K. and Bruno, R. and Carlsson, M. and Fludra, A. and Harra, L. and Hassler, D. M. and Livi, S. and Louarn, P. and Peter, H. and Schühle, U. and Teriaca, L. and Del Toro Iniesta, J. C. and Wimmer-Schweingruber, R. F. and Marsch, E. and Velli, M. and De Groof, A. and Walsh, A. and Williams, D.},
	month = oct,
	year = {2020},
	note = {867 citations (NASA ADS/DOI) [2025-10-01]},
	pages = {A1},
}

@article{marirrodriga2021solar,
  title={Solar Orbiter: Mission and spacecraft design},
  author={Marirrodriga, C Garc{\'\i}a and Pacros, A and Strandmoe, S and Arcioni, M and Arts, A and Ashcroft, C and Ayache, L and Bonnefous, Y and Brahimi, N and Cipriani, F and others},
  journal={Astronomy \& Astrophysics},
  volume={646},
  pages={A121},
  year={2021},
  publisher={EDP Sciences}
}

@article{tripathi2022aditya,
  title={The Aditya-L1 mission of ISRO},
  author={Tripathi, Durgesh and Chakrabarty, Dibyendu and Nandi, A and Prasad, B Raghvendra and Ramaprakash, AN and Shaji, Nigar and Sankarasubramanian, K and Thampi, R Satheesh and Yadav, VK},
  journal={Proceedings of the International Astronomical Union},
  volume={18},
  number={S372},
  pages={17--27},
  year={2022},
  publisher={Cambridge University Press}
}

@article{hanser2011eps,
  title={EPS/HEPAD calibration and data handbook},
  author={Hanser, FA},
  journal={Tech. Rep. GOESN-ENG-048D},
  year={2011}
}

@inproceedings{chamberlin2009next,
  title={Next generation X-ray sensor (XRS) for the NOAA GOES-R satellite series},
  author={Chamberlin, Phillip C and Woods, Thomas N and Eparvier, Francis G and Jones, Andrew R},
  booktitle={Solar physics and space weather instrumentation III},
  volume={7438},
  pages={11--20},
  year={2009},
  organization={SPIE}
}

@inproceedings{joselyn1985space,
  title={The space environment monitors onboard GOES},
  author={JOSELYN, J and GRUBB, R},
  booktitle={23rd Aerospace Sciences Meeting},
  pages={238},
  year={1985}
}

@article{bougeret1995waves,
  title={WAVES: The radio and plasma wave investigation on the Wind spacecraft},
  author={Bougeret, J-L and Kaiser, M Lꎬ and Kellogg, Paul J and Manning, Robert and Goetz, K and Monson, SJ and Monge, N and Friel, L and Meetre, CA and Perche, C and others},
  journal={Space Science Reviews},
  volume={71},
  number={1},
  pages={231--263},
  year={1995},
  publisher={Springer}
}

@article{panigrahi2018rank,
  title={Rank allocation to J48 group of decision tree classifiers using binary and multiclass intrusion detection datasets},
  author={Panigrahi, Ranjit and Borah, Samarjeet},
  journal={Procedia computer science},
  volume={132},
  pages={323--332},
  year={2018},
  publisher={Elsevier}
}

@article{ryan2012thermal,
  title={The thermal properties of solar flares over three solar cycles using GOES X-ray observations},
  author={Ryan, Daniel F and Milligan, Ryan O and Gallagher, Peter T and Dennis, Brian R and Kim Tolbert, A and Schwartz, Richard A and Alex Young, C},
  journal={The Astrophysical Journal Supplement Series},
  volume={202},
  number={2},
  pages={11},
  year={2012},
  publisher={The American Astronomical Society}
}

@article{ryan2013tebbs,
  title={TEBBS: A New Automatic Method for Calculating Background-Subtracted Thermal Flare Properties Using GOES/XRS},
  author={Ryan, Daniel F and Milligan, Ryan O and Gallagher, Peter T and Dennis, Brian R and Tolbert, A Kim and Schwartz, Richard A and Young, C Alex},
  journal={SDO-3: Exploring the Network of SDO Science},
  pages={143},
  year={2013}
}

@article{delaboudiniere1995eit,
  title={EIT: extreme-ultraviolet imaging telescope for the SOHO mission},
  author={Delaboudiniere, J-P and Artzner, GE and Brunaud, J and Gabriel, Alan H and Hochedez, Jean-Francois and Millier, F and Song, XY and Au, B and Dere, KP and Howard, Russell A and others},
  journal={Solar Physics},
  volume={162},
  number={1},
  pages={291--312},
  year={1995},
  publisher={Springer}
}

@inproceedings{meyer2021alternative,
  title={An alternative probabilistic interpretation of the huber loss},
  author={Meyer, Gregory P},
  booktitle={Proceedings of the ieee/cvf conference on computer vision and pattern recognition},
  pages={5261--5269},
  year={2021}
}

@article{mccomas2016integrated,
  title={Integrated Science Investigation of the Sun (ISIS): Design of the energetic particle investigation},
  author={McComas, DJ and Alexander, N and Angold, N and Bale, S and Beebe, C and Birdwell, B and Boyle, M and Burgum, JM and Burnham, JA and Christian, ER and others},
  journal={Space Science Reviews},
  volume={204},
  pages={187--256},
  year={2016},
  publisher={Springer}
}

@ARTICLE{HillEA2017JGRA_PSP_ISOIS_EPILo,
       author = {{Hill}, M.~E. and {Mitchell}, D.~G. and {Andrews}, G.~B. and {Cooper}, S.~A. and {Gurnee}, R.~S. and {Hayes}, J.~R. and {Layman}, R.~S. and {McNutt}, R.~L. and {Nelson}, K.~S. and {Parker}, C.~W. and {Schlemm}, C.~E. and {Stokes}, M.~R. and {Begley}, S.~M. and {Boyle}, M.~P. and {Burgum}, J.~M. and {Do}, D.~H. and {Dupont}, A.~R. and {Gold}, R.~E. and {Haggerty}, D.~K. and {Hoffer}, E.~M. and {Hutcheson}, J.~C. and {Jaskulek}, S.~E. and {Krimigis}, S.~M. and {Liang}, S.~X. and {London}, S.~M. and {Noble}, M.~W. and {Roelof}, E.~C. and {Seifert}, H. and {Strohbehn}, K. and {Vandegriff}, J.~D. and {Westlake}, J.~H.},
        title = "{The Mushroom: A half-sky energetic ion and electron detector}",
      journal = {Journal of Geophysical Research (Space Physics)},
         year = 2017,
        month = feb,
       volume = {122},
       number = {2},
        pages = {1513-1530},
          doi = {10.1002/2016JA022614},
       adsurl = {https://ui.adsabs.harvard.edu/abs/2017JGRA..122.1513H}
}

@article{cuesta2025comparing,
  title={Comparing Methods for Calculating Solar Energetic Particle Intensities: Rebinning versus Spectral Binning},
  author={Cuesta, ME and Khoo, LY and Livadiotis, G and Shen, MM and Szalay, JR and McComas, DJ and Rankin, JS and Bandyopadhyay, R and Farooki, HA and Niehof, JT and others},
  journal={The Astrophysical Journal},
  volume={980},
  number={2},
  pages={235},
  year={2025},
  publisher={IOP Publishing}
}

@article{kasper2016solar,
  title={Solar wind electrons alphas and protons (SWEAP) investigation: Design of the solar wind and coronal plasma instrument suite for solar probe plus},
  author={Kasper, Justin C and Abiad, Robert and Austin, Gerry and Balat-Pichelin, Marianne and Bale, Stuart D and Belcher, John W and Berg, Peter and Bergner, Henry and Berthomier, Matthieu and Bookbinder, Jay and others},
  journal={Space Science Reviews},
  volume={204},
  number={1},
  pages={131--186},
  year={2016},
  publisher={Springer}
}

@article{livi2022solar,
  title={The solar probe analyzer—ions on the Parker Solar Probe},
  author={Livi, Roberto and Larson, Davin E and Kasper, Justin C and Abiad, Robert and Case, Anthony W and Klein, Kristopher G and Curtis, David W and Dalton, Gregory and Stevens, Michael and Korreck, Kelly E and others},
  journal={The Astrophysical Journal},
  volume={938},
  number={2},
  pages={138},
  year={2022},
  publisher={IOP Publishing}
}

@article{queipo2005surrogate,
  title={Surrogate-based analysis and optimization},
  author={Queipo, Nestor V and Haftka, Raphael T and Shyy, Wei and Goel, Tushar and Vaidyanathan, Rajkumar and Tucker, P Kevin},
  journal={Progress in aerospace sciences},
  volume={41},
  number={1},
  pages={1--28},
  year={2005},
  publisher={Elsevier}
}

@book{forrester2008engineering,
  title={Engineering design via surrogate modelling: a practical guide},
  author={Forrester, Alexander and Sobester, Andras and Keane, Andy},
  year={2008},
  publisher={John Wiley \& Sons}
}

@article{yu2025solar,
  title={Solar Energetic Particle Forecasting with Multi-Task Deep Learning: SEPNET},
  author={Yu, Yian and Chen, Yang and Zhao, Lulu and Whitman, Kathryn and Manchester, Ward and Gombosi, Tamas},
  journal={arXiv preprint arXiv:2512.12786},
  year={2025}
}

@article{cohen2026imap,
  title={IMAP’s role in understanding particle injection and energization throughout the heliosphere},
  author={Cohen, CMS and Alterman, Benjamin L and Baker, Daniel N and Bruno, Alessandro and Bzowski, Maciej and Christian, Eric R and Cohen, IJ and Dalla, Silvia and Dayeh, MA and Desai, MI and others},
  journal={Space Science Reviews},
  volume={222},
  number={1},
  pages={6},
  year={2026},
  publisher={Springer}
}

@incollection{benesty2009pearson,
  title={Pearson correlation coefficient},
  author={Benesty, Jacob and Chen, Jingdong and Huang, Yiteng and Cohen, Israel},
  booktitle={Noise reduction in speech processing},
  pages={1--4},
  year={2009},
  publisher={Springer}
}

@article{kasapis2023turning,
  title={Turning noise into data: Characterization of the Van Allen radiation belt using SDO spikes data},
  author={Kasapis, Spiridon and Thompson, Barbara J and Rodriguez, Juan V and Attie, Raphael and Cucho-Padin, Gonzalo and Da Silva, Daniel and Jin, Meng and Pesnell, William D},
  journal={Space Weather},
  volume={21},
  number={3},
  pages={e2022SW003310},
  year={2023},
  publisher={Wiley Online Library}
}

@article{hodson2022root,
  title={Root mean square error (RMSE) or mean absolute error (MAE): When to use them or not},
  author={Hodson, Timothy O},
  journal={Geoscientific Model Development Discussions},
  volume={2022},
  pages={1--10},
  year={2022},
  publisher={G{\"o}ttingen, Germany}
}

@article{ash1999r2,
  title={R2: a useful measure of model performance when predicting a dichotomous outcome},
  author={Ash, Arlene and Shwartz, Michael},
  journal={Statistics in medicine},
  volume={18},
  number={4},
  pages={375--384},
  year={1999},
  publisher={Wiley Online Library}
}

@article{goel2010understanding,
  title={Understanding survival analysis: Kaplan-Meier estimate},
  author={Goel, Manish Kumar and Khanna, Pardeep and Kishore, Jugal},
  journal={International journal of Ayurveda research},
  volume={1},
  number={4},
  pages={274},
  year={2010}
}

@article{regnault202020,
  title={20 years of ACE data: How superposed epoch analyses reveal generic features in interplanetary CME profiles},
  author={Regnault, F and Janvier, M and D{\'e}moulin, Pascal and Auch{\`e}re, F and Strugarek, Antoine and Dasso, S and No{\^u}s, C},
  journal={Journal of Geophysical Research: Space Physics},
  volume={125},
  number={11},
  pages={e2020JA028150},
  year={2020},
  publisher={Wiley Online Library}
}

@article{richardson201425,
  title={> 25 MeV proton events observed by the high energy telescopes on the STEREO A and B spacecraft and/or at Earth during the first~ seven years of the STEREO mission},
  author={Richardson, IG and Von Rosenvinge, TT and Cane, HV and Christian, ER and Cohen, CMS and Labrador, AW and Leske, RA and Mewaldt, RA and Wiedenbeck, ME and Stone, EC},
  journal={Solar Physics},
  volume={289},
  number={8},
  pages={3059--3107},
  year={2014},
  publisher={Springer}
}

@article{bemporad2021possible,
  title={Possible advantages of a twin spacecraft Heliospheric mission at the Sun-Earth Lagrangian points L4 and L5},
  author={Bemporad, Alessandro},
  journal={Frontiers in Astronomy and Space Sciences},
  volume={8},
  pages={627576},
  year={2021},
  publisher={Frontiers Media SA}
}

@article{cho2023opening,
  title={Opening new horizons with the L4 mission: vision and plan},
  author={Cho, Kyung-Suk and Hwang, Junga and Han, Jeong-Yeol and Choi, Seong-Hwan and Park, Sung-Hong and Lim, Eun-Kyung and Kim, Rok-Soon and Seough, Jungjoon and Sohn, Jong-Dae and Song, Donguk and others},
  journal={Journal of the Korean Astronomical Society},
  volume={56},
  number={2},
  pages={263--275},
  year={2023},
  publisher={Korean Astronomical Society}
}

@article{posner2021multi,
  title={A multi-purpose heliophysics L4 mission},
  author={Posner, Arik and Arge, Charles Nickolos and Staub, Jan and StCyr, Orville C and Folta, David and Solanki, Sami K and Strauss, Roelf Du Toit and Effenberger, Frederic and Gandorfer, Achim and Heber, Bernd and others},
  journal={Space Weather},
  volume={19},
  number={9},
  pages={e2021SW002777},
  year={2021},
  publisher={Wiley Online Library}
}
\bibliographystyle{aasjournal}

\pagebreak


\appendix 


\section{Description of ML Models} \label{app:Description_of_Models}

The descriptions of the models presented here are based on the modelers' contributions and their answers to the questionnaire presented in Appendix~\ref{app:Questionnaire}. The models here appear in order of complexity, from less complex to deeper, as they are summarized in Table~\ref{tab:all_models}. Each of the following subsections (Sections \ref{sec:XGBoost}-\ref{sec:EPREM-S}) presents a single-page summary of the ML models that predict SEP events, along with a table of quantitative and qualitative characteristics of each models, as summarized in Table~\ref{tab:Qualitative_Table} and Figures~\ref{fig:Quantitative_Plot} and \ref{fig:Qualitative_Plot}.

\subsection{eXtreme Gradient Boosting (XGBoost) Model} \label{sec:XGBoost}

\textbf{Model Developers and Relevant Citation}: Aatiya Ali, Viacheslav Sadykov, Alexander Kosovichev, Irina N. Kitiashvili, Vincent Oria, Gelu M. Nita, Egor Illarionov, Patrick M. O'Keefe, Fraila Francis, Chun-Jie Chong, Paul Kosovich, and Russell D. Marroquin; \cite{ali2024predicting}.

\begin{table}[h] 
\caption{Model, Input and Output Specification Table for the XGBoost model.}
\label{tab:XGBoost}
\centering
\begin{tabular}{|lc|}
\hline
\multicolumn{2}{|c|}{Model}                                            \\ \hline
\multicolumn{1}{|l|}{Type}            & Decision Tree                  \\
\multicolumn{1}{|l|}{Complexity}      & 2                              \\ \hline
\multicolumn{2}{|c|}{Input}                                            \\ \hline
\multicolumn{1}{|l|}{Shape}           & Time Series (1D)               \\
\multicolumn{1}{|l|}{Type}            & Soft X-ray, Proton Flux        \\
\multicolumn{1}{|l|}{History}         & 33 years (1986-2019)           \\
\multicolumn{1}{|l|}{Diversity}       & 12484 samples                  \\
\multicolumn{1}{|l|}{Imbalance}       & 0.045 positive                 \\
\multicolumn{1}{|l|}{Sample Size}     & 240 bytes                      \\
\multicolumn{1}{|l|}{Sample Coverage} & 24 hours                       \\ \hline
\multicolumn{2}{|c|}{Output}                                           \\ \hline
\multicolumn{1}{|l|}{Prediction}      & Classification                 \\
\multicolumn{1}{|l|}{Type}            & Continuous                     \\
\multicolumn{1}{|l|}{Forecast Window} & 23 hours                       \\ \hline
\end{tabular}
\end{table}

\textbf{Summary}: The eXtreme Gradient Boosting (XGBoost; Figure~\ref{fig:S3_XGBoost}) model generates binary predictions of SPEs for the following day at Geostationary Earth Orbit (GEO). Comparing the performance of an SVM and XGBoost for these predictions, we find that XGBoost significantly outperforms SVM in most training-testing configurations based on metrics such as TSS, HSS, and recall. Using GOES proton and soft X-ray flux data spanning from 1986 to 2019. Data from SCs 22-24 are treated separately for training and testing. To address class imbalance, various oversampling and weight-balancing methods were tested, as well as model cross-cycle transferability.

\textbf{Model Description}: XGBoost ---an ensemble classifier based on gradient boosting, presents better results compared to SVMs (supervised classifiers using decision surfaces), across all tests: using default parameters, applying imbalance-handling class weights, and using data oversampled by standard (positive-class duplication), ADAptive SYNthetic (ADASYN), and Synthetic Minority Oversampling TEchniques (SMOTE) separately. Flux feature importance is determined using Gini importance, Fisher scoring, and the inherent feature ranking provided by XGBoost. Because our primary goal is not to parameterize a specific algorithm with minute detail, only parameters related to classification are optimized using Grid Search Cross-Validation (\textit{GridSearchCV}\footnote{\url{https://scikit-learn.org/stable/modules/generated/sklearn.model_selection.GridSearchCV.html}}).

\textbf{Inputs}: Model input features include statistics of (1-8 \AA) soft X-ray fluxes and proton fluxes $\geq 10$ MeV from the Energetic Particle Sensors \citep[EPS;][]{hanser2011eps} and Energetic Proton, Electron and Alpha Detectors \citep[EPEAD;][]{bruno2017} onboard the GOES missions. These features include daily flux mean, median, minimum, maximum, standard deviation, skewness, kurtosis, and the last measured flux of the previous day. The dataset exhibits a significant class imbalance, with 11,946 days classified as negative (no SPEs) and only 538 days classified as positive (SPEs detected).

\textbf{Outputs}: The model produces a daily binary flag, indicating whether an SPE is expected at GEO in the next 23 hours.

\textbf{Model Configuration}: The XGBoost model specifies two default parameters: booster = gbtree (to use tree-based models for ensemble building) and scale\_pos\_weight (to balance data classes).

\textbf{Model Validation and Results}: On average, XGBoost outperformed SVM by approximately +0.10 in TSS, +0.20 in HSS, and +0.10 in recall. While XGBoost showed higher recall values, it also produced a significant number of false positives. Evaluation of the XGBoost model across long (two SCs) and short (a single SC) training timescales shows that TSS and HSS were comparable for both timescales. The cross-cycle transferability studies the dependence of the results on properties of a solar cycle. Nonetheless, when compared to baseline models, such as SWPC daily probabilistic forecasts and a persistence model, XGBoost (optimized for TSS) outperformed these models in both TSS and recall. Overall, our results suggest that with proper tuning, XGBoost can enhance SPE prediction accuracy, particularly in refining all-clear predictions.

\textbf{Access to Model Data and Forecasts}: The SPE catalogs developed during this study are archived at: \url{https://sun.njit.edu/SEP3/datasets.html}. The GOES proton and soft X-ray flux data are available at \url{http://www.ncei.noaa.gov/data/goes-space-environment-monitor/access/avg/}. The XGBoost python package is accessible via \url{https://xgboost.readthedocs.io/en/stable/python/python\_intro.html}.

\textbf{Limitations, Caveats and Discussion}: This work builds on NOAA's classification of an S1 SPE where protons $\geq 10$ MeV exceed 10 pfu. The model results were tested and validated with GOES flux data from only SCs 22-24 individually across different training and testing configurations. Further work is needed to improve the model's ability to generalize across multiple SCs with varying levels of solar activity.


\subsection{Supervised Time Series Forest (STSF) Model} \label{sec:STSF}

\textbf{Model Developers and Relevant Citation}: Sumanth A. Rotti, Berkay Aydin, and Petrus C. Martens; \cite{rotti2024short, rotti2024precise} and \cite{loning2019sktime}.

\begin{table}[h]
\caption{Model, Input and Output Specification Table for the STSF model.}
\label{tab:STSF}
\centering
\begin{tabular}{|lc|}
\hline
\multicolumn{2}{|c|}{Model}                                            \\ \hline
\multicolumn{1}{|l|}{Type}            &  Forest                        \\
\multicolumn{1}{|l|}{Complexity}      &  4                             \\ \hline
\multicolumn{2}{|c|}{Input}                                            \\ \hline
\multicolumn{1}{|l|}{Shape}           &  Time Series (1D)              \\
\multicolumn{1}{|l|}{Type}            &  X-ray, Energetic Protons      \\
\multicolumn{1}{|l|}{History}         &  32 years (1986-2018)          \\
\multicolumn{1}{|l|}{Diversity}       &  998 samples                   \\
\multicolumn{1}{|l|}{Imbalance}       &  0.09 positive                 \\
\multicolumn{1}{|l|}{Sample Size}     &  25000 bytes                   \\
\multicolumn{1}{|l|}{Sample Coverage} &  11 hours                      \\ \hline
\multicolumn{2}{|c|}{Output}                                           \\ \hline
\multicolumn{1}{|l|}{Prediction}      &  Classification                \\
\multicolumn{1}{|l|}{Type}            &  Continuous                    \\
\multicolumn{1}{|l|}{Forecast Window} &  1 hour                        \\ \hline
\end{tabular}
\end{table}

\textbf{Summary}: In \cite{rotti2024short}, an ensemble modeling methodology is introduced, consisting of a feature-based multivariate variant of univariate time series classifiers to classify between strong and weak SEP events in the GSEP data set \citep{gsep_2022, rotti2022integrated, rotti2023analysis} covering SCs 22-24. The same model architecture was implemented in \cite{rotti2024precise} on an extended dataset comprising SEP-quiet samples. There are 2893 samples, of which 244 are strong SEP events (those crossing the SWPC's S1 threshold). Furthermore, a fixed input length of the time series (11 hours) is considered and the model is assessed for a prediction window of up to 60 minutes. The model's performance on an expanded dataset is promising, obtaining high skill scores.

\textbf{Model Description}: Both studies utilized a binary classification framework (SEPs vs. weak/non-SEPs) using an ensemble of univariate time series classifiers. The best-performing model is the Supervised Time Series Forest \cite[STSF][]{cabello2020fast}, which is described in this review. The STSF model employs three representations (time, frequency, and derivative) of the input time series and uses a supervised learning approach to find discriminatory intervals. The model computes the region of interest to highlight the location of discriminatory intervals, defined as the intersection of such intervals. Furthermore, it extracts seven statistical features, including mean, median, standard deviation, slope, minimum, maximum, and interquartile range, from each interval. The ranking of the interval feature is determined by a scoring function that indicates how effectively the feature distinguishes one class of time series from the other classes. The final set of intervals is obtained in a top-down approach to represent the entire series. The feature set is concatenated to form a new dataset upon which decision trees are built. The final output is based on the majority vote of averaged probability estimates from the individual estimators in the ensemble.

\textbf{Inputs}: In our approach, strong SEPs (244 samples) correspond to the positive class, while weak and non-SEPs (2,649 samples) are negatives. Here, a strong SEP-event indicates the GOES $\geq 10$ MeV proton fluxes crossing 10 pfu. The data set contains long-band (1-8\AA) X-ray measurements from the XRS instrument and proton fluxes from the Space Environment Monitor \citep[SEM;][]{joselyn1985space} instrument onboard the GOES missions. The model is trained using these four physical parameters as input.

\textbf{Outputs}: The output of the ensemble STSF model is a binary flag (yes/no SEP). That is, it indicates whether a strong SEP event will occur within the next 60 minutes, based on the observed X-ray and proton enhancements. 

\textbf{Model Configuration}: The STSF model is available in the \texttt{sktime} library \citep{loning2019sktime, markussktime} for Python. We use the training set to train the model and perform a grid search for hyperparameter optimization. The best hyperparameters for STSF  were in the default model settings, with the number of estimators set to 200.

\textbf{Model Validation and Results}: The model utilizes TSS, HSS, Gilbert Skill Score (GSS), and Matthew’s Correlation Coefficient (MCC) scores for evaluation. For a 60-minute prediction window, the scores are TSS = 0.850, HSS = 0.878, GSS = 0.783, and MCC = 0.879. This study examines periods of non-occurrence of SEPs following a flare with magnitudes $\geq C6.0$ to maintain a natural class imbalance in the sample distribution. Nonetheless, there was only a decrease of $\sim 7$ \% ($\pm$2\%) in the skill scores compared to \cite{rotti2024short}.

\textbf{Access to Model Data and Forecasts}: The GSEP data set \citep{gsep_2022} and coding methodology of our model implementation have been made available on the GitHub repository: \url{https://github.com/sumanth-ra23/SEP-Predictions}. The SEP catalog and time series data set developed as a part of this study are available on Harvard Dataverse: \url{https://doi.org/10.7910/DVN/DZYLHK}.

\textbf{Limitations, Caveats and Discussion}: Identifying and flagging strong SEP events with proton fluxes fluctuating around 10 pfu is challenging for our model, which can lead to some misses and false positives.


\subsection{SHARP-SMARP Model} \label{sec:SHARP-SMARP}

\textbf{Model Developers and Relevant Citation}: Spiridon Kasapis, Lulu Zhao, Yang Chen, Xiantong Wang, Monica Bobra, Tamas Gombosi, Irina N. Kitiashvili, Paul Kosovich, Alexander G. Kosovichev, Viacheslav M. Sadykov, Patrick O’Keefe, and Vincent Wang; \cite{kasapis2022interpretable,kasapis2024forecasting} and \cite{kosovich2024time}.

\begin{table}[h]
\caption{Model, Input and Output Specification Table for the SHARP-SMARP model.}
\centering
\begin{tabular}{|lc|}
\hline
\multicolumn{2}{|c|}{Model}                                            \\ \hline
\multicolumn{1}{|l|}{Type}            & Support Vector Machine         \\
\multicolumn{1}{|l|}{Complexity}      & 7                              \\ \hline
\multicolumn{2}{|c|}{Input}                                            \\ \hline
\multicolumn{1}{|l|}{Shape}           & Point Data (0D)                \\
\multicolumn{1}{|l|}{Type}            & Magnetic Fields                \\
\multicolumn{1}{|l|}{History}         & 26 years (1996-2022)           \\
\multicolumn{1}{|l|}{Diversity}       & 3466 samples                   \\
\multicolumn{1}{|l|}{Imbalance}       & 0.032 positive                 \\
\multicolumn{1}{|l|}{Sample Size}     & 56 bytes                       \\
\multicolumn{1}{|l|}{Sample Coverage} & 0 hours                        \\ \hline
\multicolumn{2}{|c|}{Output}                                           \\ \hline
\multicolumn{1}{|l|}{Prediction}      & Classification, Probability    \\
\multicolumn{1}{|l|}{Type}            & Triggered                      \\
\multicolumn{1}{|l|}{Forecast Window} & 14.21 hours                    \\ \hline
\multicolumn{2}{|p{8.5cm}|}{Comments: \textsuperscript{*}The forecast window is variable, defined by the model's prediction timestamp until the beginning of the SEP event. The average forecast window is 14.21 hours.}   \\ \hline
\end{tabular}
\end{table}

\textbf{Summary}: The 2022 model \citep{kasapis2022interpretable} introduced an interpretable ML framework using the SMARP (SDO/MDI) dataset to forecast whether solar flares would lead to SEP events during SC 23. The 2024 model \citep{kasapis2024forecasting} extended this work to include SHARP (SDO/HMI) data using the \cite{kosovich2024time} dataset, allowing prediction of SEPs across SCs 23 and 24 with a combined dataset of 3,869 ARs and 110 SEP events, twice as many as the 2022 study. Despite the expanded dataset, the model performance remained similar, indicating a limit to the predictive power of the selected SHARP and SMARP features.

\textbf{Model Description}: Both studies \citep{kasapis2022interpretable, kasapis2024forecasting} utilized a binary classification framework (SEP and non-SEP warnings after a flare occurrence) by using SVMs and Regression models. Different kernels were explored for the SVMs such as a) the linear kernel, b) the polynomial kernel, c) the radial basis function, and d) the sigmoid kernel, while both logistic and linear regression models were also tested. The best performing model mentioned in this review is the linear SVM.

\textbf{Inputs}: The approach defines flares associated with SEPs as positive cases (110 positive flares), and flare-only events (3356 negative flares) as negatives. SMARP and SHARP \cite{bobra2021smarps} physical parameters are used as predictive features, spanning 26 years (1996-2022). The model was trained using five physical features from the SMARP-SHARP dataset: the total line-of-sight unsigned flux, the mean value of line-of-sight magnetic field gradient, the unsigned flux $R$ near polarity inversion lines, the Vertical component of the total unsigned magnetic flux, and the mean value of the vertical field gradient. Two more parameters are calculated using the AR coordinates, the AR area, and the AR’s angular distance between the associated flare and Earth's magnetic footpoint. From the full SMARP-SHARP timelines for the aforementioned physical parameters provided by \cite{kosovich2024time}, only those recorded right before each flare occurrence were selected. Therefore, for each positive or negative flare instance, a 7-dimensional vector (56 bytes) is used for training and testing the SVM model. 

\textbf{Outputs}: The output of the model is a probability of SEP occurrence based on a flare trigger. The probability is converted to a binary label (True/False) using a threshold of 0.5, informing us whether a flare will produce an SEP event.

\textbf{Model Configuration}: The model that yielded the best results in our analysis is an SVM that uses a linear kernel. For the 7-dimensional data case, the SVM uses 7 trainable parameters (weights). The Python Scikit Learn (\textit{sklearn}\footnote{\url{https://scikit-learn.org/stable/}}) library was used for implementation along with the \textit{GridSearchCV} in order to explore the best regularization parameter C=2.4 using the standard l2 penalty.

\textbf{Model Validation and Results}: The study shows that despite the augmented volume of data compared to \cite{kasapis2022interpretable}, the prediction accuracy reaches $0.7 \pm 0.1$ (experimental setting / balanced dataset), which aligns with but does not exceed these published benchmarks. A linear SVM model with training and testing configurations that mimic an operational setting (original positive–negative imbalance) reveals a slight increase ($0.04 \pm 0.05$) in the accuracy of a 14-hour SEP forecast compared to \cite{kasapis2022interpretable}. Other metrics used in the study are: TSS, HSS, FAR and F1. 

\textbf{Access to Model Data and Forecasts}: Data and preprocessing scripts are available at \url{github.com/skasapis/SEP\_Pred\_SMARP-SHARP}. The SEP list used in this study is provided by NOAA at \url{https:/ngdc.noaa.gov/stp/satellite/goes/doc/SPE.txt} while the SMARP-SHARP dataset, although not publicly available yet, can be obtained by contacting the authors of \cite{kosovich2024time}.

\textbf{Limitations, Caveats and Discussion}: A major limitation of both the SMARP-SHARP models is the inherent class imbalance in SEP prediction: only a small fraction of flaring ARs actually produce SEPs. This imbalance can bias the models toward predicting non-events unless careful sampling or penalization strategies are employed. While physical parameters like magnetic flux and magnetic field gradients show some predictive capability, they are not sufficient to achieve high skill scores, even when expanding the dataset from one to two SCs. The plateau in model performance between the two studies suggests a ceiling on what can be achieved with current inputs and models alone. 


\subsection{AA Model} \label{sec:AA}

\textbf{Model Developers and Relevant Citation}: Eleni Lavasa, Giorgos Giannopoulos, Athanasios Papaioannou, Anastasios Anastasiadis; \cite{lavasa2021assessing}.

\begin{table}[h]
\caption{Model, Input and Output Specification Table for the AA model.}
\centering
\begin{tabular}{|lc|}
\hline
\multicolumn{2}{|c|}{Model}                                            \\ \hline
\multicolumn{1}{|l|}{Type}            & Forest\textsuperscript{*}      \\
\multicolumn{1}{|l|}{Complexity}      & 8                              \\ \hline
\multicolumn{2}{|c|}{Input}                                            \\ \hline
\multicolumn{1}{|l|}{Shape}           & Point data (0D)                \\
\multicolumn{1}{|l|}{Type}            & Soft X-ray, Coronagraphs       \\
\multicolumn{1}{|l|}{History}         & 15 years (1998–2013)           \\
\multicolumn{1}{|l|}{Diversity}       & 3,307 samples                  \\
\multicolumn{1}{|l|}{Imbalance}       & 0.039 positive                 \\
\multicolumn{1}{|l|}{Sample Size}     & 64 bytes                       \\
\multicolumn{1}{|l|}{Sample Coverage} & 0 hours                        \\ \hline
\multicolumn{2}{|c|}{Output}                                           \\ \hline
\multicolumn{1}{|l|}{Prediction}      & Classification                 \\
\multicolumn{1}{|l|}{Type}            & Triggered                      \\
\multicolumn{1}{|l|}{Forecast Window} & 24 hours                       \\ \hline
\multicolumn{2}{|p{8.5cm}|}{Comments: \textsuperscript{*}The sklearn.ensemble.RandomForestClassifier class from the \textit{sklearn} library was used here.}   \\ \hline
\end{tabular}
\end{table}

\textbf{Summary}: The Random Forest (RF) model generates binary predictions of solar energetic particle (SEP) events (integral proton flux $\geq 10$ pfu at $E \geq 10$ MeV), for the following day at GEO. Comparing the performance of several ML algorithms (logistic regression, SVM, decision tree, random forest, extremely randomized trees, XGBoost and NNs), we find that random forests show both high performance in terms of F1-score, TSS, HSS (achieving high POD with relatively low FAR) and generalization (i.e. low variance), in the mean test scores of a 5-fold nested cross-validation scheme. Random forests also show a small difference in performance between the validation and test sets. The model is trained, validated and tested on flare soft X-ray data from GOES and CME data from SOHO/LASCO spanning from 1998 to 2013 (SC 23 and the rising phase of SC 24). Weight-adjusting to more heavily penalize misclassified positive SEP events is applied to address class imbalance. 

\textbf{Model Description}: Random Forests ---a bagging ensemble classifier built on decision trees--- are the method of choice among other supervised classifiers (logistic regression, SVM, decision tree, extremely randomized trees, XGBoost and NNs) in this evaluation setting. A nested 5-fold cross-validation scheme is used, with inner folds for hyperparameter tuning (under \textit{RandomizedGridSearchCV}\footnote{\url{https://scikit-learn.org/stable/modules/generated/sklearn.model_selection.RandomizedSearchCV.html}} across 1000 configurations) and outer folds for the evaluation of model performance and generalization. Data are split randomly without replacement to prevent information leakage and stratified to preserve class imbalance across all partitions. Permutation feature importance, measuring the decrease in model performance when a feature’s values are randomly shuffled, is applied to assess the importance of input features to the prediction target.

\textbf{Inputs}: Solar eruptive flare and CME events associated with SEPs are defined as positive cases (1 label) and eruptive events without SEP association as negatives (0 label). Model input features are extracted from solar flare identifications (classes C, M, X) in the 1-8 \AA soft X-ray flux from the XRS instrument\footnote{\url{ftp://ftp.ngdc.noaa.gov/STP/space-weather/solar-data/solar-features/solar-flares/x-rays/goes/}} onboard the GOES missions, complemented with locations extracted in Ha, as well as CME recordings in white light by the SOHO/LASCO coronagraph. These features include the peak flux, fluence (time-integrated flux), heliographic longitude, duration, and the rise time of solar flares, as well as the sky-projected linear speed and width of the CME. An additional Cycle index is used as indicative to the magnitude of solar activity. The dataset includes a significant class imbalance ($\sim$3.9\%), with 3181 negative (no SEP) and only 126 positive (SEP) class events, thus, being representative of the real distribution of events.  

\textbf{Outputs}: Given parent solar event triggers ($\geq C1$ flare, CME), the model produces a binary flag, indicating whether an SEP event is expected at Earth in the next 24 hours, exceeding integral proton flux $\geq 10$ pfu at $E \geq 10$ MeV.

\textbf{Model Configuration}: The following hyper-parameters of the random forest classifier are optimized for F1-score t: i) n\_estimators = total number of decision trees in the ensemble, ii) criterion = split criterion in individual decision trees, iii) min\_samples\_split = minimum number of samples required to perform a split, iv) min\_samples\_leaf = min. number of samples in decision tree leaf nodes, v) max\_depth = maximum depth of decision trees, vi) class\_weight controls the strength of penalization applied to wrong predictions of positive and negative class events, vii) max\_features = maximum number of features in constructing individual decision trees. 

\textbf{Model Validation and Results}: On the realistic, imbalanced test folds, random forests emerged as the top-scoring classifier. Using the combined flare and CME feature set without imputed gaps, it achieved TSS = $0.75 \pm 0.05$, HSS = $0.69 \pm 0.04$, POD = $0.76 \pm 0.06$ and FAR = $0.34 \pm 0.10$, yielding an overall F1 $\sim 0.70 \pm 0.04$. As compared to e.g. XGBoost, random forests' ensemble captures $\sim 11$\% more true SEP events, lifting F1 ($\sim$1\%), TSS ($\sim$11\%) and HSS ($\sim$1\%), but at the cost of $\sim 7$\% more false alarms. Variance is lower too ($\sim$2\%), hinting at slightly steadier behavior across cross-validation splits. Random forests show a clear improvement over the SWPC legacy probabilistic baseline. The model has been validated against the independent SEPVAL event sample, and the performance scores were POD = 0.75 and FAR = 0.22. Permutation importance scores reveal that flare soft X-ray fluence and CME speed mostly affect the model's predictions.  

\textbf{Access to Model Data and Forecasts}: The flare and CME dataset (SEP-labeled) without imputed gaps, as well as the model training and evaluation pipeline developed during this study, are available at: \url{https://github.com/SolarML/SEP-ML}

\textbf{Limitations, Caveats and Discussion}: Despite its strong skill, the random forests show four operational weaknesses: i) they are prone to over-fitting, with training scores $\sim 90$ \% but validation $\sim 10–15$ \% lower, revealing high variance; ii) they depend on having both flare and CME inputs, since dropping either source pushes all metrics below service thresholds; iii) their performance degrades when gaps are median-filled (F1 falls from 0.70 to 0.63) underscoring sensitivity to data quality; iv) any alert is bounded by the telemetry lag of CME speed and width measurements (e.g. $\sim 6$ hours for SOHO/LASCO), limiting real-time lead time for prediction. 


\subsection{Empirical model for Solar Proton Event Real Time Alert (ESPERTA) Model} \label{sec:ESPERTA}

\textbf{Model Developers and Relevant Citation}: Monica Laurenza, Edward W. Cliver, Alan G. Ling, Tommaso Alberti, Mirko Stumpo, Simone Benella; \cite{laurenza2009technique,laurenza2018short,laurenza2024upgrades}, \cite{alberti2017solar, alberti2019forecasting}, \cite{stumpo2021open} and \cite{benella2023statistical}.

\begin{table}[h]
\caption{Model, Input and Output Specification Table for the ESPERTA model.}
\centering
\begin{tabular}{|lc|}
\hline
\multicolumn{2}{|c|}{Model}                                                            \\ \hline
\multicolumn{1}{|l|}{Type}            & Logistic Regression                            \\
\multicolumn{1}{|l|}{Complexity}      & 12                                             \\ \hline
\multicolumn{2}{|c|}{Input}                                                            \\ \hline
\multicolumn{1}{|l|}{Shape}           & Point data (0D)                                \\
\multicolumn{1}{|l|}{Type}            & Soft X-ray, Flare Location, Space-Based or Ground-Based Radio  \\
\multicolumn{1}{|l|}{History}         & 23 years (1995–2017)\textsuperscript{*}        \\
\multicolumn{1}{|l|}{Diversity}       & 989 samples                                    \\
\multicolumn{1}{|l|}{Imbalance}       & 0.1 positive                                   \\
\multicolumn{1}{|l|}{Sample Size}     & 44 bytes                                       \\
\multicolumn{1}{|l|}{Sample Coverage} & 0 hours                                        \\ \hline
\multicolumn{2}{|c|}{Output}                                                           \\ \hline
\multicolumn{1}{|l|}{Prediction}      & Classification, Probability                    \\
\multicolumn{1}{|l|}{Type}            & Triggered                                      \\
\multicolumn{1}{|l|}{Forecast Window} & 7 hours                                        \\ \hline
\multicolumn{2}{|p{8.5cm}|}{Comments: \textsuperscript{*}Extended with recent data where available. \textsuperscript{**}Average for S1, since the forecast window is $\approx 6–8$ hours for $\geq S1$ and $1.7–4$ hours for $\geq S2$.}   \\ \hline
\end{tabular}
\end{table} 

\textbf{Summary}: The ESPERTA model introduced a logistic regression approach \citep{laurenza2009technique} for the prediction of solar proton events (defined as $\geq S1$ in the NOAA scale) following $\geq M2$ flares, and using three solar parameters: flare heliolongitude, soft X-ray fluence, and radio fluence ---originally at 1 MHz from WIND/WAVES \citep{bougeret1995waves}. The \cite{laurenza2024upgrades} ESPERTA upgrades include: a) an ML approach with stratified cross-validation \citep{stumpo2021open}; b) a binary classification algorithm to forecast the occurrence of $\geq S1$ events and $\geq S2$ ones (defined as those reaching a peak flux of $\geq 100$ pfu), to give an indication of the storm severity and c) replacement of space-based radio data with ground-based low-frequency observations (30 MHz from LOFAR) to allow for real-time operations. The upgraded ESPERTA maintains or improves performance compared to earlier versions, with POD up to $79\%$ for $\geq S2$ events. Note here that NOAA categorizes solar radiation storms using the NOAA Space Weather Scale\footnote{\url{https://www.swpc.noaa.gov/noaa-space-weather-scales}} on a scale from S1 - S5.

\textbf{Model Description}: Original ESPERTA uses logistic regression to estimate the probability of $\geq S1$ SPE occurrence based on the three input parameters. Predictions for $\geq S1$ events are issued 10 minutes after the $\geq M2$ soft X-ray flare peak time. The upgraded ML version of ESPERTA within a supervised learning framework, provides forecasting also for $\geq S2$ events at the time of $\geq S1$ threshold crossing. The $\geq S2$ proton events were identified for the period 1995–April 2017, extending the list in \cite{laurenza2018short}. A supervised ML approach was then applied, treating the $\geq S1$ and $\geq S2$ events as separate classes. The model was calibrated by determining the optimal threshold that maximizes the Critical Success Index (CSI), representing a balance between maximizing the POD and minimizing the FAR. The optimal thresholds were found to be 0.36 and 0.37 for $\geq S1$ and $\geq S2$ events, respectively.

\textbf{Inputs}: The three ESPERTA features are: a) flare heliolongitude; b) time-integrated soft X-ray flux ($1–8 \AA$, GOES); c) time-integrated radio flux (originally 1 MHz WIND/WAVES, but can be replaced with 30 MHz LOFAR).

\textbf{Outputs}: The output of the model, based on a $\geq M2$ flare trigger, are: probability of $\geq S1$ SPE occurrence 10 minutes after flare peak and probability of $\geq S2$ SPE occurrence after S1 onset. The probability is converted to a binary label (True/False) using the aforementioned optimal thresholds, informing us whether a flare will produce an SEP event.

\textbf{Model Configuration}: A logistic regression model with three features with a threshold optimized via CSI and for the upgraded ESPERTA \citep{laurenza2024upgrades} a supervised ML classifier for two-class prediction ($\geq S1$ and $\geq S2$ severity levels).

\textbf{Model Validation and Results}: The model has been validates by using N-1 observations in the training set and 1 event in the test set and repeated this for N-1 times. The optimization led to the following scores. For the upgraded ESPERTA and for $\geq S2$ events: theoretical POD = 0.88 (operational 0.79), FAR = 0.32 and median warning time $\approx 2$ hours. Using 30 MHz ground-based data for $\geq S1$ events (between 1995–2017) the upgraded ESPERTA achieved POD = 0.69 and FAR = 0.33.

\textbf{Access to Model Data and Forecasts}: Original ESPERTA uses WIND/WAVES and GOES data (publicly available from NASA/NOAA). Ground-based 30 MHz data available from LOFAR \citep{van2013lofar} archives. The whole ESPERTA dataset, although not publicly available yet, can be obtaining by contacting the authors of \cite{laurenza2024upgrades}.

\textbf{Limitations, Caveats and Discussion}: Class imbalance ---especially for $\geq S2$ events--- affects model calibration and FAR, and there is limited real-time availability of global low-frequency radio coverage as the single LOFAR station limits ESPERTA to 24-hour operation.


\subsection{University of MAlaga Solar particle Event Predictor (UMASEP) Model} \label{sec:UMASEP}

\textbf{Model Developers and Relevant Citation}: Marlon Nunez and Daniel Paul-Pena; \cite{nunez2011predicting}.

\begin{table}[h] 
\caption{Model, Input and Output Specification Table for the UMASEP model. Please note that due to the authors of this work not participating in this effort, the values of the below table are estimates derived from \cite{nunez2011predicting}.}
\label{tab:UMASEP}
\centering
\begin{tabular}{|lc|}
\hline
\multicolumn{2}{|c|}{Model}                                                         \\ \hline
\multicolumn{1}{|l|}{Type}            & Decision Tree, Linear Regression, Ensemble  \\
\multicolumn{1}{|l|}{Complexity}      & 20                                          \\ \hline
\multicolumn{2}{|c|}{Input}                                                         \\ \hline
\multicolumn{1}{|l|}{Shape}           & Time Series (1D)                            \\
\multicolumn{1}{|l|}{Type}            & Soft X-Ray, Proton Flux                     \\
\multicolumn{1}{|l|}{History}         & 19 years (1987-2006)                        \\
\multicolumn{1}{|l|}{Diversity}       & 166 samples                                 \\
\multicolumn{1}{|l|}{Imbalance}       & 0.45 positive                               \\
\multicolumn{1}{|l|}{Sample Size}     & 10,000 bytes                                \\
\multicolumn{1}{|l|}{Sample Coverage} & 24 hours                                    \\ \hline
\multicolumn{2}{|c|}{Output}                                                        \\ \hline
\multicolumn{1}{|l|}{Prediction}      & Classification, Regression                  \\
\multicolumn{1}{|l|}{Type      }      & Continuous                                  \\
\multicolumn{1}{|l|}{Forecast Window} & 5.17 hours                                  \\ \hline
\end{tabular}
\end{table}

\textbf{Summary}: The UMASEP (University of Málaga Solar Energetic Proton) model introduces a dual-model ML framework for forecasting $>10$ MeV SEP events by distinguishing between well-connected and poorly connected magnetic configurations between the Sun and Earth. The first model identifies well-connected events by detecting temporal correlations between GOES soft X-ray and differential proton flux time series, empirically estimating magnetic connectivity and associating flares of C7 or greater class, with potential proton enhancements. The second model targets poorly connected events using an ensemble of nonlinear regression trees trained on historical proton flux profiles to recognize patterns preceding gradual flux increases. An additional high-level analysis module filters inconsistent forecasts and estimates the expected intensity during the first 7 hours of the predicted event. Validated on SC 22–23, UMASEP achieved POD = 0.81, FAR = 0.40 and an average warning time of 5 hours and 10 minutes (1 hour and 5 minutes for well-connected and 8 hours and 28 minutes for poorly connected events), outperforming earlier automatic forecasters such as ESPERTA.

\textbf{Model Description}: The UMASEP model is based on an empirical dual-predictor design that analyzes the magnetic connectivity between the Sun and Earth to forecast $\geq 10$ MeV SEP events. It includes two complementary components: one for well-connected events, which identifies time correlations between GOES soft X-ray flux and proton flux increases to infer flare–particle linkages, and another for poorly connected events, which applies a set of nonlinear regression trees to proton flux time series to detect gradual rises indicative of eastern limb or backside origins. Both predictors operate continuously and independently, generating forecasts when predefined thresholds of correlation or regression output are reached. A final decision module integrates both predictors, filters contradictory results, and estimates the expected proton intensity during the first 7 hours after onset. This architecture allows UMASEP to automatically detect the solar origin of each event type and provide early, interpretable warnings using only near-real-time GOES data.

\textbf{Inputs}: The inputs to UMASEP consist entirely of near–real-time measurements from GOES satellites, specifically the soft X-ray flux (0.1–0.8 nm channel) and differential proton flux in the $>10$ MeV energy range. These two time-series form the basis of both predictors within the model. For the well-connected predictor, UMASEP computes the evolving correlation between short time windows of X-ray and proton flux data to infer the magnetic linkage between the flare site and Earth. For the poorly connected predictor, the model uses only the proton flux data to train regression trees that recognize characteristic patterns preceding gradual proton enhancements. The temporal cadence of the inputs is typically 5 minutes, ensuring sufficient resolution for real-time SEP forecasting. 

\textbf{Outputs}: The outputs of UMASEP are binary SEP occurrence forecasts indicating whether a $>10$ MeV proton event will occur, along with the estimated onset time and predicted intensity profile for the first 7 hours after detection. Each of the two predictors (well-connected and poorly connected) independently issues a probability-based SEP warning, which is then integrated by the decision module into a single operational forecast. The model also identifies the most likely solar origin—classifying events as well- or poorly connected based on the timing and strength of X-ray–proton correlations. UMASEP’s forecasts are generated automatically and continuously in real-time, producing both the categorical SEP warning and quantitative estimates of expected proton flux intensity.

\textbf{Model Configuration}: The UMASEP model operates as a real-time forecasting system built around two independently configured predictors. The first predictor, for well-connected events, continuously computes correlation coefficients between short sliding windows of soft X-ray and proton flux to detect magnetically connected flare–particle pairs. The second predictor, for poorly connected events, uses an ensemble of nonlinear regression trees trained on historical GOES proton flux profiles from SCs 22 and 23. Both components are configured to update every 5 minutes, applying adaptive thresholds for correlation and flux variation that trigger a forecast when exceeded. A decision layer then combines the outputs, giving priority to well-connected predictions when both models issue simultaneous warnings. The entire framework was calibrated using a multi-cycle dataset and validated against NOAA’s operational criteria for $>10$ MeV SEP event detection.

\textbf{Model Validation and Results}: The UMASEP model was validated using data from SCs 22 and 23 (January 1986 – December 2009), encompassing 166 $>10$ MeV SEP events listed in NOAA’s catalog. Validation involved running the model in a simulated real-time mode using historical GOES X-ray and proton flux data. UMASEP achieved a POD of 0.81, a FAR\textsuperscript{*} of 0.40, and an average warning time of 5 hours and 10 minutes. The model’s separate predictors yielded mean lead times of 1 hour and 5 minutes for well-connected SEPs and 8 hours 28 minutes for poorly connected events. These results outperform previous automatic forecasters such as ESPERTA, particularly in detecting eastern-hemisphere events. The study also showed that UMASEP maintained stable performance across both SCs, demonstrating its robustness and operational readiness for real-time SEP forecasting.

\textbf{Access to Model Data and Forecasts}: All input data are publicly available through the NOAA/GOES archive and require no additional preprocessing beyond smoothing and normalization applied internally by the model. The model's performance can be found on the CCMC SEP Scoreboard by following this link: \url{https://ccmc.gsfc.nasa.gov/scoreboards/sep/}.

\textbf{Limitations, Caveats and Discussion}: A main limitation of UMASEP is its dependence on GOES satellite data availability and quality, which makes the model sensitive to telemetry gaps or delayed data streams that can interrupt real-time forecasting. The approach also assumes that soft X-ray–proton correlations accurately represent magnetic connectivity, which may not always hold for complex or multi-source events. While the dual-predictor architecture improves detection of both well- and poorly connected SEPs, it occasionally produces false alarms for proton flux fluctuations unrelated to solar flares. UMASEP’s average warning time of about five hours is constrained by the onset of measurable X-ray or proton signatures, limiting its utility for extremely rapid SEP events. Additionally, the model does not incorporate data from coronagraphs, radio bursts, or CME kinematics, which could enhance prediction accuracy. Despite these caveats, UMASEP remains a reliable operational tool that balances interpretability and automation, demonstrating robust performance across two SCs and forming the foundation for later real-time forecasting systems such as UMASEP-10.


\subsection{University of Malaga predictor from Solar Data (UMASOD) Model} \label{sec:UMASOD}

\textbf{Model Developers and Relevant Citation}: Marlon Nunez and Daniel Paul-Pena; \cite{nunez2020predicting}.

\begin{table}[h]
\caption{Model, Input and Output Specification Table for the UMASOD model. Please note that due to the authors of this work not participating in this effort, the values of the below table are estimates derived from \cite{nunez2020predicting}.}
\centering
\begin{tabular}{|lc|}
\hline
\multicolumn{2}{|c|}{Model}                                            \\ \hline
\multicolumn{1}{|l|}{Type}            & Decision Tree                  \\
\multicolumn{1}{|l|}{Complexity}      & 90                             \\ \hline
\multicolumn{2}{|c|}{Input}                                            \\ \hline
\multicolumn{1}{|l|}{Shape}           & Time Series (1D)               \\
\multicolumn{1}{|l|}{Type}            & Soft X-Ray, Space-Based Radio  \\
\multicolumn{1}{|l|}{History}         & 17 years (1997-2014)           \\
\multicolumn{1}{|l|}{Diversity}       & 502 samples                    \\
\multicolumn{1}{|l|}{Imbalance}       & 0.14 positive                  \\
\multicolumn{1}{|l|}{Sample Size}     & 1,000 bytes                    \\
\multicolumn{1}{|l|}{Sample Coverage} & 0.5 hours                      \\ \hline
\multicolumn{2}{|c|}{Output}                                           \\ \hline
\multicolumn{1}{|l|}{Prediction}      & Classification                 \\
\multicolumn{1}{|l|}{Type}            & Triggered                      \\
\multicolumn{1}{|l|}{Forecast Window} & 9.87 hours                     \\ \hline
\end{tabular}
\end{table}

\textbf{Summary}: The UMASOD (University of Málaga Solar Data Predictor) model introduced an interpretable ML framework to forecast $>10$ MeV SEP events from solar flare and space-based radio burst data obtained from NOAA/SWPC event lists (updated every 30 min). Using a J48 decision tree \citep{panigrahi2018rank}, the model was trained on 502 events (75 SEP-producing and 427 non-SEP) spanning November 1997 to February 2014 ($\sim 16$ years). Each event corresponds to a compact record ($\sim 0.001$ MB) of flare and radio parameters such as soft X-ray flux, duration, and radio type III intensity and frequency range. 

\textbf{Model Description}: UMASOD uses a decision tree (J48) classifier trained on flare and radio burst parameters from NOAA/SWPC event lists to forecast $>10$ MeV SEP events. The model combines features such as soft X-ray peak and integrated flux, flare duration and rise time, heliographic coordinates, and radio burst intensity and frequency range to identify pre-SEP scenarios. During training, the algorithm found that the most relevant predictors were the integral of soft X-ray flux, flare rise time, and radio type III maximum frequency. The resulting tree provides interpretable decision paths for distinguishing SEP-producing from non-SEP events and was optimized using the CSI to balance POD and FAR\textsuperscript{*}, achieving performance comparable to the ESPERTA model.

\textbf{Inputs}: The inputs to the UMASOD model consist of flare and radio burst parameters derived from NOAA/SWPC’s Solar Edited Event Lists (updated every 30 minutes) and cross-referenced with the NOAA/NASA SEP list to label events as SEP or non-SEP. Each record includes flare information such as start, peak, and end times, heliolongitude and heliolatitude, X-ray peak flux (logarithmic scale), duration, and rise time, as well as radio burst characteristics like type (II–V), intensity, duration, frequency range, and integrated flux. The preprocessing step discretized flare coordinates into 12 regions, converted the X-ray flux scale to a linearized logarithmic form, and calculated additional variables such as the integral of soft X-ray flux, flare rise time, and product of soft X-ray and type III integrals, which served as combined predictive attributes. Events with a flare peak $\geq M2$ were retained, resulting in 502 labeled instances (75 SEP, 427 non-SEP) used to train the J48 decision tree.

\textbf{Outputs}: The output of UMASOD is a binary classification indicating whether a solar flare and radio burst event will result in a $>10$ MeV SEP occurrence. The model predicts either an SEP-producing (positive) or non-SEP (negative) outcome based on flare and radio parameters. The outputs are derived from decision tree rules optimized for operational use, providing interpretable “if–then” conditions that identify pre-SEP configurations using only solar data available at the time of flare and radio emission.

\textbf{Model Configuration}: The UMASOD model was implemented using the J48 decision tree algorithm within the Weka ML environment. The model was trained on 502 labeled events (75 SEP, 427 non-SEP) and optimized through 20-fold cross-validation, varying the minimum number of instances per leaf to maximize the CSI. The optimal configuration was found with eight instances per leaf, yielding a balance between Probability of Detection and False Alarm Ratio. Events with flare classes below M2 were filtered out, and the model trained exclusively on greater than or equal to M2 flare–associated cases. All computations were based on preprocessed flare and radio attributes, and the final decision tree, shown in the paper, reflects the interpretable rule-based structure generated by Weka for operational use.

\textbf{Model Validation and Results}: The UMASOD model was validated using historical flare and radio burst events spanning November 1997 to February 2014, employing 20-fold cross-validation during training and an independent evaluation of 104 flare-associated SEP events for performance assessment. Performance evaluation produced a POD of 0.70, a FAR\textsuperscript{*} of 0.40, and an average warning time of 9 hours and 52 minutes, closely matching the empirical ESPERTA model. When optimized using the CSI, the model reached POD = 0.85 and FAR = 0.55 during cross-validation. These results confirm that UMASOD provides comparable accuracy to existing empirical predictors while maintaining interpretability and real-time operational capability based solely on solar flare and radio burst observations.


\textbf{Limitations, Caveats and Discussion}: A key limitation of the UMASOD model lies in its event-triggered and empirical design, which restricts forecasts to periods following flare and radio detections, preventing continuous monitoring of the solar environment. The class imbalance between SEP-producing (75) and non-SEP (427) events introduces bias toward negative predictions, though the use of the CSI helps mitigate this effect. While the decision-tree structure enhances interpretability, it may oversimplify nonlinear relationships among flare and radio parameters, limiting generalization to unseen or extreme events. The model also depends entirely on NOAA/SWPC event lists that update every 30 minutes; any delay or data gap directly impacts its real-time applicability. Furthermore, the training data end in 2014, leaving performance under recent SCs unverified. Despite these caveats, UMASOD demonstrates that combining flare and radio burst parameters can yield operationally useful, physically interpretable forecasts for $>10$ MeV SEP events, achieving skill scores comparable to more complex empirical systems such as ESPERTA.


\subsection{MS-SEP Model} \label{sec:MS_SEP}

\textbf{Model Developers and Relevant Citation}: Mohammed AbuBakr Ali, Ali G. A. Abdelkawy, Abdelrazek M. K. Shaltout, and M. M. Beheary; \cite{ali2025forecasting}.

\begin{table}[h] 
\caption{Model, Input and Output Specification Table for the MS-SEP model.}
\label{tab:MS-SEP}
\centering
\begin{tabular}{|lc|}
\hline
\multicolumn{2}{|c|}{Model}                                            \\ \hline
\multicolumn{1}{|l|}{Type}            & Random Forest                  \\
\multicolumn{1}{|l|}{Complexity}      & 52                             \\ \hline
\multicolumn{2}{|c|}{Input}                                            \\ \hline
\multicolumn{1}{|l|}{Shape}           & Time Series (1D)               \\
\multicolumn{1}{|l|}{Type}            & Coronagraph, Soft X-Ray, Space-Based Radio  \\
\multicolumn{1}{|l|}{History}         & 25 years (1997-2022)           \\
\multicolumn{1}{|l|}{Diversity}       & 740 samples                    \\
\multicolumn{1}{|l|}{Imbalance}       & 0.108                          \\
\multicolumn{1}{|l|}{Sample Size}     & 416 bytes                      \\
\multicolumn{1}{|l|}{Sample Coverage} & 6.5 hours                      \\ \hline
\multicolumn{2}{|c|}{Output}                                           \\ \hline
\multicolumn{1}{|l|}{Prediction}      & Classification                 \\
\multicolumn{1}{|l|}{Type}            & Triggered                      \\
\multicolumn{1}{|l|}{Forecast Window} & 4.6 hours                      \\ \hline
\multicolumn{2}{|p{8.5cm}|}{Comments: \textsuperscript{*}These values are for the fixed frequency model. The values for the sweep frequency are a) Complexity: 29, b) Diversity: 534, c) Imbalance: 0.15, d) Sample Size: 232 and Forecast Window: 4.23 hours.}   \\ \hline
\end{tabular}
\end{table}

\textbf{Summary}: This study developed an interpretable machine-learning framework to forecast $>10$ MeV SEP events associated with M2.0 and stronger solar flares by integrating flare, space-based radio-burst, and CME data from NOAA/SWPC, NOAA/NASA and SOHO/LASCO catalogs. The system combines multi-source observations, including solar flare and radio bursts updated every 30 minutes and CME reports issued every 6 hours, covering 1997–2022. Random forest, decision tree, and SVM classifiers were trained and validated using both sweep-frequency and fixed-frequency radio datasets. The final dataset links each solar event to a compact feature record containing flare intensity, soft X-ray flux, CME speed and angular width, and type II/III radio-burst characteristics. Incorporating CME and radio parameters improved predictive skill, while restricting the analysis to stronger flares enhanced class balance. Nested cross-validation ensured robust and unbiased evaluation. 

\textbf{Model Description}: This study implemented a binary classification framework to issue SEP and non-SEP warnings following $\geq M2.0$ solar flares with associated CME and radio-burst activity. The system integrates flare, CME, and radio-burst parameters into compact event-based records and evaluates multiple ML classifiers, including ramdom forest, Decision Trees, and linear and non-linear SVMs. Input features comprise flare soft X-ray flux and duration, CME speed and angular width, and type II/III radio-burst intensity and frequency characteristics to identify pre-SEP conditions. During training, CME kinematics and flare intensity consistently emerged as the most informative predictors. Model performance was assessed under imbalanced, balanced, and hybrid sampling strategies using nested cross-validation to ensure robust and unbiased estimates. Among the tested approaches, the random forest provided the most reliable and stable forecasting skill.

\textbf{Inputs}: The input dataset comprises solar flares associated with SEP events labeled as positive cases (80 SEP-producing flares) and flare-only events with concurrent CMEs and radio bursts labeled as negative cases (454 in the sweep-frequency set and 660 in the fixed-frequency set). Observations were compiled from NOAA/SWPC Solar Event Lists for flare and radio-burst parameters, the SOHO/LASCO Coordinated Data Analysis Workshop (CDAW) catalog\footnote{\url{https://cdaw.gsfc.nasa.gov/CME_list/}} for CME characteristics, and the NOAA/NASA $\geq 10$ MeV proton flux list for SEP labeling, covering the period 1997–2022. Two complementary event-based datasets were constructed. The sweep-frequency dataset (534 events) includes features derived from Type II and Type III dynamic radio spectra together with flare and CME properties such as flare rise time, duration, soft X-ray flux and intensity, heliographic location, and CME speed and angular width. The fixed-frequency dataset (740 events) incorporates radio-burst intensity, duration, and flux measurements at multiple discrete frequencies, combined with the same flare and CME parameters. All variables were standardized using min-max scaling, and events with missing data were excluded to ensure consistency and reliable model training.

\textbf{Outputs}: The model produces a binary classification indicating whether a flare–CME–radio event is expected to generate a $>10$ MeV SEP occurrence, labeling each case as SEP-producing (positive) or non-SEP (negative). Model performance is evaluated using standard skill metrics, including POD, FAR\textsuperscript{*}, TSS, and HSS.

\textbf{Model Configuration}: The machine-learning framework was implemented in Python using the \textit{sklearn} library and included random forest, decision trees, and linear and non-linear SVM classifiers. Model performance and hyperparameters were optimized using a nested cross-validation scheme with five outer and five inner folds to obtain robust and unbiased estimates. Hyperparameter tuning was conducted through randomized search, and class-weight adjustments were applied to mitigate dataset imbalance. All features were scaled using min-max normalization, and a fixed random seed ensured reproducibility. Model selection prioritized configurations that balanced detection capability and false alarms based on precision, Recall, POD, FAR, TSS, and HSS. Final performance statistics were computed as the mean and standard deviation across the outer folds.

\textbf{Model Validation and Results}: The framework was validated using historical flare, CME, and radio-burst events spanning 1997–2022, with performance assessed through nested cross-validation under imbalanced, balanced, and hybrid sampling conditions. Among the evaluated classifiers, the Random Forest consistently delivered the strongest performance across both datasets. For the sweep-frequency dataset, the model achieved a POD of $0.85\pm0.08$, a FAR of $0.30\pm0.05$, a TSS of $0.78\pm0.07$, a HSS of $0.71\pm0.03$, and an average warning time of approximately 5 hours. For the fixed-frequency dataset, corresponding values were POD $=0.76\pm0.12$, FAR $=0.31\pm0.08$, TSS $=0.71\pm0.11$, HSS $=0.67\pm0.06$, and an average warning time of about 4.5 hours. The results indicate stable generalization with no evidence of overfitting and improved detection capability compared with earlier empirical and machine-learning approaches. Feature importance analysis showed that CME speed and angular width were the dominant predictors, followed by flare intensity, soft X-ray flux, and key radio-burst characteristics.

\textbf{Access to Model Data and Forecasts}: All datasets and source files used in this study are publicly accessible through established solar data repositories. Solar flare and radio-burst data were obtained from the \href{ftp://ftp.swpc.noaa.gov/pub/indices/events/}{NOAA Space Weather Prediction Center (SWPC) Solar Event List}, the CME data were retrieved from the \href{https://cdaw.gsfc.nasa.gov/CME_list/}{SOHO/LASCO Coordinated Data Analysis Web (CDAW) catalog} and the SEP event data were collected from the \href{ftp://ftp.swpc.noaa.gov/pub/indices/SPE.txt}{NOAA/NASA SEP list}.
The SolarML/SEP-ML codebase was adapted from \cite{lavasa2021assessing}, applying modifications to the weighting scheme and restricting the analysis to four models: random forest, decision tree, SVM, and linear SVM. The adapted implementation is available at: \url{https://github.com/SolarML/SEP-ML}. Processed datasets and model scripts supporting this study are available upon reasonable request from the corresponding authors \citep{ali2025forecasting}.

\textbf{Limitations, Caveats and Discussion}: The proposed framework focuses on SEP events associated with strong ($\geq M2.0$) flares, which excludes weaker-flare–driven events and restricts applicability to high-activity scenarios. The requirement for complete flare, CME, and radio-burst observations further reduces the available sample size and limits event diversity. Class imbalance between SEP and non-SEP cases remains an inherent challenge that may bias predictions despite mitigation strategies. Operationally, the model depends on external event catalogs, and delays in CME reporting—particularly the $\sim 6$ hour latency of SOHO/LASCO detections and manual CDAW updates—constrain real-time forecasting and shorten the effective warning window. These factors limit continuous monitoring capability and may affect generalization to unseen or extreme solar conditions.


\subsection{Classification and Regression Tree (CART) Model} \label{sec:CART}

\textbf{Model Developers and Relevant Citation}: Soukaina Filali Boubrahimi, Berkay Aydin, Petrus Martens, and Rafal Angryk; \cite{boubrahimi2017prediction}.

\begin{table}[h]
\caption{Model, Input and Output Specification Table for the CART model.}
\label{tab:CART}
\centering
\begin{tabular}{|lc|}
\hline
\multicolumn{2}{|c|}{Model}                                            \\ \hline
\multicolumn{1}{|l|}{Type}            & Decision Tree                  \\
\multicolumn{1}{|l|}{Complexity}      & 61                             \\ \hline
\multicolumn{2}{|c|}{Input}                                            \\ \hline
\multicolumn{1}{|l|}{Shape}           & Time Series (1D)               \\
\multicolumn{1}{|l|}{Type}            & Soft X-ray, Energetic Protons  \\
\multicolumn{1}{|l|}{History}         & 16 years (1997-2013)           \\
\multicolumn{1}{|l|}{Diversity}       & 94 samples                     \\
\multicolumn{1}{|l|}{Imbalance}       & 0.5                            \\
\multicolumn{1}{|l|}{Sample Size}     & 59,451 bytes                   \\
\multicolumn{1}{|l|}{Sample Coverage} & 10 hours                       \\ \hline
\multicolumn{2}{|c|}{Output}                                           \\ \hline
\multicolumn{1}{|l|}{Prediction}      & Classification                 \\
\multicolumn{1}{|l|}{Type}            & Triggered                      \\
\multicolumn{1}{|l|}{Forecast Window} & 0 hours\textsuperscript{*}     \\ \hline
\multicolumn{2}{|p{8.5cm}|}{Comments: \textsuperscript{*}A forecast window of 0 hours indicates that the model performs event-level classification rather than temporal forecasting. These models rely on the GSEP catalog to label whether an SEP event occurs, without predicting its onset time or lead interval. Consequently, the output reflects the presence or absence of an SEP event conditioned on the input observations, not a forward-looking warning horizon.}   \\ \hline
\end{tabular}
\end{table} 

\textbf{Summary}: The paper introduces a method for predicting $\geq 100$ MeV SEP events using GOES satellite data, focusing on time series from both X-ray and proton flux channels. It uses a Vector Autoregression (VAR) model to capture cross-channel correlations, including interactions among proton channels and between X-ray and proton data. Features extracted from these time series are used to train interpretable decision tree models on a balanced dataset of SEP and non-SEP events. The results show that certain correlations, especially involving proton channel P6 and the long X-ray channel, are strong indicators of upcoming SEP events. The proposed method achieves similar accuracy to existing systems like UMASEP while offering clear interpretability.

\textbf{Model Description}: The model predicts $>100$ MeV SEP events using multivariate time series data from GOES satellite proton and X-ray channels. It applies a VAR model to capture linear dependencies among time series, focusing on how proton channel fluctuations relate to past values of both themselves and the X-ray channels (short and long wavelength). Each time series window (spanning up to 30 hours before an X-ray flare) is represented by a feature vector of VAR coefficients, expressing how proton responses are influenced by earlier activity. These features are used to train interpretable Classification and Regression Tree (CART) models. The decision trees use splitting criteria based on Gini impurity or information gain to identify feature thresholds that best separate SEP and non-SEP classes. The model is trained and evaluated on a balanced dataset of 47 SEP and 47 non-SEP events, using stratified 10-fold cross-validation. Results highlight that features such as the correlation between proton channel P6 and the long X-ray channel are among the most predictive for SEP event occurrence.

\textbf{Inputs}: The model takes as input multivariate time series data consisting of GOES X-ray and proton flux measurements. Specifically, it uses two X-ray channels—short (0.05–0.3 nm) and long (0.1–0.8 nm) ---along with six proton channels: P6 and P7 from the EPS instrument (covering 80–500 MeV) and P8 to P11 from the High Energy Proton and Alpha Detector \citep[HEPAD;][]{hanser2011eps} instrument (covering 350 MeV to $>700$ MeV). Each input sequence spans a fixed observation window, up to 30 hours before the onset of an X-ray event, sampled at a 5-minute cadence. These raw time series are then transformed into feature vectors using a VAR model that captures the dependencies of each proton channel on its own past values and those of the X-ray channels.

\textbf{Outputs}: The output of the model is a probability of SEP occurrence based on a flare trigger. The probability is converted to a binary label (True/False), informing us whether a flare will produce an SEP event.

\textbf{Model Configuration}: The model is configured as a CART decision tree classifier, trained using feature vectors derived from VAR applied to multivariate time series data. Two key parameters are tuned: the observation window span (ranging from 3 to 30 hours) and the VAR lag order (tested for values 1, 3, 5, 7, and 9), which controls how far back in time dependencies are modeled. The best performance was achieved with a 30-hour span and a lag of 5, balancing model complexity and predictive accuracy. The tree uses either Gini impurity or information gain as the splitting criterion, and training is done using stratified 10-fold cross-validation to ensure balanced class distribution and robust evaluation.

\textbf{Model Validation and Results}: The model is validated using stratified 10-fold cross-validation on a balanced dataset of 47 SEP and 47 non-SEP events, ensuring equal representation of both classes in each fold. Performance is assessed using standard classification metrics including accuracy, precision, recall, F1-score, and AUC. The best results are achieved with a 30-hour span and lag value of 5, using information gain as the splitting criterion. Under this configuration, the model reaches an Accuracy of 0.78, Precision of 0.86, POD of 0.73, F1-score of 0.82, and AUC of 0.77, indicating strong predictive power. These results are comparable to or slightly better than existing systems like UMASEP, with the added advantage of interpretability through decision tree rules.


\textbf{Limitations, Caveats and Discussion}: The model relies on historical correlations and may not generalize well to unseen solar conditions or rare event types. Missing data, especially in channels P6 and P7 during GOES-12, poses a risk of bias despite balancing. It also assumes a flare-based trigger, limiting applicability to flare-independent SEP events.


\subsection{Random Hivemind (RH) Model} \label{sec:RH}

\textbf{Model Developers and Relevant Citation}: Patrick M. O’Keefe, Viacheslav Sadykov, Alexander Kosovichev, Irina N. Kitiashvili, Vincent Oria, Gelu M. Nita, Fraila Francis, Chun-Jie Chong, Paul Kosovich, Aatiya Ali, Russell D. Marroquin; \cite{o2024random}.

\begin{table}[h]
\caption{Model, Input and Output Specification Table for the RH model.}
\centering
\begin{tabular}{|lc|}
\hline
\multicolumn{2}{|c|}{Model}                                            \\ \hline
\multicolumn{1}{|l|}{Type}            & Neural Networks, Ensemble      \\
\multicolumn{1}{|l|}{Complexity}      & 202\textsuperscript{**}        \\ \hline
\multicolumn{2}{|c|}{Input}                                            \\ \hline
\multicolumn{1}{|l|}{Shape}           & Point Data (0D)                \\
\multicolumn{1}{|l|}{Type}            & Soft X-ray, Flare Location     \\
\multicolumn{1}{|l|}{History}         & 15 years (2002-2017)           \\
\multicolumn{1}{|l|}{Diversity}       & 18,311 samples                 \\ 
\multicolumn{1}{|l|}{Imbalance}       & 0.0035 positive                \\
\multicolumn{1}{|l|}{Sample Size}     & 48 bytes                       \\
\multicolumn{1}{|l|}{Sample Coverage} & Varies (length of the associated flare)          \\ \hline
\multicolumn{2}{|c|}{Output}                                           \\   \hline
\multicolumn{1}{|l|}{Prediction}      & Triggered                      \\
\multicolumn{1}{|l|}{Type}            & Classification                 \\
\multicolumn{1}{|l|}{Forecast Window} & 0 hours                        \\ \hline
\multicolumn{2}{|p{8.5cm}|}{Comments: \textsuperscript{*}No window used in this study as it associates flares with SEP events. \textsuperscript{**}The values in this table concern the RHv2 model.}   \\ \hline
\end{tabular}
\end{table}

\textbf{Summary}: In this study, the considered problem is whether the particular soft X-ray flare event on the Sun is associated with the $\geq 10$ MeV $\geq 10$ pfu SEP event sometime in the future (with no particular forecasting time window). The ML model implemented is the Random Hivemind (RH) model, which represents the ensemble of individually-trained NNs, each considering a randomized set of features and voting proportionally to their importance. The input to the model is the soft X-ray properties of solar flares coming from the Temperature and Emission measure-Based Background Subtraction \citep[TEBBS;][]{ryan2012thermal, ryan2013tebbs} algorithm \citep{sadykov2019statistical}, along with their locations. The RH has been compared to the conventional NN approach (by keeping about the same architecture as for the ensemble members) and to the committee approach (identical NNs trained individually for a majority vote). It was demonstrated that RH has a comparable or better performance with respect to the models it has been compared to, has a lesser spread of the scores for individual train-validation subsets, and captures almost all SEP events (making it a promising solution for all-clear predictions).

\textbf{Model Description}: The Random Hivemind (RH) model represents the ensemble NN model. Unlike the traditional approach, where the identical NN architectures are trained individually and issue a majority vote, the RH utilizes a) a randomized set of features propagating into each ensemble member, b) an adaptive learning rate that depends on the importance of features propagating to the ensemble member, and c) the output vote weighted for each ensemble member based on the importance of the features propagated into it. The basis network architecture includes two linear layers (10 neurons each), one dropout layer (20\% rate), and one linear layer (2 neurons). The feature importances are estimated using the combination of $\chi^2$ and mutual information gain statistics. Two versions of RH are considered using different numbers of features as inputs, and different approaches for progressing learning rates. 

\textbf{Inputs}: The model utilizes the soft X-ray properties of solar flares computed from the 1-8\,$\AA$ and 0.5-4\,$\AA$ emission observed by the GOES XRS instrument. The properties are computed using the updated TEBBS algorithm. These properties include peak temperature, peak emission measure, background-subtracted flare class, and flare duration. The times of the temperature, emission measure, and soft X-ray flux peaks relative to the flare start and end times are also computed. Together with the flare coordinates, this results in 12 features per flare. For the flare-SEP association, the list of Solar Proton Events Affecting the Earth's Environment has been used and can be found here: \url{https://www.ngdc.noaa.gov/stp/space-weather/interplanetary-data/solar-proton-events/SEP\%20page\%20code.html}.

\textbf{Outputs}: The model produces a binary prediction of whether a particular flare is associated with the $\geq 10$ MeV $\geq 10$ pfu SEP event sometime in the future.

\textbf{Model Configuration}: Overall, one can configure the number of ensemble members, the number of features propagating to the estimators, the way of progressing the learning rate and other. One of the advantages of the model is that the newly trained ensemble member can be added without adjusting the previously trained members.

\textbf{Model Validation and Results}: The model has been validated using a variety of metrics, including widely-used TSS, HSS, and AUC. The model has been tested against the conventional NN of approximately the same architecture as RH ensemble members and having all features as inputs, and the committee ensemble of the identical networks. The performance of RH was found to be comparable or better than the competing approaches (TSS = $0.944 \pm 0.023$ and HSS = $0.168 \pm 0.013$ for RH version 2). The RH also typically demonstrated lower standard deviations for the scores, resulting in being less dependent on the particular train-test subdivision. In addition, the RH resulted in a very few false-negative predictions, demonstrating that it captures almost every SEP event, which would be desirable for all-clear purposes.

\textbf{Access to Model Data and Forecasts}: The SPE catalogs developed during this study are archived at the Solar Energetic Particle Prediction Portal which can be found here: \url{https://sun.njit.edu/SEP3/datasets.html}.

\textbf{Limitations, Caveats and Discussion}: This work has several important limitations, including a) a limited-span dataset that included only 64 unique SEP events; the validation on a larger dataset is desirable, and b) the non-operational nature of the currently implemented TEBBS algorithm, which requires the presence of the entire soft X-ray profile of the solar flare before producing the flare properties.


\subsection{Survival SEP (SSEP) Model} \label{sec:SSEP-Survival}

\textbf{Model Developers and Relevant Citation}: India Jackson, Petrus Martens; \cite{martens2024advancing} and \cite{DVN/GXY9MZ_2024}.

\begin{table}[h]
\caption{Model, Input and Output Specification Table for the SSEP model.}
\centering
\begin{tabular}{|lc|}
\hline
\multicolumn{2}{|c|}{Model}                                            \\ \hline
\multicolumn{1}{|l|}{Type}            & Forest, Decision Tree          \\
\multicolumn{1}{|l|}{Complexity}      & 300                            \\ \hline 
\multicolumn{2}{|c|}{Input}                                            \\ \hline
\multicolumn{1}{|l|}{Shape}           & Point Data (0D)                \\
\multicolumn{1}{|l|}{Type}            & Flare Location                 \\
\multicolumn{1}{|l|}{History}         & 31 years (1986-2017)           \\
\multicolumn{1}{|l|}{Diversity}       & 293 samples                    \\
\multicolumn{1}{|l|}{Imbalance}       & 0.9044                         \\
\multicolumn{1}{|l|}{Sample Size}     & 47 bytes                       \\
\multicolumn{1}{|l|}{Sample Coverage} & 0 hours                        \\ \hline
\multicolumn{2}{|c|}{Output}                                           \\ \hline
\multicolumn{1}{|l|}{Prediction}      & Probability                    \\
\multicolumn{1}{|l|}{Type}            & Triggered                      \\
\multicolumn{1}{|l|}{Forecast Window} & 22.46 hours\textsuperscript{*} \\ \hline
\multicolumn{2}{|p{8.5cm}|}{Comments: \textsuperscript{*}The forecast window is variable, defined by the model's prediction of the time until SEPs exceed 10 MeV following a solar flare. The average forecast window is 22.46 hours.}   \\ \hline
\end{tabular}
\end{table}

\textbf{Summary}: The Survival SEP (SSEP) model applies survival analysis techniques to estimate the time-to-detection of SEP events following solar flares, using flare latitude, longitude, and GOES class as input features. Built on a curated dataset of flare-associated SEP events, the model implements Kaplan–Meier estimation and Cox PH modeling, with additional evaluation of survival trees and random survival forests. The output is a survival function \( S(t) \), representing the probability that an SEP event has not occurred by time \( t \).

\textbf{Model Description}: Five feature sets were tested, including combinations of flare latitude, longitude, and GOES class, as well as subsets selected based on Cox PH significance. No NN was used; instead, classical survival analysis models and tree-based ensembles were evaluated. Grid search and 5-fold cross-validation were used for hyperparameter tuning, although the models themselves contain no trainable weights.

\textbf{Inputs}: The model uses point-based input data, where each solar flare event is independently represented as a set of features (time-to-detection, longitude, latitude, and GOES class) with no temporal or spatial relationship between samples. This structure qualifies as Point Data (0D features). The physical quantity represented is energetic protons detected at $\geq10$ MeV. The dataset spans 31 years, from 1986 to 2017, and includes a total of 293 labeled flare events: 265 positive cases (flare followed by an SEP) and 28 negative cases (flare not followed by an SEP), yielding a class imbalance of approximately 9.56\% negative cases (or 90.44\% positive). Each event consists of 5 features, resulting in a data size of approximately 40 bytes per sample. The time coverage of a single input sample, defined as the duration between flare onset and SEP detection, ranges from 89 minutes to 5,886 minutes across the dataset.

\textbf{Outputs}: Each trained survival model outputs a survival function \( S(t) \), which indicates the probability that an SEP event has not occurred at a given time post-flare. The random survival forest version enhances interpretability and captures non-linear relationships in the flare–SEP timing data.

\textbf{Model Configuration}: The models were implemented in Python using \textit{scikit-survival} and \textit{sklearn}. Hyperparameters were tuned using \textit{GridSearchCV}, but no learnable weights exist in these models. For the random survival forests, 300 estimators were used. Tree depth and split criteria were optimized for performance via log-rank test statistics.

\textbf{Model Validation and Results}: Performance was assessed using the concordance index (C-index) on held-out validation sets. The best-performing Cox PH model achieved a C-index of $\sim 0.82$, indicating strong predictive capability for time-to-detection of SEPs following solar flares. Feature importance analysis identified flare longitude as the most influential predictor, followed by GOES class and flare latitude. Models were validated using 5-fold cross-validation to ensure generalizability, and survival curves were compared between SEP and non-SEP events to evaluate separation. Survival tree and random survival forests methods yielded similar trends in predictor importance, with added benefits in capturing nonlinear interactions. Results support the viability of survival analysis as a practical forecasting framework for operational space weather applications.

\textbf{Access to Model Data and Forecasts}: All data and code are publicly available. The model dataset is hosted on Harvard Dataverse at \url{https://doi.org/10.7910/DVN/GXY9MZ}, and the codebase is maintained on GitHub at \url{https://github.com/indiajacksonphd}.

\textbf{Limitations, Caveats and Discussion}: This model assumes that SEP onset is directly related to flare timing, which excludes CME-only or shock-driven SEP events. Additionally, the flare-to-SEP association window may introduce uncertainty due to event overlap. Future work may expand to include CME parameters and solar wind context for increased accuracy.


\subsection{SEP-C Model} \label{sec:SEP-C}

\textbf{Model Developers and Relevant Citation}: Jesse Torres, Philip K. Chan, Lulu Zhao, and Ming Zhang; \cite{torres2022machine}.

\begin{table}[h]
\caption{Model, Input and Output Specification Table for the SEP-C model.}
\centering
\begin{tabular}{|lc|}
\hline
\multicolumn{2}{|c|}{Model}                                       \\ \hline
\multicolumn{1}{|l|}{Type}            & Neural network            \\
\multicolumn{1}{|l|}{Complexity}      & 780                       \\ \hline
\multicolumn{2}{|c|}{Input}                                       \\ \hline
\multicolumn{1}{|l|}{Shape}           & Point Data (0D Features)  \\
\multicolumn{1}{|l|}{Type}            & Proton Flux, Coronagraphs\textsuperscript{*}, Solar Wind  \\
\multicolumn{1}{|l|}{History}         & 21 years (1996-2017)      \\
\multicolumn{1}{|l|}{Diversity}       & 20,210 samples            \\
\multicolumn{1}{|l|}{Imbalance}       & 0.0046 positive           \\
\multicolumn{1}{|l|}{Sample Size}     & 100 bytes                 \\
\multicolumn{1}{|l|}{Sample Coverage} & 2 hours                   \\ \hline
\multicolumn{2}{|c|}{Output}                                      \\ \hline
\multicolumn{1}{|l|}{Prediction}      & Classification, Probability, Regression \\
\multicolumn{1}{|l|}{Type}            & Continuous                \\
\multicolumn{1}{|l|}{Forecast Window} & 24 hours                  \\ \hline
\multicolumn{2}{|p{8.5cm}|}{Comments: \textsuperscript{*}This study uses properties of CMEs (Point Data) derived by the SOHO/LASCO coronagraphs.}   \\ \hline
\end{tabular}
\end{table}

\textbf{Summary}: The SEP-C model is a NN based on characteristics of CMEs and Type II radio waves from the CDAW catalog, solar wind speed, and sunspot number, to forecast whether CMEs would lead to SEP events. 

\textbf{Model Description}: The NN is a Multi-Layer Perceptron (MLP) with one hidden layer of 30 units and Rectified Linear Unit (ReLU) activations.  The loss function is binary cross entropy with L2 regularization to reduce overfitting. 

\textbf{Inputs}: The input features are based on characteristics of CMEs, Type II radio waves, and sunspot number. The basic CME features include linear speed, width, acceleration, second order speed initial, second order speed final, second order speed at 20 solar radii, central position angle, measurement position angle. Extended features include the number of CMEs with the past month and past 9 hours, the maximum speed of all CMEs within the past day, the number of CMEs with a speed greater than 1,000 km/s, $v*log(v)$ (where $v$ is the linear speed), halo, and particle intensity based on Diffusive Shock Acceleration \citep{drury1983introduction}. Other features are the solar wind speed at Earth, area in the spectrogram (duration $\times$ frequency range) of a Type II burst, sunspot number, and empirical SEP intensity prediction formula based on Richardson et al. \cite{richardson2018prediction}. The imbalance ratio is about 1:300 and by varying oversampling a 1:3 ratio was found to be desirable. 

\textbf{Outputs}: The output of the model is a score, which can roughly be interpreted as a probability, of whether a CME is associated with an SEP event. The natural logarithm of proton intensity 30 (or 60) minutes in the future.

\textbf{Model Configuration}: Training the model uses a learning rate of 0.1, momentum coefficient of 0.9, batch size of 200, and an L2 regularization term of 0.1. The model is allowed to train for up to 2,000 iterations before stopping unless the loss does not decrease by $10^{-4}$ within 10 consecutive iterations.

\textbf{Model Validation and Results}: This study indicates that the model with all the features can achieve 0.906 in TSS, 0.245 in HSS, and 0.246 in F1. The features were divided into 5 groups (speed, size, location, history, and others) and found that speed-related and other (e.g. Type II burst, sunspot) features are relatively more important. For forecasting intensity, our study indicates that this model can achieve 0.379 in Mean Absolute Error (MAE) for 30-minute forecast and 0.599 in MAE for 60-min forecast. For forecasting the start of SEP events exceeding 10 pfu intensity threshold of $\geq 10$ MeV protons, periods of advanced and extended warnings are incorporated. The SEP-C model can achieve 0.76 in F1 for 30-minute forecast and 0.85 in F1 for 60-minute forecast.

\textbf{Access to Model Data and Forecasts} The CME list used in this study can be found in \url{https://cdaw.gsfc.nasa.gov/CME_list/} and \url{https://cdaw.gsfc.nasa.gov/CME_list/radio/waves_type2.html} and the repository with the data and code can be found here \url{https://doi.org/10.5281/zenodo.12832882}.

\textbf{Limitations, Caveats and Discussion}: Most false alarms occur when CMEs are not fast and large. Future work will continue to reduce the false alarms by adding more features, particularly those that can distinguish SEPs with features that are similar to non-SEPs, such as low speed. The CME speed and size data from the CDAW catalog contain observation limitations such as projection effects. Including properties of CME size, location, and speed listed in the DONKI database can reduce the false alarm rate. The data in this study cover a partial SC. Particularly, the test data set in the evaluation is near the solar activity maximum. Extending the data to cover two SCs could help improve the model.


\subsection{Custom Architecture Neural Network (CANN) Model} \label{sec:CANN}

\textbf{Model Developers and Relevant Citation}; Viacheslav M. Sadykov, Alexander G. Kosovichev, Irina N. Kitiashvili, Vincent Oria, Gelu M. Nita, Egor Illarionov, Patrick M. O’Keefe, Yucheng Jiang, Sheldon H. Fereira, Aatiya Ali; \cite{sadykov2021prediction}.

\begin{table}[h]
\caption{Model, Input and Output Specification Table for the CANN model.}
\centering
\begin{tabular}{|lc|}
\hline
\multicolumn{2}{|c|}{Model}                                            \\ \hline
\multicolumn{1}{|l|}{Type}            & Neural Network                 \\
\multicolumn{1}{|l|}{Complexity}      & 1,243                          \\ \hline
\multicolumn{2}{|c|}{Input}                                            \\ \hline
\multicolumn{1}{|l|}{Shape}           & Point Data (0D)                \\
\multicolumn{1}{|l|}{Type}            & Soft X-ray, Proton Flux, Magnetic Fields and Ground-Based Radio \\
\multicolumn{1}{|l|}{History}         & 9 years (2010–2019)            \\
\multicolumn{1}{|l|}{Diversity}       & 2,288 samples                  \\
\multicolumn{1}{|l|}{Imbalance}       & 0.0294                         \\
\multicolumn{1}{|l|}{Sample Size}     & 1,244 bytes                    \\
\multicolumn{1}{|l|}{Sample Coverage} & 24 hours                       \\ \hline
\multicolumn{2}{|c|}{Output}                                           \\   \hline
\multicolumn{1}{|l|}{Prediction}      & Classification, Probability    \\
\multicolumn{1}{|l|}{Type}            & Continuous                     \\
\multicolumn{1}{|l|}{Forecast Window} & 24 hours                       \\ \hline
\end{tabular}
\end{table}

\textbf{Summary}: The daily whole-Sun binary prediction of SPEs ($\geq 10$ MeV, $\geq 10$ pfu) is considered, along with the related all-clear problem. The network is a custom-architected NN that utilizes the magnetic field properties of the solar ARs, the statistical properties of the preceding soft X-ray and proton fluxes, as well as records of solar radio bursts, as an input. The model has been evaluated on the SC 24 data, and demonstrated the performance comparable to or better than the daily probabilistic SWPC NOAA forecasts, especially in situations when missing the events is undesirable (all-clear regime).

\textbf{Model Description}: The ML model considered in this work is a custom-architected NN. The key idea of the custom architecture is to pre-process the magnetic field properties of individual ARs in the identical way first (in so-called AR blocks), then sum up the output from these blocks, and propagate it into the fully-connected part to concatenate with the whole-Sun statistical features from soft X-ray and proton flux observations and daily counts of radio bursts. Since each AR block shares the same weights and biases updated synchronously during the network training, the network has significantly fewer free parameters compared to the fully-connected analog, which makes it less prone to overfitting. The network implemented by \cite{sadykov2021prediction} has 30–15–8–4–2 neurons for AR blocks and 21–15–10–5–2 for the main network, resulting in 1243 free parameters.

\textbf{Inputs}: The model takes the median SHARP \citep{bobra2014helioseismic} properties of 10 solar ARs with the largest unsigned magnetic flux (including AR locations and quality parameters) as inputs into the AR blocks. The ARs that are rotated behind the 68-degree longitude are assumed to have their SHARP properties unchanged for 11 days, and are propagated with the Carrington rotation rate. The model then concatenates the AR block outputs with daily soft X-ray properties (mean, standard deviation, median, minimum, maximum), daily $\geq 10$ MeV proton flux properties (mean, standard deviation, median, minimum, maximum, last value), and daily counts of type II, III, and IV radio bursts.

\textbf{Outputs}: The model produces a daily binary and probabilistic forecast for $\geq 10$ MeV $\geq 10$ pfu proton events.

\textbf{Model Configuration}: While a certain model architecture has been considered by \cite{sadykov2021prediction}, the network can be adjusted with respect to any additional AR or whole-Sun properties (requires the model retraining).

\textbf{Model Validation and Results}: The model has been trained and tested on the SC 24 proton events. The train data set contained 2,222 non-SPE days and 66 SPE days (the days when the flux of $\geq 10$ MeV protons has exceeded 10 pfus). These numbers are 1178 and 35 for the test data set, respectively. The model has been evaluated using the standard metrics of TSS, two variations of the HSS, AUC, and the new metric of weighted TSS introduced in \cite{sadykov2021prediction}. The results have been compared with the performance of the SWPC NOAA daily probabilistic models. It was found that, overall, ML-driven prediction outperforms SWPC NOAA forecasts in the all-clear regimes (when missing an SEP is undesirable), and has a competitive TSS = $0.82 \pm 0.01$ and HSS = $0.38 \pm 0.03$ overall.

\textbf{Access to Model Data and Forecasts}: The SPE catalogs and the Jupyter notebooks developed during this study are archived at the Solar Energetic Particle Prediction Portal, which can be found here: \url{https://sun.njit.edu/SEP3/datasets.html}.

\textbf{Limitations, Caveats and Discussion}: The major limitation of the current work is the training and test data set sizes, which have included only the SC 24 data (and, therefore, a very limited number of unique SEPs). Other limitations include the consideration of relatively weak S1 NOAA events only ($\geq 10$ MeV $\geq 10$ pfu), reliance on the science-quality SHARP and GOES data instead of the operational data streams.


\subsection{SEP-E Model} \label{sec:SEP-E}

\textbf{Model Developers and Relevant Citation}: Jesse Torres, Philip K. Chan, Lulu Zhao, and Ming Zhang; \cite{torres2025machine}.

\begin{table}[h]
\caption{Model, Input and Output Specification Table for the SEP-E model.}
\centering
\begin{tabular}{|lc|}
\hline
\multicolumn{2}{|c|}{Model}                                           \\ \hline
\multicolumn{1}{|l|}{Type}            & Neural network                \\
\multicolumn{1}{|l|}{Complexity}      & 1,530                         \\ \hline
\multicolumn{2}{|c|}{Input}                                           \\ \hline
\multicolumn{1}{|l|}{Shape}           & Time Series (1D)              \\
\multicolumn{1}{|l|}{Type}            & Proton Flux, Electron Flux    \\
\multicolumn{1}{|l|}{History}         & 7 years (1995-2002)           \\
\multicolumn{1}{|l|}{Diversity}       & 517,769 samples               \\
\multicolumn{1}{|l|}{Imbalance}       & 0.0103                        \\
\multicolumn{1}{|l|}{Sample Size}     & 200 bytes                     \\
\multicolumn{1}{|l|}{Sample Coverage} & 2 hours                       \\ \hline
\multicolumn{2}{|c|}{Output}                                          \\ \hline
\multicolumn{1}{|l|}{Prediction}      & Classification, Probability, Regression \\
\multicolumn{1}{|l|}{Type}            & Continuous                    \\
\multicolumn{1}{|l|}{Forecast Window} & 0.75 hours                    \\ \hline
\multicolumn{2}{|p{8.5cm}|}{Comments: \textsuperscript{*}This value reflects the average of the forecast window range of 30 - 60 minutes.}   \\ \hline
\end{tabular}
\end{table}

\textbf{Summary}: The SEP-E  model is a NN based on electron and proton intensities to forecast proton intensities 30 (or 60) minutes in the future every 5 minutes.

\textbf{Model Description}: The model is an RNN with one hidden layer of 30 gated recurrent units. Mean squared error (MSE) is used for the loss function.

\textbf{Inputs}: Two hours of electron intensities from the $\geq 0.25$ and $\geq 0.67$ MeV channels and proton intensities $\geq 10$ Mev channels at 5-minute intervals. Input to the model is the time series of natural logarithm of measured particle intensities in pfu. solar X-ray emission data have also been used in the SEP-E model, but they did not help improve the prediction performance.

\textbf{Outputs}: The natural logarithm of proton intensity 30 (or 60) minutes in the future.

\textbf{Model Configuration}: Weights are updated using the Adam optimizer, and up to 1,000 iterations are allowed unless the network converges before then. The NN converges if the loss function does not change by more than $1{0}^{-4}$ over 20 iterations.

\textbf{Model Validation and Results}: For forecasting intensity, our study indicates that our model can achieve 0.379 in MAE for 30-minute forecast and 0.599 in MAE for 60-minute forecast. For forecasting the start of SEP events exceeding 10 pfu intensity threshold of $\geq 10$ MeV protons, periods of advanced and extended warnings are incorporated. SEP-E can achieve 0.76 in F1 for 30-minute forecast and 0.85 in F1 for 60-minute forecast.

\textbf{Access to model data and Forecasts}: The code and data for this work can be found here: \url{https://doi.org/10.5281/zenodo.12832882}.

\textbf{Limitations, Caveats and Discussion}: The data in this study cover a partial SC. Particularly, the test data set in the evaluation is near the solar activity maximum. Extending the data to cover two SCs could help improve the model.


\subsection{Space Radiation Intelligence System (SPRINTS) Model} \label{sec:SPRINTS}

\textbf{Model Developers and Relevant Citation}: Alec Engell, Brianna Maze, Harold Farmer; \cite{engell2017sprints}. 

\begin{table}[h]
\caption{Model, Input and Output Specification Table for the SPRINTS model.}
\centering
\begin{tabular}{|lc|}
\hline
\multicolumn{2}{|c|}{Model}                                           \\ \hline
\multicolumn{1}{|l|}{Type}            & Neural Network                \\
\multicolumn{1}{|l|}{Complexity}      & 5,401\textsuperscript{*}      \\ \hline
\multicolumn{2}{|c|}{Input}                                           \\ \hline
\multicolumn{1}{|l|}{Shape}           & Point Data (0D)               \\
\multicolumn{1}{|l|}{Type}            & Soft X-ray, Flare Location    \\
\multicolumn{1}{|l|}{History}         & 31 years (1986–2017)          \\
\multicolumn{1}{|l|}{Diversity}       & 2,263 samples                 \\
\multicolumn{1}{|l|}{Imbalance}       & 0.067 \textsuperscript{*}     \\
\multicolumn{1}{|l|}{Sample Size}     & 32 bytes                      \\
\multicolumn{1}{|l|}{Sample Coverage} & 0 hours                       \\ \hline
\multicolumn{2}{|c|}{Output}                                          \\ \hline
\multicolumn{1}{|l|}{Prediction}      & Classification, Probability   \\
\multicolumn{1}{|l|}{Type}            & Triggered                     \\
\multicolumn{1}{|l|}{Forecast Window} & 24 hours \textsuperscript{**} \\ \hline
\multicolumn{2}{|p{8cm}|}{Comments: \textsuperscript{*}These values correspond to the 10\_10\_24 model \textsuperscript{**}SPRINT can have prediction windows that range from 0 to 96 hours, the results presented in this document is for the 24 hours prediction model.}   \\ \hline
\end{tabular}
\end{table}

\begin{figure}[h]  
    \centering 
    \includegraphics[width=0.75\textwidth]{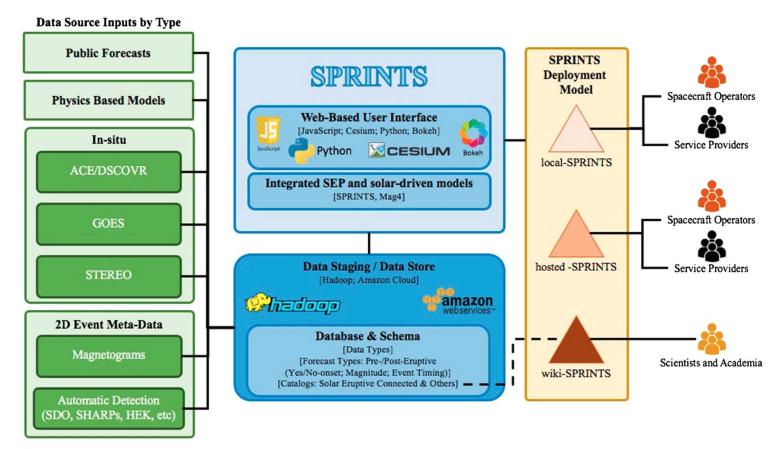} 
    \caption{SPRINTS logical architecture organized by data sources, data staging/store, user interface, forecast models (integrated MAG4 and SPRINTS models), deployment models, and users.} 
    \label{fig:SPRINTS}
\end{figure}

\textbf{Summary}: The Space Radiation Intelligence System (SPRINTS; Figure~\ref{fig:SPRINTS}) is a modeling framework for data-driven forecasting of solar-driven events. It applies event catalogs and databased observations including flares, SEPs, CMEs, and radio bursts as well as associated catalogs such as flare events that are associated to SEP events. The developed SPRINTS SEP forecasting model uses an MLP model based on flare parameters including flare flux, flare fluence, flare decay phase, flare long/short X-ray ratio, and flare longitude.  

\textbf{Model Description}: The MLP model is used for binary classification to distinguish between SEP and non-SEP events. Specifically, the model outputs a probability whether an SEP will occur within a given time window at a given proton energy channel. A separate MLP model is trained to predict SEP occurrences at user-defined proton energy channels (e.g., 1, 5, 10, 30, 50, 100 MeV), flux thresholds, and time resolutions (e.g., 12-hour) up to 96 hours in advance. One model is trained per combination of energy channel and time resolution, resulting in 20 independent MLP models. 

\textbf{Inputs}: There are 252 flare-SEP events and 19,959 flare-only events that are cataloged in the SPRINTS database. To reduce the number of flare-only events used during training, flares that had a fluence below the minimum fluence observed in any SEP event were excluded. This filtering step reduced the flare-only set to 10,084 events, ensuring that the model was trained on more challenging negative samples. To further address the class imbalance between SEP and flare-only events, a random subset comprising 20\% of the remaining flare-only events was selected for inclusion in the training set. This resulted in 2,017 flare-only events. Because the models were designed to predict whether an SEP would occur at a given proton energy channel and flux threshold, it was necessary to calculate whether each of the 252 events exceeded the threshold at each given energy channel. An SEP event was labeled as positive (1 label) for a given energy-threshold bin if it exceeded that threshold; otherwise, it was labeled negative (0 label). Since not all SEP events exceeded every threshold, the number of positive samples varied across bins, further reducing the positive samples down from the original 252 for each bin. 

Each model was trained using flare characteristics including X-ray flux, X-ray fluence, X-ray peak ratio (long vs. short channel), and flare longitude. Challenge events defined by CCMC were completely removed from the dataset and used for a hold-out test set. The remaining dataset was split into training and testing sets using a 90/10 ratio, with stratification to preserve the distribution of positive and negative samples. If the resulting test set did not include any SEP events, one SEP event was randomly moved from the training set to the test set to ensure representation. Feature normalization was performed using the \textit{StandardScaler}\footnote{\url{https://scikit-learn.org/stable/modules/generated/sklearn.preprocessing.StandardScaler.html}} from Python’s \textit{sklearn} library, which standardizes features by removing the mean and scaling to unit variance. It’s important to note the mean and standard deviation were computed from the training data alone and then applied to both training and testing sets to prevent data leakage.

\textbf{Outputs}: The model outputs a probability representing the likelihood that a given solar flare will produce an SEP event. This probability is converted into a binary classification (True/False) using a threshold of 0.5, indicating whether an SEP event is expected to occur given the flare metadata. The system is configurable to any desired GOES energy channel and flux threshold and is currently deployed in real-time for the following thresholds: 10 MeV at 10 pfu, 10 MeV at 40 pfu, 30 MeV at 10 pfu, 50 MeV at 10 pfu, 100 MeV at 1 pfu.

\textbf{Model Configuration}: A separate MLP model was trained for each energy channel and threshold requirement, resulting in a total of 20 models. To optimize performance, each model underwent a two-stage hyperparameter tuning process. Initially, randomized grid search with k-fold cross-validation using 3-folds was employed to efficiently explore a broad hyperparameter space. Based on the results of this preliminary search, a refined parameter space was defined for a full grid search, again using 3-fold cross-validation. The use of k-fold cross-validation helped ensure that the models were not overfitting to the training data. Hyperparameters tuned during this process included the number and size of hidden layers, activation functions, learning rate, and maximum number of training iterations. Because tuning was performed individually for each model, the final configurations varied across energy channels and time resolutions. Once the optimized parameters were found using cross-validation, the models were re-trained on the entire training dataset and saved.

\textbf{Model Validation and Results}: Model performance was evaluated using the HSS, POD, and False Detection Rate. HSS results of the models after training on the train set and evaluating on the hold-out test set range from 0.17 – 0.65 for the 24-hour time window across the different energy channels. It is important to note that as the forecast time window increases, in particular at 72 and 96 hours, the skill scores drop dramatically or exhibit large variability. This is primarily due to the substantial class imbalance between SEP and flare-only events at longer lead times, which results in poor representation of positive (SEP) samples for both training and evaluation.

\textbf{Access to Model Data and Forecasts}: Model outputs are publicly available on the CCMC SEP scoreboard \url{https://ccmc.gsfc.nasa.gov/scoreboards/sep/} via the SPRINTS REST API.

\textbf{Limitations, Caveats and Discussion}: Because each MLP model is trained independently for a specific combination of temporal bin and proton energy/flux threshold, the resulting probability outputs are not inherently consistent across models. This means that the predicted probability of SEP occurrence at a higher energy channel may exceed that of a lower energy channel, even though such behavior may appear counterintuitive from a physical standpoint. This is a direct consequence of training a separate MLP model for each unique combination of temporal bin and energy/flux threshold. The disproportionate number of flare-only events relative to SEP events presented challenges during model training. Without proper handling, this imbalance can lead to biased models that underperform in predicting the minority class (SEPs). Careful attention was paid to create representative and balanced training and testing datasets. This includes strategies such as selective sampling of flare-only events and stratified data splitting, ensuring that each model is trained and evaluated on data that reflect the operational scenario as closely as possible.


\subsection{Time-Series Forest (TSF) Model} \label{sec:TSF}

\textbf{Model Developers and Relevant Citation}: Pouya Hosseinzadeh, Soukaina Filali Boubrahimi, and Shah Muhammad Hamdi; \cite{hosseinzadeh2024improving}.

\begin{table}[h]
\caption{Model, Input and Output Specification Table for the TSF model.}
\label{tab:TSF}
\centering
\begin{tabular}{|lc|}
\hline
\multicolumn{2}{|c|}{Model}                                            \\ \hline
\multicolumn{1}{|l|}{Type}            & Forest, Ensemble               \\
\multicolumn{1}{|l|}{Complexity}      & 15,000                         \\ \hline
\multicolumn{2}{|c|}{Input}                                            \\ \hline
\multicolumn{1}{|l|}{Shape}           & Time Series (1D)               \\
\multicolumn{1}{|l|}{Type}            & Energetic Protons              \\
\multicolumn{1}{|l|}{History}         & 25 years (1986-2011)           \\
\multicolumn{1}{|l|}{Diversity}       & 141 samples                    \\
\multicolumn{1}{|l|}{Imbalance}       & 0.0746 positive                \\
\multicolumn{1}{|l|}{Sample Size}     & 11.52 bytes                    \\
\multicolumn{1}{|l|}{Sample Coverage} & 5 hours                        \\ \hline
\multicolumn{2}{|c|}{Output}                                           \\ \hline
\multicolumn{1}{|l|}{Prediction}      & Classification                 \\
\multicolumn{1}{|l|}{Type}            & Triggered                      \\
\multicolumn{1}{|l|}{Forecast Window} & 0 hours\textsuperscript{*}     \\ \hline
\multicolumn{2}{|p{8.5cm}|}{Comments: \textsuperscript{*}A forecast window of 0 hours indicates that the model performs event-level classification rather than temporal forecasting, similar to Table~\ref{tab:CART}.}    \\ \hline
\end{tabular}
\end{table}

\begin{figure}[h]
    \centering
    \begin{subfigure}{0.48\textwidth}
        \centering
        \includegraphics[width=\textwidth]{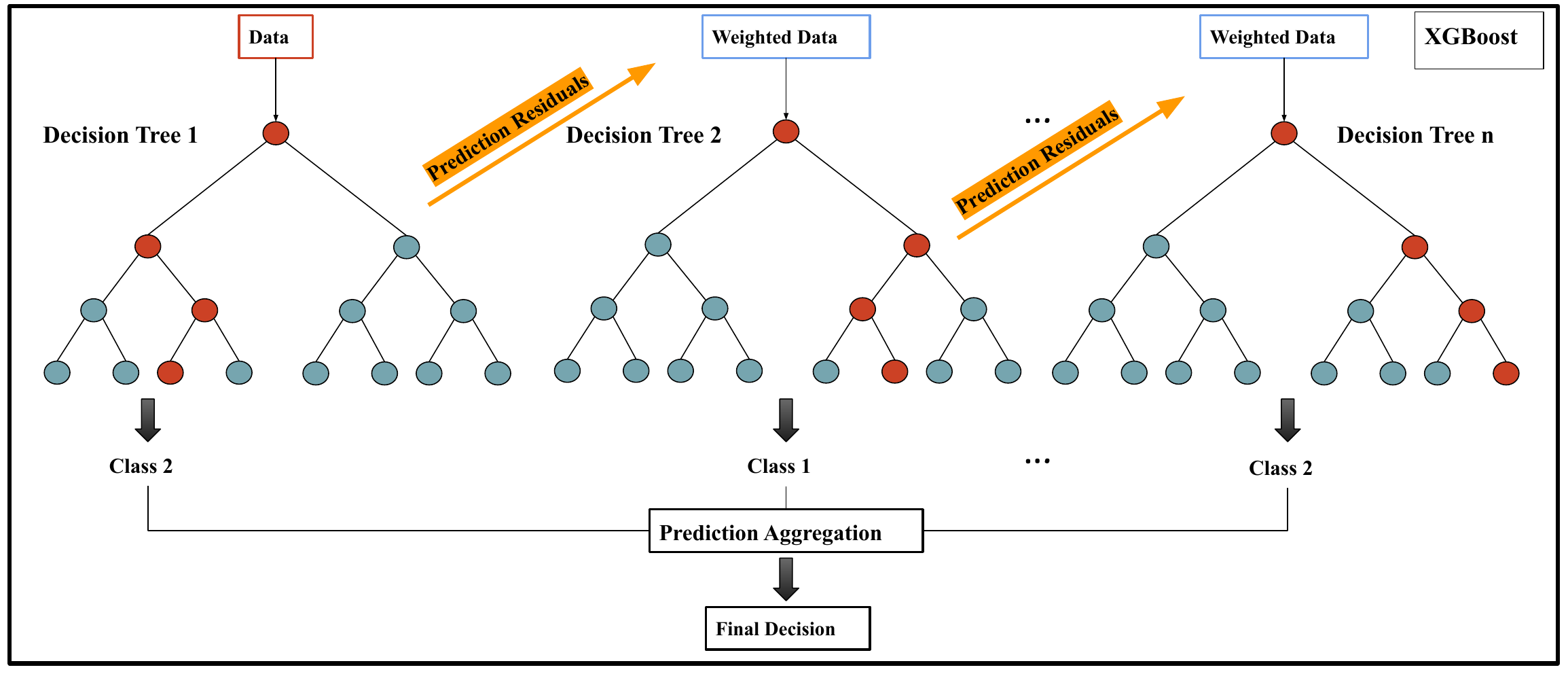}
        \caption{XGBoost-style boosted decision-tree ensemble.}
        \label{fig:S3_XGBoost}
    \end{subfigure}
    \hfill
    \begin{subfigure}{0.48\textwidth}
        \centering
        \includegraphics[width=\textwidth]{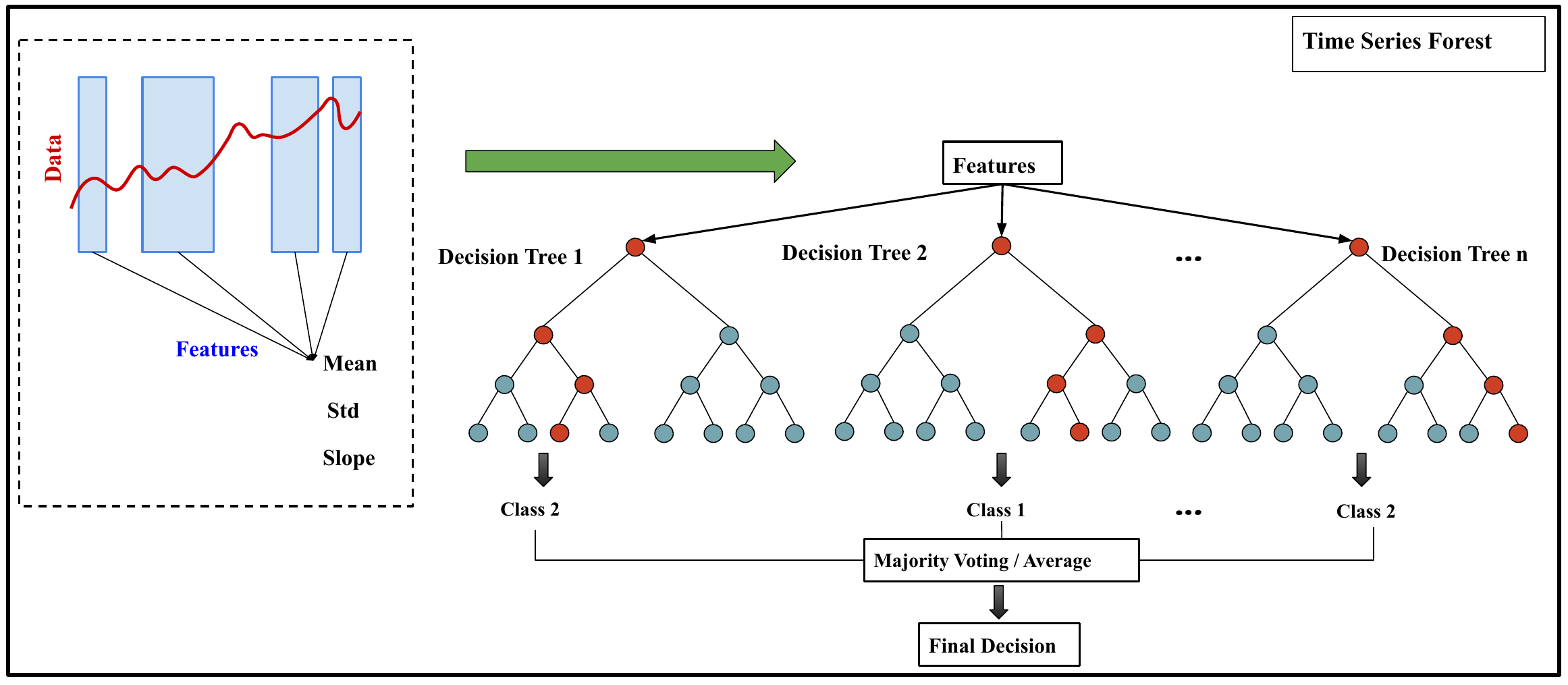}
        \caption{Time Series Forest (TSF) pipeline.}
        \label{fig:S3_TSF}
    \end{subfigure}

    \vspace{0.6em}

    \begin{subfigure}{0.48\textwidth}
        \centering
        \includegraphics[width=\textwidth]{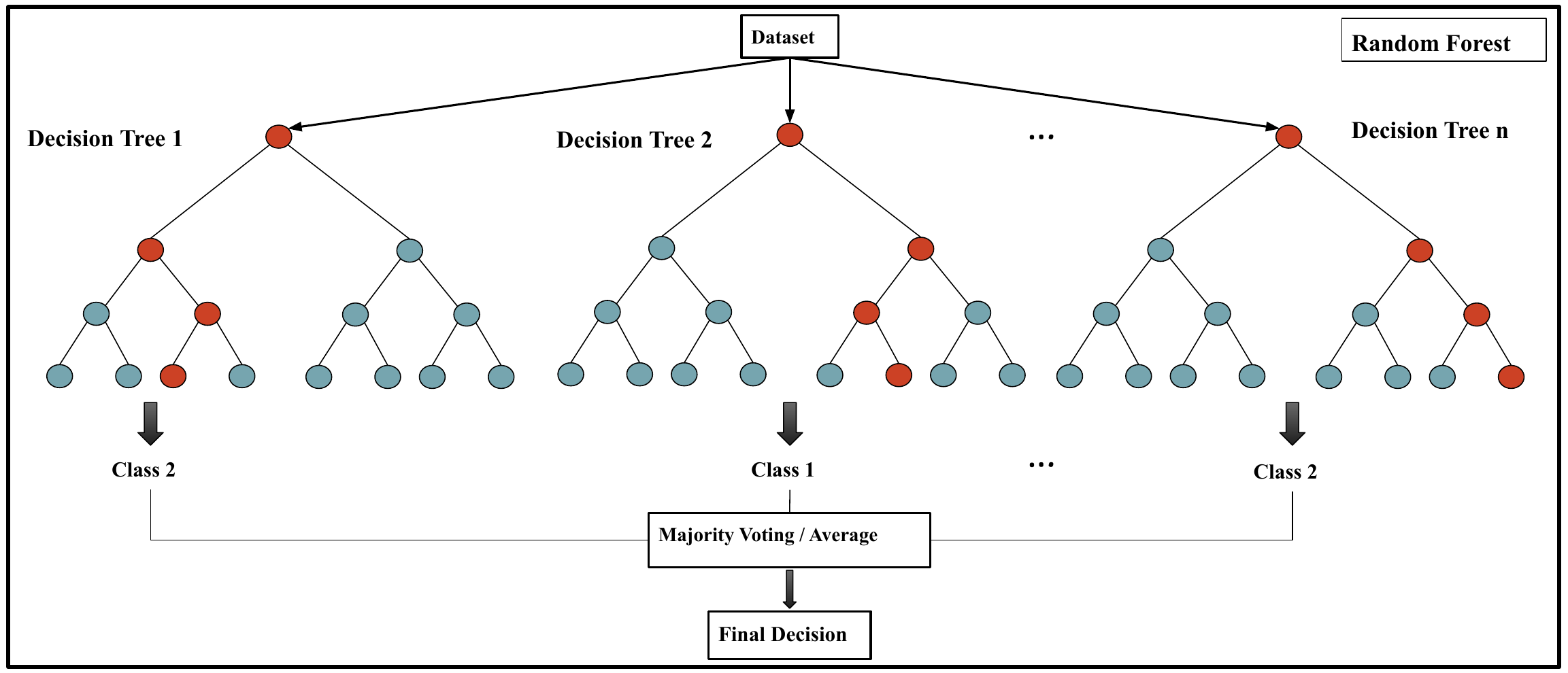}
        \caption{Random forest ensemble classifier.}
        \label{fig:S3_RF}
    \end{subfigure}
    \caption{Tree-based ensemble architectures used for SEP classification, including gradient boosting, Time Series Forest, and Random Forest models.}
    \label{fig:S3_Tree_Ensembles}
\end{figure}

\textbf{Summary}: This work introduces a time series data augmentation approach to improve SEP event prediction for $\sim 30$, $\sim 60$, and $\sim 100$ MeV energy bands using GOES proton flux. Three time series classification models ---Time Series Forest (TSF; Figure~\ref{fig:S3_TSF}), ROCKET, and SHAPELET--- are evaluated. The model performance is significantly enhanced by applying SMOTE, ADASYN, and Gaussian noise. Among all models, TSF consistently outperforms others, reaching up to 90\% accuracy in the $\sim 100$ MeV prediction task.

\textbf{Model Description}: The model uses time series classification techniques to distinguish between SEP and non-SEP events using GOES proton flux data. TSF, a random forest-based ensemble method (Figure~\ref{fig:S3_RF}), is applied to 1D time series segments extracted from 5-hour observation windows. TSF randomly selects intervals from each time series and calculates statistical features such as mean, standard deviation, and slope. The model is trained using both real and synthetically generated SEP samples to balance the class distribution. This approach supports both binary classification (SEP vs. non-SEP) and hierarchical multi-class classification across energy levels ($\sim 30$, $\sim 60$, $\sim 100$ MeV).

\textbf{Inputs}: We used GOES proton flux time series data from three channels: P4 ($\sim 30$ MeV), P5 ($\sim 60$ MeV), and P6 ($\sim 100$ MeV), with 5-hour observation windows prior to solar flares. SEP event lists were extracted from the GSEP catalog, while non-SEP events were collected using the Heliophysics Event Knowledgebase (HEK\footnote{\url{https://www.lmsal.com/hek/}}). The SEP labels correspond to the increase in energetic protons detected after solar flare activity. Multivariate and univariate versions of these time series were created and used as input for classification models, with data augmentation techniques applied to SEP samples to address class imbalance.

\textbf{Outputs}: The output of the model is a probability of SEP occurrence based on a flare trigger. The probability is converted to a binary label (True/False), informing us whether a flare will produce an SEP event.

\textbf{Model Configuration}: The best-performing model is the TSF, which uses an ensemble of decision trees built on random intervals of the time series. Each tree is trained on statistical features (mean, standard deviation, slope) extracted from selected time intervals of the GOES proton flux. Data augmentation is applied using Gaussian noise, SMOTE, and ADASYN methods to balance the SEP vs. non-SEP classes. Models are evaluated using k-fold cross-validation, and both univariate and multivariate time series inputs are considered. Hierarchical modeling predicts high-energy SEPs first ($\sim 100$ MeV), then medium ($\sim 60$ MeV), and finally low-energy ($\sim 30$ MeV) events.

\textbf{Model Validation and Results}: The TSF model achieved significant improvements with data augmentation. For $\sim 100$ MeV SEP classification, accuracy increased from 70\% to 90\% using Gaussian noise. For $\sim 60$ and $\sim 30$ MeV, Gaussian yielded the best accuracy (~90\%). The model was evaluated using k-fold cross-validation with 10 folds for 100 MeV, and 7 folds for 60 and 30 MeV due to limited data. TSS and HSS scores increased to ~0.8–0.9 with augmentation. SMOTE and ADASYN also improved performance, especially for 100 MeV classification.

\textbf{Access to Model Data and Forecasts}: All input data used in the TSF model are derived from publicly accessible heliophysics repositories. Time-series proton flux measurements are obtained from the GOES Space Environment Monitor (SEM) archives provided by NOAA National Centers for Environmental Information (NCEI), with SEP-producing events sourced from the GSEP catalog and corresponding non-SEP flare events obtained from the HEK. These publicly available time-series datasets enable reproducible SEP classification experiments for high-energy ($\sim 100$ MeV) event prediction under fixed 6-hour observation window configurations.

\textbf{Limitations, Caveats and Discussion}: A key limitation of SEP prediction is the extreme class imbalance due to the rarity of high-energy events, especially $\sim 100$ MeV. Although data augmentation significantly improves classification performance, synthetic samples may not capture all physical characteristics of real SEP events. Additionally, performance varies by energy band ($\sim 100$ MeV classification remains more challenging than $\sim 30$ or $\sim 60$ MeV). While TSF achieves high accuracy, further improvement may require incorporating additional data modalities or physics-based constraints to better generalize to unseen SEP scenarios.


\subsection{Univariate Deep Merge (UDM) Model} \label{sec:UDM}

\textbf{Model Developers and Relevant Citation}: Pouya Hosseinzadeh, Soukaina Filali Boubrahimi, and Shah Muhammad Hamdi; \cite{hosseinzadeh2024toward}.

\begin{table}[h]
\caption{Model, Input and Output Specification Table for the UDM model.}
\centering
\begin{tabular}{|lc|}
\hline
\multicolumn{2}{|c|}{Model}                                            \\ \hline
\multicolumn{1}{|l|}{Type}            & Time Series Forest             \\
\multicolumn{1}{|l|}{Complexity}      & 15,548                         \\ \hline
\multicolumn{2}{|c|}{Input}                                            \\ \hline
\multicolumn{1}{|l|}{Shape}           & Time Series (1D), Vectors (1D) \\
\multicolumn{1}{|l|}{Type}            & EUV Imagery, Energetic Protons \\
\multicolumn{1}{|l|}{History}         & 15 years (1997-2012)           \\
\multicolumn{1}{|l|}{Diversity}       & 59 samples                     \\
\multicolumn{1}{|l|}{Imbalance}       & 0.2655 positive                \\
\multicolumn{1}{|l|}{Sample Size}     & 11.52 bytes                    \\
\multicolumn{1}{|l|}{Sample Coverage} & 6 hours                        \\ \hline
\multicolumn{2}{|c|}{Output}                                           \\ \hline
\multicolumn{1}{|l|}{Prediction}      & Classification                 \\
\multicolumn{1}{|l|}{Type}            & Triggered                      \\
\multicolumn{1}{|l|}{Forecast Window} & 0 hours\textsuperscript{*}                        \\ \hline
\multicolumn{2}{|p{8.5cm}|}{Comments: \textsuperscript{*}A forecast window of 0 hours indicates that the model performs event-level classification rather than temporal forecasting, similar to Table~\ref{tab:CART}.}   \\ \hline
\end{tabular}
\end{table}

\textbf{Summary}: This study presents a multimodal time series data fusion framework for predicting high-energy ($\sim 100$ MeV) SEP events by combining GOES proton flux data and solar EUV images. Six ML models are evaluated: two unimodal models—UTS (time series only) and Image (image only), and four fusion models (UFC, UDC, UDM, and USC). Among these, the Univariate Deep Merge (UDM) achieves the highest performance, reaching 0.80 Accuracy and 0.81 Precision under the balanced setting. The results highlight the importance of both temporal and spatial information for SEP classification, and show how optimal observation window sizes and image vector lengths significantly impact prediction accuracy.

\textbf{Model Description}: The best-performing model in this study is the UDM model, a multimodal data fusion architecture that integrates time series proton flux data and solar EUV image (converted to vectors). The model applies a deep learning-based strategy in which separate neural branches process each modality independently—one branch for 5-hour time series segments from the GOES P6 proton channel ($\sim 100$ MeV), and another for 200-dimensional vectors extracted from EUV images using an autoencoder. These representations are merged using element-wise operations within the network to capture both shared and complementary features. The fused output is then passed through dense layers for final classification. UDM effectively captures temporal dynamics and spatial patterns, enabling robust prediction of high-energy SEP events.

\textbf{Inputs}: The model uses two data modalities: a) time series proton flux data from the GOES P6 channel ($\sim 100$ MeV) over a 6-hour observation window preceding solar flares, and b) solar single images from SOHO’s Extreme-ultraviolet Imaging Telescope \citep[EIT;][]{delaboudiniere1995eit} 304\AA  \, channel, captured within 4 hours before the flare start time. The time series data captures temporal dynamics of energetic particle activity, while the EUV images are transformed into 200-dimensional latent vectors using an autoencoder to extract spatial features. SEP events are sourced from the GSEP catalog, and non-SEP events are selected from the HEK flare records with peak intensity $\geq C1.3$ that did not lead to SEP events.

\textbf{Outputs}: The output of the model is a probability of SEP occurrence based on a flare trigger. The probability is converted to a binary label (True/False), informing us whether a flare will produce an SEP event.

\textbf{Model Configuration}: The UDM model consists of two parallel neural branches: one processes 6-hour GOES P6 time series using a sequence of dense and dropout layers, and the other processes 200-dimensional EUV image vectors extracted via an autoencoder. Features from both branches are merged using element-wise operations (e.g., addition), followed by fully connected layers and a softmax output for classification. The model is implemented using \textit{TensorFlow}\footnote{\url{https://www.tensorflow.org/}} and trained using the Adam optimizer. Hyperparameter tuning determined the optimal configuration, including 150 estimators for the base TSF model and 5-fold stratified cross-validation for both balanced and imbalanced data settings.

\textbf{Model Validation and Results}: Model performance was evaluated using 5-fold cross-validation on both balanced and imbalanced datasets. The univariate UDM model consistently outperformed all other models across multiple metrics. On the balanced dataset, UDM achieved 0.80 Accuracy, 0.79 F1-score, and 0.81 Precision. It also showed strong Recall, TSS, and HSS scores, indicating its reliability in detecting high-energy SEP events. The model's robustness was further confirmed through noise sensitivity analysis, where UDM maintained the highest F1-scores under varying levels of Gaussian noise. Compared to state-of-the-art baselines such as XGBoost, SVM, RF, and decision trees, UDM showed superior average F1 and TSS scores, validating its effectiveness for $\sim 100$ MeV SEP prediction using multimodal fusion.

\textbf{Access to Model Data and Forecasts}:

The implementation of the UDM model, including preprocessing and training configurations, is available at \url{https://github.com/pouyahosseinzadeh/High-Impact-SEP-Prediction---Space-Weather}. All input data used in this study are derived from publicly accessible heliophysics repositories. Time-series proton flux measurements are obtained from the GOES SEM archives provided by NOAA NCEI, while solar EUV imagery is accessed via the SOHO instrument through the \textit{Helioviewer}\footnote{\url{https://gs671-suske.ndc.nasa.gov/}} platform. SEP event lists are sourced from the GSEP catalog, and non-SEP events are collected from the HEK. These multimodal inputs enable reproducible SEP classification experiments under fixed observation windows for high-energy ($\sim 100$ MeV) event prediction.

\textbf{Limitations, Caveats and Discussion}: A key limitation of this study is the limited number of high-energy ($\sim 100$ MeV) SEP events available, which restricts the size and diversity of the dataset. Although balanced and imbalanced settings were both evaluated, real-world scenarios involve extreme class imbalance that may impact generalization. The SOHO EUV images, used as part of the multimodal input, are infrequent and sometimes captured hours before flare onset, introducing temporal variability and potential misalignment with the time series data. Additionally, the model focuses solely on classification (SEP vs. non-SEP) without addressing the timing or intensity of SEP events. Future work should explore advanced augmentation, domain adaptation, and real-time integration strategies to improve robustness and operational readiness.


\subsection{UNifying Solar Particle Event modeLLing (UNSPELL) Model} \label{sec:UNSPELL}

\textbf{Model Developers and Relevant Citation}: Sigiava Aminalragia-Giamini, Constantinos Papadimitriou, Ingmar Sandberg, Savvas Raptis; \cite{aminalragia2021solar}.

\begin{table}[h]
\caption{Model, Input and Output Specification Table for the UNSPELL model.}
\centering
\begin{tabular}{|lc|}
\hline
\multicolumn{2}{|c|}{Model}                                            \\ \hline
\multicolumn{1}{|l|}{Type}            & Neural Network                 \\
\multicolumn{1}{|l|}{Complexity}      & 81,120\textsuperscript{*}  \\ \hline
\multicolumn{2}{|c|}{Input}                                            \\ \hline
\multicolumn{1}{|l|}{Shape}           & Point Data (0D), Spectra (1D)  \\
\multicolumn{1}{|l|}{Type}            & Soft X-ray, Flare Location     \\
\multicolumn{1}{|l|}{History}         & 25 years (1988-2013)           \\
\multicolumn{1}{|l|}{Diversity}       & 18,025 samples                 \\
\multicolumn{1}{|l|}{Imbalance}       & 0.0128 positive                \\
\multicolumn{1}{|l|}{Sample Size}     & 208 bytes                      \\
\multicolumn{1}{|l|}{Sample Coverage} & 0.4 hours\textsuperscript{**}  \\ \hline
\multicolumn{2}{|c|}{Output}                                           \\ \hline
\multicolumn{1}{|l|}{Prediction}      & Classification, Probability    \\
\multicolumn{1}{|l|}{Type}            & Triggered                      \\
\multicolumn{1}{|l|}{Forecast Window} & 24 hours                       \\ \hline
\multicolumn{2}{|p{8.5cm}|}{Comments: \textsuperscript{*}UNSPELL is an ensemble of 1000 NNs, each one of which has a complexity of 81120 trainable parameters. \textsuperscript{**}The sample coverage varies, with an average of 25 minutes.} \\ \hline
\end{tabular}
\end{table}

\textbf{Summary}: A ML approach was introduced in \cite{aminalragia2021solar} which uses as the main input X-ray measurements data to forecast whether solar flares will lead to SEP events. This model approach and rationale is focused on real-conditions applicability and the production of forecasts during and immediately after the occurrence of solar flares. The model uses publicly and real-time available data which entail GOES X-ray fluxes and the flare detection/definition provided by NOAA SWPC, as well as the flare heliolongitude of the solar flare, when available. The latter is retrieved from \textit{Solar Demon}\footnote{\url{https://www.sidc.be/solardemon/}} provided by the Royal Observatory of Belgium. Subsequent work further expanded on the methodology presented in \citep{aminalragia2021solar} and lead to the currently operational version integrated in the UNSPELL system as the solar flare module.

\textbf{Model Description}: The model uses a binary classification for SEP and non-SEP occurrence and provides a probability $P \in [0,1]$ that a flare will lead to an event. The core of the model are NNs which are run in parallel with the same inputs, each providing an independent estimation of the SEP probability occurrence. The final output consists of the mean probability as well as the probability standard deviation resulting from the ensemble outputs. To collapse the probabilistic output to a categorical one, these two outputs are compared against defined thresholds and if both are above or below defined values an alert that an SEP will occur is issued. 

\textbf{Inputs}: Model training used 228 samples labeled as positive ---flares associated with subsequent SEP events, and 17797 samples labeled as negative--- flares not associated with SEP events. The flare X-ray timeseries are used to derive 24 features for a flare which are used as input. If the heliolongitude is also available, its cosine and sine are also used as inputs. The flares used for training are detailed in the NOAA GOES solar flare catalogue and the training dataset spans the years 1988-2013 covering the largest part of SC22, the whole of SC23, and the rising phase of SC24. 

\textbf{Outputs}: The output of the model is the ensemble mean probability of SEP occurrence and the ensemble standard deviation of probabilities. The output is converted to a categorical binary label (True/False) using a threshold of 0.7 (above for True) for the average probability, and 0.06 (below for True) for the standard deviation.

\textbf{Model Configuration}: The model that is currently operational is based on the methodology described in \cite{aminalragia2021solar} with two substantial differences. The NNs used in the development and validation were deep networks with several layers and the ensemble members numbers $N$ that were tested were $N = 3$ or $N = 10$. Subsequent investigations showed that ensembles with shallow feed-forward networks with three layers and fewer neurons were able to match the performance previously achieved having the benefit of lower complexity and much faster training times. At the same time the ensemble number N was increased to 1000. While this is a high number, each NN was trained using a different subset of the total available training data, an approach bearing similarities to that of random forests and its variants. This approach was selected in order to achieve good generalization on the ensemble level, avoid any potential overfitting with a singular or few NNs, and have a large enough ensemble from which to derive a meaningful standard deviation of outputs to be used in the subsequent thresholding process.

\textbf{Model Validation and Results}: Our investigation on the operational module reproduced the findings of the original publication for all flares above C1 with true positive rates of 0.86 and true negative rates of 0.92 with a resulting TSS of $\sim 0.78$. The model has participated in SEPVAL and further validations on the operational outputs will be performed in the near future during the current SC25.

\textbf{Access to Model Data and Forecasts}: The X-ray data and flare list are provided by NOAA at online repositories and the SEP list used is detailed in \cite{pacheco2019analysis}. The forecasts of the model will be publicly available in 2026 through the ESA \textit{Space Weather Service Network}\footnote{\url{https://swe.ssa.esa.int/}} portal and specifically the \textit{Space Radiation Expert Service Centre}\footnote{\url{https://swe.ssa.esa.int/space-radiation}}, or can contact the authors of \cite{aminalragia2021solar} directly. For more information follow the links bellow: 

\vspace{-0.75em}
\begin{itemize} 
    \item \url{https://www.ncei.noaa.gov/data/goes-space-environment-monitor/access/science/xrs/}
    \vspace{-0.75em}
    \item \url{https://www.ngdc.noaa.gov/stp/space-weather/solar-data/solar-features/solar-flares/x-rays/goes/xrs/}
\end{itemize}

\textbf{Limitations, Caveats and Discussion}: A limitation of the model is that it relies solely on solar flare inputs to forecast the SEP occurrence and it does not use CME data. This is an inherent limitation in the model and it was a choice made in its development so that it is able to provide as accurate as possible forecasts, as quickly as possible, since real-time CME data can have large uncertainties and be delayed for several hours. Another limitation is of course the large class imbalance between the positive and negative categories, flares associated and not associated with SEP events. While there is no perfect substitute for real data in the training of ML models, this caveat was addressed to a large degree by taking into account this imbalance in the very training of the model itself. 


\subsection{Time Series-Histogram of Oriented Gradients-TaBular (TS-HOG-TB) Model} \label{sec:TS-HOG-TB}

\textbf{Model Developers and Relevant Citation}: Pouya Hosseinzadeh, Soukaina Filali Boubrahimi, and Shah Muhammad Hamdi; \cite{hosseinzadeh2025end}.

\begin{table}[h]
\caption{Model, Input and Output Specification Table for the TS-HOG-TB model.}
\centering
\begin{tabular}{|lc|}
\hline
\multicolumn{2}{|c|}{Model}                                            \\ \hline
\multicolumn{1}{|l|}{Type}            & Ensemble Method                \\
\multicolumn{1}{|l|}{Complexity}      & 100,000                        \\ \hline
\multicolumn{2}{|c|}{Input}                                            \\ \hline
\multicolumn{1}{|l|}{Shape}           & Time Series (1D), Vectors (1D) \\
\multicolumn{1}{|l|}{Type}            & EUV Imagery, Energetic Protons  \\
\multicolumn{1}{|l|}{History}         & 15 years (1997-2012)                       \\
\multicolumn{1}{|l|}{Diversity}       & 207 samples                    \\
\multicolumn{1}{|l|}{Imbalance}       & 0.1449 positive                \\
\multicolumn{1}{|l|}{Sample Size}     & 16.90 bytes                    \\
\multicolumn{1}{|l|}{Sample Coverage} & 6 hours                        \\ \hline
\multicolumn{2}{|c|}{Output}                                           \\ \hline
\multicolumn{1}{|l|}{Prediction}      & Classification                 \\
\multicolumn{1}{|l|}{Type}            & Triggered                      \\
\multicolumn{1}{|l|}{Forecast Window} & 0 hours                        \\ \hline
\multicolumn{2}{|p{8.5cm}|}{Comments: \textsuperscript{*}A forecast window of 0 hours indicates that the model performs event-level classification rather than temporal forecasting. These models rely on the GSEP catalog to label whether an SEP event occurs, without predicting its onset time or lead interval. Consequently, the output reflects the presence or absence of an SEP event conditioned on the input observations, not a forward-looking warning horizon.}   \\ \hline
\end{tabular}
\end{table}

\textbf{Summary}: This study introduces an end-to-end ensemble ML framework for predicting high-impact ($\sim 100$ MeV) SEP events by integrating multimodal data. The proposed model combines three key data modalities: GOES proton flux time series, AR polygons extracted from SOHO EUV images, and solar flare-related tabular data. Each modality is independently evaluated using specialized models, and the best-performing classifiers are combined using ensemble strategies. The final model, the Time Series-Histogram of Oriented Gradients-TaBular (TS-HOG-TB), which integrates all three modalities, achieves strong results with a recall of 0.80 (for balanced) and 0.75 (for imbalanced). The framework shows robustness under noise and across various temporal settings, highlighting the advantage of multimodal fusion for reliable SEP forecasting.

\textbf{Model Description}: The best-performing model, TS-HOG-TB, is an ensemble ML framework that combines predictions from three specialized unimodal models, each trained on a distinct data modality: a) time-series proton flux data from the GOES P6 channel processed using the TSF classifier; b) AR polygon data extracted from SOHO EUV images, encoded with HOG features and classified using random forests; and c) tabular data including sunspot counts, AR counts, and flare class, modeled using an SVM. Each unimodal classifier outputs probabilistic predictions, which are averaged in the ensemble to produce the final classification. This late-fusion strategy captures complementary temporal, spatial, and statistical patterns of solar activity, resulting in improved accuracy and robustness in predicting $\sim 100$ MeV SEP events. 

\textbf{Inputs}: The model takes three distinct data modalities as input: a) GOES time-series proton flux data from the P6 channel ($\sim 100$ MeV), using a 6-hour observation window prior to the associated solar flare; b) EUV images from SOHO’s EIT 304 \AA \, channel, processed to extract AR polygons using thresholding, contour detection, and binary masking; c) tabular features including sunspot counts, AR counts, and flare class (C, M, X), computed from the 6-hour period preceding each flare. All SEP events are sourced from the GSEP catalog, while non-SEP events are selected from the HEK database. Events with incomplete data in any modality are excluded.

\textbf{Outputs}: The output of the model is a probability of SEP occurrence based on a flare trigger. The probability is converted to a binary label (SEP/no-SEP), informing us whether a flare will produce an SEP event.

\textbf{Model Configuration}: For time-series input, the model uses TSF, which extracts summary statistics (mean, variance, slope) over random intervals from the 6-hour GOES P6 proton flux data. For image-based input, EUV AR polygons are processed with HOG features and classified using random forest. For tabular input (sunspots, AR counts, flare class), an SVM is used. The final ensemble model (TS-HOG-TB) averages probabilistic outputs from each unimodal classifier. All models are trained with 5-fold cross-validation. Z-score normalization is applied to time-series data; AR images are resized to $200\times200$ pixels before feature extraction.

\textbf{Model Validation and Results}: Model performance was evaluated using 5-fold cross-validation under both balanced and imbalanced settings. For the balanced setting (37 SEP, 37 non-SEP), the TS-HOG-TB ensemble achieved 0.81 recall, 0.80 F1-score, and the highest TSS and HSS scores among all models. In the imbalanced setting (37 SEP, 104 non-SEP), the model maintained strong Recall (0.75), indicating robustness under real-world class imbalance. Sensitivity analysis with Gaussian noise (mean=0, std=0.5) confirmed that TS-HOG-TB outperforms unimodal models (TS-only, AR-only) under noisy conditions. The model also showed stable performance with reduced training data and achieved best results with a 6–7 hour observation window.

\textbf{Access to Model Data and Forecasts}: The implementation of the TS-HOG-TB ensemble model, including preprocessing scripts and training configurations, is publicly available at \url{https://github.com/pouyahosseinzadeh/Solar-Energetic-Particle-Event-Prediction-Ensemble-TS-HOG-TB}. All input data modalities used in this study are obtained from publicly accessible heliophysics repositories. Time-series proton flux measurements are retrieved from the GOES SEM archives provided by NOAA NCEI, EUV AR imagery is accessed via the SOHO/EIT instrument through the \textit{Helioviewer} platform, and tabular solar activity parameters (e.g., sunspot counts, AR counts, flare class) are obtained from the HEK. These resources enable reproducible multimodal SEP forecasting experiments using combined temporal, spatial, and tabular predictors.

\textbf{Limitations, Caveats and Discussion}: A primary limitation is the small number of high-energy ($\sim 100$ MeV) SEP events with complete multimodal data, which restricts dataset size. While the model performs well in balanced settings, performance varies under class imbalance. Non-SEP event selection required manual filtering to avoid temporal overlap, limiting scalability. EUV image resolution and timing may also introduce uncertainty, as AR segmentation depends on pre-flare image availability. Despite these constraints, the ensemble model demonstrated robustness to noise and reduced training data. Future work will address real-time prediction, expand the non-SEP set, and explore physics-informed ensemble methods.


\subsection{Solar Energetic Particle Network (SEPNET) Model} \label{sec:SEPNET}

\textbf{Model Developers and Relevant Citation}: Yian Yu, Yang Chen, Lulu Zhao, Kathryn Whitman, Ward Manchester, Tamas Gombosi; \cite{yu2025solar}.

\begin{table}[h]
\caption{Model, Input and Output Specification Table for the SEPNET model.}
\centering
\begin{tabular}{|lc|}
\hline
\multicolumn{2}{|c|}{Model}                                            \\ \hline
\multicolumn{1}{|l|}{Type}            & Neural Network                 \\
\multicolumn{1}{|l|}{Complexity}      & 130,000                        \\ \hline
\multicolumn{2}{|c|}{Input}                                            \\ \hline
\multicolumn{1}{|l|}{Shape}           & Time Series (1D)               \\
\multicolumn{1}{|l|}{Type}            & Magnetic Fields, Soft X-ray    \\
\multicolumn{1}{|l|}{History}         & 40 years (1986-2025)           \\
\multicolumn{1}{|l|}{Diversity}       & 11,773 samples                 \\
\multicolumn{1}{|l|}{Imbalance}       & 0.3004 positive                \\
\multicolumn{1}{|l|}{Sample Size}     & 105000 bytes                   \\
\multicolumn{1}{|l|}{Sample Coverage} & 24 hours                       \\ \hline
\multicolumn{2}{|c|}{Output}                                           \\ \hline
\multicolumn{1}{|l|}{Prediction}      & Classification, Probability    \\
\multicolumn{1}{|l|}{Type}            & Continuous                     \\
\multicolumn{1}{|l|}{Forecast Window} & 24 hours                       \\ \hline
\end{tabular}
\end{table}

\textbf{Summary}: We introduce SEPNET (and its extensions, SEPNET-TS and SEPNET-O), an innovative multi-task NN that integrates forecasting of solar flares and CME summary statistics into the SEP prediction model, leveraging their shared dependence on SHARP magnetic field parameters. 

\textbf{Model Description}: SEPNET incorporates long short-term memory and transformer architectures to capture contextual dependencies in temporally evolving features for SEP forecasting. 

\textbf{Inputs}: For each sample, the input consists of a set of min-max normalized features derived from solar flare, CME, and SHARP magnetic field data. 

\textbf{Outputs}: The {SEPNET} model, together with {SEPNET-TS}, is trained using all SEP event enhancements above GOES background, indicated by a proton flux threshold of $10^{-6}$ pfu in the CLEAR SEP benchmark dataset. For operational deployment ({SEPNET-O}), samples labeled as operational SEP events ($\geq 10$ MeV proton flux $\geq 10$ pfu) are used as a validation set to fine-tune the classification threshold for distinguishing operational SEP events.

\textbf{Model Configuration}: The input features are processed through three shared fully connected (dense) layers with gradually reduced feature dimensionality (from $256$ to $128$, $64$, and $16$). Each dense layer is followed by layer normalization, ReLU activation, and dropout. The shared embedding is then fed into two distinct output heads to implement multi-task learning: a regression head that predicts the counts of future flare and CME events, and a classification head that outputs the predicted probability of a future SEP event. To better capture temporal dependencies and complex sequential patterns in the input data, the updated model \texttt{SEPNET-TS} integrates recurrent and attention mechanisms by combining a unidirectional LSTM layer with a transformer encoder.

\textbf{Model Validation and Results}: The performance of SEPNET is evaluated on the state-of-the-art SEPVAL SEP dataset and compared with classical ML methods and current state-of-the-art pre-eruptive SEP prediction models. The results show that SEPNET achieves higher detection rates and skill scores while being suitable for real-time space weather alert operations. 

\textbf{Access to Model Data and Forecasts}: The operational forecasting is available in real-time (updated every hour) through the MLSW website at \url{https://mlsw.engin.umich.edu/apps/runSEP}. The code and data are available at: \url{https://github.com/yuyian/SEP-Prediction.git}.

\textbf{Limitations, Caveats and Discussion}: In our follow-up ongoing work, we extend the forecasting to incorporate X-ray and proton flux history, in predicting both SEP occurrence and corresponding proton flux values. 


\subsection{Bidirectional Long Short-Term Memory (BiLSTM-SEP) Model} \label{sec:BiLSTM-SEP}

\textbf{Model Developers and Relevant Citation}: Mohamed Nedal, Kamen Kozarev, Nestor Arsenov, and Peijin Zhang; \cite{nedal2023forecasting}.

\begin{table}[h]
\caption{Model, Input and Output Specification Table for the BiLSTM-SEP model.}
\centering
\begin{tabular}{|l|l|}
\hline
\multicolumn{2}{|c|}{Model}                                                  \\ \hline
\multicolumn{1}{|l|}{Type}            & Neural Network                       \\
\multicolumn{1}{|l|}{Complexity}      & 333,699                              \\ \hline
\multicolumn{2}{|c|}{Input}                                                  \\ \hline
\multicolumn{1}{|l|}{Shape}           & Time Series (1D)                     \\
\multicolumn{1}{|l|}{Type}            & Magnetic Fields, Soft X-ray, Proton Flux, Ground-Based Radio, Solar Wind  \\ 
\multicolumn{1}{|l|}{History}         & 43 years (43)                        \\
\multicolumn{1}{|l|}{Diversity}       & 15,558 samples\textsuperscript{*}    \\
\multicolumn{1}{|l|}{Imbalance}       & 0.0146 positive                      \\
\multicolumn{1}{|l|}{Sample Size}     & 15,120 bytes                         \\ 
\multicolumn{1}{|l|}{Sample Coverage} & 6,480 hours\textsuperscript{**}      \\ \hline 
\multicolumn{2}{|c|}{Output}                                                 \\ \hline
\multicolumn{1}{|l|}{Prediction}      & Classification (All Clear), Probability, Regression \\
\multicolumn{1}{|l|}{Type}            & Continuous                           \\
\multicolumn{1}{|l|}{Forecast Window} & 24 hours\textsuperscript{***}        \\ \hline 
\multicolumn{2}{|p{11cm}|}{Comments: \textsuperscript{*} BiLSTM-SEP does time series forecast therefore diversity reflects the number of days in the training samples. The total number of days is 15,558, from December 25th 1976 to July 30th 2019. \textsuperscript{**} For every sample, time-series history of 270 days (6480 hours) is used. \textsuperscript{***} A forecast of 24 hours ahead produces the best results. Forecast windows of 48 and 72 hours (2 and 3 days) are explored too.}   \\ \hline
\end{tabular}
\end{table}

\textbf{Summary}: The Bidirectional Long Short-Term Memory (BiLSTM-SEP) is a deep learning model for forecasting SEP fluxes across three GOES energy channels using bidirectional LSTM networks. The model processes multi-decadal, multivariate time series of solar and heliospheric data to generate 3-day forecasts of log-integral proton flux. Designed to capture both short and long-term dependencies in space weather data, it performs competitively with other forecasting approaches while maintaining low FAR. Its outputs are suited for both operational radiation hazard mitigation and downstream scientific analysis.

\textbf{Model Description}: The model consists of four BiLSTM layers with 64 neurons each, followed by a dense layer that outputs a 3-day sequence of predictions. Its bidirectional architecture allows the model to learn from both past and forward temporal context, which improves performance on highly nonlinear and non-stationary SEP data. Each GOES energy channel is modeled independently. Training uses early stopping, adaptive learning rate reduction, and the Huber loss function \citep{meyer2021alternative} for robustness to outliers.

\textbf{Inputs}: Seven input features were chosen for their physical relevance and correlation with SEP flux: sunspot number, F10.7 index, long and short-band X-ray fluxes (log-transformed), solar wind speed, interplanetary magnetic field magnitude, and prior log-SEP fluxes in each energy band. All features were daily averaged, normalized, and linearly interpolated to fill gaps. The model uses a sliding window of 270 days of inputs to forecast the next 3 days.

\textbf{Outputs}: Each trained model predicts the next three days of log-integral proton flux for a specific energy channel ($\geq 10$ MeV, $\geq 30$ MeV, or $\geq 60$ MeV). Outputs are real-valued (not categorical) and can be used in radiation exposure calculations or space weather alert systems. The multi-input ulti-output strategy avoids recursive input-feedback and allows efficient, simultaneous multi-step prediction.

\textbf{Model Configuration}: The model was trained using a batch size of 30 days (approximately a Carrington rotation), with an input window of 270 days. A training set (74.29\%), validation set (16.2\%), and test set (9.51\%) were carved out using a fixed 9:2:1 month-based strategy from each year. The Adam optimizer and Huber loss were used, with learning rate reduction and early stopping. Separate models were trained per energy band to reduce cross-channel interference.

\textbf{Model Validation and Results}: On validation and test sets, the model achieved $R \geq 0.9$ for all energy bands at a 1-day lead time. MAE ranged from 0.045 to 0.125 across channels and lead times. Mean Absolute Percentage Error ranged from 12.36 to 49.14. Performance declined slightly with increasing forecast horizon, as expected. The $\geq 60$ MeV model showed the highest consistency, while the $\geq 30$ MeV model showed a somewhat larger discrepancy between predictions and observations. Confusion-matrix-based skill scores showed a low FAR and competitive POD compared to prior models such as UMASEP \cite{nunez2011predicting} and Relativistic Electron Alert System for Exploration \cite[REleASE;][]{malandraki2018solar}.

\textbf{Access to Model Data and Forecasts}: Data and preprocessing scripts are available at \url{https://gitlab.com/iahelio/mosaiics/sep-lstm/}. All data used are publicly available from OMNIWeb at \url{https://omniweb.gsfc.nasa.gov}, GOES SEM archives at \url{https://satdat.ngdc.noaa.gov/sem/goes/data/avg}, and Solar Influences Data Analysis Center (SILSO; sunspot number) at \url{https://www.sidc.be/silso/home}. Forecast near-real-time model outputs are under development and will be made available in a future public repository, as part of an effort to support operational space weather forecasting.

\textbf{Limitations, Caveats and Discussion}: While overall results are encouraging, the model still struggles to forecast rare high-flux events (e.g., SEP $\geq 10$ pfu) due to the inherent data imbalance ---these events are underrepresented in the dataset, often leading to underestimation of their peak flux. Additionally, forecasts for the $\geq 30$ MeV energy channel exhibited more pronounced deviations from observations than the $\geq 10$ and $\geq 60$ MeV models. Forecast performance also decreased during solar minimum and quiet periods, where the predictive signal in inputs is weak. Addressing these limitations will require higher-resolution (e.g., hourly) inputs, more sophisticated feature engineering, and additional data from SC 25.


\subsection{Models for Probabilistic Forecast of Solar Energetic Particles (MEMPSEP)} \label{sec:MEMPSEP}

\textbf{Model Developers and Relevant Citation}: Subhamoy Chatterjee, Maher Dayeh, Andrés Muñoz-Jaramillo, Kim Moreland, Hazel Bain, Samuel Hart, Michael Starkey; \cite{chatterjee2024mempsep}, \cite{dayeh2024mempsep} and \cite{moreland2024mempsep}.

\begin{table}[h]
\caption{Model, Input and Output Specification Table for the MEMPSEP model.}
\centering
\begin{tabular}{|l|p{10cm}|}
\hline
\multicolumn{2}{|c|}{Model}                                                 \\ \hline
\multicolumn{1}{|l|}{Type}            & CNN                                 \\
\multicolumn{1}{|l|}{Complexity}      & 6,092,617                           \\ \hline
\multicolumn{2}{|c|}{Input}                                                 \\ \hline
\multicolumn{1}{|l|}{Shape}           & Point data (0D), Time-series (1D), Images (2D)\textsuperscript{*} \\
\multicolumn{1}{|l|}{Type}            & Magnetic Field, Soft X-ray, Electron Flux, EUV Imagery, Coronagraphs, Space-Based Radio, Solar Wind \\
\multicolumn{1}{|l|}{History}         & 15 years (1998–2013)                \\
\multicolumn{1}{|l|}{Diversity}       & 13,850 samples                      \\
\multicolumn{1}{|l|}{Imbalance}       & 0.07 positive                       \\
\multicolumn{1}{|l|}{Sample Size}     & 17,000,000 bytes\textsuperscript{**}\\
\multicolumn{1}{|l|}{Sample Coverage} & 72 hours\textsuperscript{***}       \\ \hline
\multicolumn{2}{|c|}{Output}                                                \\ \hline
\multicolumn{1}{|l|}{Prediction}      & Classification, Probability, Regression (Time-series) \\
\multicolumn{1}{|l|}{Type}            & Triggered                           \\
\multicolumn{1}{|l|}{Forecast Window} & 6 hours                             \\ \hline
\multicolumn{2}{|p{14cm}|}{Comments: \textsuperscript{*} Time Series of magnetograms (3D), Time Series of X-ray flux(1D), Time series of Electron (1D), Time-frequency map of Radio bursts (2D), Scalar solar wind and suprathermal particle properties (0D) \textsuperscript{**} For each datapoint-1.7 MB for magnetograms, 272KB for wind/waves time-freq map,  60 KB for X-ray time-series, 500 KB for electron time-series \textsuperscript{***} 3 days for magnetogram sequence and Wind/Waves time-frequency map prior to flare onset, 1 day for X-ray and electron time series. }   \\ \hline
\end{tabular}
\end{table}

\begin{figure}[h]    
    \centering 
    \includegraphics[width=\textwidth]{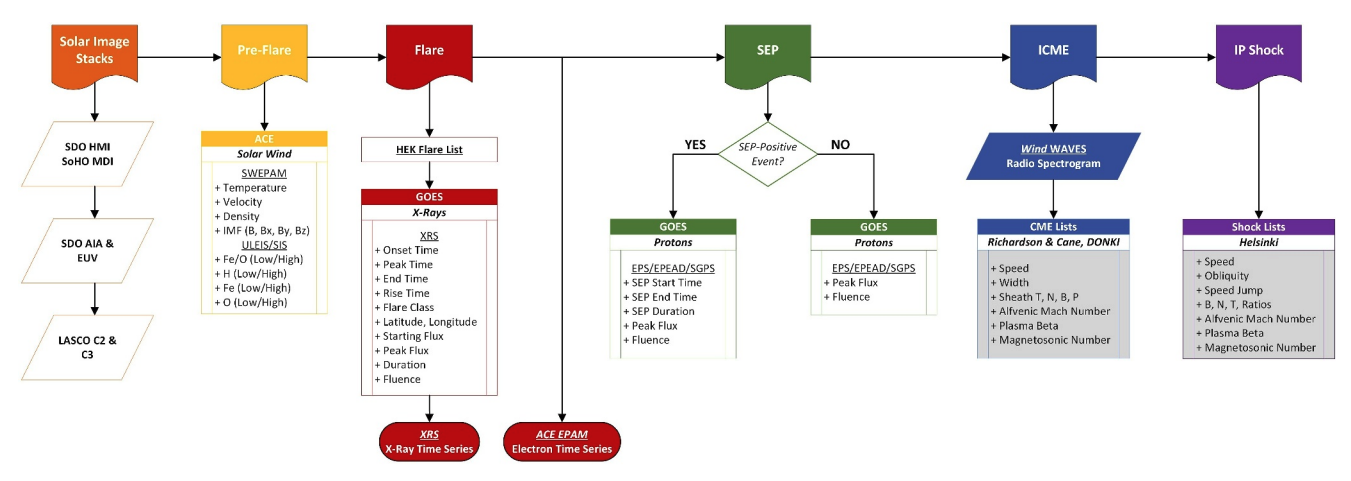} 
    \caption{The complete data set flowchart shows all incorporated observations from remote imaging to in‐situ measurements. Each section relates to a time frame in the event series: Solar images (pre‐flare, flare), solar wind conditions (pre‐flare), X‐ray properties, and time series (flare). Post‐flare properties for the SEP event, interplanetary CME parameters with WIND/WAVES radio spectrogram, interplanetary shock properties, and published shock lists. We note the observation (italics) and the instrumentation (underlined) used to obtain in‐situ data along with each parameter observed or calculated parameter.} 
    \label{fig:S3_MEMPSEP}
\end{figure}

\textbf{Summary}: The Multivariate Ensemble of Models for Probabilistic Forecast of Solar Energetic Particles (MEMPSEP; Figure~\ref{fig:S3_MEMPSEP}) introduced an ensemble of ML models to predict if a solar flare would lead to SEP events and, if so, what the properties of that event would be with associated uncertainty. The model ingests both remote-sensing (SoHO/MDI and SDO/HMI) and in-situ data over 1998-2013 to make predictions.

\textbf{Model Description}: A CNN architecture was built that ingests multiple inputs such as images, scalar parameters, and time series to make SEP event/non-event classification and regression of SEP properties. The CNN processes the input image sequences with repeated 2D convolution, batch-normalization, non-linear activation, and max-pooling. The flattened layers are concatenated with scalar parameters and features extracted from time series data through 1D convolution. The concatenated features are passed through dense (fully connected) layers to reproduce the classification and regression targets. First, the classification model is trained, followed by the regression model in two different ways: gated and non-gated. For the gated approach, the regression is trained by coupling the classification model outcome to the regression loss function, and for the non-gated approach, the regression model is trained independently.

\textbf{Inputs}: The GOES integrated ($\geq 10$ MeV) particle flux is checked for whether it crosses a threshold of 5 pfu within 6 hours of flare onset. If yes, it is then labeled as an SEP event and otherwise as a non-event. The inputs to the model are acquired over 1-3 days prior to the flare onset. A 3-day preflare magnetogram time sequence is used as remote sensing input. For in-situ, WIND/WAVES radio burst time-freq map over 3 days, X-ray, and L1 Electron time series over 1 day before flare, and scalar properties related to solar wind and suprathermal population are used. A total of 10200 non-events and 675 events for training the model ensemble are utilized.

\textbf{Outputs}: MEPSEP produces SEP occurrence probability and SEP properties such as proton peak fluxes ($\geq 5$,  $\geq 10$, $\geq 30$, $\geq 60$, $\geq 100$ MeV), event onset, and duration. 

\textbf{Model Configuration}: A large class imbalance of SEP events and non-events poses difficulty in the success of ML-based prediction models. We made use of the class imbalance, creating multiple training and validation sets through stratified undersampling of non-events and the same set of events. With each of those training and validation sets, a CNN is trained with the architecture described previously. The sets generated an ensemble of models for both the classification and regression tasks. The CNN classifier is turned into a predictor of true probability using a non-parametric calibration technique called Bayesian Binning Quantile. An ensemble of BBQ calibrators was derived, providing uncertainty estimates on the SEP occurrence probability. Among the non-gated and gated regression models, MEMPSEP achieved better results with the gated model. 

\textbf{Model Validation and Results}: A test set is designed including events/non-events that are not seen by the model during the training phase for estimating both probabilistic and deterministic skill scores. A Brier Score (BS) of 0.14 is achieved and an Expected Calibration Error (ECE) of 0.07 on the ensemble median probability. Applying a threshold of 0.5, a POD, False Positive Rate (FPR), TSS, and HSS of 0.83, 0.2, 0.63, and 0.6, respectively is achieved. For the regression model, an $R^2$ score of $\geq 0.63$, $0.07$, and $0.01$ for SEP peak, onset, and duration with occurrence probability $\geq 0.5$ is achieved. MEMPSEP participated in the Solar Heliospheric and INterplanetary Environment (SHINE) 2022 independent validation as well as SEPVAL campaign, providing a BS of 0.2, and predicting 6 out of 8 events and 11 out of 14 non-events correctly with a probability threshold of 0.5.

\textbf{Access to Model Data and Forecasts}: The MEMPSEP training data and codes have been made publicly available. They can be accessed via the following links: \url{https://zenodo.org/records/10044865} and \url{https://doi.org/10.5281/zenodo.11201195}.

\textbf{Limitations, Caveats and Discussion}: The model ensemble is currently flare-triggered and cannot forecast the full temporal profile of the SEP fluxes. Although the model currently ingests Magnetograms as imaging data, the MEMPSEP dataset defined in \citet{moreland2024mempsep} also consists of EUV and coronagraph imagery. The impact of those additional inputs on MEMPSEP performance is being tested. We are currently developing MEMPSEP further to eliminate the flare trigger requirement and enable rolling predictions of time series through the course of an SEP event using near-real-time data \citep{dayeh2025machine}. An effort is also ongoing to add the energetic storm particle sudden enhancements, often associated with SEPs. 


\subsection{Parker Solar Probe SEP Prediction (PSPSP) Model} \label{sec:PSPSP}

\textbf{Model Developers and Relevant Citation}: Tate Hutchins, Spiridon Kasapis, Hameedullah Farooki, Manuel Cuesta, Lengying Khoo2, Sungmin Pak, Robert Czarnota, Jamie Rankin, Jamey Szelay, Georgios Livadiotis, Xiaoyan Li, David McComas, Zigong Xu, Nikolaos Sarlis; \cite{hutchins2026}.

\begin{table}[h]
\caption{Model, Input and Output Specification Table for the PSPSP Model.}
\centering
\begin{tabular}{|lc|}
\hline
\multicolumn{2}{|c|}{Model}                                            \\ \hline
\multicolumn{1}{|l|}{Type}            & Neural Network                 \\
\multicolumn{1}{|l|}{Complexity}      & 13,814,081                     \\ \hline
\multicolumn{2}{|c|}{Input}                                            \\ \hline
\multicolumn{1}{|l|}{Shape}           & Time Series (1D), Images (2D)  \\
\multicolumn{1}{|l|}{Type}            & EUV Imagery, Proton Flux, Solar Wind       \\
\multicolumn{1}{|l|}{History}         & 6 years (2019-2025)            \\
\multicolumn{1}{|l|}{Diversity}       & 1,015 samples                  \\
\multicolumn{1}{|l|}{Imbalance}       & 0.1547 positive                \\
\multicolumn{1}{|l|}{Sample Size}     & 6,300,000 bytes                \\
\multicolumn{1}{|l|}{Sample Coverage} & 240 hours                      \\ \hline
\multicolumn{2}{|c|}{Output}                                           \\ \hline
\multicolumn{1}{|l|}{Prediction}      & Classification, Regression     \\
\multicolumn{1}{|l|}{Type}            & Continuous                     \\
\multicolumn{1}{|l|}{Forecast Window} & 35 hours                       \\ \hline
\multicolumn{2}{|p{7cm}|}{Comments: \textsuperscript{*}The forecast window varies from 2 hours to 3 days, with a median value of 35 hours.}                                            \\ \hline
\end{tabular}
\end{table}

\textbf{Summary}: Most ML models in this document aim to predict geoeffective SEP events. The PSPSP model uses EUV images from the AIA onboard the SDO to predict the intensity values within any given point in ecliptic plane between Sun and Earth. To train the model, PSP Integrated Science Investigation of the Sun \citep[IS$\odot$IS;][]{mccomas2016integrated} Energetic Particle Instruments - Low \citep[EPI-Lo;][]{HillEA2017JGRA_PSP_ISOIS_EPILo, wiedenbeck_capabilities_2017} particle intensity measurements were used as targets. A series of convolutional layers extract image features, which are combined with information about the location of PSP and passed through a set of dense layers in order to provide a binary classification about whether the particle population in the given space is $\geq 0.1$.

\textbf{Model Description}: The study focuses on a snapshot model focusing on single image input and a better performing video model focusing on a chronological sequence of images input. Each model consists of a configurable-depth CNN that extracts spatial features from the AIA $171\AA$ inputs using successive Convolution, Batch Normalization, ReLU and Maximum Pooling blocks. Global average pooling is applied to obtain a fixed-length feature vector, which is concatenated with the PSP magnetic footpoint information.  For the video model, a transformer encoder layer learns and appends a non-local temporal embedding to learn the progression of SEP events. The combined representation is passed through a fully connected MLP with ReLU activations and dropout regularization to produce a single output prediction of intensity.

\textbf{Inputs}: The inputs to the model are a) SDO AIA $171\AA$ images that are reduced in dimension ($512\times512$ pixels) and b) the solar footprint longitude and distance values taken from the PSP ephemeris data. The solar footpoint longitude is calculating assuming the Parker spiral utilizing the ephermeris data and the PSP solar-wind speed, which is derived from measurements of the Solar Wind Electrons Alphas and Protons \citep[SWEAP;][]{kasper2016solar} instrument. Here, we use the level 3 Solar Probe Analyzer for Ions \citep[SPAN-I;][]{livi2022solar} solar wind proton speed magnitude. As targets to the model we use the energy weighted average \citep[$J_\mathrm{linlin}$,][]{cuesta2025comparing} particle intensity measured by the PSP IS\(\odot\)IS EPI-Lo instrument. The inputs span 6 years, from the beginning of the PSP mission to December 2025. Both inputs are log-normalized and then min-max normalized too.

\textbf{Outputs}: The model outputs a prediction for the $J_{linlin}$ particle intensity on the given point of the PSP orbit. To provide SEP event classification predictions, a threshold is set at $10^{-1}$ such that any $J_{linlin}$ reading above is defined as a positive event and anything else is defined as a negative non-event. Models were trained on targets of the actual $J_{linlin}$ values as well as modified and trained on the 0/1 classification values.

\textbf{Model Configuration}: The best performing model has a batch size of 32 images, dropout rate of 0.25, 4 attention heads, 3 attention blocks, hidden head size of 256, and image embedding size of 256. It was trained using a learning rate of 0.0001 and achieved the best loss after 31 epochs.

\textbf{Model Validation and Results}: With a 0.1 $J_{linlin}$ threshold, the Video model trained on classification targets achieves an Accuracy of 0.7705, Precision of 0.3803, Recall of 0.8039, FAR of 0.2355, TSS of 0.5684, and HSS of 0.3902. 

\textbf{Access to Model Data and Forecasts}: All relevant code can be found at \url{https://github.com/thutch17/PSP-SEP-Event-Prediction}.

\textbf{Limitations, Caveats and Discussion}: The main limitation of this study is that the model does not effectively capture temporal relationships, considering the video model only performed marginally better than the snapshot model. Future studies should use model architectures that can capture temporal dependencies not only in the $J_{linlin}$ timelines but also in the solar disc progression captured in the SDO AIA image series.


\subsection{Energetic Particle Radiation Environment Module - S (EPREM-S) Model} \label{sec:EPREM-S}

\textbf{Model Developers and Relevant Citation}: Atilum Gunes Baydin, Bala Poduval; \cite{baydin2023surrogate}.

\begin{table}[h]
\caption{Model, Input and Output Specification Table for the \cite{baydin2023surrogate} Model.}
\label{tab:EPREM-S}
\centering
\begin{tabular}{|lc|}
\hline
\multicolumn{2}{|c|}{Model}                                            \\ \hline
\multicolumn{1}{|l|}{Type}            & Neural Networks                \\
\multicolumn{1}{|l|}{Complexity}      & 285,881,344                    \\ \hline
\multicolumn{2}{|c|}{Input}                                            \\ \hline
\multicolumn{1}{|l|}{Shape}           & Time Series (1D)               \\
\multicolumn{1}{|l|}{Type}            & Proton Flux                    \\
\multicolumn{1}{|l|}{History}         & N/A\textsuperscript{*}         \\
\multicolumn{1}{|l|}{Diversity}       & 32000 samples                  \\
\multicolumn{1}{|l|}{Imbalance}       & 1\textsuperscript{**}          \\
\multicolumn{1}{|l|}{Sample Size}     & 1,600,00 bytes                 \\
\multicolumn{1}{|l|}{Sample Coverage} & 96                             \\ \hline
\multicolumn{2}{|c|}{Output}                                           \\ \hline
\multicolumn{1}{|l|}{Prediction}      & N/A\textsuperscript{***}       \\
\multicolumn{1}{|l|}{Type}            & Continuous                     \\
\multicolumn{1}{|l|}{Forecast Window} & N/A\textsuperscript{***}       \\ \hline
\multicolumn{2}{|p{8cm}|}{Comments: \textsuperscript{*}This study does not use observational data but rather synthetic (simulated) data that contained 32,000 SEP events. \textsuperscript{**}Only positive events are used in this simulation study. \textsuperscript{***}This is a surrogate model capable of SEP prediction but no prediction accuracy analysis carried out.} \\ \hline
\end{tabular}
\end{table}

\begin{figure}[h] 
    \centering 
    \includegraphics[width=0.75\textwidth]{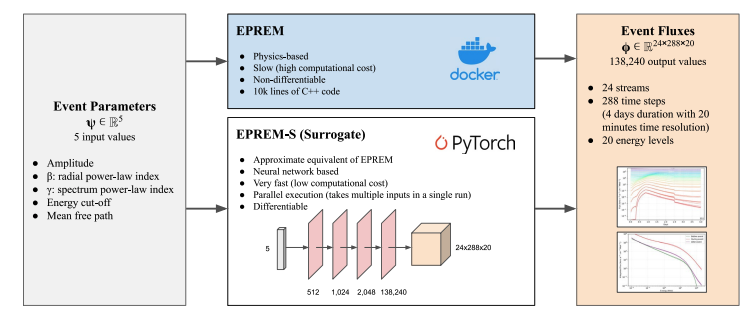} 
    \caption{Overview of EPREM and EPREM-S models. Both map input parameters $\psi$ to output fluxes $\phi$. EPREM is a physics-based model for generating SEP events implemented in C++. EPREM-S is a NN trained with a data set $D = \{\psi_i,\phi_i = EPREM(\psi_i)\}^{N}_{i=1}$ obtained by running EPREM with inputs $\psi_i ~ p(\psi)$ sampled from a continuous uniform prior distribution.} 
    \label{fig:EPREM-S} 
\end{figure}

\textbf{Summary}: The challenge in developing an ML model for the prediction of SEPs is the lack of sufficient number of SEP events for training and validation of the ML model ---the class imbalance problem--- despite decades of SEP observations by spacecraft. Availability of synthetic (or simulated) SEP events created using first principles models will solve the class imbalance problem to a large extent. However, many first principles models are computationally intensive and takes tens of minutes to hours for completing one simulation. Therefore, simulating hundreds of thousands of SEP events for training the ML model becomes rather impractical within a reasonable time-frame. This difficulty can be overcome with the method of emulation or surrogate models where a NN model trained on the simulated out put (SEPs, for example) of a first principle model is developed to perform the exact same function as that of the original model with acceptable accuracy but much faster. \cite{baydin2023surrogate} describes the method and the results of the NN surrogate model of the Energetic Particle Radiation Environment Model (EPREM) developed by \cite{schwadron2010earth}. For this, 32,000 SEP events are generated to train a feed-forward NN, EPREM-S (Figure~\ref{fig:EPREM-S}), with 4 hidden layers and ReLU nonlinearities after each layer except the last one. It should be noted that the EPREM and EPREM-S outputs are in remarkable agreement, the MSE being 0.07 as it is found during validation of the surrogate model. Analysis of an event previously unseen by the surrogate model as a Bayesian inference problem revealed that the parameters were correctly inferred as their ground truth values (unknown to the inference algorithm) are contained within the resulting posterior distributions. By measuring the runtime costs of EPREM and EPREM-S, it is found that EPREM-S is tens to hundreds of thousands times ($10^4$ - $10^6$) faster than EPREM.

\textbf{Model Description}: The results are based on a feed-forward NN with four hidden layers of sizes 512, 1,024, 2,048, and 138,240, and ReLU nonlinearities after each layer except the last one, giving rise to a total number of 285,881,344 trainable parameters. The output of the last layer is reshaped into a cube with shape $24 \times 288 \times 20$ (24 streams, 288 time steps, and 20 energy levels). In order to provide uncertainty quantification when running trained EPREM-S models, a deep ensemble \citep{lakshminarayanan2017simple} approach is used, where multiple independently trained EPREM-S instances are involved. Following standard practice, these model instances are trained with the same data, but using a different random number seed leading to different model weight initialization and course of stochastic optimization for each instance. Given the set of pretrained surrogate models $S_i = 1,2,..., M$, and a new event parameter $\psi$, the mean and standard deviation of the flux predictions were estimated as

\begin{equation}
    \mu_{\phi} = \dfrac{1}{M} \Sigma S_i(\psi)
\end{equation}

\begin{equation}
    \sigma_{\phi} = \sqrt{\dfrac{1}{M} \Sigma S_i(\psi) - \mu_{\phi}}
\end{equation}

\textbf{Inputs}: EPREM simulates the SEPs by solving the focused transport equation numerically in a Lagrangian frame of reference \citep{schwadron2010earth}. Nested cubes the surfaces of which are subdivided into square cells with the grid nodes at their centers make up the EPREM grid. The grids at the inner boundary corotates with the Sun at each time step and the nested shells advance radially outward with the solar wind. EPREM will compute the distribution function of a given source of particles such as SEPs or pickup ions, anywhere in the heliosphere at individual nodes advecting with the speed of solar wind, naturally tracing the Parker spirals. For this, EPREM makes use of a seed particle spectrum that is a function of energy and heliocentric distance. Five core parameters of the initial seed function are selected, namely, boundary function amplitude, energy spectrum power-law index $\gamma$, radial scaling index $\beta$, boundary function cut-off energy, also called roll over energy or knee cut-off, and the mean free path, as the variable input parameters, that is, the parameters which EPREM-S takes as input.

\textbf{Outputs}: The model output is time series of particle fluxes as a function of time, heliocentric distance and energy. The simulation domain consists of 4 days and there are 20 energy levels from 0.01 - 200 MeV.

\textbf{Model Configuration}: A feed forward NN with four layers and about 286 million learnable parameters is used.

\textbf{Model Validation and Results}: Surrogate modeling or emulation is the creation of fast and simple models that approximate the behavior of complex analytical models that are computationally expensive to evaluate \citep{queipo2005surrogate, forrester2008engineering}. An ensemble of five independently trained EPREM-S models is trained, where each model is a feed-forward NN with four layers and approximately 286 million learnable parameters. Using the EPREM-S ensemble, several thousand SEP events are generated for a range of values for five physical parameters of the initial seed spectrum in EPREM. A comparison of the outputs of EPREM and EPREM-S showed remarkable agreement supported by the fact that the MSE was 0.07 during validation of the surrogate model. EPREM-S was used for simulation-based inference by treating the range of values for the selected five parameters of the initial seed spectrum as the prior distribution for inferring the posterior distribution. This has been demonstrated using a set of events unseen by EPREM-S during training where in all the cases the ground truth values of the selected events were found to be contained within the posterior distributions.

\textbf{Access to Model Data and Forecasts}: The 32,000 SEP events simulated using EPREM and the codes are available at \url{https://zenodo.org/records/10109868}, and \url{https://zenodo.org/records/10038847}. 

\textbf{Limitations, Caveats and Discussion}: EPREM-S is a generative model which can be used to predict SEP events and also for simulation-based inference of observed events. Since EPREM-S is designed and trained on five specific parameters of the initial seed spectrum, the inferences will primarily be based on the influence of the seed population in the time profiles and evolution of SEP events. Surrogates trained on more physical parameters of the first principle model EPREM, giving rise to more general-purpose surrogates, will be the continuation of the emulation work presented here.


\section{Description of Datasets for SEP Prediction} \label{app:Datasets}

\subsection{MEMPSEP-III Dataset} \label{sec:MEMPSEP-III_Dataset}

The MEMPSEP-III dataset \citep{moreland2024mempsep} is an ML-oriented multivariate dataset specifically designed for forecasting the occurrence and properties of SEP events. It integrates both in-situ and remote sensing observations from multiple spacecraft, including GOES, ACE, SDO, SOHO, and WIND/WAVES, covering SC 23 and part of SC 24 (1998–2013). The dataset comprises 252 SEP-producing solar flare events and 17,542 non-SEP events, identified using the GOES flare event list. For each event, MEMPSEP-III includes a rich set of features such as energetic proton and electron fluxes, upstream solar wind parameters, interplanetary magnetic field vectors, and remote solar imaging and radio observations. This multivariate structure enables flexible input configurations for ML models and supports both classification and regression tasks related to SEP forecasting. The MEMPSEP-III dataset has been carefully curated, validated, and cleaned to ensure reliability for ML applications. It has been used to train the MEMPSEP, as described in a series of accompanying papers \citep{chatterjee2024mempsep, dayeh2024mempsep}. Its design facilitates experimentation with different feature sets and model architectures, making it a valuable resource for advancing data-driven SEP prediction. 

\subsection{MTS-SEP Dataset} \label{sec:MTS-SEP_Dataset}

\cite{hosseinzadeh2024improving} present a dataset and methodology aimed at improving the prediction of high-energy SEP events, particularly those involving $\sim 30$, $\sim 60$, and $\sim 100$ MeV protons. A key challenge addressed in this work is the scarcity of SEP events, which limits the effectiveness of ML models. To overcome this, the authors apply data augmentation techniques to synthetically increase the number of SEP samples, thereby enhancing model performance. The dataset consists of univariate and multivariate time series of proton flux measurements, spanning SCs 22 to 24. These time series are used as input to ML classifiers, with a particular focus on the TSF algorithm. The authors demonstrate that using multivariate time series data significantly improves prediction accuracy, especially for the $\sim 100$ MeV SEP events. By applying the SMOTE, they report a 20\% increase in average accuracy, reaching approximately 90\% for the highest energy SEP prediction task. In addition to the dataset, the authors develop a pipeline framework for hierarchical classification of SEP and non-SEP events across different energy thresholds. This work highlights the importance of data augmentation and multivariate temporal features in enhancing the predictive capabilities of ML models for SEP forecasting. 

\subsection{GSEP Dataset} \label{sec:GSEP_Dataset}

\cite{rotti2022integrated} introduce the integrated Geostationary Solar Energetic Particle Events Catalog (GSEP), a homogenized dataset of SEP events spanning SCs 22 to 24. The catalog is constructed by correlating and integrating three existing SEP datasets based on GOES integral proton flux measurements. Each event in the catalog has been visually verified and labeled to ensure consistency and reliability. It has been revised in \cite{rotti2023analysis} to include additional weak SEP events. The latest GSEP catalog identifies a total of 433 SEP events, of which 244 exceed the SWPC threshold for significant proton events. For each event, the dataset includes sliced time-series data of proton flux intensity profiles across multiple energy bands, along with metadata describing associated solar eruptions such as flares and CMEs. This dataset is publicly available and designed to support ML and statistical analyses of SEP events and their solar sources. Its structured format and validated event labeling make it a valuable resource for developing predictive models and improving space weather forecasting capabilities.

\subsection{SMARP-SHARP Dataset} \label{sec:SMARP-SHARP_Dataset}

\cite{kosovich2024time} present a merged dataset of magnetic field parameters derived from two major solar AR data products: the SMARPs \citep{bobra2021smarps} from the MDI instrument onboard SOHO and the SHARPs \citep{bobra2014helioseismic} from the HMI instrument onboard SDO. This unified dataset spans from 1996 April 4 to 2022 December 13 and is designed to support solar flare and SEP forecasting and broader space weather applications. The merging process involves filtering, rescaling, and combining SMARP and SHARP parameters into uniform multivariate time series representations of solar ARs. These time series can be spatially reduced and correlated with other space weather indicators, such as the daily solar flare index and soft X-ray flux measurements from GOES satellites. Preliminary statistical analysis using time-lagged cross-correlation and rolling-window techniques reveal that certain magnetic field properties of ARs may precede flare activity, suggesting potential predictive relationships. The dataset enables exploration of these dynamics across multiple solar cycles and provides a foundation for developing ML models that incorporate magnetic field evolution as a predictive feature for solar flares and SEP events.

\subsection{CLEAR Dataset} \label{sec:CLEAR_Dataset}

The Center for All-Clear Solar Energetic Particle Forecasts (CLEAR) Space Weather Center of Excellence\footnote{\url{https://ccmc.gsfc.nasa.gov/swxcoe/}} has developed the CLEAR SEP Benchmark Dataset derived from GOES data between January 1986 and September 2025. This dataset emphasizes consistent and automated identification of SEP enhancements using the \texttt{fetchsep}\footnote{\url{https://github.com/ktindiana/fetchsep}} tool. The proton time series for all GOES satellites from GOES-06 to GOES-18 were independently analyzed with \texttt{fetchsep} to calculate mean background levels and identify all SEP enhancements by applying the event definitions specified in Table~\ref{tab:CLEAR}. The full benchmark dataset package ($\sim 30$ GB) consists of SEP and non-event (quiet-time period) lists for each GOES satellite, along with the complete time series of the original GOES fluxes, calculated mean background, plots, and other information. The final SEP Benchmark List ($\sim 2$ MB) was compiled by selecting SEP event information from the primary GOES satellite at the time of the event. The benchmark list includes associated flare, CME, radio, and solar wind information extracted from SEP lists maintained by A. Steve Johnson (NASA SRAG) and Ian Richardson (University of Maryland, NASA GSFC) spanning SCs 22 to 25. Two sub-lists are provided in the Benchmark dataset ---the Operational List and the Energy-Bin Calibrated List. The Operational List (1986-2025) uses the archived GOES integral fluxes without background subtraction or modification. The proton values in this list represent data streams used by operational end-users for decision-making. The Energy-Bin Calibrated list (2010-2017) is derived from GOES-13 and GOES-15 background-subtracted uncorrected differential fluxes, with calibrated energy bins provided by \cite{sandberg2014} and \cite{bruno2017}. Integral fluxes were then estimated from the differential channels. The calibrated list provides a better representation of the SEP energy spectrum. The CLEAR SEP Benchmark Dataset is hosted by CCMC and may be downloaded from \url{https://ccmc.gsfc.nasa.gov/swxcoe/clear/}. To promote transparency, reproducibility, and continued maintenance of this dataset, the \texttt{fetchsep} repository includes scripts that may be used to generate the benchmark dataset from scratch. 

\begin{table}[h]
\centering
\caption{Number of SEP events in the CLEAR Benchmark Dataset}
\begin{tabular}{| l |c | c |}
    \hline
     Event Definition & Operational List & Energy-Bin Calibrated List\\
     \hline
     $\geq 10$ MeV above background & 565  & 98 \\
     $\geq 10$ MeV $\geq 10$ pfu  &  265 & 47 \\
     $\geq 30$ MeV above background & 358  & 43 \\
     $\geq 30$ MeV $\geq 1$ pfu  &  258 &  33\\
     $\geq 50$ MeV above background & 292  & 24 \\
     $\geq 50$ MeV $\geq 1$ pfu  &  161 & 22 \\
     $\geq 100$ MeV above background & 163  & 24 \\
     $\geq 100$ MeV $\geq 1$ pfu  & 88  &  8 \\
     \hline
\end{tabular}
\label{tab:CLEAR}
\end{table}


\section{Questionnaire} \label{app:Questionnaire}

Below we present the ML model taxonomy form that was filled out by the SEP modelers of the 23 different publications included in this review.

\vspace{10pt}

\noindent \textbf{ML Models for SEP Prediction - Taxonomy Form}

Dear co-authors, thank you once again for agreeing to participate in our effort to put together a review paper for the ML models that predict SEP events. As a first step of our collaboration, I would like to ask you to fill out the following form which will be used for comparing the different models in the manuscript and will also help us map the research field of SEP prediction using ML. 

The form contains three parts: \textit{Architecture}, \textit{Input} and \textit{Output} Information. Every section has two different types of questions/classifications: quantifiable (you should provide a number) and categorical (you should provide text/choose options). Each section provides more detailed information about the questions asked. Please take your time to fill out the form. Many questions will require some research from your side in order to answer. Please try to answer all questions to the best of your ability. 

If your work/published manuscript includes more than one model (for example works that test multiple different ML models), please provide information for the best performing model. If you have trained models that are substantially different from each other and therefore you would like to submit more than one pages for the review manuscript, please submit this form again, one time for each model. This also applies to cases where the authors have published different manuscripts for each model. 

\vspace{8pt}

\noindent \textbf{Model Information}

\noindent In this Section we would like you to provide information about the ML model you have trained. 
\vspace{-8pt}
\begin{itemize}
    \item \textbf{Model Type - ML Class:} \\
    What is the type/architecture of your model? Some examples would be SVMs, Regression Models, Deep NNs, Random Forests, LSTM architectures etc. \\
    \textit{Categorical Response:} Open.
    \vspace{-8pt}
    \item \textbf{Model Complexity - Number of Trainable Parameters (n):} \\
    How deep is your model? Here you need to answer with the number (scalar) of free parameters  that your model includes. Do not confuse this number with the number of hyperparameters (parameters chosen by user before training), here we are looking for the number of trainable weights included within your model. \\ 
    \textit{Quantifiable Response:} Open.
\end{itemize}

\noindent \textbf{Input Information}

\noindent Here we would like you to provide information about the inputs that you have used to train the model. Most questions are related to the training input, but there are also some questions that are relevant to the model inputs during validation.
\vspace{-8pt}
\begin{itemize}
    \item \textbf{Shape of Input Data:} \\
    Here we would like you to select what is the shape (1D/2D/3D) of your input data. Note that here we are looking for the shape of a single occurrence/event, i.e. if your model is trained on multiple time series events, therefore constructing a 2D input matrix, the shape of your input data is still 1D (time series rather than a matrix). You can select multiple choices in the case where you train your model with data of multiple sizes (ex. you input during training both time series and image data to the model). \\
    \textit{Categorical Response:} Point Data (0D Features), Time Series (1D), Images (2D or 3D), Spectra (1D), Other.
    \vspace{-8pt}
    \item \textbf{Type of Data - Physical Quantity:} \\
    Here we would like you to select the physical quantity/ies that your input training data represents. For more information on the input type categories please check \cite{whitman2023review} (Table 10). \\
    \textit{Categorical Response:} UV / EUV Imagery, Magnetic Fields / Magnetograms, Electric Fields, X-Ray / Soft X-Ray Intensity, White Light / Optical Imaging, Ground-Based Radio, Space-Based Radio, Coronagraph, Solar Wind (n, T, p, V), Suprathermal Particles, Energetic Protons, Other.
    \vspace{-8pt}
    \item \textbf{Input History - Time Coverage of Training Set:} \\
    Here we would like you to type in the number of months/years your training data covers. For example, if your very first training event is in 2010 and the last one occurred in 2020, this would be 10 years worth of data. Some studies use multiple solar cycles worth of data, therefore the answer here would be a value greater than 11 years. \\
    \textit{Quantifiable Response:} Open.
    \vspace{-8pt}
    \item \textbf{Input Diversity - Number of Events - Total Training Samples:} \\
    Here we would like to know the total amount of events you use as a training input (or targets in some applications). For example, if you predict SEPs based on solar flare occurrences, we would like to know the total number of positive (SEP producing) and negative (non-SEP producing) flare events you have used. For many studies this can be simply the total number of SEP events used as targets. Please do not take in account here the number of events you use to validate your model. Positive Samples + Negative Samples = Total Training Samples \\
    \textit{Quantifiable Response:} Open.
    \vspace{-8pt}
    \item \textbf{Class Imbalance - Percentage of Positive Samples in Training Set:} \\
    Here we need you to answer with a number between 0-100. For example, if you perform prediction by training on a dataset that includes 3000 negative flares and 100 positive, the answer to this question would be 100/3100 = 0.0323. Percentage of Positive Samples = Positive Samples/Total Training Samples \\ 
    \textit{Quantifiable Response:} Open.
    \vspace{-8pt}
    \item \textbf{Input Sample Size - Single Event Size:} \\
    Here you should answer with the information size (in bytes) for a single event. For example, if one positive or negative event is represented in your input dataset as a solar EUV image, then the answer to this question is the size of that image. \\ 
    \textit{Quantifiable Response:} Open.
    \vspace{-8pt}
    \item \textbf{Time Coverage of Single Input Sample:} \\ 
    Here you should answer with the time (in hours) coverage of the single event you considered in the previous question. For example, if each one of your events is represented by a timeline, the answer to this question would be the amount of time this timeline covers. There might be works/models for which the answer to this question is zero, as they use for a single event only one solar imagery frame, or a single data point. \\
    \textit{Quantifiable Response:} Open.
\end{itemize}

\noindent \textbf{Output Information}

\noindent Here we would like you to provide information about the testing/validation output your model provides. This section is related to the performance of your model and the type of predictions it offers.
\vspace{-8pt}
\begin{itemize}
    \item \textbf{Output Type - ML Prediction Category:} \\
    Here we would like to know what is the type of prediction your model performs. For example, some studies provide a binary prediction (Classification) whereas there are others that predict whether there are other studies that predict non-activity for the next X hours (All-Clear). Note here that a model can fall into multiple categories. For example, an LSTM/Regression model might predict a physical quantity's values (such as magnetic flux) for the next $x$ hours, and then converts this predicted timeline to a probability of a positive event or an all-clear flag. In such a case, the Regression, Physical Quantity and Probability or All-Clear checkboxes must be selected. For more information on the output type categories please check Table 11 of \cite{whitman2023review}. \\ 
    \textit{Categorical Response:} Classification, Regression (Time-Series), Probability, Time Prediction (Onset Time/ Peak Time/ End Time), Physical Quantity Prediction (Peak/Fluence), All Clear, Other.
    \vspace{-8pt}
    \item \textbf{Triggered vs. Continuous Prediction:} \\
    This question is related to the previous question. Here we need to know whether the model outcome/prediction is Triggered or Continuous. A Triggered prediction means that something happens to the sun (flare, CME etc.) and there are parameters available that were not there before, and therefore prediction happens based on this event's parameters. An example would be a flare-based prediction. If your model relies on information from an event such as a flare or a CME, then it falls under the triggered category. A Continuous prediction model issues a warning at any time regardless of a solar precursor event as it relies on parameters available at all times. An example of a continuous prediction would be a regression model which provides continuous time-series prediction of a physical quantity, no matter whether there are flares or CMEs erupting. \\ 
    \textit{Categorical Response:} Triggered, Continuous, Other.
    \vspace{-8pt}
    \item \textbf{Output Time Resolution - Forecast Window:} \\
    We define as forecast window the time period for which the forecast is valid, e.g. all clear for the next 24 hours. From another perspective, when a forecast is issued, the forecast window indicates the time period in which the predicted phenomenon is expected to occur, e.g. SEP threshold crossing in the next 7 hours. \\
    \textit{Quantifiable Response:} Open.
\end{itemize}

\noindent \textbf{Other Information}
\vspace{-8pt}
\begin{itemize}
    \item \textbf{Relevant Publication:} \\
    Please add the DOI/Link to the publication this model is presented in (if published). \\
    \textit{Categorical Response:} Open.
    \vspace{-8pt}
    \item \textbf{Developers:} \\
    Please add your name and any team members (if any) you would like me to include as a co-author. \\
    \textit{Categorical Response:} Open.
    \vspace{-8pt}
    \item \textbf{SEP Event Definition:} \\
    Open ended. Please describe what is your SEP definition (ex. $\geq 10$ MeV protons). \\
    \textit{Categorical Response:} Open.
    \vspace{-8pt}
    \item \textbf{Validated using SEPVAL?} \\
    Let us know if you have used \href{https://ccmc.gsfc.nasa.gov/community-workshops/ccmc-sepval-2023/}{SEPVAL} for the validation of your model or whether you would be interested in validating it for the review paper. The absence of common validation methods and therefore the difficulty of comparing the results of different models, will be discussed in the manuscript. The SEPVAL challenge time periods for 33 SEP events and 30 non-event periods can be downloaded from this \href{https://doi.org/10.5281/zenodo.15020585}{Zenodo repository}. \\
    \textit{Categorical Response:} Yes, No, Other.
\end{itemize}

\pagebreak

\section{Acronyms} \label{app:Acronyms}

\begin{table}[h]
\begin{tabular}{ll}
\textbf{Acronym} & \textbf{Meaning }                                  \\
AA        & Not defined (Model~\ref{sec:AA})                          \\
AAS       & American Astronomical Society                             \\
ACC       & Accuracy                                                  \\
ACE       & Advanced Composition Explorer                             \\
ADASYN    & ADAptive SYNthetic                                        \\
AR        & Active Region                                             \\
AUC       & Area Under the Curve                                      \\
AWT       & Average Warning Time                                      \\ 

BA        & Balanced Accuracy                                         \\
BBQ       & Bayesian Binning Quantile                                 \\
BiLSTM-SEP& Bidirectional LSTM - SEP (Model~\ref{sec:BiLSTM-SEP})     \\
BS        & Brier Score                                               \\

CANN      & Custom Architecture Neural Network (Model~\ref{sec:CANN}) \\
CART      & Classification and Regression Tree (Model~\ref{sec:CART}) \\ 
CCMC      & Community Coordinated Modeling Center                     \\ 
CDAW      & Coordinated Data Analysis Workshop                        \\
CLEAR     & Center for All-Clear Solar Energetic Particle Forecasts   \\
CME       & Coronal Mass Ejection                                     \\
CNN       & Convolutional Neural Networks                             \\
Cox PH    & Cox Proportional Hazards                                  \\
CSI       & Critical Success Index                                    \\

DOI       & Digital Object Identifier                                 \\ 
DONKI     & Database Of Notifications, Knowledge, Information         \\ 
DSCOVR    & Deep Space Climate Observatory                            \\  

ECE       & Expected Calibration Error                                \\
EIT       & Extreme-ultraviolet Imaging Telescope                     \\
EPI-Lo    & Energetic Particle Instrument - Low                       \\
EPEAD     & Energetic Proton, Electron and Alpha Detectors            \\
EPREM-S   & Energetic Particle Radiation Environment Module - S (Model~\ref{sec:EPREM-S} \\
EPS       & Energetic Particles Sensors                               \\
ESA       & European Space Agency                                     \\
ESPERTA   & Empirical model for Solar Proton Event Real Time Alert (Model~\ref{sec:ESPERTA}) \\
EUV       & Extreme Ultra Violet                                      \\ 

FAR       & False Alarm Rate                                          \\ 
FAR\textsuperscript{*}     & False Alarm Ratio                        \\
FM        & Foundation Model                                          \\
FPR       & False Positive Rate                                       \\
F1        & F1 Score                                                  \\

GEO       & Geostationary Earth Orbit                                 \\
GeV       & giga-electron Volt (unit)                                 \\
GOES      & Geostationary Operational Environmental Satellite         \\ 
GridSearchCV & Grid Search Cross-Validation                           \\
GSS       & Gilbert Skill Score                                       \\

HEK       & Heliophysics Event Knowledgebase                          \\

\end{tabular}
\end{table}

\begin{table}[h]
\begin{tabular}{ll}

HMI       & Helioseismic Magnetic Imager                              \\
HSS       & Heidke Skill Score                                        \\
IMAP      & Interstellar Mapping and Acceleration Probe               \\

IMP       & Interplanetary Monitoring Platform                        \\ 
ISEP      & Integrated Solar Energetic Proton Event Alert/Warning System \\
ISRO      & Indian Space Research Organisation                        \\
IS\(\odot\)IS & Integrated Science Investigation of the Sun           \\
I-ALiRT   & IMAP Active Link for Real-Time                            \\

keV       & kilo-electron Volt (unit)                                 \\
KM        & Kaplan–Meier                                              \\

JSC       & Johnson Space Center                                      \\

LASCO     & Large Angle and Spectrometric Coronagraph                 \\
LLM       & Large Language Model                                      \\
LOFAR     & Low-Frequency Array                                       \\ 
LSTM      & Long Short-Term Memory                                    \\ 
L1        & Lagrange 1                                                \\
L4        & Lagrange 4                                                \\

MAE       & Mean Absolute Error                                       \\
MCC       & Matthew’s Correlation Coefficient                         \\
MDI       & Michelson Doppler Imager                                  \\
MEMPSEP   & Models for Probabilistic Forecast of Solar Energetic Particles (Model~\ref{sec:MEMPSEP}) \\
ML        & Machine Learning                                          \\
MLP       & Multi-Layer Perceptron                                    \\
MLSW      & (University of Michigan) Machine Learning for Space Weather\footnote{\url{https://mlsw.engin.umich.edu/apps/runSEP}} \\
MS-SEP    & Not defined (Model~\ref{sec:MS_SEP})                      \\

NASA      & National Aeronautics and Space Administration             \\ 
NCEI      & National Centers for Environmental Information            \\
NDA       & Nançay Decameter Array                                    \\
NN        & Neural Network                                            \\   
NOAA      & National Oceanic and Atmospheric Administration           \\   

PCC       & Pearson Correlation Coefficient                           \\
pfu       & particle flux unit ($counts/(cm^{2} s sr)$)               \\
PINN      & Physics-Informed Neural Network                           \\
POD       & Probability of Detection                                  \\
PSP       & Parker Solar Probe                                        \\
PSPSP     & PSP SEP Prediction (Model~\ref{sec:PSPSP})                \\
PUNCH     & Polarimeter to Unify the Corona and Heliosphere           \\

REleASE   & Relativistic Electron Alert System for Exploration        \\
ReLU      & Rectified Linear Unit                                     \\
RH        & Random Hivemind (Model~\ref{sec:RH})                      \\
RMSE      & Root Mean Squared Error                                   \\
RNN       & Recurrent Neural Network                                  \\ 
$R^2$        & $R^2$ Score (Coefficient of Determination)             \\
R2O       & Research-to-Operations                                    \\

SC        & Solar Cycle                                               \\ 
SDO       & Solar Dynamics Observatory                                \\
SEM       & Space Environment Monitor                                 \\
SEP       & Solar Energetic Particles                                 \\
SEPEM     & Solar Energetic Particle Environment Modeling             \\
SEPVAL    & Solar Energetic Particle Model Validation                 \\

\end{tabular}
\end{table}

\begin{table}[h]
\begin{tabular}{ll}

SEP-C     & Not defined (Model~\ref{sec:SEP-C})                       \\
SEP-E     & Not defined (Model~\ref{sec:SEP-E})                       \\ 
SHARP     & Space Weather HMI Active Region Patches                   \\ 
SMARP     & Space-Weather MDI Active Region Patches                   \\ 
SHINE     & Solar Heliospheric and INterplanetary Environment         \\
SILSO     & Solar Influences Data Analysis Center                     \\
sklearn   & Scikit Learn                                              \\
SMOTE     & Synthetic Minority Oversampling TEchniques                \\ 
SOHO      & Solar and Heliospheric Observatory                        \\
SolO      & Solar Orbiter                                             \\
SPAN-I    & Solar Probe Analyzer for Ions                             \\
SPD       & Solar Physics Division (refers to AAS)                    \\
SPE       & Solar Proton Event                                        \\
SPRINTS   & Space Radiation Intelligence System (Model~\ref{sec:SPRINTS}) \\
SRAG      & Space Radiation Analysis Group                            \\
SSEP      & Survival SEP (Model~\ref{sec:SSEP-Survival})              \\
STEREO    & Solar TErrestrial RElations Observatory                   \\
STSF      & Supervised Time Series Forest (Model~\ref{sec:STSF})      \\
SWEAP     & Solar Wind Electrons Alphas and Protons                   \\
SWFO-L1   & Space Weather Follow On – Lagrange 1                      \\
SWPC      & Space Weather Prediction Center                           \\ 
S1        & Minor Solar Radiation Storm as defined by NOAA\footnote{\url{https://www.swpc.noaa.gov/noaa-scales-explanation}}\\
S2        & Moderate Solar Radiation Storm as defined by NOAA         \\
S3        & Strong Solar Radiation Storm as defined by NOAA           \\
S4        & Severe Solar Radiation Storm as defined by NOAA           \\
S5        & Extreme Solar Radiation Storm as defined by NOAA          \\

TEBBS     & Temperature and Emission measure-Based Background Subtraction \\
TSF       & Time-Series Forest                                        \\ 
TSS       & True Skill Score                                          \\
TS-HOG-TB & Time Series - Histogram of Oriented Gradients - TaBular (Model~\ref{sec:TS-HOG-TB}) \\

UDM       & Univariate Deep Merge (Model~\ref{sec:UDM})               \\
UMASEP    & University of MAlaga Solar particle Event Predictor (Model~\ref{sec:UMASEP}) \\
UMASOD    & University of Malaga predictor from Solar Data (Model~\ref{sec:UMASOD}) \\
UNSPELL   & UNifying Solar Particle Event modeLLing (Model~\ref{sec:UNSPELL}) \\

VAR       & Vector Autoregression                                     \\

XGBoost   & eXtreme Gradient Boosting (Model~\ref{sec:XGBoost})       \\
XRS       & X-Ray Sensor                                              \\

1D        & 1 Dimension                                               \\ 
2D        & 2 Dimensions                                              \\ 
3D        & 3 Dimensions                                              \\ 

\end{tabular}
\end{table}

\end{document}